\documentclass[12pt,a4paper]{book}

\usepackage{epsfig}
\usepackage{graphicx}
\usepackage{epstopdf}
\usepackage{bm}
\usepackage{amsmath}
\usepackage{amssymb}
\usepackage{eufrak}
\usepackage{subfigure}

\begin{document}

\evensidemargin=4ex
\topmargin=-3ex
\textwidth=80ex
\textheight=120ex
\baselineskip=3ex

\thispagestyle{empty}

\begin{center}

\vspace{50mm}

{\Large {Daniel Marco Philip Bozi}}

\vspace{15mm}
                         
{\bf {\LARGE {Finite Energy Electronic Correlations}}}

\vspace{\baselineskip}

{\bf {\LARGE {in Low-Dimensional Systems}}}

\vspace{\baselineskip}

\vspace{40mm}

\begin{figure}[h!]
{ \par\centering \includegraphics[scale=0.15]{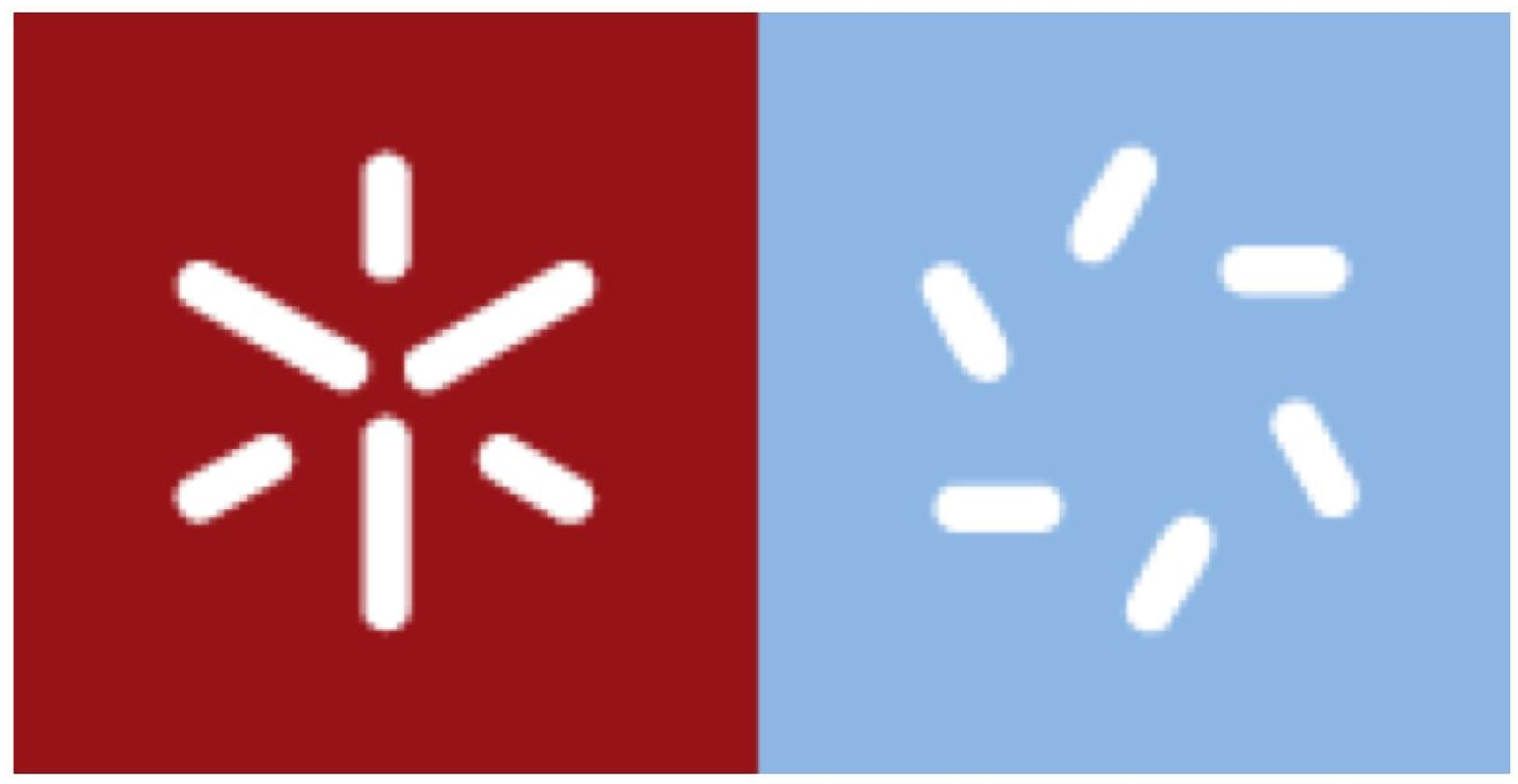} \par }
\end{figure}

\vspace{30mm}

{\Large {Dissertion submitted for the degree of Ph.D. in Physics}}

{\Large {Universidade do Minho, Braga (Portugal)}}

\vspace{15mm}

{\Large {Braga 2006}}

\end{center}

\newpage
\thispagestyle{empty}

\vspace{3.5cm}

\ \ \ \ 
\newpage
\thispagestyle{empty}

 \ \ \ 
 
\vspace{10cm}

\begin{flushright}
\hfill {\Large {\it Till Erik, min b\"asta v\"an}} \linebreak[4] \\ \hfill {\Large {\it Till Irma, min b\"asta lillasyster}}
\end{flushright}

\newpage
\thispagestyle{empty}

\vspace{3.5cm}

\ \ \ \ 
\newpage
\thispagestyle{empty}

 \ \ \ 
 
\vspace{10cm}

\begin{flushright}
\hfill {\normalsize {\it Science cannot solve the ultimate mystery of nature. And that is because, in the last analysis, we ourselves are a part of the mystery that we are trying to solve.}} \vspace{1cm} \\ \hfill {\normalsize Max Planck} 
\end{flushright}

\newpage
\thispagestyle{empty}

\vspace{3.5cm}

{\bf {\Large Abstract}}
\vspace{0.5cm}

This thesis report deals with the one dimensional Hubbard model. Specifically, we describe the quantum objects that diagonalize the normal ordered Hubbard hamiltonian, among those the so called {\it pseudofermions}. These pseudofermions, $\eta$-spin and spin zero objects, are the scatterers and the scattering centers of the representation developed here. They have no residual energy interactions. The S-matrix of this representation can be written as a simple phase factor, which involves the phase shifts of the zero energy forward momentum scattering events. The form of the pseudofermion S-matrix constitutes an important new result of this thesis report. In contrast to the usual low-energy Luttinger liquid theory, the theory reported here allows us to categorize a separation of the charge and spin degrees of freedom at a finite energy excitation scale, of quantum objects called rotated electrons. The rotated electrons are linked to the electrons by a mere unitary transformation.

Furthermore, we develop a pseudofermion dynamical theory and apply it to the evaluation of the spectral function in the one-electron removal and one-electron lower Hubbard band addition cases. For any value of the on-site effective Coloumb repulsion and electronic density, and in the limit of zero magnetization, we derive closed form expressions for these spectral functions, showing explicitly the emergence of the characteristic power-law type behavior of correlation functions of Luttinger liquids. We note however, that our derived expressions are valid for the entire elementary excitation energy bandwidth, and not just the linear regime. We are able to identify practically all features of the spectral weight of the 1D Hubbard model, in terms of pseudofermion or pseudofermion hole excitations. This brings new light to the understanding of the spectral properties of the one dimensional Hubbard model.

The singular behavior of the theoretical spectral weight, as predicted by the explicitly calculated values of the relevant exponents and pre-factors, leads to a spectral weight distribution which is detectable by photo emission and photo absorption experiments on quasi one dimensional materials. As an important contribution to the understanding of these materials, we are able to reproduce for the whole energy bandwidth, the experimental spectral distributions recently found for the organic compound TTF-TCNQ by high-resolution photo emission spectroscopy. This confirms the validity of the pseudofermion dynamical theory, and provides a deeper understanding of low dimensional correlated systems.
\newpage
\thispagestyle{empty}

{\bf {\Large Acknowledgements}}

\vspace{0.5cm}

First and foremost, I wish to thank my two wonderful supervisors, Prof. Jos\'e M.P. Carmelo at Universidade do Minho, Braga, and Prof. Pedro D. Sacramento at Instituto Superior T\'ecnico, Lisbon (both in Portugal). I owe many thanks to them for having proposed such an interesting research topic to me. I have learned a great deal on correlated electron systems in general and on the exciting universe of the 1D Hubbard model in specific. During these four years, they have also become very good friends. I would further like to thank Dr. Karlo Penc at the Research Institute for Solid State Physics and Optics in Budapest, Hungary, for receiving me during four months (in total), during my PhD candidacy. My visits to Budapest were very educative and exciting, and I treasure them very much. Also, I would like to thank Prof. Patrick A. Lee for receiving me during during the first half of 2005 at the Massachusets Institute of Technology, Cambridge, USA. Thanks to the generosity of Prof. Patrick A. Lee, I was able to profit from the resources and the positive research environment in one of the best universities in Physics. Furthermore, I wish to thank Prof. Nuno M.R. Peres for introducing me to the Hubbard model in two and three spatial dimensions. Unfortunately, the work done with him, resulting in the article Phys. Rev. B. $\bm{70}$, 195122 (2004), did not make it into this thesis report. Last, but definitely not least, I would like to thank the kind financial support of the Portuguese "Funda\c{c}\~ao para Ci\^{e}ncia e a Tecnologia" (FCT), under the scholarship grant SFRH/BD/6930/2001. Needless to say, without such a financial support, this PhD would have been impossible to even start.

\vspace{0.5cm}

Erik and Anouk, my true friends, thank you for being the listeners in times of need. Four years in wonderful Portugal couldn't tear our strong friendship apart, now that's something!

Carlinha, you will always have a special place in my heart. I wish you all the happiness you deserve.

Thank you Portugal, a country made of delicious food, wonderful coffee, lovely spontaneous people and sunny beaches.

It was the country in which I learned how to Love.

\tableofcontents
\setcounter{chapter}{0}
\setcounter{section}{0}

\chapter{Introduction}

\section{The one dimensional many-body problem}

This thesis deals with a model used to study strongly correlated electrons in solids, called the Hubbard model. The model was introduced by J. Hubbard in the article series on "Electron correlations in narrow energy bands" \cite{Hubbar1}-\cite{Hubbar6}. We will exclusively focus on the case of one spatial dimension. 

In general, the Hubbard model is viewed as one of the most important models of strongly correlated electron systems in solids, and is used frequently in various applications both in one, two and three spatial dimensions. Formally, it presupposes the existence of an atomic-like static lattice, upon which valence electrons exist in a single energy band. 

The Hubbard model is obtained for the approximative scheme where, due to the screening of the long-range Coloumb repulsion, its on-site Coloumb interaction term is large as compared to the inter-atomic interactions however not rendering the inter-atomic hopping amplitudes negligible. Thus, the Hubbard model is loosely described as a dynamical model of electrons, where the Coloumb interaction ($\sim$ potential energy) competes with the transfer integral between neighboring lattice sites ($\sim$ kinetic energy), under the influence of the Pauli exclusion principle. Despite its conceptual simplicity, the Hubbard model is indeed a very difficult model to solve exactly. It has only been solved in one dimension so far, by use of the Bethe ansatz technique, originally due to H. Bethe \cite{betheorig}, where it was first applied to the isotropic Heisenberg chain. 

Even though no "true" one dimensional materials exist, many materials have been observed with a {\it quasi one dimensional} behavior. For example, some materials exhibit very strong anisotropies, for instance in the electrical conductivity, where the motion of the electrons is confined to certain specific directions along the Brillouin zone. As an example, "tetrathiafulvalene-tetracyanoquinodimethane", abbreviated TTF-TCNQ \cite{claessen1} \cite{carmclaes} \cite{ttftcnq1}-\cite{ttftcnq10}, is a {\it charge transfer salt} consisting of linear stacks of planar TTF molecules and planar TCNQ molecules, which in the metallic phase exhibit an intra-stack conductivity three orders of magnitude larger than the inter-stack conductivity.

Some other examples of materials that are adequately described by the Hubbard model include $\text{V}_2 \text{O}_3$ \cite{v2o3}, strontium-copper oxide compounds (also known as "chain cuprates") such as $\text{Sr}_2\text{CuO}_3$ \cite{cupper1} \cite{cupper2} and $\text{Sr}_2\text{CuO}_2$ \cite{cupper3} \cite{cupper4}, the Bechgaard salts (organic conductors) \cite{bechsalt1}-\cite{bechsalt3} and $\pi$-conjugated polymers \cite{piconjug1} \cite{piconjug2}.

Historically, one example of the motivation to study such a model refers to the unusual metal-insulator transition theoretically predicted by N.F. Mott \cite{mott1} \cite{mott2}. This transition can not be explained by standard band theory \cite{mott}. One of the first materials observed to undergo a Mott-Hubbard metal-insulator transition, was the Cr-doped  $\text{V}_2 \text{O}_3$ \cite{mott3}-\cite{mott5}. The transition is characterized by the strong mutual interactions between the charge carriers, forcing them to be localized. This can intuitively be understood in the case of {\it half filling} (i.e. exactly one valence electron per lattice site). In this limit, it is not energetically favorable for electrons to be delocalized since hopping onto a neighboring lattice site, where another electron is present, costs more energy than staying put. In this way, a solid with an odd number of valence electrons per lattice site may exhibit insulating behavior, contrary to the predictions of standard band theory.

Quantum systems exhibiting Mott related features continues to be one of the main topics in low dimensional strongly correlated electron systems, with some recent applications involving 1D cuprates \cite{mott51}, spin frustrated organic conductors \cite{mott6}, "orbital-selective Mott systems" (modeled by a multi-band hamiltonian) \cite{mott7} and dispersionless-boson interaction (modeled with a Holstein-Hubbard hamiltonian) \cite{mott8}.

From a theoretical point of view, the strongly correlated one dimensional systems have spectral properties not explainable by the usual {\it Fermi liquid theory}. One of the trademarks of this theory, is the description of the low lying excitations in terms of "quasi particles", whose interaction is described by the scattering $f$-functions of Landau \cite{Fermi1}-\cite{Fermi4}. The ground state distribution in a Fermi liquid is given by the usual Fermi distribution, becoming a step distribution at zero temperature, where the step occurs at the Fermi level. The basic assumption of Fermi liquid theory is that there exists a one-to-one correspondence between the particle states of the original system and the quasiparticle states of the interacting system. Thus, there exists a strong $\delta$-function coherent peak in the spectral function at the quasi particle energy (at the Fermi level). This picture breaks down if for example the interacting system produces bound states, as in a superconductor. Indeed, according to the usual BCS theory, the formation of Cooper pairs ultimately destroys the above mentioned one-to-one correspondence \cite{CBS}. 

However, in the case of one spatial dimension, the breakdown of the Fermi liquid picture is more general as 1D metals are characterized by the absence of Fermi liquid type quasi particles. Instead, the low excitation energy dynamics is described by charge and spin elementary excitations propagating independently of each other. These properties of interacting 1D metals signal the advent of a new type of quantum liquid, known as the Luttinger liquid \cite{luttliq1}-\cite{luttliq4}.  A characteristic of a Luttinger liquid is that it exhibits no coherent quasi particle peaks, as all low energy contributions to the one-particle spectral weight are of a non coherent origin and characterizeable in terms of the separated charge and spin degrees of freedom \cite{luttliq5}. Moreover, correlation functions decay algebraically as the Fermi level is approached, by an interaction dependent exponent. For example, in the case of the 1D Hubbard model, the density of states decays with an exponent assuming values between $0$ and $\frac1 8$, where the latter value corresponds to the strong coupling limit  \cite{luttliq6}.

Using conformal invariance and finite-size scaling, H. Frahm and V. Korepin obtained the low lying spectral properties for the Hubbard model, evaluating the asymptotics of correlation functions \cite{crit6} \cite{crit61}. Since for the low lying elementary excitations the Hubbard model belongs to the same universality class as the Tomonaga-Luttinger model, the exponents obtained in these references are the same as those of a Luttinger liquid.

$\newline$
{\bf Examples of recent experiments related to 1D Luttinger liquids}

Recently, there has been a surge of exciting new low dimensional materials, which commonly share the traits of Luttinger liquids. For example, in Ref. \cite{sexy1}, the conductive properties of carbon nanotubes are investigated and compared with the theoretical predictions of a Luttinger liquid. The authors find a power-law type scaling behavior of the conductance, with a measured value for the exponent in good agreement with the theoretical value. 

The material $\text{La}_{1.4-x}\text{Nd}_{0.6}\text{Sr}_x\text{CuO}_4$ is studied for $x=0.12$ in Refs. \cite{sexy2} and \cite{sexy3}, and below and above this value in Ref. \cite{sexy4}. The reason for this material attracting attention is due to the quest for a better understanding of the copper-oxide superconductors. Unlike conventional metals, the charge carriers are confined to one dimensional "domain lines", but where the electronic spins in the region between these lines order themselves antiferromagnetically. This charge- and spin-ordered state is in these references called a "stripe phase". This exotic charge transport suffers a dimensional crossover 1D $\rightarrow$ 2D as the critical concentration grows beyond $x=0.12$. Numerical calculations employing the Hubbard model confirmed the interpretation of "stripe phases" in this material. 

Ref. \cite{sexy5} describes a one dimensional optical lattice of ultracold fermions in a harmonical trap in order to study the Mott transition at micro Kelvin temperatures. In Ref. \cite{sexy51}, such a lattice is set up by using "thousands" of parallel atomic waveguides, to form a nearly perfect atomic quantum wire. This new method provides an unprecedented control over the study of strongly correlated electron systems, and will be of great interest for anyone aspiring to work with low dimensional systems. Related numerical work is presented in Ref. \cite{sexy52}, with the survival of spin-charge separation far outside the low excitation energy (Luttinger) regime, as one of the key results.

Mentioned briefly, Ref. \cite{sexy6} studies the angle resolved photoelectron spectra on metallic nano-wires. Ref. \cite{sexy7} observes a spin-charge separation in quantum wires, which is readily cast into the Luttinger liquid scheme. Ref. \cite{sexy8} studies the electronic transfer between various structures of the DNA double helix.

$\newline$
{\bf The aim of this Thesis Report:}

In light of recent high-resolution photo emission experiments on various materials, and in light of significantly improved experimental techniques for the study of strongly correlated electron systems in general, the {\it absence} of a dynamical theory for the theoretical model that is expected to describe the vast majority of the properties of these new materials - the Hubbard model - is a serious drawback for the complete understanding of many low dimensional quantum systems. 

Indeed, apart from the limit of infinite Coloumb interaction strength, so far it is only the properties of the low excitation energy regime that is understood to a sufficient degree of satisfaction, by employing theoretical tools only valid for that regime. 

However, many recent experiments (see for example the organic conductor TTF-TCNQ cited above) suggest the existence of elementary excitations at all energy scales, and not just the low lying ones. Thus, the need for a dynamical theory for the Hubbard model, capable of capturing the finite energy physics of the model hamiltonian in terms of a consistent scattering theory, is urgent. The recent high-resolution photo emission studies of the organic conductor TTF-TCNQ \cite{claessen1} \cite{carmclaes} is a good example of this. Naturally, with the fine tuning of photo emission and photo absorption techniques, as well as with the development of new experimental procedures, such as the trapped ultracold atoms technique, the need for such a dynamical theory will most likely grow in the future.

With this thesis report, we aim to derive and apply such a dynamical theory.

For the derivation part, we will aim at calculating the spectral functions for the cases of one electron removal and one electron lower Hubbard band addition. This derivation will actually be intertwined with the development of the dynamical theory itself, and therefore constitutes the main part of this work, chapters (\ref{dynamics}) and (\ref{onelecspec}). The aim of this disposition is to describe how a one electron spectral function is derived, and at the same time gain a significant insight into the general physics of the model. However before this, in chapter (\ref{themodel}), we characterize the symmetries of the model and introduce the quantum objects that diagonalize the normal ordered hamiltonian, the so-called {\it pseudofermions}, related to the likewise introduced {\it pseudoparticles}. A scattering theory for theses pseudofermions is also developed in this chapter. In chapter (\ref{theorweight}), we use numerical studies to evaluate the one-electron spectral functions from the general expressions of the pseudofermion dynamical theory. The results of this chapter are then compared to recent photo emission experimental results (using the ARPES technique) for the organic conductor TTF-TCNQ, chapter (\ref{applica}). A short discussion with a summary of the main conclusions is presented in chapter (\ref{conclusion}).

\section{The model hamiltonian}

For the following, and indeed throughout this thesis report, we will assume a vanishing magnetization $m \rightarrow 0$. Consider a solid whose ions are arranged in a crystalline structure. Since the ions of the solids are much heavier than the electrons, it is reasonable to assume the ions to be static, i.e. not participating in the dynamics of the solid. A general hamiltonian governing the dynamics of electrons with mass $m_e$ and charge $e$ can be written as

\begin{equation}
H=\sum_{j=1}^N \Big[ \frac {\bm{p}_j^2} {2m_e} + V(\bm{x}_j) \Big] + \sum_{1 \leq j < j' \leq N} \frac {e^2} {\vert \bm{x}_j-\bm{x}_{j'} \vert} \label{modelham}
\end{equation}
where $\bm{x}_j$ denotes the position and $\bm{p}_j$ the momentum of electron $j$ (there are a total of $N$ electrons present in the system). $V(\bm{x_j})$ is a potential with the periodicity of the lattice structure. We will assume a maximum of two valence electrons per lattice site (with opposite spin projection). This means that we are considering atoms where all the other electrons are strongly bound to the ions and hence do not contribute to the dynamics. 

The eigenfunctions $\chi_{\bm{k}}(\bm{x})$ of the one-body term of the hamiltonian (\ref{modelham}) are just Bloch functions, i.e. $\chi_{\bm{k}}(\bm{x})=e^{i \bm{k} \cdot \bm{x}} u_{\bm{k}}(\bm{x})$, where $u_{\bm{k}}(\bm{x})$ is a function with the periodicity of the lattice. These can in turn be expressed in another basis, namely the Wannier basis, comprising the Wannier eigenfunctions $\psi (\bm{x}-\bm{R}_j)$ that are centered on the lattice site at position $\bm{R}_j$. The relation between the functions of the two bases is:

\begin{equation}
\chi_{\bm{k}}(\bm{x}) = \frac 1 {\sqrt{L}} \sum_{j=1}^N e^{i \bm{k} \cdot \bm{R}_j} \psi (\bm{x}-\bm{R}_j)
\end{equation}
where $L$ is the total length of the lattice (many times, units such that the lattice constant $a$ is equal to $1$ are employed, and hence $L=N$).

By using wave functions centered on the lattice sites, we may express the second quantization annihilation field operator $\Psi_{\sigma} (\bm{x})$, as a sum over the entire lattice:

\begin{equation}
\Psi_{\sigma} (\bm{x}) = \sum_{j=i}^N \psi (\bm{x}-\bm{R}_j) c_{j\sigma} \label{2ndquant}
\end{equation}
where we introduced the second quantization electron annihilation operator  $c_{j\sigma}$, that annihilates an electron on lattice site $j$ with spin projection $\sigma$. The creation field operator $\Psi_{\sigma}^{\dag} (\bm{x})$ is nothing but the hermitian conjugate of the operator given in equation (\ref{2ndquant}).

In terms of the field operators, the first-quantization hamiltonian above may be rewritten as

\begin{equation}
H=\sum_{\sigma} \int dx \ \Psi_{\sigma}^{\dag} (\bm{x}) \: \hat{T} \: \Psi_{\sigma} (\bm{x}) + \sum_{\sigma \bar{\sigma}} \int dx \: dy \ \Psi_{\sigma}^{\dag} (\bm{x})  \Psi_{\bar{\sigma}}^{\dag} (\bm{y}) \: \hat{U} \: \Psi_{\bar{\sigma}} (\bm{y}) \Psi_{\sigma} (\bm{x})
\end{equation}
where $\hat{T}$ and $\hat{U}$ correspond to the first-quantization one-body (kinetic) and two-body (Coloumb interaction) operators, respectively. Thus, in terms of Wannier states:

\begin{equation}
H=\sum_{ij,\sigma} t_{ij} c_{i\sigma}^{\dag} c_{j\sigma} + \frac 1 2 \sum_{ijkl} \sum_{\sigma \bar{\sigma}} U_{ijkl} c_{i\sigma}^{\dag} c_{j\bar{\sigma}}^{\dag} c_{k\bar{\sigma}} c_{l\sigma} \label{snartslut}
\end{equation}
where we have defined the kinetic and the potential matrix elements according to:

\begin{eqnarray}
t_{ij} &=& \int dx \  \psi^* (\bm{x}-\bm{R}_i)  \: \hat{T} \: \psi (\bm{x}-\bm{R}_j) \\
U_{ijkl} &=& \int dx \: dy \  \psi^* (\bm{x}-\bm{R}_i) \psi^* (\bm{y}-\bm{R}_j)  \: \hat{U} \:  \psi (\bm{y}-\bm{R}_k) \psi (\bm{x}-\bm{R}_l) \nonumber
\end{eqnarray}

Note that in this derivation we have all along assumed a single band model. For multi-band solids we would have to use an extended model, except for the special case of weak inter-band interactions. 

Let us assume the Wannier states to be "strongly localized", i.e. that $\psi (\bm{x}-\bm{R}_j)$ is centered on $\bm{R}_j$. This means that the Wannier states of two neighboring lattice sites may have a finite overlap with each other, but lattice sites further apart may not. Hence, in the first term of Eq. (\ref{snartslut}), the summation over the lattice sites reduces to a summation over {\it nearest neighbor pairs}, denoted $\langle ij \rangle$ (where $i=j\pm1$). 

Due to the "strong localization" of the Wannier states, we can assume that for the two-body term, only matrix elements referring to the {\it same} site become non-negligable. This makes the Coloumb interaction term become an {\it effective potential}: the Coloumb repulsion between two electrons on the same lattice site dominates completely over the repulsion of two electrons on different lattice sites, due to the {\it screening} within the actual material sample.  

Introducing these assumptions into Eq. (\ref{snartslut}), we obtain an effective hamiltonian where $t_{ij}=t_{ij} \delta_{i,j\pm1}$ and $U_{ijkl}=U_{iiii} \delta_{ijkl}$:

\begin{equation}
H=\sum_{\langle ij \rangle \sigma} t_{ij} c_{i\sigma}^{\dag} c_{j\sigma} +  \frac 1 2 \sum_{i} \sum_{\sigma \bar{\sigma}} U_{iiii} c_{i\sigma}^{\dag} c_{i\bar{\sigma}}^{\dag} c_{i\bar{\sigma}} c_{i\sigma}
\end{equation}

Since we are considering systems with identical lattice sites, we assume isotropic nearest neighbor transfer amplitudes $t_{ij}=-t$ and effective Coloumb interaction strengths $U_{iiii}=2U$ (where the factor of 2 comes from the spin summation), which finally leads us to the Hubbard model hamiltonian:

\begin{equation}
H=-t \sum_{\langle ij \rangle \sigma} c_{i\sigma}^{\dag} c_{j\sigma} + U \sum_{i} n_{i\uparrow} n_{i\downarrow}
\end{equation}
where $n_{i\sigma}=c_{i\sigma}^{\dag} c_{i\sigma}$ is the electron number operator. As we will see in the next chapter, the interaction term is written in a different way in order to comprise useful symmetries, such as the particle-hole symmetry. Note that the $i=j$ term in the kinetic part only counts the number of electrons, and can hence be absorbed by the chemical potential in a grand canonical ensemble description.

The hamiltonian derived above allows for some characterization merely by "hand waving" arguments. For example, one can see that in the limit $(U/t) \rightarrow 0$, the electrons do not interact and hence are delocalized for almost any value of the band filling $n=N/N_a$, where $N$ is the number of electrons and $N_a$ the number of lattice sites (in subsequent chapters, however, we will define $n=N/L$, where $L=aN_a$ is the {\it length} of the lattice chain, where $a$ is the lattice constant with dimension of length). Exceptions are the fully polarized half filled case, and the fully occupied case $n=2$, respectively. This is the {\it tight-binding model}, for which the electrons disperse in a cosine-like band \cite{Mahan}.

In the opposite limit, $(U/t) \rightarrow \infty$, the electrons are localized. At half filling, the ground state of the model describes an antiferromagnetic insulator \cite{ggallan}. Actually, at half filling the Hubbard model is insulating for all values of  $(U/t) > 0$ \cite{LiebWu}. Moreover, in this limit, any eigenstate of the model can be factorized into two eigenstates: one describing spinless fermions and another describing chargeless spins \cite{explicitwavefcn}. It can be shown that the effective hamiltonian for the spin part, is nothing but the 1D antiferromagnetic Heisenberg spin hamiltonian \cite{collus}, with an effective coupling constant equal to $n(4t^2/U)(1-\sin [2\pi n]/[2\pi n])$.

This result can be understood by second order perturbation theory around $(U/t) = \infty$. Since the spin configuration is antiferromagnetic, one can think of virtual states in which an electron hop to one of its neighbors, with the amplitude proportional to $t$. However, due to the large repulsion $U$, one of the electrons immediately hops back to the originating site, again with an amplitude $t$. This gives rise to an antiferromagnetic exchange of strength $\sim t \cdot (1/U) \cdot t = t^2 / U $. When away from half filling, the value of the effective interaction of the spins will then be modified due to the presence of holes. 

Due to the insulating properties of the Hubbard model at half filling, we will always stay away from this value of the density $n$ in this thesis report.

\setcounter{chapter}{1}
\setcounter{section}{1}

\chapter{The Hubbard Model}
\label{themodel}
\section{Symmetries and exact solution}

\subsection{SO(4) symmetry}
\label{so4sect}
By rewriting the model hamiltonian of the previous chapter, it is possible to arrive at the following general hamiltonian for electrons on a one dimensional chain consisting of $N_a$ equally spaced lattice sites:
\begin{equation}
\hat{H}=\hat{T}+\hat{U}=-t\sum_{\left< ij \right> \sigma} c_{i\sigma}^{\dag}c_{j\sigma}+U\sum_i (c_{i\uparrow}^{\dag}c_{i\uparrow}-\frac 1 2)(c_{i\downarrow}^{\dag}c_{i\downarrow}-\frac 1 2)
\label{H}
\end{equation}
where $c_{i\sigma}^{\dag}$ is the usual electronic creation operator and $c_{i\sigma}$ is the usual electronic annihilation operator. We will assume $N_a$ to be an even number. The symbol $\left< ij \right>$ indicates that the summation is done over neighbouring sites only, i.e. $j=i\pm1$ and $\sigma$ stands for the electronic spin projection. We consider periodic boundary conditions. The transfer integral $t$ and the effective on-site coloumb interaction strength $U$ were introduced in the previous chapter. The hamiltonian conserves the total number of electrons, as well as the total number of $\uparrow$-spin electrons and the total number of $\downarrow$-spin electrons even though, in the following, we will concentrate on somewhat more subtle and interesting symmetries. Any one site can either be empty of electrons, singly occupied by a $\uparrow$-spin or a $\downarrow$-spin electron, or "doubly occupied", i.e. occupied by two electrons (with opposing spin projection due to the Pauli principle). The number of electrons $N$, can be written as $N=N_{\uparrow}+N_{\downarrow}$, where $N_{\sigma}$ is the total number of $\sigma$-spin electrons. The creation and annihilation operators have the usual Fourier representation on a lattice:
\begin{eqnarray}
c_{k\sigma}^{\dag}&=&\frac 1 {\sqrt{N_a}} \sum_{j=1}^{N_a} e^{-ikja} c_{j\sigma}^{\dag} \nonumber \\
c_{k\sigma}&=&\frac 1 {\sqrt{N_a}} \sum_{j=1}^{N_a} e^{ikja} c_{j\sigma} \label{Fourierelectr}
\end{eqnarray}
The total momentum operator can thus most easily be written:
\begin{equation}
\hat{P}=\sum_{\sigma=\uparrow,\downarrow} \sum_k \hat{N}_{\sigma} (k) k
\end{equation}
where $\hat{N}_{\sigma} (k)=c_{k\sigma}^{\dag}c_{k\sigma}$ is the Fourier transformed number operator, and $L=aN_a$ is the physical length of the lattice chain ($a$ is the lattice constant).
This hamiltonian is many times written as $\hat{H}_{SO(4)}$ due to the fact that it is invariant under the $SO(4)$ symmetry group \cite{SO4} \cite{SO4energy}. The $SO(4)$ group is isometric with the $SU(2) \otimes SU(2)$ group and differs in that only half of the irreducible representations of $SU(2) \otimes SU(2)$ correspond to energy eigenstates, as further explained below. Therefore, symmetry properties of the model are usually explained in terms of the two related independent $SU(2)$ symmetries, namely the $SU(2)$ spin algebra and the $SU(2)$ $\eta$-spin algebra. The generators of the spin algebra (subscript $s$) and of the $\eta$-spin algebra (subscript $c$) are
\begin{eqnarray}
\hat{S}_s^z &=& \frac 1 2 \left( \hat{N}-N_a \right) \hspace{2.20cm}  \hat{S}_c^z = \frac 1 2 \left( \hat{N}_\downarrow-\hat{N}_\uparrow \right) \nonumber \\
\hat{S}_s^{\dag} &=& \sum_i c_{i\downarrow}^{\dag}c_{i\uparrow}  \hspace{3.0cm} \hat{S}_c^{\dag} =\sum_i (-1)^i c_{i\downarrow}^{\dag}c_{i\uparrow}^{\dag} \label{gensu2} \\
\hat{S}_s &=& \sum_i c_{i\uparrow}^{\dag}c_{i\downarrow}  \hspace{3.0cm} \hat{S}_c = \sum_i (-1)^i c_{i\uparrow}c_{i\downarrow} \nonumber
\end{eqnarray}
where $\hat{N}=\sum_{\sigma} \hat{N}_{\sigma}$ and $\hat{N}_{\sigma}=\sum_i c_{i\sigma}^{\dag}c_{i\sigma}$ is the total electronic number operator and the $\sigma$-spin electronic number operator, respectively. They satisfy the usual $SU(2)$ commutation relations:
\begin{equation}
\left[ \hat{S}_{\alpha}^{\dag} , \hat{S}_{\alpha} \right] = 2\hat{S}_{\alpha}^z \hspace{1.0cm} \alpha=c,s
\end{equation}

Apart from commuting with the hamiltonian (\ref{H}), all the generators of the spin algebra commute with all the generators of the $\eta$-spin algebra. Together with the square of the total spin and total $\eta$-spin operator, $(\hat{\bm{S}}^{\alpha})^2$ (where $\alpha=c$ or $s$), the diagonal generators $\hat{S}_{\alpha}^z$ and the hamiltonian form a set of commuting operators. The transformations associated with the $SU(2)$ symmetries are the spin flip ($c_{i\sigma}^{\dag} \rightarrow c_{i-\sigma}^{\dag}$) and the $\eta$-spin flip ($c_{i\sigma}^{\dag} \rightarrow (-1)^i c_{i\sigma}$), respectively (the latter transformation is associated with the so called {\it particle-hole} invariance). As is trivially understood by physical reasoning, these symmetries are easily broken, for example by applying an external magnetic field or by introducing a chemical potential, since in these cases the full hamiltonian will not commute with the off diagonal generators and thus the system will prefer a certain spin projection and a certain $\eta$-spin projection, respectively. Mathematically, this comes about by adding a term to the hamiltonian equal to $\mu_{\alpha}\hat{S}_{\alpha}^z$, where in the case of $\eta$-spin ($\alpha=c$) $\mu_c=2\mu$ where $\mu$ is the chemical potential and in the case of spin ($\alpha=s$) $\mu_s=2\mu_0h$ where $\mu_0$ is the Bohr magneton and $h$ the strength of the applied external magnetic field. 

Applying any of the ladder operators, $\hat{S}_{\alpha}^{\dag}$ or $\hat{S}_{\alpha}$ (where $\alpha=c$ or $s$), to an eigenstate of the model (\ref{H}), leads either to zero or to a new state for which the total spin projection ($\alpha=s$) or the total $\eta$-spin projection ($\alpha=c$) has changed by one. One can then successively apply this ladder operator until we reach either the "top" of the ladder (by repeatedly applying $\hat{S}_{\alpha}^{\dag}$) or the "bottom" of the ladder (by repeatedly applying $\hat{S}_{\alpha}$). The state reached by successive application of the operator $\hat{S}_{\alpha}$ is usually called a {\it lowest weight state} (LWS), whilst the state reached by successive application of the operator $\hat{S}_{\alpha}^{\dag}$ is usually called a {\it highest weight state} (HWS), respectively. Trying to climb lower than the bottom of the ladder, as well as trying to climb higher than the top of it, yields zero: $\hat{S}_{\alpha} \vert \text{LWS} \rangle=0$ and $\hat{S}_{\alpha}^{\dag} \vert \text{HWS} \rangle=0$. The eigenvalue of $\hat{S}_{\alpha}^z$, denoted by $S_{\alpha}^z$, ranges from $-S_{\alpha},-S_{\alpha}+1$ , $\ldots$ , $S_{\alpha}-1,S_{\alpha}$, where $(\hat{\bm{S}}^{\alpha})^2$ has eigenvalue $S_{\alpha}(S_{\alpha}+1)$. 

Because of the above found symmetries, it will be sufficient to consider electronic densities $n=\frac N {L}$ and magnetization $m=\frac {N_{\uparrow}-N_{\downarrow}} {L}$, where $N=N_{\uparrow}+N_{\downarrow}$ is the number of electrons, such that $0 < na < 1$ and $0 < ma < na$. For example, considering electronic densities larger than one is equivalent to considering {\it hole concentrations} smaller than one, and thus we are back at the original mathematical formulation. In the remainder of this thesis report, $na$ and $ma$ will always be assumed to obey these inequalities. There is one final remark to be made, originally due to the discoveries of Ref \cite{SO4symmetry}: the hamiltonian would not have commuted with the generators of the $\eta$-spin algebra had we chosen the number of lattice sites $N_a$ to be an odd number (this is easily verified by explicit calculation of the commutators), which is why $N_a$ is assumed even. 

When forming $\hat{S}_s^z+\hat{S}_c^z = \hat{N}_{\downarrow}-\frac {N_a} 2$, we see that $\hat{S}_s^z+\hat{S}_c^z$ only takes integer values. This implies some restrictions in the types of multiplets that are allowed by the model: states for which both $\hat{S}_s^z$ and $\hat{S}_c^z$ are integers, or for which both are half-odd integers, are allowed, whilst states for which one is integer and the other is a half-odd integer, are prohibited. This is the reason for why the hamiltonian (\ref{H}) does not possess a full $SU(2) \otimes SU(2)$ symmetry.

\subsection{The Bethe ansatz solution}

By using an ansatz wave-function, Lieb and Wu managed, in their famous paper of 1968 \cite{LiebWu}, to reduce the problem of diagonalizing the hamiltonian (\ref{H}) into solving a set of coupled non-linear equations. The variables in these equations are two sets of numbers usually referred to as {\it charge momenta} $\{{ k_j} \}$ and {\it spin rapidities} $\{{ \lambda_{\l} }\}$. These numbers can be finite or infinite and are in general complex. However, it is these two sets of numbers that characterize the eigenfunctions of the model. As a side note, it is worth to mention that this eigenfunction, the "Bethe ansatz wave function" was explicitly presented by F. Woynarovich in 1982, for $(U/t) \gg 1$ \cite{explicitwavefcn}. Unfortunately, for finite values of ($U/t$), it is not suitable for direct calculation of correlation functions, due to its complexity. The coupled non-linear equations, also called the "Lieb-Wu equations", are:
\begin{eqnarray}
&{\displaystyle e^{k_j L} = \prod_{l=1}^M \frac {\lambda_l-\sin k_j a -i \frac U 4}  {\lambda_l-\sin k_j a+i \frac U 4}} &j=1,\ldots,N \nonumber \\
&{\displaystyle \prod_{j=1}^N \frac {\lambda_l-\sin k_j a -i \frac U 4}  {\lambda_l-\sin k_j a+i \frac U 4} = \prod_{\substack { \bar{m}=1 \\ \bar{m} \neq l}}^M \frac {\lambda_l-\lambda_{\bar{m}} -i \frac U 2}  {\lambda_l-\lambda_{\bar{m}} +i \frac U 2}} \hspace{0.6cm} &l=1,\ldots,M\label{LiebWueqn}
\end{eqnarray}
where $M$ is the number of $\downarrow$-spin electrons. Here and in the following, we use the notation of T. Deguchi et al \cite{TDeguchi} (with the exception of $n$ and $m$ in that reference, which we here denote by $\bar{n}$ and $\bar{m}$, on order to separate them from the density and the magnetization). Taking logarithms on both sides of the two equations (\ref{LiebWueqn}), introduces the quantum numbers, which are integers or half-odd integers. Even though the solution provides us with the ground state energy as well as the energy spectrum, it does not provide us with an association of the quantum numbers with the electrons, nor to the configurations of electrons on a lattice.

Moving on, in 1972, Takahashi reformulated the original solution using the so called {\it string hypothesis} \cite{takahashi}. In this paper, the hypothesis is used to classify the finite quantum numbers of the problems into "strings", which are valid as the system size $N_a$ becomes very large. Basically, the numbers $k_j$ and $\lambda_l$ are grouped into strings according to the value of their real parts, and are distributed symmetrically with respect to the real axis. Like this, Takahashi arrives at three different groupings of numbers; one that only involves real charge momenta $k_j$ (type I), one that only involves (complex) spin rapidities $\lambda_l$ (type II), and finally one that involves both complex charge momenta $k_j$ and complex spin rapidities $\lambda_l$ (type III). By using these relationships between the numbers of equation (\ref{LiebWueqn}), we arrive after some algebra to the following set of equations, one for each type, usually called the "discrete thermodynamic Takahashi equations":
\begin{eqnarray}
\label{Takahashi}
&& k_j L = 2\pi I_j - 2 \sum_{\bar{n}=1}^{\infty} \sum_{\alpha=1}^{M_{\bar{n}}} \arctan \left(4 \frac {\sin k_j a -\Lambda_{\alpha}^{\bar{n}}} {\bar{n}U} \right) - 2 \sum_{\bar{n}=1}^{\infty} \sum_{\alpha=1}^{M'_{\bar{n}}}  \arctan \left( 4 \frac {\sin k_j a-{\Lambda'}_{\alpha}^{\bar{n}}} {\bar{n}U} \right) \nonumber \\
&& -2 \sum_{j=1}^{N-2M'} \arctan \left( 4 \frac {\sin k_j a-\Lambda_{\alpha}^{\bar{n}}} {\bar{n}U} \right) = 2\pi J_{\alpha}^{\bar{n}} + \sum_{\bar{m}=1}^{\infty} \sum_{\beta=1}^{M_{\bar{m}}}  \Theta_{\bar{n}\bar{m}} \left( 4 \frac {\Lambda_{\alpha}^{\bar{n}}-\Lambda_{\beta}^{\bar{m}}} U \right) \hspace{2.3cm} \nonumber \\
&& \frac L a \left( \arcsin ({\Lambda'}_{\alpha}^{\bar{n}}+i\frac {\bar{n}U} 4)+\arcsin ({\Lambda'}_{\alpha}^{\bar{n}}-i\frac {\bar{n}U} 4) \right) = \hspace{5.6cm} \label{stringhyp} \\
&&=2\pi {J'}_{\alpha}^{\bar{n}}-2 \sum_{j=1}^{N-2M'} \arctan \left( 4 \frac {\sin k_j a-{\Lambda'}_{\alpha}^{\bar{n}}} {\bar{n}U} \right)+\sum_{\bar{m}=1}^{\infty} \sum_{\beta=1}^{M'_{\bar{m}}}  \Theta_{\bar{n}\bar{m}} \left( 4 \frac {{\Lambda'}_{\alpha}^{\bar{n}}-{\Lambda'}_{\beta}^{\bar{m}}} U \right) \nonumber
\end{eqnarray}

Here $\alpha$ enumerates the string of length $\bar{m}$ (where "length" translates into "amount of numbers on a given string"). $\Lambda_{\alpha}^{\bar{m}}$ and ${\Lambda'}_{\alpha}^{\bar{m}}$ are the purely real midpoints of strings of type II and type III, respectively. $M_{\bar{n}}$ and $M'_{\bar{m}}$ are the total numbers of strings of type II (with length $\bar{n}$) and of type III (with length $\bar{m}$), respectively. Finally, ${M'}$ is the total number of charge momenta numbers involved in a string of type III, $M'=\sum_{\bar{n}=1}^{\infty} \bar{n}M'_{\bar{n}}$. The number of $\sigma$-spin electrons are related to the numbers of strings through: 
\begin{eqnarray}
N_{\downarrow}&=&\sum_{\bar{n}=1}^{\infty} \bar{n}M_{\bar{n}} + M' \nonumber \\
N_{\uparrow}&=&N-N_{\downarrow} \label{takaspinupdown}
\end{eqnarray}
and the function $\Theta_{\bar{n}\bar{m}}(x)$ is given by:
\begin{equation}
\Theta_{\bar{n}\bar{m}}(x)= \left\{
\begin{array}{c}
2\arctan \left( \frac x {\vert \bar{n}-\bar{m} \vert} \right)+4\arctan \left( \frac x {\vert \bar{n}-\bar{m} \vert +2} \right)+\ldots \hspace{1.0cm} \\
\ldots+4\arctan\left( \frac x {\bar{n}+\bar{m}-2} \right)+2\arctan \left( \frac x {\bar{n}+\bar{m}} \right), \hspace{0.2cm} \bar{n} \neq \bar{m}\\
4\arctan \left( \frac x 2 \right)+4\arctan\left( \frac x 4 \right)+\ldots \hspace{2.7cm} \\
\ldots+4\arctan\left( \frac x {2\bar{n}-2} \right)+2\arctan \left( \frac x {2\bar{n}} \right), \hspace{0.8cm} \bar{n} = \bar{m}
\end{array}
\right.
\label{thetafunc}
\end{equation}

It is important to note that these equations introduce the quantum numbers of the Bethe ansatz. These are a set of purely real numbers: $I_j$, $J_{\alpha}^{\bar{n}}$ and ${J'}_{\alpha}^{\bar{n}}$. $I_j$ is an integer if $\sum_{\bar{m}} (M_{\bar{m}}+{M'}_{\bar{m}})$ is even and a half-odd integer if odd, $J_{\alpha}^{\bar{n}}$ is an integer if $N-M_{\bar{n}}$ is odd and a half-odd integer if even, ${J'}_{\alpha}^{\bar{n}}$ is an integer if $N_a-N+M'_{\bar{n}}$ is odd and a half-odd integer if even. They obey the following inequalities:
\begin{eqnarray}
\vert 2I_j \vert &\leq& N_a \nonumber \\
\vert 2J_{\alpha}^{\bar{n}} \vert &\leq& N-2M'-1-\sum_{\bar{m}=1}^\infty t_{\bar{m}\bar{n}}M_{\bar{m}} \label{qnrange} \\
\vert 2{J'}_{\alpha}^{\bar{n}} \vert &\leq& N_a-N+2M'-1-\sum_{\bar{m}=1}^\infty t_{\bar{m}\bar{n}}M'_{\bar{m}} \nonumber
\end{eqnarray}
where $t_{\bar{m}\bar{n}}=2\min (\bar{m},\bar{n}) -\delta_{\bar{m}\bar{n}}$. The quantum numbers are equidistant from each other, for example $I_{j+1}-I_{j}=1$. With these numbers specified, one can solve Eqs. (\ref{stringhyp}) for the numbers $k_j$, $\Lambda_{\alpha}^{\bar{n}}$ and ${\Lambda'}_{\alpha}^{\bar{n}}$. The specification of the occupancy configurations of the above quantum numbers allows for the construction of the ground state and any excited state of the original hamiltonian \cite{takahashi}.

Some very important insights into the Bethe-ansatz solution was given by F.H.L. E{\ss}ler, V.E, Korepin and K. Schoutens in 1992 \cite{LWSBethe0} - \cite{completesolution}. In Ref. \cite{LWSBethe0} and \cite{LWSBethe} it was shown that the Bethe ansatz solution only accounts for either {\it lowest weight} or {\it highest weight} states of the one dimensional Hubbard model. However, in Ref. \cite{completesolution}, it was shown that, after taking into account all the states reached by the off-diagonal generators of the spin- and the $\eta$-spin algebras, the Bethe ansatz solution is indeed complete, in that the total number of states present in the solution, gives the accurate dimension of the Hilbert space of the original model. In this notation, the total energy and momentum (modulo $2\pi$) are expressed as:
\begin{eqnarray}
E&=&-2t \sum_{j=1}^{N-2M'} \cos k_j a +4t \sum_{\bar{n}=1}^{\infty}\sum_{\alpha=1}^{M'_{\bar{n}}} \text{Re} \sqrt {1-\left( {\Lambda'}_{\alpha}^{\bar{n}}+i\frac {\bar{n}U} 4 \right)^2}+U\bar{N} \label{Etaka} \\
P&=&\sum_{j=1}^{N-2M'} k_j -\frac 2 a \sum_{\bar{n}=1}^{\infty}\sum_{\alpha=1}^{M'_{\bar{n}}} \left[ \text{Re} \left\{ \arcsin \left({\Lambda'}_{\alpha}^{\bar{n}}+i\frac {\bar{n}U} 4 \right) \right\}-(\bar{n}+1) \frac {\pi} a \right] \label{Ptaka}
\end{eqnarray}
where $\bar{N}=(N_a-2N)/4$ and $\text{Re}$ denotes the real part.

\newpage
\section{Pseudoparticles and rotated electrons}

\subsection{Pseudoparticles - historical overview}

It was mentioned earlier that the Bethe ansatz does not provide information about the connection of its quantum numbers to the original electrons of the problem. However, these numbers have been associated with various quantum objects, different from the electrons, in various descriptions valid for some strict subspace of the Hilbert space of the model. Some examples include:  {\it charge pseudoparticles} and {\it spin pseudoparticles}, introduced in Refs. \cite{lowlying1} - \cite{lowlying12}, in the study of low-lying excitations.

In 1990, M. Ogata and H. Shiba used the Bethe ansatz wave-function factorization to calculate physical quantities, such as the momentum distribution function and the spin correlation function, for the case of strong coupling, $U/t \rightarrow \infty$ \cite{OgShib}.  They used the fact that, in this limit, the Bethe ansatz ground state wave function factorizes into a charge part and a spin part. The charge degrees of freedom are then calculated via a Slater determinant of "spinless fermions" and the spin degrees of freedom are described by the one dimensional $S=\frac 1 2$ Heisenberg model. This framework, with spinless fermions and a "squeezed" spin wave taken from the $S=\frac 1 2$ Heisenberg model, was subsequently used by K. Penc et al in various publications \cite{Karlo1} - \cite{Karlo3}, in the study of the exact one electron spectral function for $(U/t) \rightarrow \infty$. K. Penc and B.S. Shastry then adopted a similar technique for the Schultz-Shastry model \cite{Karlo4}. A related representation valid in the $(U/t) \rightarrow \infty$ limit was presented by R.G. Dias and J.M.B. Lopes dos Santos \cite{santos}. 

In Ref. \cite{holonspinon1} the $SO(4)$ symmetry of the model is used in case of exact half filling $na=1$, to describe the excitation spectrum and an S-matrix in terms of spinon and holon excitations. In this work, the elementary scatterers carry either spin but no charge (thus, they are dubbed spin $+\frac 1 2$ and spin $-\frac 1 2$ spinons, respectively) or charge but no spin (consequently dubbed $\eta$-spin $+\frac 1 2$ and $\eta$-spin $-\frac 1 2$ holons, respectively). Note that these "holons" and "spinons" are not the same quantum objects as those which will be introduced in this thesis report.

The common ground between these works is the fact that they are only valid for some strict subspace of the model, and that the introduced quantum object description, either derived from the numbers introduced by the string hypothesis or by considering the symmetries of the model, accounts for the famous separation of the electronic degrees of freedom: the spin and the charge degrees of freedom of the electron are described by different quantum objects that propagate through the system with different velocities. 

In 1997, J.M.P. Carmelo and N.M.R. Peres managed to generalize the previous pseudoparticle picture, to be valid for the {\it entire} Hilbert space \cite{CarmeloNuno}. The energy bands and the residual interactions of these pseudoparticles were explicitly presented. Furthermore, a normal ordered (relative to the ground state) formulation of the problem allowed the hamiltonian to be rewritten solely in terms of momentum configuration distribution operators. However, the connection between these pseudoparticles and the original electrons remained unknown until 2004, when the electrons were "re-connected" to the problem by relating them to quantum objects baptized {\it rotated electrons}, obtainable from the electrons by a mere unitary transformation for all values of $U/t$ \cite{NPB04}. The pseudoparticles are then "constructed" in terms of the decoupled spin- and charge degrees of freedom of these rotated electrons.

To give a flavour of how the pseudoparticles of Ref. \cite{CarmeloNuno} were born out of the quantum numbers of Takahashi, and also to conform with the notation that we will use from now on, we let $\alpha \rightarrow j $ and $\bar{n} \rightarrow \nu$ in the Takahashi string hypothesis, Eq. (\ref{stringhyp}). Furthermore, we will denote the numbers $I_j$, $J_{\alpha}^{\bar{n}}$ and ${J'}_{\alpha}^{\bar{n}}$ by $I_j^{c0}$, $I_j^{s\nu}$ and $I_j^{c\nu}$ respectively. Correspondingly,  let $k_j$, $\Lambda_{\alpha}^{\bar{n}}$ and ${\Lambda'}_{\alpha}^{\bar{n}}$ be re-baptized into $k_j$ (unchanged), $\Lambda_{s\nu}$ and $\Lambda_{c\nu}$ respectively. Since these quantities depend on each other, according to $j \rightarrow I_j^{c0} \rightarrow k_j$ (and similarly for the other two quantum numbers $I_j^{s\nu}$ and $I_j^{c\nu}$), we actually have by Eq. (\ref{stringhyp}), that $k_j=k(I_j^{c0})$, $\Lambda_{s\nu}=\Lambda_{s\nu}(I_j^{s\nu})$ and $\Lambda_{c\nu}=\Lambda_{c\nu}(I_j^{c\nu})$. Associating the quantum numbers with {\it pseudoparticle momenta} $q_j$, according to $q_j=(2\pi/L)I_j^{\alpha \nu}$ (where $I_j^{\alpha\nu}$ is shorthand for any of the three types of quantum numbers), we can finally write $k=k(q_j)$, $\Lambda_{s\nu}=\Lambda_{s\nu}(q_j)$ and $\Lambda_{c\nu}=\Lambda_{c\nu}(q_j)$ respectively. One should not confuse the different $q_j$'s with each other: they are different in that they are equal to $(2\pi/L)I_j^{c0}$, $(2\pi/L)I_j^{s\nu}$ and $(2\pi/L)I_j^{c\nu}$ respectively. In short:
\begin{eqnarray}
I_j &\rightarrow& I_j^{c0} \hspace{1.5cm} \hspace{0.15cm}  k_j \rightarrow k_j \nonumber \\
J_{\alpha}^{\bar{n}} &\rightarrow& I_j^{s\nu} \hspace{1.5cm} \Lambda_{\alpha}^{\bar{n}} \rightarrow \Lambda_{s\nu} \\
{J'}_{\alpha}^{\bar{n}} &\rightarrow& I_j^{c\nu} \hspace{1.4cm} {\Lambda'}_{\alpha}^{\bar{n}} \rightarrow \Lambda_{c\nu} \nonumber
\end{eqnarray}
and
\begin{eqnarray}
k&=&k(q_j) \hspace{2.0cm} \hspace{0.35cm} q_j=\frac {2\pi} {L} I_j^{c0} \nonumber \\
\Lambda_{s\nu}&=&\Lambda_{s\nu}(q_j) \hspace{2.0cm} q_j=\frac {2\pi} {L} I_j^{s\nu} \\
\Lambda_{c\nu}&=&\Lambda_{c\nu}(q_j) \hspace{2.0cm} q_j=\frac {2\pi} {L} I_j^{c\nu} \nonumber
\end{eqnarray}

In the following, we will write $\alpha \nu$ as a collective symbol standing for all of the $c0,c\nu$ and $s\nu$ branches (unless otherwise specified). This change of notation helps us to identify the sets of quantum numbers as occupational configurations of {\it pseudoparticles}, whose discrete momentum spacing is the usual $2\pi/L$. Furthermore, it identifies some functions of these momenta, collectively referred to as {\it rapidities}, namely $k$, $\Lambda_{s\nu}$ and $\Lambda_{c\nu}$, that are related to each other via Eq. (\ref{stringhyp}). The inequalities of Eq. (\ref{qnrange}) define the minimum and maximum values of the quantum numbers, and thus defines the occupancies inside the effective Brillouin zone of the pseudoparticles. For the ground state, the functions $k(q_j)$ and $\Lambda_{\alpha\nu}(q_j)$ are odd functions of their arguments \cite{wavefcnfact}.

\subsection{Rotated electrons - historical overview}
\label{sectrotE}

The mapping that transforms the electrons into their rotated counterparts, the rotated electrons, follows the previous work by A.B. Harris and R.V. Lange \cite{HarrisLange1} and the work of A.H. MacDonald et al \cite{HarrisLange2}. In these publications, a unitary transformation from now on denoted $\hat{V}(U/t)$, is introduced for large values of the on-site coloumb repulsion $U\!$, that cancels all terms in the original hamiltonian that change the number of doubly occupied sites. Later, the same transformation is successfully used in Ref. \cite{Karlo3}, in order to calculate the one electron spectral function in the limit $U/t \rightarrow \infty$. Because of the "large $U$ history" behind the introduction of this transformation, it will be our starting point as well. In this limit, $\hat{V}(U/t)$ can be written as an expansion in powers of $(t/U)$, but we should note that formally, this expansion is {\it not} the definition of $\hat{V}(U/t)$, which we will give later. 

The basic consideration behind this transformation is the fact that double occupancy is a good quantum number in the limit $U/t \rightarrow \infty$. This is easily seen by investigating the Coulomb interaction term of the hamiltonian (\ref{H}): a doubly occupied site gives a contribution to the total energy of the system equal to $U$, whilst other types of occupancies does not. Thus, states with $j$ and $j\pm1$ number of doubly occupied sites have an energy difference equal to $U$, so that this difference goes to infinity as $U$.

The double occupancy quantum number $D$, is nothing but  the expectation value of the double occupancy operator, $D=\langle\hat{D\rangle}$, where
\begin{equation}
\hat{D}=\sum_{i=1}^{N_a} n_{i\uparrow}n_{i\downarrow}, \hspace{1.0cm} n_{i\sigma}=c_{i\sigma}^{\dag}c_{i\sigma}
\end{equation}
In the following, we will adopt the following notation for an arbitrary operator $\hat{X}$, transformed by $\hat{V}(U/t)$:
\begin{equation}
\tilde{X}=\hat{V}^{\dag}(U/t)\hat{X}\hat{V}(U/t)
\end{equation}
where $\tilde{X}$ is the corresponding "rotated operator". Later we will see that $\hat{V}(U/t)$ is in fact unitary, which means that $\hat{X}$ and $\tilde{X}$ share the same set of eigenvalues and preserves the norm of the eigenstates, which will turn out to be very useful. 

The kinetic term of (\ref{H}), responsible for the dynamics of the model, can be written as a sum of three terms, $\hat{T}=\hat{T_0}+\hat{T}_U+\hat{T}_{-U}$, according to the energy difference that the hopping results in ("before" the hopping as compared to "after" the hopping). For example, a lonely electron on site $i$ that hops to a neighbouring site $i\pm1$, where one electron already is present, will have increased the total energy of the system by $U$. Thus, the job that $\hat{V}(U/t)$ has to accomplish is to cancel $\hat{T}_U$ and $\hat{T}_{-U}$. It is then possible to rewrite the hamiltonian according to 
\begin{equation}
\hat{H}=\hat{H}^{(0)}+\hat{H}^{(1)}+\hat{H}^{(2)}+\ldots
\label{Hexpansion}
\end{equation}
where $\hat{H}^{(j)}$ allows hopping with a total number of $j$ doubly occupied sites. One should note that Eq. (\ref{Hexpansion}) is actually nothing but a large - $U$ expansion of the original hamiltonian, where $\hat{H}^{(0)}$ is the $U=\infty$ term, and the following terms are corrections of order $(t/U)^j$. The first eight terms in the series (\ref{Hexpansion}) were explicitly calculated and presented in Ref. \cite{HarrisLange2}. 

The question arises, then, {\it how} one obtains the operator expressions that constitute the terms of (\ref{Hexpansion}). To answer this, we remember the fact that any unitary operator can be written as the exponential of an anti-hermitian operator, which here will be called $\hat{Y}(U/t)$:
\begin{equation}
\left.
\begin{array} {c}
\hat{H}=\hat{V}(U/t)\tilde{H}\hat{V}^{\dag}(U/t) \\
\hat{V}(U/t)=e^{\hat{Y}(U/t)}
\label{rotation}
\end{array}
\right\} \implies \hat{H}=e^{\hat{Y}(U/t)}\tilde{H}e^{-\hat{Y}(U/t)}
\end{equation}
Now, using the Baker-Hausdorff Lemma, we see that this can be rewritten as
\begin{equation}
\hat{H}=\tilde{H}+\left[ \hat{Y}(U/t),\tilde{H} \right]+\frac 1 2 \left[ \hat{Y}(U/t),\left[ \hat{Y}(U/t),\tilde{H} \right] \right]+\ldots
\label{BakerHaus}
\end{equation}
which immediately produces very many terms as we need to introduce $\tilde{H}=\tilde{T_0}+\tilde{T}_U+\tilde{T}_{-U}+\tilde{U}$ (note that $\tilde{U}=U\tilde{D}$). Requiring $\langle \tilde{D} \rangle$ to be a good quantum number, we need to have 
\begin{equation}
\left[ \tilde{D} , \hat{H} \right] = 0
\label{zeroD}
\end{equation}
Now, assuming that the following series expansion exists
\begin{equation}
\hat{Y}=\hat{Y}^{(1)}+\hat{Y}^{(2)}+\hat{Y}^{(3)}+\ldots
\label{Sseries}
\end{equation}
where $\hat{Y}^{(j)} \sim (t/U)^j$, we can express Eq. (\ref{BakerHaus}) by using Eq. (\ref{Sseries}) together with $\tilde{H}=\tilde{T_0}+\tilde{T}_U+\tilde{T}_{-U}+\tilde{U}$. The assumption of the existence of Eq. (\ref{Sseries}) is actually not such a gamble as it may seem: just like an ansatz solution to a differential equation "if it works it works, if it does not it does not". In this case "works" translates into "deriving a closed form expression for $\hat{Y}^{(j)}$".
By retaining only the first term of the expansion of $\hat{Y}$ in the above mentioned scheme, one can deduce that in order for Eq. (\ref{zeroD}) to be valid, we need to have:
\begin{equation}
\hat{Y}^{(1)}=\frac 1 U \left( \tilde{T}_U - \tilde{T}_{-U} \right)
\end{equation}
Inserting into Eq. (\ref{BakerHaus}) with an added unknown $\hat{Y}^{(2)}$, we find after some algebra that in order for Eq. (\ref{zeroD}) to hold, we must have
\begin{equation}
\hat{Y}^{(2)}=\frac 1 {U^2} \left[ \tilde{T}_U + \tilde{T}_{-U} , \tilde{T_0} \right]
\end{equation}
and so forth. In this way, the first terms of Eq. (\ref{Hexpansion}) becomes:
\begin{equation}
\hat{H}=\tilde{T}_0+\tilde{U}+\frac 1 U \left[  \tilde{T}_U , \tilde{T}_{-U} \right]\ + \mathcal{O} (\frac {t^3} {U^2})
\label{HoneoverU}
\end{equation}
This expansion can be continued to higher orders in $(t/U)$, finding closed form expressions of $\hat{V}(U/t)$ that are successively valid for a larger range of $(t/U)$ values, but it is actually {\it not needed} in order to create a valid theory for {\it arbitrary} values of this parameter. What is enough is the overall definition of the transformation, which is a combination of Eqs. (\ref{rotation}), (\ref{zeroD}) and (\ref{BakerHaus}):
\begin{eqnarray}
&&\hat{V}(U/t)=e^{\hat{Y}(U/t)} \nonumber \\
&&\left[ \tilde{D} , \hat{H} \right]=0 \label{deftrans} \\
&&\hat{H}=\tilde{H}+\left[ \hat{Y}(U/t),\tilde{H} \right]+\frac 1 2 \left[ \hat{Y}(U/t),\left[ \hat{Y}(U/t),\tilde{H} \right] \right]+\ldots \nonumber
\end{eqnarray}
Even though this definition seems quite abstract, there are some things that can be said about $\hat{V(U/t)}$ by pure physical reasoning. Since electronic double occupancy is a good quantum number for $(U/t)=\infty$, electronic $\uparrow$-spin and electronic $\downarrow$-spin single occupancies, as well as electronic no occupancy are also good quantum numbers in this limit (we will come back to this in section \ref{bridge}). For rotated electrons, however, these numbers are {\it always} good quantum numbers. This means that when a finite-$(U/t)$ energy eigenstate is acted upon by the operator $\hat{V}(U/t)$, the resulting state must bear some similarities with the corresponding $(U/t)=\infty$ eigenstate. This also implies that $\hat{V}(U/t)\rightarrow \bm{1}$ when $(U/t)\rightarrow \infty$ due to the simple fact that in this limit, the rotated electrons are identical to the electrons, or in other words, $\hat{Y}=\bm{0}$ in the expansion given by Eq. (\ref{Sseries}). The formal proof of the existence, uniqueness and unitariness of $\hat{V}(U/t)$ is given in Ref. \cite{NPB04}. Lastly, since $\hat{V}(U/t)$ does not change the lattice (neither the amount of lattice sites nor the lattice constant), we must have that 
\begin{equation}
\left[ \hat{V}(U/t) , \hat{P} \right] = 0
\end{equation}
considering that the momentum operator is the generator of lattice translations \cite{NPB04}.

\subsection{Connecting rotated electrons to pseudoparticles}
\label{bridge}

By the definition (\ref{deftrans}), we have actually defined a new quantum object, dubbed the {\it rotated electron}, that has the same spin and charge and that exists in the same lattice as the original electron, but for which double occupancy is a good quantum number for {\it all values of $U/t$}. Let us look a little bit deeper into this claim. 

Define the number operators for {\it electronic} double occupancy, no occupancy, single $\uparrow$-spin and single $\downarrow$-spin occupancy, respectively: 
\begin{eqnarray}
\hat{R}_{c-}=\sum_i c_{i\uparrow}^{\dag}c_{i\uparrow}c_{i\downarrow}^{\dag}c_{i\downarrow} \nonumber \\
\hat{R}_{c+}=\sum_i c_{i\uparrow}c_{i\uparrow}^{\dag}c_{i\downarrow}c_{i\downarrow}^{\dag} \nonumber \\
\hat{R}_{s-}=\sum_i c_{i\downarrow}^{\dag}c_{i\uparrow}c_{i\uparrow}^{\dag}c_{i\downarrow} \label{occnumbers} \\
\hat{R}_{s+}=\sum_i c_{i\uparrow}^{\dag}c_{i\downarrow}c_{i\downarrow}^{\dag}c_{i\uparrow} \nonumber
\end{eqnarray}

These operators are not independent. In fact, they can all be expressed in terms of the electronic double occupancy operator $\hat{D}$:
\begin{eqnarray}
\hat{R}_{c-}&=&\hat{D} \nonumber \\
\hat{R}_{c+}&=&N_a-\hat{N}+\hat{D} \nonumber \\
\hat{R}_{s-}&=&\hat{N}_{\downarrow}-\hat{D} \label{occnumbers2} \\
\hat{R}_{s+}&=&\hat{N}-\hat{N}_{\downarrow}-\hat{D} \nonumber
\end{eqnarray}
Note that this description of the different types of electronic occupations will be very important for the framework that we are going to use. Indeed, it accounts for all the electrons in the system, $2\hat{R}_{c-}+\hat{R}_{s+}+\hat{R}_{s-}=\hat{N}$. Now, let us define the {\it rotated electronic creation} and {\it annihilation operators}:
\begin{equation}
\begin{array} {c}
c_{i\sigma}^{\dag}= \hat{V}(U/t)\tilde{c}_{i\sigma}^{\dag}\hat{V}^{\dag}(U/t) \\
c_{i\sigma}=\hat{V}(U/t)\tilde{c}_{i\sigma}\hat{V}^{\dag}(U/t)
\end{array}
\label{rotops1}
\end{equation}
The operators $\tilde{c}_{i\sigma}^{\dag}$ and $\tilde{c}_{i\sigma}$ create and annihilate some quantum objects whose double occupancy number, single occupancy $\uparrow$-spin and single occupancy $\downarrow$-spin number, as well as its no occupancy number, are all good quantum numbers for any value of ($U/t$). This is seen by inserting Eq. (\ref{rotops1}) into Eq. (\ref{occnumbers}), and thus forming the corresponding rotated number operators, $\tilde{R}_{\alpha l}$, where $\alpha=c$ or $s$ and $l=-,+$. Like this we obtain by definition (\ref{deftrans}) and by Eq. (\ref{occnumbers2}) that:
\begin{equation}
\left[ \hat{H} , \tilde{R}_{c-} \right] =\left[ \hat{H} , \tilde{R}_{c+} \right] =\left[ \hat{H} , \tilde{R}_{s-} \right] =\left[ \hat{H} , \tilde{R}_{s+} \right] =0 \label{commutators}
\end{equation}
According to the studies of Ref. \cite{NPB04}, it is the separated charge and spin degrees of freedom of the {\it rotated} electrons that constitute some exotic quantum objects called {\it pseudoparticles}. The occupancy configurations of these pseudoparticles are described by the quantum numbers given by the Takahashi string hypothesis \cite{takahashi}. The connection to the original electrons was not reached because of the fact that the relevant electronic occupational numbers were not good quantum numbers, so that any separation of the original electronic degrees of freedom is bound to lead to quantum objects consisting of several very involved superpositions of the original electronic occupational configurations. The picture becomes much more elegant if, when describing the various excitations of the model, the starting point consists of quantum objects whose occupancy number operators commute with the original hamiltonian, like those of the rotated electron. Like this, it will be much more simple to describe the properties of the {\it new} quantum objects, here called the pseudoparticles, in terms of the occupancy configurations of the previous ones, the rotated electrons.

To proceed any further, it is necessary to explain what the pseudoparticles are constituted of. Since $\hat{V}(U/t)$ is unitary, all eigenvalues of the electronic number operators of Eq. (\ref{occnumbers}) are equal to the eigenvalues of the corresponding rotated operators, $\tilde{R}_{\alpha l}$. Now, there are $N$ electrons in the system together with $N^h=2N_a - N$ electronic holes. Let the number of sites that are {\it singly occupied by rotated electrons} be equal to $N_c$. This means that there are $N-N_c$ rotated electrons on doubly occupied sites, and hence a number of $(N-N_c)/2$ doubly occupied sites. By the same reasoning, we have $(N^h-N_c)/2$ sites doubly occupied by rotated holes ("empty sites"). Next we baptize the $\uparrow$-spin $\downarrow$-spin pair, on the sites doubly occupied by rotated electrons, $-\frac 1 2$ holons. There is a total number of $M_{c,-\frac 1 2}=(N-N_c)/2$ of such quantum objects. Whilst the $-\frac 1 2$ holons are spin zero objects, the value of its $\eta$-spin projection is $-\frac 1 2$, which justifies the choice of name. Equivalently, we form a total number $M_{c,+\frac 1 2}=(N^h-N_c)/2$ of $+\frac 1 2$ holons, from the sites doubly occupied by rotated holes. The $N_c$ rotated electrons on the singly occupied sites decouple into $N_c$ {\it chargeons} (with the same charge as the rotated electron but with no spin degrees of freedom) and $N_c$ {\it spinons} (with the same spin as the rotated electron but with no charge degrees of freedom). Now, the total number of $\uparrow$-spin spinons is $M_{s,+\frac 1 2}=N_{\uparrow}-(N-N_c)/2$, which is nothing but the total number of $\uparrow$-spin rotated electrons of the system, minus the $\uparrow$-spins of the doubly occupied sites. Equivalently, $M_{s,-\frac 1 2}=N_{\downarrow}-(N-N_c)/2$ is the total number of $\downarrow$-spin spinons. Below, we will make use of the number $N_c^h$ defined as the number of lattice sites {\it not} singly occupied, i.e. either doubly occupied or empty: $N_c^h=N_a-N_c$. Lastly, we note that the $N_c$ sites singly occupied by rotated electrons also carry $N_c$ rotated electronic holes. The charge part of these rotated holes, living on singly occupied sites, will be called {\it antichargeons}.
To summarize, we have the following amount of quantum objects:
\begin{eqnarray}
\text{$\downarrow$ $\eta$-spin holons} \hspace{1.0cm} M_{c,-\frac 1 2}&=&(N-N_c)/2 \nonumber \\
\text{$\uparrow$ $\eta$-spin holons} \hspace{1.0cm} M_{c,+\frac 1 2}&=&(N^h-N_c)/2 \nonumber \\
\text{$\downarrow$-spin spinons} \hspace{1.0cm} M_{s,-\frac 1 2}&=&N_{\downarrow}-(N-N_c)/2 \\
\text{$\uparrow$-spin spinons} \hspace{1.0cm} M_{s,+\frac 1 2}&=&N_{\uparrow}-(N-N_c)/2 \nonumber
\end{eqnarray}
The following should be observed:
\begin{equation}
\left.
\begin{array} {c}
M_{c,-\frac 1 2} + M_{c,+\frac 1 2}= N_c^h \\
M_{s,-\frac 1 2} + M_{s,+\frac 1 2}= N_c
\end{array}
\right \} \implies \sum_{\alpha=c,s} \sum_{\sigma= \pm \frac 1 2} M_{\alpha,\sigma} = N_c+N_c^h = N_a
\label{sumparticles}
\end{equation}

So far we have only regrouped the rotated electrons, giving them new names according to the occupancies of the lattice sites. Following the interpretation of Ref. \cite{NPB04}, let now the $N_c$ chargeons and the $N_c$ antichargeons recombine into $N_{c0}=N_c$ {\it c0-pseudoparticles}. Let further a total number of $\nu \hspace{0.2cm}$ $+\frac 1 2$ spinons and $\nu \hspace{0.2cm}$ $-\frac 1 2$ spinons form one {\it s$\nu$ pseudoparticle}, where the number $\nu=1,2,\ldots$ . Equivalently, let a total number of $\nu \hspace{0.2cm}$ $+\frac 1 2$ holons and $\nu \hspace{0.2cm}$ $-\frac 1 2$ holons form one {\it c$\nu$ pseudoparticle}, where the number $\nu=1,2,\ldots$ .

The claim of Ref. \cite{CarmeloNuno} is that the quantum numbers describing the momentum occupancies of these created pseudoparticles, are nothing but the Takahashi numbers, given by the string hypothesis. Thus an $\alpha \nu$ pseudoparticle (where $\alpha=c$ or $s$ and $\nu=1,2,\ldots$), that contains a number $\nu$ of holon ($\alpha=c$) or spinon ($\alpha=s$) pairs, corresponds to a string with length $\nu$ of type II ($\alpha=s$) or of type III ($\alpha=c$). There will be a total number of $N_{s\nu}$ $s\nu$ pseudoparticles, and a total number of $N_{c\nu}$ $c\nu$ pseudoparticles.

\begin{figure}
\begin{center}
\includegraphics[height=17cm,width=17cm]{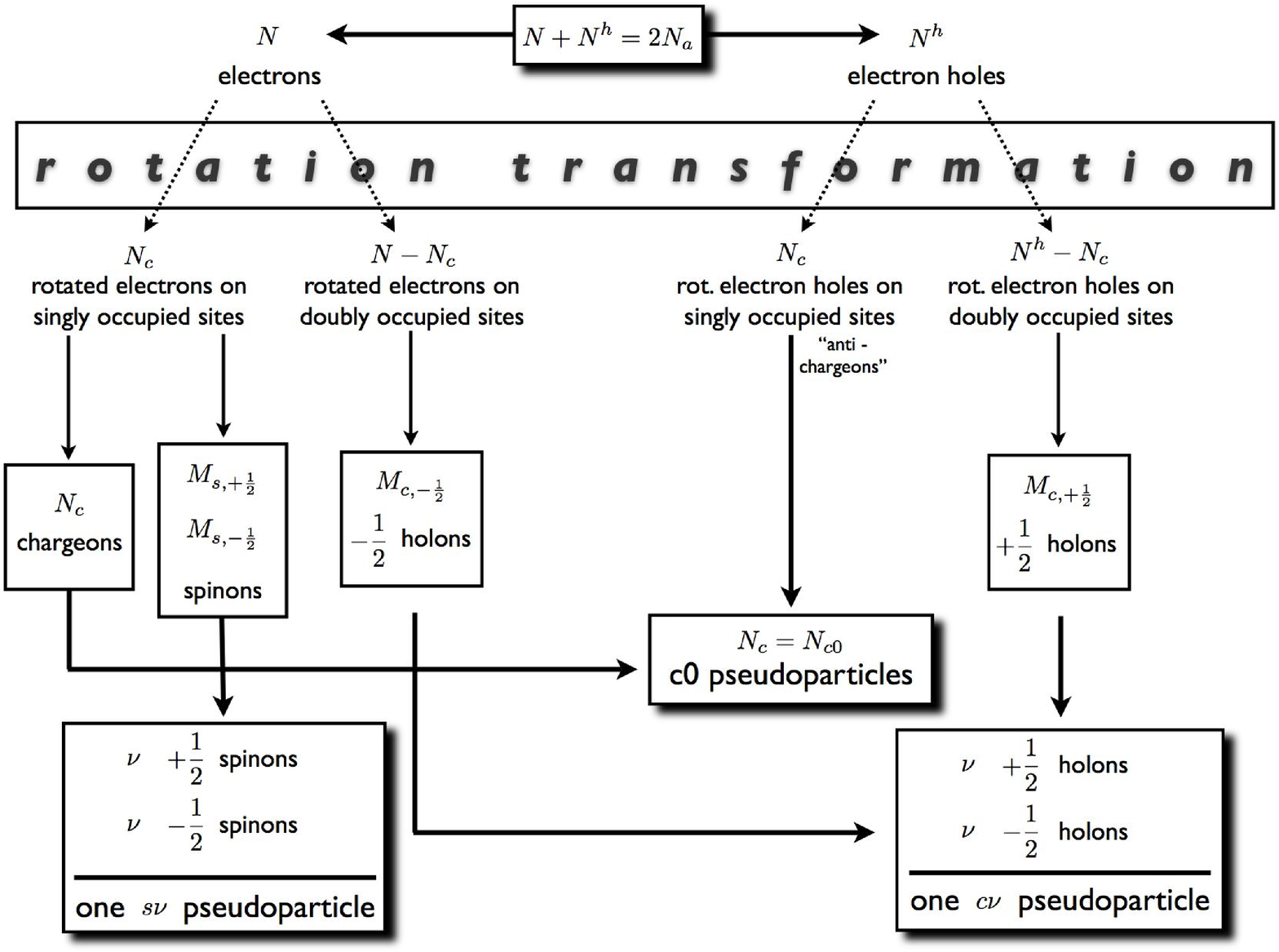}
\caption{\label{figflow} Flow chart describing how the electrons and the electron holes, due to the rotation transformation given by $\hat{V}(U/t)$, are described in terms of rotated electrons and rotated electron holes. These, in turn, are then forming the holons and spinons. These holons and spinons combine in a way described in section (\ref{bridge}), to form the pseudoparticles. \vspace{0.5cm}}
\end{center}
\end{figure}

These here introduced quantum objects cannot, however, account for the entire Hilbert space, as all of them are $\eta$-spin zero and spin zero objects (remember that we always combine an equal number of "$+\frac 1 2$" and "$-\frac 1 2$" to form one $\alpha \nu$ pseudoparticle). Since we have electronic densities $na$ and magnetization $ma$ in the ranges $0<na<1$ and $0<ma<na$, this means, for example in the spinon case, that some spins would be "left out" in the pairing process of constructing the $s\nu$ pseudoparticles (because in general $N_{\uparrow} > N_{\downarrow}$). The physical properties of the holons and spinons that were "not able to make it" into any of the $\alpha \nu$ pairs are quite different from those of the pseudoparticles, as further discussed in section (\ref{ppyhhls}).

The description of pseudoparticles in terms of holons and spinons, rotated electrons and electrons, respectively, is given in Fig. (\ref{figflow}).

\subsection{Pseudoparticles, Yang holons and HL spinons}
\label{ppyhhls}

Starting from the total number of electrons $N=\sum_{\sigma} N_{\sigma}$, where $N_{\sigma}$ is the total number of $\sigma$-spin electrons, we can derive expressions for the corresponding numbers of the pseudoparticles $N_{c0}$, $N_{s\nu}$ and $N_{c\nu}$ ($N_{\alpha \nu}$ gives the number of $\alpha \nu$ pseudoparticles: $N_{c2}=3$ means that we have 3 c2-pseudoparticles consisting of 4 holons each, 2 with $\eta$-spin projection $+\frac 1 2$ and 2 with $\eta$-spin projection $-\frac 1 2$). We will see that, in order to account for all the electrons in the system, we need to define some new objects that are inherently linked to the pseudoparticle description, but that do not contribute to the dynamics of the Hubbard model.

Our starting point is the symmetry consideration that led us to conclude that it is sufficient to study the region of the parameter space where $0 < na < 1$ and $0 < ma < na$. This means that:
\begin{eqnarray}
M_{c,+\frac 1 2}-M_{c,-\frac 1 2}&=&N_a-N > 0 \nonumber \\
M_{s,+\frac 1 2}-M_{s,-\frac 1 2}&=&N_{\uparrow}-N_{\downarrow} > 0
\end{eqnarray}
Since the total number of $\pm \frac 1 2$ holons ($\alpha=c$) and $\pm \frac 1 2$ spinons ($\alpha=s$) that take part in the $\alpha \nu$ pseudoparticles can be written as $\sum_{\nu=1}^{\infty} \nu N_{\alpha \nu}$ , we can write the difference between the total number of holons and spinons and those paired up in $\alpha \nu$ pseudoparticles, as:

\begin{equation}
L_{\alpha, \pm \frac 1 2}=M_{\alpha, \pm \frac 1 2}-\sum_{\nu=1}^{\infty} \nu N_{\alpha \nu}
\label{defLparts}
\end{equation}

These remaining particles, there are a number of $L_{\alpha, \pm \frac 1 2}$ of them, have been given the names $\pm \frac 1 2$ {\it Yang holons} (for $\alpha=c$) and $\pm \frac 1 2$ {\it HL spinons} (for $\alpha=s$) \cite{NPB04}, \cite{CarmeloPedro}. "Yang" stands for C.N. Yang who authored Ref. \cite{SO4}, whilst "HL" stands for Heilmann and Lieb, who authored Ref \cite{HL}. It is not difficult to hint a relationship between the total $\eta$-spin and spin on the one hand, and the total number of Yang holons and HL spinons on the other. In this case of a LWS, we have $L_{c,+\frac 1 2}^{LWS}=N_a-N=2S_c$ and $L_{c,-\frac 1 2}^{LWS}=0$ in the holon case, and $L_{s,+\frac 1 2}^{LWS}=N_{\uparrow}-N_{\downarrow}=2S_s$ and $L_{s,-\frac 1 2}^{LWS}=0$ in the spinon case. Now, by acting onto the LWS by $\hat{S}_{\alpha}^{\dag}$, we increase the number of $L_{\alpha,-\frac 1 2}$ by one, on the expense of $L_{\alpha,+\frac 1 2}$ which decreases by one (the z-component of the $\eta$-spin and/or spin changes accordingly). Like this we can continue until we reach the HWS. Hence $L_{\alpha,\pm \frac 1 2}=S_{\alpha}-2(\pm \frac 1 2)S_{\alpha}^z=0,1, \ldots , 2S_{\alpha}$ and furthermore $L_{\alpha}=L_{\alpha,+\frac 1 2}+L_{\alpha,-\frac 1 2}=2S_{\alpha}$, where $S_{\alpha}$ is the total $\eta$-spin ($\alpha=c$) or spin ($\alpha=s$) of the system.

These HL spinons and Yang holons behave quite differently inside the many-body system, than the $c0$-pseudoparticles and the $\alpha \nu$ pseudoparticles. Firstly, the creation and annihilation operators of the former objects do not commute with the generators of the $\eta$-spin and the spin algebras (as already noted above), whilst the corresponding operators of the latter objects indeed do. This is easily seen by understanding that the $c0$-pseudoparticles have no $\eta$-spin or spin degrees of freedom, and that the $\alpha \nu$ pseudoparticles are singlet $\eta$-spin ($\alpha$=c) and spin ($\alpha=s$) quantum objects, and thus yield zero when acted upon by any of the $\eta$-spin and spin generators, respectively (this is easily confirmed mathematically by forming a singlet state of rotated electrons corresponding to a certain pseudoparticle, and then letting any of the $\eta$-spin or spin generators act upon this state). 

The physical implication is that all ($2S_{\alpha}+1$) energy eigenstates inside any given ladder have the {\it same} occupancies of pseudoparticles: "climbing up" the ladder, from a LWS towards a HWS, will only change the numbers $L_{\alpha ,\pm \frac 1 2}$. Due to the fact that the occupancies of pseudoparticles are insensitive to the application of the generators of the two $SU(2)$ algebras, the discrete momentum values and hence their corresponding rapidity numbers stay unaltered as well. If we define the number operators $\hat{N}_{\alpha \nu}=\sum_q \hat{N}_{\alpha \nu}(q)$, with eigenvalues $N_{\alpha \nu}$, we have that
\begin{equation}
\left[ \hat{N}_{\alpha \nu} , \hat{S}_{\alpha'}^{\dag} \right] = \left[ \hat{N}_{\alpha \nu} , \hat{S}_{\alpha'} \right] = \left[ \hat{N}_{\alpha \nu} , \hat{S}_{\alpha'}^z \right] = 0 \hspace{1.5cm} \alpha'=c,s 
\end{equation}
valid for all $\alpha \nu$ branches. Next, we claim that all the six generators of the two $SU(2)$ algebras commute with the unitary operator $\hat{V}(U/t)$, whilst the operators $\hat{N}_{\alpha \nu}(q)$ do not. Actually, the last claim is trivial: if they would commute, then the pseudoparticles as described by the rotated electrons would be the same objects as described by the original electrons. However, since double occupancy is not a good quantum number for the original electrons, we know that the pseudoparticles can {\it not} be described as simple occupancy configurations of these objects. For the generators of the two $SU(2)$ algebras, we have the opposite: these operators create the {\it same} quantum objects in the unrotated as well as in the rotated frame. Unlike the operators for the pseudoparticles, the generators of the two $SU(2)$ algebras can be easily expressed both in terms of electronic as well as rotated electronic creation and annihilation operators:
\begin{equation}
\hat{S}_c^{\dag} = \sum_i (-1)^i c_{i\downarrow}^{\dag}c_{i\uparrow} = \hat{V}(U/t)^{\dag}\hat{S}_c^{\dag}\hat{V}(U/t) =  \sum_i (-1)^i \tilde{c}_{i\downarrow}^{\dag}\tilde{c}_{i\uparrow} 
\end{equation}
where $\hat{S}_c^{\dag}$ serves as an example of a typical $SU(2)$ generator. The point is that $\hat{S}_c^{\dag}$ has the same expression in terms of electrons and rotated electrons. This was studied in detail in Ref. \cite{CarmeloPedro}, where the electronic double occupancy expectation value $D_{c1}(q)$ was obtained as a function of the occupancies of various pseudoparticles. It was found, for example, that when creating a $c1$-pseudoparticle, the electronic average double occupancy did not in general increase by one (which would be the naive guess), but rather, it was found to depend on the momenta $q$ and on the value of $(U/t)$, according to
\begin{equation}
D_{c1}(0) \leq D_{c1} (q) \leq D_{c1} (\pi-2k_F) = 1 \nonumber
\end{equation}

In Fig. 4 of that reference, we see that $D_{c1}(0) \rightarrow 1$ as $(U/t) \rightarrow \infty$. This can actually serve as a "measure" for how the electron - rotated electron unitary transformation depends on $(U/t)$; the closer $D_{c1}(0)$ gets to $1$, the closer $\hat{V}(U/t)$ gets to unity. The fact that the generators of the two $SU(2)$ algebras commute with $\hat{V}(U/t)$ implies that the Yang holons and HL spinons are the same quantum objects in terms of electrons as they are in terms of rotated electrons, due to the fact that the creation and annihilation operators for these objects are nothing but the off-diagonal generators themselves, multiplied by a normalizing constant.  Indeed, it was found in Ref. \cite{CarmeloPedro}, that creating a $-\frac 1 2$ Yang holon always yields a double occupancy increase by one, $D \rightarrow D+1$, independently of $(U/t)$. Furthermore it was found that creation of a Yang holon or a HL spinon does not change the expectation value of the kinetic operator, thus deeming these objects "localized" in terms of (lack of) charge and spin transport. Summarized:
\begin{equation}
\left[ \hat{V}(U/t) , \hat{S}_{\alpha}^{\dag} \right] = \left[ \hat{V}(U/t) , \hat{S}_{\alpha} \right] = \left[ \hat{V}(U/t) , \hat{S}_{\alpha}^z \right] = 0 \hspace{0.7cm} \alpha=c,s 
\end{equation}
which implies that
\begin{equation}
\left[ \hat{V}(U/t) , \hat{L}_{c,\pm \frac 1 2} \right] = \left[ \hat{V}(U/t) , \hat{L}_{s,\pm \frac 1 2} \right]  = 0 \hspace{1.20cm}
\end{equation}
where $\hat{L}_{\alpha,\pm \frac 1 2}$ is the $\pm \frac 1 2$ Yang holon ($\alpha=c$) and $\pm \frac 1 2$ HL spinon ($\alpha=s$) number operator, respectively. 

Finally, the proof of the consistency of this pseudoparticle, Yang holon and HL spinon picture, with the Bethe-Ansatz solution, in terms of the counting of the states, was given in Ref. \cite{NPB04}. In this reference, it is shown that this representation accounts for all the $4^{N_a}$ eigenstates of the model, as well as showing that the number of states in a subspace consisting of a fixed number of $M_{\alpha}=\sum_{\sigma} M_{\alpha, \sigma}$ holons or spinons with a fixed $\eta$-spin or spin value $S_{\alpha}$, equals the number of states obtained by $\alpha \nu$ pseudoparticle occupancy configurations (where $\alpha \nu=c0,c\nu,s\nu$), according to the previously described scheme of the recombination of rotated electrons into pseudoparticles.

\newpage
\subsection{Occupational configurations of the pseudoparticles}
\label{occpp}

Since the pseudoparticles are derived from the Takahashi string hypothesis, with the occupational configurations given by the quantum numbers, we can define the functions $N_{c0}(q_j)$, $N_{s \nu}(q_j)$ and $N_{c \nu}(q_j)$ respectively, where $N_{\alpha \nu}(q_j)=1$ means that the discrete pseudoparticle momentum $q_j$ is occupied and $N_{\alpha \nu}(q_j)=0$ means that the discrete momentum $q_j$ is unoccupied. To each set of allowed quantum numbers correspond a unique set of rapidity numbers, which in turn correspond to a unique energy eigenstate \cite{TDeguchi} \cite{takahashi}. The pseudoparticles obey a generalized Pauli principle, known as {\it Haldane statistics} \cite{haldane}. In short, the Haldane particles affect the number of states available to any other particle in the many-body system (whilst in the case of exact fermions, only the state occupied by one fermion is forbidden to the next fermion). 

Let us define pseudoparticle creation and annihilation operators: $b_{q,\alpha \nu}^{\dag}$ and $b_{q,\alpha \nu}$ respectively, where $q$ is the momentum and $\alpha \nu = c0,c\nu,s\nu$. 
The pseudoparticle number operator can be written as:
\begin{equation}
\hat{N}_{\alpha \nu}(q)=b_{q,\alpha \nu}^{\dag}b_{q,\alpha \nu}
\end{equation}
which when summing over all momenta gives the total number of $\alpha \nu$ pseudoparticles
\begin{equation}
\hat{N}_{\alpha \nu}=\sum_q \hat{N}_{\alpha \nu}(q)
\end{equation}
The statistics obeyed by the pseudoparticles is \cite{wavefcnfact}
\begin{equation}
\{ b_{q,\alpha \nu}^{\dag} , b_{q',\alpha' \nu'} \} = \left\{
\begin{array} {c}
\delta_{\alpha\nu,\alpha'\nu'} \delta_{q,q'} \hspace{6.0cm} Q_{\alpha\nu}^0=0\\
i \delta_{\alpha\nu,\alpha'\nu'} e^{ia_{\alpha\nu} \left( q'-q \right) /2} \left[ N_{\alpha\nu}^* \sin \left(  \frac {a_{\alpha\nu} \left(q-q' \right)} 2 \right) \right]^{-1} \hspace{0.8cm} Q_{\alpha\nu}^0=\pm \pi
\end{array}
\right. \label{anticompseudo}
\end{equation}
where $N_{\alpha\nu}^*$ is defined below and $Q_{\alpha\nu}^0$ measures the quantum shake-up effect, and is introduced in section (\ref{pseudofermionII}). $Q_{\alpha\nu}^0$ is zero for the ground state (by construction) and nonzero if the actual excited state is described by quantum numbers shifted as compared to those of the ground state (from integers to half-odd integers or vice-versa). The momentum dependent creation and annihilation operators are formally defined locally on an effective $\alpha \nu$ lattice, with lattice constant $a_{\alpha \nu}$, defined so that the length of such a lattice is $\alpha \nu$ independent and equal to $L$:
\begin{equation}
a_{\alpha \nu}=a \frac {N_a} {N_{\alpha \nu}^*}
\end{equation}
where $N_{\alpha \nu}^*$ is the number of $\alpha \nu$ lattice sites. That in general $N_{\alpha \nu}^* \neq N_a$ stems originally from the upper and lower bounds on the quantum numbers from the Takahashi string hypothesis (\ref{qnrange}), which controls the value of the largest possible occupied momentum (the positive boundary of the effective Brillouin zone, the largest quantum number) and the smallest possible occupied momentum (the negative boundary of the effective Brillouin zone, the smallest quantum number). Only in one case (for the $c0$-pseudoparticles), does the total number of allowed discrete momenta equal the number of "real" lattice sites $N_a$. The sum of the number of occupied momentum values and the number of unoccupied momentum values, must  always equal the number of $\alpha \nu$ effective lattice sites.
\begin{equation}
N_{\alpha \nu}^*=N_{\alpha \nu} + N_{\alpha \nu}^h
\end{equation}
This gives us instantly that $N_{c0}^*=N_a$ due to Eq. (\ref{sumparticles}). Ref. \cite{NPB04} provides us with an expression for the total number of $\alpha \nu$ pseudoparticle holes:
\begin{equation}
N_{\alpha \nu}^h=L_{\alpha} + 2 \sum_{\nu'=\nu+1}^{\infty} \left( \nu' - \nu \right) N_{\alpha \nu'}
\label{pseudoholes}
\end{equation}
which also can be expressed as:
\begin{eqnarray}
N_{c0}^h&=&N_a-N_{c0} \nonumber \\
N_{c\nu}^h&=&N_a - N_{c0} - \sum_{\nu'=1}^{\infty} \left( \nu+\nu' - \vert \nu - \nu' \vert \right) N_{c\nu'} \hspace{0.5cm} \nu \geq 1\label{pseudoholesII} \\
N_{s\nu}^h&=&N_{c0} - \sum_{\nu'=1}^{\infty} \left( \nu+\nu' - \vert \nu - \nu' \vert \right) N_{s\nu'} \hspace{0.5cm} \nu \geq 1 \nonumber
\end{eqnarray}

The number of allowed momentum values in the ground state reads: 
\begin{eqnarray}
N_{c0}^{*,0}&=&N_a \nonumber \\
N_{c\nu}^{*,0}&=&N_a-N \hspace{1.1cm} \nu \geq 1\nonumber \\ 
N_{s1}^{*,0}&=&N_{\uparrow} \label{plattice} \\
N_{s\nu}^{*,0}&=&N_{\uparrow}-N_{\downarrow} \hspace{1.0cm} \nu \geq 2 \nonumber
\end{eqnarray}
(the corresponding numbers for an arbitrary excited energy eigenstate will be given in section (\ref{specdevnum})). With these numbers well defined, we can then relate the local $\alpha \nu$ pseudoparticle operators with the Fourier transformed momentum $\alpha \nu$ pseudoparticle operators:
\begin{eqnarray}
b_{q,\alpha \nu}^{\dag}&=&\frac 1 {\sqrt{N_{\alpha \nu}^*}} \sum_{j=1}^{N_{\alpha \nu}^*} e^{iqja_{\alpha \nu}} b_{j,\alpha \nu}^{\dag} \nonumber \\
b_{q,\alpha \nu}&=&\frac 1 {\sqrt{N_{\alpha \nu}^*}} \sum_{j=1}^{N_{\alpha \nu}^*} e^{-iqja_{\alpha \nu}} b_{j,\alpha \nu}
\end{eqnarray}
where $b_{j,\alpha \nu}^{\dag}$ creates a $\alpha \nu$ pseudoparticle on lattice site position $j$ (with space coordinate $x_j=ja_{\alpha \nu}$) and $b_{j,\alpha \nu}$ annihilates a $\alpha \nu$ pseudoparticle on lattice site position $j$. From the equalities of Eq. (\ref{commutators}), we have that $\hat{N}_{\alpha \nu}$ commutes with the original hamiltonian.
  
As can be deduced from Eqs. (\ref{pseudoholes}) and (\ref{plattice}) (together with the total number of Yang holons and HL spinons), we find that the ground state is completely void of $s \nu$ pseudoparticles for $\nu=2,3,\ldots$ and of $c \nu$ pseudoparticles for all $\nu=1,2,\ldots$ \cite{CarmeloNuno}. Moreover, the two branches that have finite occupancies in the ground state, as given by $N_{c0}(q)=N_{c0}^0 (q)$ and $N_{s1}(q)=N_{s1}^0 (q)$, are both densely packed around a minimum energy - zero momentum point with well defined left and right Fermi points. In the finite ground state system, this means that the {\it occupied} quantum numbers are symmetrically distributed around zero, with the exception of ($1/L$) corrections (see below). We can define the smallest possible quantum number for the $\alpha \nu$ branch as $I_-^{\alpha\nu}$ and the largest as $I_+^{\alpha\nu}$, defining the limiting momenta for the effective Brillouin zone, and similarily for the occupied momenta, the negative (left) Fermi point $I_{F-}^{\alpha\nu}$ and the positive (right) Fermi point $I_{F+}^{\alpha\nu}$. The "true" limiting momenta for the effective Brillouin zone and Fermi momenta will be shown to be ($1/L$) corrections to the momenta given by these numbers. In our new notation, we can reformulate the conditions on when the quantum numbers of the Takahashi string hypothesis must be integers and when they must be half-odd integers. For all $\alpha \nu \neq c0$ branches, we have that
\begin{eqnarray}
N_{\alpha\nu}^* \text{       even} &\implies& I_j^{\alpha\nu} \text{       half-odd integer} \nonumber \\
N_{\alpha\nu}^* \text{       odd  } &\implies& I_j^{\alpha\nu} \text{       integer} \nonumber
\end{eqnarray}
whilst for the $c0$ branch,
\begin{eqnarray}
\frac {N_a} 2 - \sum_{\alpha\nu=c\nu,s\nu }N_{\alpha\nu} \hspace{0.7cm} \text{even} &\implies& I_j^{c0} \text{       half-odd integer} \nonumber \\
\frac {N_a} 2 - \sum_{\alpha\nu=c\nu,s\nu }N_{\alpha\nu} \hspace{0.7cm} \text{odd  } &\implies& I_j^{c0} \text{       integer}  \label{c0numbers}
\end{eqnarray}
For all $\alpha\nu$ branches, let us define the quantum numbers introduced above according to
\begin{eqnarray}
&I_-^{\alpha\nu}={\displaystyle - \frac {N_{\alpha\nu}^*-1} 2} \hspace{1.6cm} &I_+^{\alpha\nu}=-I_-^{\alpha\nu}= {\displaystyle \frac {N_{\alpha\nu}^*-1} 2} \nonumber \\
&I_{F-}^{\alpha\nu}={\displaystyle - \frac {N_{\alpha\nu}-1} 2} \hspace{1.6cm} &I_{F+}^{\alpha\nu}=-I_{F-}^{\alpha\nu}= {\displaystyle \frac {N_{\alpha\nu}-1} 2} \label{brillouinfermi}
\end{eqnarray}
To obtain the limiting momenta for the effective Brillouin zone as well as the Fermi momenta, let us define some temporary variables $\kappa_{\alpha \nu}^+$, $\kappa_{\alpha \nu}^-$, $\kappa_{F\alpha \nu}^+$ and $\kappa_{F\alpha \nu}^-$ respectively, according to:
\begin{eqnarray}
&\kappa_{\alpha \nu}^- = {\displaystyle \frac {2\pi} L I_-^{\alpha \nu}}  \hspace{1.5cm} &\kappa_{\alpha \nu}^+ = {\displaystyle \frac {2\pi} L I_+^{\alpha \nu} } \nonumber \\
&\kappa_{F\alpha \nu}^- = {\displaystyle \frac {2\pi} L I_{F-}^{\alpha \nu} } \hspace{1.5cm} &\kappa_{F\alpha \nu}^+ = {\displaystyle \frac {2\pi} L I_{F+}^{\alpha \nu}}
\end{eqnarray}
Next, let us consider the {\it ground state} configuration. To make matters simple, we will only consider LWS ground states, such that $L_{\alpha.-\frac 1 2}=0$ which translates into $N_{s1}=N_{\downarrow}$ since there are no other $\downarrow$-spins in the system. Furthermore, we have that $N_{c\nu}=N_{s\nu}=0$ for all $\nu=1,2,\ldots$ in the case of the $c\nu$ branches and $\nu=2,3,\ldots$ in the case of the $s\nu$ branches. The only relevant $\kappa_{\alpha\nu}^{\pm}$ and $\kappa_{F\alpha\nu}^{\pm}$ becomes
\begin{eqnarray}
\kappa_{c0}^- &=&-\frac {\pi N_a} L + \frac {\pi} L \hspace{1.5cm} \kappa_{c0}^+ = -\kappa_{c0}^- \nonumber \\
\kappa_{Fc0}^- &=&- \pi n + \frac {\pi} L  \hspace{1.58cm} \kappa_{Fc0}^+ = -\kappa_{Fc0}^- \nonumber \\
\kappa_{s1}^- &=&- \pi n_{\uparrow}+\frac {\pi} L  \hspace{1.65cm} \kappa_{s1}^+ = -\kappa_{s1}^- \label{kappaeqs}\\
\kappa_{Fs1}^- &=&- \pi n_{\downarrow}+ \frac {\pi} L \hspace{1.45cm} \kappa_{Fs1}^+ = -\kappa_{Fs1}^- \nonumber
\end{eqnarray}
where we have used Eqs. (\ref{pseudolattice}) and (\ref{brillouinfermi}) together with $N_{s1}=N_{\downarrow}$. The different cases of limiting momenta for the effective Brillouin zone $q_{\alpha\nu}^{\pm}$ and of Fermi momenta $q_{F\alpha\nu}^{\pm}$ are, according to Ref. \cite{NPB04}, in the $\alpha\nu=c0$ case:
\begin{eqnarray}
\alpha\nu=c0 \hspace{1.0cm} \text{Limiting momenta for the effective Brillouin zone:} \hspace{2.00cm}\nonumber \\
\frac {N_a} 2 - \sum_{\alpha\nu=c\nu,s\nu }N_{\alpha\nu} \hspace{0.5cm} \text{even  } \implies q_{c0}^{\pm}=\kappa_{c0}^{\pm} \nonumber \\
\frac {N_a} 2 - \sum_{\alpha\nu=c\nu,s\nu }N_{\alpha\nu} \hspace{0.5cm} \text{odd  } \implies q_{c0}^{\pm}=\kappa_{c0}^{\pm} \pm \frac {\pi} L \nonumber \\
\nonumber \\
\alpha\nu=c0 \hspace{1.0cm} \text{Fermi momenta (LWS):} \hspace{6.80cm} \nonumber \\
\frac {N_a} 2 - \sum_{\alpha\nu=c\nu,s\nu }N_{\alpha\nu} \hspace{0.5cm} \text{and  } N \text{ both odd or both even } \implies q_{Fc0}^{\pm}=\kappa_{Fc0}^{\pm} \nonumber \\
\frac {N_a} 2 - \sum_{\alpha\nu=c\nu,s\nu }N_{\alpha\nu} \hspace{0.5cm} \text{odd and  } N \text{ even } \implies q_{Fc0}^{\pm}=\kappa_{Fc0}^{\pm}-\frac {\pi} L \nonumber \\
\frac {N_a} 2 - \sum_{\alpha\nu=c\nu,s\nu }N_{\alpha\nu} \hspace{0.5cm} \text{even and  } N \text{ odd } \implies q_{Fc0}^{\pm}=\kappa_{Fc0}^{\pm}+\frac {\pi} L \nonumber 
\end{eqnarray}
and in the $\alpha\nu=s1$ case:
\begin{eqnarray}
\alpha\nu=s1 \hspace{1.0cm} \text{Limiting momenta for the effective Brillouin zone:} \hspace{2.00cm}\nonumber \\
N_{s1}^*=N_{\uparrow} \text{ even or odd (i.e. always) } \implies q_{s1}^{\pm}=\kappa_{s1}^{\pm} \nonumber \\
\nonumber \\
\alpha\nu=s1 \hspace{1.0cm} \text{Fermi momenta (LWS):} \hspace{6.80cm} \nonumber \\
N_{s1}^*=N_{\uparrow} \text{ and } N_{s1}=N_{\downarrow} \text{ both even or both odd } \implies q_{Fs1}^{\pm}=\kappa_{Fs1}^{\pm} \nonumber \\
N_{s1}^*=N_{\uparrow} \text{ even and } N_{s1}=N_{\downarrow} \text{ odd } \implies q_{Fs1}^{\pm}=\kappa_{Fs1}^{\pm}+\frac {\pi} L \nonumber \\
N_{s1}^*=N_{\uparrow} \text{ odd and } N_{s1}=N_{\downarrow} \text{ even } \implies q_{Fs1}^{\pm}=\kappa_{Fs1}^{\pm}-\frac {\pi} L \nonumber
\end{eqnarray}
where the $\kappa_{\alpha\nu}^{\pm}$'s and the $\kappa_{F\alpha\nu}^{\pm}$'s are defined in Eq. (\ref{kappaeqs}).

So far everything has been described for the finite system, however, later we will frequently use the same quantities in the thermodynamic limit. In this limit, the notation becomes less heavy due to the fact that we neglect ($1/L$) terms. Let us define $q_{\alpha\nu}^0=\lim_{L\rightarrow \infty} q_{\alpha\nu}^+$ and $q_{F\alpha\nu}=\lim_{L\rightarrow \infty} q_{F\alpha\nu}^+$. The ground state occupancy configurations of the pseudoparticles become:
\begin{eqnarray}
N_{c0}^0 (q)&=&\theta (q_{Fc0}-\vert q \vert) \hspace{1.0cm} \vert q \vert \leq q_{c0}^0 \nonumber \\
N_{s1}^0 (q)&=&\theta (q_{Fs1}-\vert q \vert) \hspace{1.0cm} \vert q \vert \leq q_{s1}^0 \label{GSpp} \\
N_{\alpha \nu}^0 (q)&=&0 \hspace{3.0cm} \vert q \vert \leq q_{\alpha \nu}^0 \nonumber
\end{eqnarray}
where here $\alpha \nu$ stands for all other pseudoparticles $\neq c0,s1$ and the Fermi momenta $q_{Fc0}$ and $q_{Fs1}$ and the limiting momentum values of the effective Brillouin zone $q_{c0}^0$, $q_{s1}^0$ and $q_{\alpha \nu}^0$, respectively, become:
\begin{eqnarray}
q_{Fc0}&=&2k_F \hspace{1.45cm} q_{c0}^0=\frac {\pi} a \nonumber \\
q_{Fs1}&=&k_{F\downarrow} \hspace{1.5cm } q_{s1}^0=k_{F\uparrow}
\end{eqnarray}
and
\begin{eqnarray}
q_{s \nu}^0&=&k_{F\uparrow}-k_{F\downarrow} \hspace{2.80cm} \nu=2,3,\ldots \nonumber \\
q_{c \nu}^0&=&\frac {\pi} a-2k_F=\pi (\frac 1 a-n) \hspace{1.0cm} \nu=1,2,\ldots
\end{eqnarray}
where $k_F=(k_{F\uparrow}+k_{F\downarrow})/2=(\pi n_{\uparrow}+\pi n_{\downarrow})/2=\pi n/2$ is the usual Fermi momentum. Note that we cannot define a corresponding Fermi momentum for the $c\nu$ and the $s\nu$ ($\nu=2,3,\ldots$) bands due to the absence of these pseudoparticles in the ground state. However, the well defined Fermi points and effective Brillouin zones of the $c0$ and $s1$ momenta, allows us to define some typical ranges and values for the ground state rapidities:
\begin{eqnarray}
&\vert q \vert \leq 2k_F  &\implies \vert k^0(q) \vert \leq k^0(2k_F)=Q \nonumber \\
&\vert q \vert \leq k_{F\downarrow}  &\implies \vert \Lambda_{s1}^0(q) \vert \leq \Lambda_{s1}^0(k_{F\downarrow})=B \label{QB}
\end{eqnarray}
where actually $k^0(-2k_F)=-k^0(2k_F)=-Q$ and $ \Lambda_{s1}^0(-k_{F\downarrow})=-\Lambda_{s1}^0(k_{F\downarrow})=-B$, defining the quantities $Q$ and $B$. The occupancy configuration functions in momentum space will be important when deriving expressions for the pseudoparticle energy bands and phase shifts, amongst other quantities. Some well known limiting values of these entities include: 
\begin{eqnarray}
k(q) \rightarrow &q& \hspace{1.5cm} U/t \rightarrow \infty \nonumber \\
B \rightarrow &\infty& \hspace{1.5cm} ma \rightarrow 0 \nonumber \\
q_{s \nu}^0 \rightarrow &0& \hspace{1.5cm} ma \rightarrow 0 \hspace{1.0cm} (\nu \geq 2) \\
q_{c \nu}^0 \rightarrow &0& \hspace{1.5cm} na \rightarrow 1 \hspace{1.1cm} (\nu \geq 1) \nonumber
\end{eqnarray}
where the two last limits are particularly interesting: the actual bands vanish and thus the entire dynamics of the system is described by the $c0$ and the $s1$ bands. The limit $na \rightarrow 1$ is thus the limit where the Fourier momentum space of the $c\nu$ ($\nu \geq 1$) pseudoparticles disappears, just as the limit $ma \rightarrow 0$ is where the Fourier momentum space of the $s\nu$ ($\nu \geq 2$) pseudoparticles disappears. Later, we will see that due to this effect, the Fermi points of the $c0$ and the $s1$ bands, are in the $c\nu$ and $s\nu$ cases simulated by the limiting values of the effective Brillouin zone, equal to $q_{\alpha\nu}^0$, being the only momenta points to survive in these limits. Half filling is a limit that we will avoid in the dynamical theory of chapter \ref{dynamics}, however the zero magnetization limit will indeed interest us.

\newpage
\subsection{Energy and momentum deviations (Introduction)}
\label{pseudoparticle}

In order to formulate a dynamical theory, which is necessary in order to obtain information about the spectral properties of the model, we need to find expressions for the deviations (from the ground state) in energy, associated with pseudoparticle excitations in the many-body system, in terms of pseudoparticle energy bands and pseudoparticle number deviations. Even though the proper representation for the dynamical theory, introduced in section (\ref{pseudofermionI}), is different from that of the pseudoparticle representation, it is similar enough to allow for the subsequent study. We will see that the quantities derived here will lead us naturally to the new representation of section (\ref{pseudofermionI}). 

In the following, when we talk about "transitions" to an excited state, we mean that the occupancies of the Takahashi quantum numbers go from the ground state distribution (which is a densely packed distribution of numbers around a minimum energy point) to some other distribution. A particle-hole excitation of a pseudoparticle means that the occupied quantum number in the ground state, becomes unoccupied at the expense of some other quantum number that in turn becomes occupied. On the other hand, when adding pseudoparticles, or when removing pseudoparticles, there will be a net increase, or decrease, in the number of occupied numbers, leading to a new excited state configuration. We will only consider excited states that differ from the ground state in the occupancy of a {\it small} number of pseudoparticles, even though the formal requirement is much more general: the number of excited electrons must remain finite \cite{spectral3} \cite{Carmspec1}. This means that, when evaluating correlation functions, we only allow operators whose expressions involve a finite number of electronic creation and annihilation operators. 

The pseudoparticle number deviations, depending on the momentum, will be a key quantity is this analysis, since all other quantities will ultimately depend on the occupation and non-occupation of pseudoparticles for different momenta. To obtain the lowest order corrections in energy, we should consider "small" deviations from the ground state pseudoparticle number configurations. This will be enough since, as studies of following sections will confirm, an overwhelmingly large portion of the total spectral weight of the one-electron addition and removal processes are generated by excitation of only a few pseudoparticles. Since the Takahashi string hypothesis is only valid when $N_a \gg 1$, we often consider the continuous momentum limit such that $q_{j+1}-q_j = (2\pi / L) \rightarrow 0$. This means that we can stop talking about discrete quantum numbers altogether, and replace the sums by integrals in the Eqs (\ref{stringhyp}). We thus arrive to the "continuous momentum Takahashi equations": 
\begin{eqnarray}
k(q)=q&-&\frac 1 {\pi} \sum_{\nu=1}^{\infty} \int_{-q_{s\nu}^0}^{q_{s\nu}^0} d{q'} \ N_{s\nu}({q'}) \arctan\left (\frac {\sin k(q)a -\Lambda_{s\nu}({q'})} {\nu u} \right ) - \nonumber \\
&-& \frac 1 {\pi} \sum_{\nu=1}^{\infty} \int_{-q_{c\nu}^0}^{q_{c\nu}^0} d{q'} \ N_{c\nu}({q'}) \arctan\left (\frac {\sin k(q)a -\Lambda_{c\nu}({q'})} {\nu u} \right ) \nonumber \\
k_{c\nu}(q)=q&+&\frac 1 {\pi} \int_{-q_{c0}^0}^{q_{c0}^0} d{q'} \ N_{c0}({q'}) \arctan\left (\frac {\Lambda_{c\nu}(q)-\sin k({q'})a} {\nu u} \right ) + \nonumber \\
&+& \frac 1 {2\pi} \sum_{{\nu'}=1}^{\infty} \int_{-q_{c{\nu'}}^0}^{q_{c{\nu'}}^0} d{q'} \ N_{c{\nu'}}({q'}) \: \Theta_{\nu {\nu'}} \! \left (\frac {\Lambda_{c\nu}(q)-\Lambda_{c{\nu'}}({q'})} u \right ) \label{takacont} \\
0=q&-&\frac 1 {\pi} \int_{-q_{c0}^0}^{q_{c0}^0} d{q'} \  N_{c0}({q'}) \arctan\left (\frac {\Lambda_{s\nu}(q)-\sin k({q'})a} {\nu u} \right ) + \nonumber \\
&+& \frac 1 {2\pi} \sum_{{\nu'}=1}^{\infty} \int_{-q_{s{\nu'}}^0}^{q_{s{\nu'}}^0} d{q'} \ N_{s{\nu'}}({q'}) \: \Theta_{\nu {\nu'}} \! \left (\frac {\Lambda_{s\nu}(q)-\Lambda_{s{\nu'}}({q'})} u \right ) \nonumber
\end{eqnarray}
where $k_{c\nu}(q)= \frac 2 a \text{Re} \left\{ \arcsin(\Lambda_{c\nu}(q)-i\nu u) \right\}$ and $u=U/4t$. The function $\Theta_{\nu {\nu'}} (x)$ can be found in Eq. (\ref{thetafunc}). 

The energy and momentum can according to the Takahashi string hypothesis, Eq. (\ref{Etaka}), easily be re-expressed in the continuous limit:
\begin{eqnarray}
E&=&4t \frac L {2\pi} \left [ -\frac 1 2 \int_{-q_{c0}^0}^{q_{c0}^0} \right. dq \ N_{c0}(q) \cos k(q)a +
\sum_{\nu=1}^{\infty} \int_{-q_{c\nu}^0}^{q_{c\nu}^0} dq \ N_{c\nu}(q) \text{Re} \left \{ \sqrt{1- \left(\Lambda_{c\nu}(q)-i\nu u \right) ^2} \right \}- \nonumber \\
&&-\frac u 2 \int_{-q_{c0}^0}^{q_{c0}^0} dq \  N_{c0}(q) -u \int_{-q_{c\nu}^0}^{q_{c\nu}^0} dq \ \nu N_{c\nu}(q) \Bigg] \nonumber \\
P&=& \frac L {2\pi} \left [ \int_{-q_{c0}^0}^{q_{c0}^0} dq \ N_{c0}(q)k(q)+\sum_{\nu=1}^{\infty} \int_{-q_{c\nu}^0}^{q_{c\nu}^0} dq \ N_{c\nu}(q) \left( \frac {\pi} a-k_{c\nu}(q) \right) \right] + \frac {\pi} a M_{c,-\frac 1 2} \label{energy} \\
P&=&  \frac L {2\pi} \left [ \int_{-q_{c0}^0}^{q_{c0}^0} dq \ N_{c0}(q)q+\sum_{\nu=1}^{\infty} \int_{-q_{c\nu}^0}^{q_{c\nu}^0} dq \ N_{c\nu}(q) \left( \frac {\pi} a-q \right)+\sum_{\nu=1}^{\infty} \int_{-q_{s\nu}^0}^{q_{s\nu}^0} dq \  N_{s\nu}(q) q \right] +\frac {\pi} a M_{c -\frac 1 2} \nonumber
\end{eqnarray}

The two equivalent expressions for the momenta are obtained by using Eqs. (\ref{takacont}). The constant term  $\frac {\pi} a M_{c,-\frac 1 2}$ is obtained by using Eq. (\ref{takaspinupdown}) together with Eq. (\ref{Ptaka}), and shows the constant momentum value of the $-\frac 1 2$ holons.The occupancy functions $N_{\alpha \nu}(q)$ have a well defined value for each energy eigenstate. 

What we will do next is to allow a small deviation $\Delta N_{\alpha \nu} (q)$ to perturb the ground state occupancy configurations,
\begin{equation}
N_{\alpha \nu}(q)=N_{\alpha \nu}^0(q)+\Delta N_{\alpha \nu} (q) \hspace{0.99cm} \alpha\nu=c0,c\nu,s\nu
\label{deviations}
\end{equation}
where the $N_{\alpha \nu}^0(q)$ are given in Eq. (\ref{GSpp}). Eq. (\ref{deviations}) then describes an excited energy eigenstate. 

\subsection{Energy deviations and the $\Phi_{\alpha\nu,\alpha'\nu'}$ functions}
\label{sectPhi}

Since it is the deviations from the ground state that interests us, we would like to shape our theory so that all quantities are expressed relative to this ground state. This is because all dynamical quantities in the following will depend on $\Delta N_{\alpha \nu}$ and not on $N_{\alpha \nu}$. Therefore, in order to capture the relevant {\it dynamics} of the problem, we will formulate all quantities in a "normal ordered relative to the ground state" fashion. 

Our starting point is to express the energy of the excited state of the system as a ground state energy plus higher order corrections due to the introduction of the pseudoparticle deviations Eq. (\ref{deviations}):
\begin{equation}
E=\sum_{j=0}^{\infty} E^{(j)}
\label{energyexp}
\end{equation}

The energy deviation $E^{(1)}$ will be expressed as proportional to some pseudoparticle energy band relative to the ground state, multiplied by the corresponding pseudoparticle occupancy first order deviation. Therefore, we define the pseudoparticle energy bands $\epsilon_{\alpha \nu}^0 (q)$ such that
\begin{equation}
E^{(1)} = \frac {L} {2\pi} \sum_{\alpha=c,s} \sum_{\nu=\delta_{\alpha,s}}^{\infty} \int_{-q_{\alpha \nu}^0}^{q_{\alpha \nu}^0} dq \  \epsilon_{\alpha \nu}^0 (q) \Delta N_{\alpha \nu} (q)
\label{energydev}
\end{equation}

The relative to the ground state pseudoparticle energy bands $\epsilon^0_{\alpha \nu}(q)$ are defined as the functional derivative of the energy with respect to the occupancy configuration deviation:
\begin{equation}
\epsilon^0_{\alpha \nu} (q) = \frac {\delta E^{(1)}} {\delta \Delta N_{\alpha \nu} (q)}
\label{funcder}
\end{equation}

The second term in the expansion of Eq. (\ref{energyexp}) would contain bilinear combinations of the $\Delta N_{\alpha \nu}$'s and would correspond to residual energy interactions between the different $\alpha \nu$ pseudoparticles.

Now, a nonzero deviation in the $\alpha \nu$ occupancy configuration yields, as can be seen in Eq. (\ref{takacont}), a corresponding deviation in the rapidity functions:
\begin{equation}
N_{\alpha \nu}(q)=N_{\alpha \nu}^0(q)+\Delta N_{\alpha \nu} (q) \implies \left \{
\begin{array} {c}
k(q)=k^0(q)+\Delta k(q) \\
\Lambda_{c\nu}(q)=\Lambda_{c\nu}^0(q)+\Delta \Lambda_{c\nu}(q) \\
\Lambda_{s\nu}(q)=\Lambda_{s\nu}^0(q)+\Delta \Lambda_{s\nu}(q)
\end{array}
\right.
\label{rapiditydev}
\end{equation}
where $\alpha \nu = c0,c\nu,s\nu$. The {\it ground state rapidity functions} come from the solution of the Takahashi equations for the particular case of having the occupied quantum numbers in their ground state configurations, i.e. such that the momenta obeys Eq. (\ref{GSpp}). They are obtainable by solving (\ref{takacont}) with $N_{\alpha \nu}(q)=N_{\alpha \nu}^0(q)$. Before deriving the energy bands, however, we will investigate the continuous momentum Takahashi equations a little bit further, in order to derive relationships between a new set of functions (denoted $\Phi_{\alpha \nu, \alpha' \nu'}(q,q')$ below) that will be used to express the energy bands in an elegant way. 

By introducing the pseudoparticle occupational deviations together with the rapidity deviations (both of them are given in (\ref{rapiditydev})), together with Eq. (\ref{GSpp}), into the Takahashi equations (\ref{takacont}), we obtain equations separable order by order. Focusing on the zeroth order contributions (the ground state) and the first order deviations, we note that as the algebra turns out, we can simplify matters a lot by applying $d/dq$ to the zeroth order equations and inserting them into the first order equations. To simplify matters even further, we define a new quantity $\Delta Q_{\alpha \nu}(q)$ according to:
\begin{eqnarray}
\Delta k(q)&=&\frac {dk^0(q)} {dq} \Delta Q_{c0}(q)  \nonumber \\
\Delta \Lambda_{\alpha \nu}(q)&=& \frac {d\Lambda_{\alpha \nu}^0(q)} {dq} \Delta Q_{\alpha \nu}(q) \hspace{0.9cm} \alpha=c,s \label{phaseshift}
\end{eqnarray}
which allows the first order contributions of the Takahashi equations be written as:
\begin{eqnarray}
\Delta Q_{c0}(q) &=& \frac 1 {\pi u} \int_{-B}^B d\Lambda \ \frac {\Delta \bar{Q}_{s1} (\Lambda)} {1+\left( \frac {\sin k^0(q)a -\Lambda} u \right)^2}+\sum_{\alpha'=c,s} \sum_{\nu'=\delta_{\alpha',s}} \int_{-q_{\alpha'  \nu'}^0}^{q_{\alpha' \nu'}^0} dq' \ z_{c0,\alpha'  \nu'}(q,q') \Delta N_{\alpha' \nu'} (q') \nonumber \\
\  \nonumber \\
\Delta Q_{c\nu}(q) &=& - \frac a {\pi \nu u} \int_{-Q}^Q dk \  \frac {\Delta \bar{Q}_{c0} (k) \cos ka} {1+\left( \frac {\Lambda_{c\nu}^0 (q) - \sin ka} {\nu u} \right)^2}+\sum_{\alpha'=c,s} \sum_{\nu'=\delta_{\alpha',s}} \int_{-q_{\alpha'  \nu'}^0}^{q_{\alpha' \nu'}^0} dq' \ z_{c\nu,\alpha' \nu'}(q,q') \Delta N_{\alpha' \nu'} (q') \nonumber \nonumber \\
\Delta Q_{s\nu}(q) &=& \frac a {\pi \nu u} \int_{-Q}^Q dk \ \frac {\Delta \bar{Q}_{c0} (k) \cos ka} {1+\left( \frac {\Lambda_{s\nu}^0 (q) - \sin ka} {\nu u} \right)^2} -\frac 1 {2\pi u} \int_{-B}^B d\Lambda \ \Delta \bar{Q}_{s1} (\Lambda) {\Theta'}_{\nu 1} \left(  \frac {\Lambda_{s\nu}^0(q)-\Lambda} u  \right) +  \nonumber \\
&&+\sum_{\alpha'=c,s} \sum_{\nu'=\delta_{\alpha',s}} \int_{-q_{\alpha'  \nu'}^0}^{q_{\alpha' \nu'}^0} dq' \ z_{s\nu,\alpha' \nu'}(q,q') \Delta N_{\alpha' \nu'} (q') \label{Qdev}
\end{eqnarray}
where we have introduced the function $\Delta \bar{Q}_{\alpha \nu}(X)=\Delta \bar{Q}_{\alpha \nu}(X(q))=\Delta Q_{\alpha \nu}(q)$ and the functions $z_{\alpha \nu, \alpha' \nu'}(q,q')$ which can be obtained explicitly and are given below. $\Theta' (x) = d\Theta (x) / dx$ from Eq. (\ref{thetafunc}) and $Q$ and $B$ are defined by Eq. (\ref{QB}). The equations (\ref{Qdev}) express relationships between the rapidity deviations, since
\begin{eqnarray}
\Delta Q_{c0}(q)&=&\frac {\Delta k (q)} {\left[   d k^0(q) /dq   \right]} \nonumber \\
\Delta Q_{\alpha \nu}(q)&=&\frac {\Delta \Lambda_{\alpha \nu} (q)} {\left[   d\Lambda_{\alpha \nu}^0(q) /dq   \right]} \label{defQ}
\end{eqnarray}
by definition. Examining the mathematical form of the relationships between the different $\Delta Q_{\alpha \nu}(q)$'s, we see that there is a possibility of expressing them as linear combinations of the pseudoparticle deviations, 
\begin{equation}
\Delta Q_{\alpha \nu} (q) = \sum_{\alpha' \nu'} \int_{-q_{\alpha' \nu'}^0}^{q_{\alpha' \nu'}^0} dq' \ \Phi_{\alpha \nu,\alpha' \nu'}(q,q') \Delta N_{\alpha' \nu'} (q')
\label{Qphase1}
\end{equation}
or equivalently,
\begin{equation}
\Phi_{\alpha \nu,\alpha' \nu'}(q,q') = \frac {\delta \Delta Q_{\alpha \nu}(q)} {\delta \Delta N_{\alpha' \nu'} (q')} \label{Qphase2}
\end{equation}
if we take the functional derivative with respect to these deviations, i.e. applying $\delta / \delta \Delta N_{\alpha' \nu'}$ to the equations (\ref{Qdev}). By doing this, and by changing variables according to
\begin{equation}
\left.
\begin{array} {c}
k(q)a \rightarrow ka \rightarrow \frac {\sin ka} u = r \text{ or } r' \\
\Lambda_{\alpha \nu} (q) \rightarrow \Lambda_{\alpha \nu} \rightarrow \frac {\Lambda} u = r \text{ or } r'
\end{array}
\right \} \implies \Phi_{\alpha \nu ,\alpha' \nu'}(q,q') \rightarrow \bar{\Phi}_{\alpha \nu ,\alpha' \nu'}(r,r')\end{equation}
where the indices of $\Phi$ always indicate what $\alpha \nu$ branch the variable belongs to (the first variable is always unprimed and the second always primed), we find after some algebra:
\begin{eqnarray}
\bar{\Phi}_{c0 ,\alpha' \nu'}(r,r')&=& \frac 1 {\pi} \int_{-B/u}^{B/u} dr'' \ \frac {\bar{\Phi}_{s1,\alpha'\nu'}(r'',r')} {1+(r-r'')^2} + \bar{z}_{c0,\alpha' \nu'}(r,r') \nonumber \\
\bar{\Phi}_{c\nu ,\alpha' \nu'}(r,r')&=& - \frac 1 {\pi \nu} \int_{-\sin Qa/u}^{\sin Qa/u} dr'' \ \frac {\bar{\Phi}_{c0,\alpha'\nu'}(r'',r')} {1+(\frac {r-r''} {\nu})^2} + \bar{z}_{c\nu,\alpha' \nu'}(r,r') \nonumber \\
\bar{\Phi}_{s\nu ,\alpha' \nu'}(r,r')&=& \frac 1 {\pi \nu} \int_{-\sin Qa/u}^{\sin Qa/u} dr'' \ \frac {\bar{\Phi}_{c0,\alpha'\nu'}(r'',r')} {1+(\frac {r-r''} {\nu})^2} - \label{Phi} \\
&&- \frac 1 {2\pi} \int_{-B/u}^{B/u} dr'' \  \bar{\Phi}_{s1,\alpha'\nu'}(r'',r') \Theta'_{\nu 1} (r-r'') + \bar{z}_{s\nu,\alpha' \nu'}(r,r') \hspace{1.0cm} \nonumber
\end{eqnarray}
where the auxiliary functions $\bar{z}_{\alpha \nu,\alpha' \nu'}$ are defined by:

$
\begin{array}{lllll}
 & & \\
\bar{z}_{c0,c0}(r,r')=0  & \bar{z}_{c0,c\nu}(r,r')=-\phi_{\nu}(r-r') & \bar{z}_{c0,s\nu}(r,r')=-\phi_{\nu}(r-r')  \\
\bar{z}_{c\nu,c0}(r,r')=\phi_{\nu}(r-r')  & \bar{z}_{c\nu,c\nu'}(r,r')=\Theta_{\nu \nu'} (r-r') / 2\pi \ \  & \bar{z}_{c\nu,s\nu'}(r,r')=0  \\
\bar{z}_{s\nu,c0}(r,r')=-\phi_{\nu}(r-r') \ \  & \bar{z}_{c\nu,c\nu'}(r,r')=0  &  \bar{z}_{s\nu,s\nu'}(r,r')=\Theta_{\nu \nu'} (r-r') / 2\pi \\
 & &
\end{array}
$
Here $\phi_{\nu}(x)=\arctan(x/\nu)/\pi$ and $\bar{z}_{\alpha \nu,\alpha' \nu'}(r,r')=z_{\alpha \nu,\alpha' \nu'}(q,q')$. 

We note that the functions defined in (\ref{Phi}) obey the following symmetry: $\bar{\Phi}_{\alpha\nu ,\alpha' \nu'}(r,r')=-\bar{\Phi}_{\alpha\nu ,\alpha' \nu'}(-r,-r')$, which together with the oddness of the ground state rapidity functions implies that $\Phi_{\alpha\nu ,\alpha' \nu'}(q,q')=-\Phi_{\alpha\nu ,\alpha' \nu'}(-q,-q')$. The same line of thought can be applied to the energy, by use of the energy expression (\ref{energy}), introducing the pseudoparticle deviations and separating contributions order by order. The newly derived $\bar{\Phi}_{\alpha \nu ,\alpha' \nu'}(r,r')$ enters the calculation via $\Delta k (q)$, and we can by comparing the different resulting first order terms with Eq. (\ref{energydev}), obtain the pseudoparticle energy bands:
\begin{eqnarray}
\epsilon^0_{c0}(q)&=&-2t\cos k^0(q)a +2ta \int_{-Q}^Q dk \ \sin k \ \tilde{\Phi}_{c0,c0}(k,k^0(q)) -\frac U 2\nonumber \\
\epsilon^0_{c\nu}(q)&=&4t \text{Re} \left \{ \sqrt{1-(\Lambda_{c\nu}^0(q)-i\nu u)^2} \right \} + 2ta \int_{-Q}^Q dk \  \sin ka \  \tilde{\Phi}_{c0,c\nu}(k,\Lambda_{c\nu}^0(q)) - \nu U \nonumber \\
\epsilon^0_{s\nu}(q)&=& 2ta \int_{-Q}^Q dk \ \sin ka \  \tilde{\Phi}_{c0,s\nu}(k,\Lambda_{c\nu}^0(q)) \label{energybands}
\end{eqnarray}
where we have used an alternative to $\Phi_{\alpha \nu ,\alpha' \nu'}(q,q') = \tilde{\Phi}_{\alpha \nu,\alpha'\nu'}(k^0 (q),\Lambda_{\alpha\nu}^0(q))$. As a bi-product of this calculation, we have that the zero order energy term, i.e. the ground state energy, can be expressed as:
\begin{equation}
E^{(0)}=-2t \frac L {2\pi} \int_{-2k_F}^{2k_F} dq \ \cos k^0(q)a \label{GSenergy}
\end{equation}
where the ground state rapidity function $k^0(q)$ satisfies the first equality of Eq. (\ref{takacont}), with $N_{\alpha\nu}(q')=N_{\alpha\nu}^0(q')$ (the ground state configuration).

\begin{figure}
\begin{center}
\includegraphics[width=9cm,height=6cm]{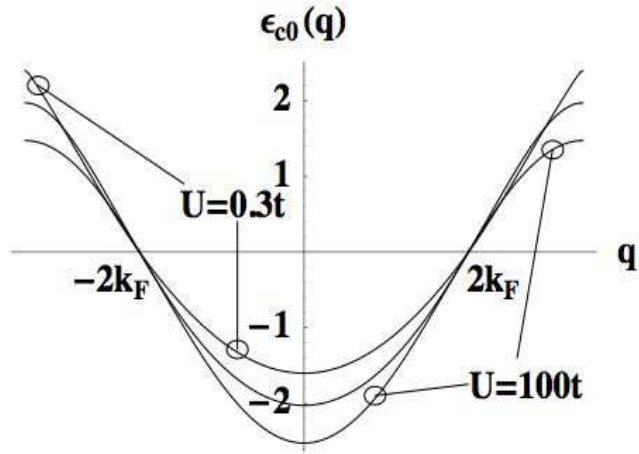}
\caption{\label{figEc0} Energy dispersion relation for the $\alpha\nu=c0$ pseudoparticle, for $(U/t)=0.3,\ 4.9,\text{ and } 100$ and for $n=0.59$ and $m \rightarrow 0$, in units of $t$. The $(U/t)=4.9$ curve is visible in between the other two curves. Note that the dispersion for $-2k_F < q < 2k_F$ becomes successively deeper for $(U/t) \rightarrow \infty$, however always keeping the bandwidth constant at $4t$. \vspace{0.5cm}}
\end{center}
\end{figure}

\begin{figure}
\begin{center}
\includegraphics[width=9cm,height=6cm]{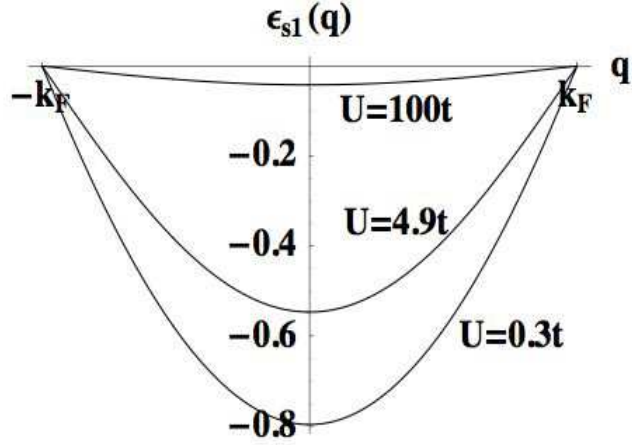}
\caption{\label{figEs1} Energy dispersion relation for the $\alpha\nu=s1$ pseudoparticle, for $(U/t)=0.3,\ 4.9,\text{ and } 100$ and for $n=0.59$ and $m \rightarrow 0$, in units of $t$. Note that the energy bandwidth is a decreasing function of ($U/t$) and that $s1$ pseudoparticle becomes dispersionless in the $(U/t) \rightarrow \infty$ limit. \vspace{0.5cm}}
\end{center}
\end{figure}

\begin{figure}
\begin{center}
\includegraphics[width=9cm,height=6cm]{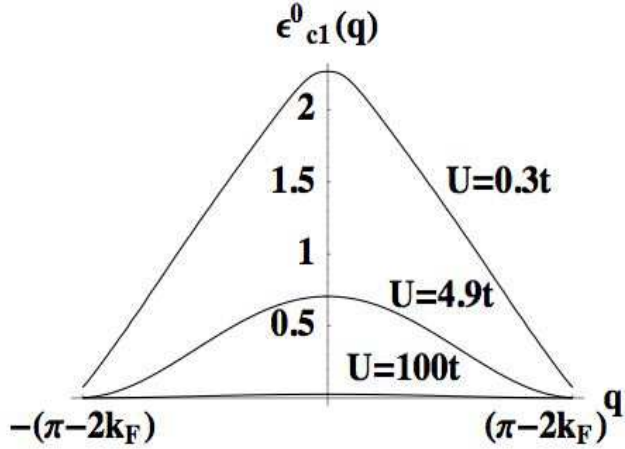}
\caption{\label{figEc1} Energy dispersion relation for the $\alpha\nu=c1$ pseudoparticle, for $(U/t)=0.3,\ 4.9,\text{ and } 100$ and for $n=0.59$ and $m \rightarrow 0$, in units of $t$. The energy bandwidth is a decreasing function of ($U/t$) and the $c1$ pseudoparticle becomes dispersionless in the $(U/t) \rightarrow \infty$ limit, and is a horizontal line along the zero energy level in the figure. \vspace{0.5cm}}
\end{center}
\end{figure}

\begin{figure}
\subfigure{\includegraphics[width=7cm,height=7cm]{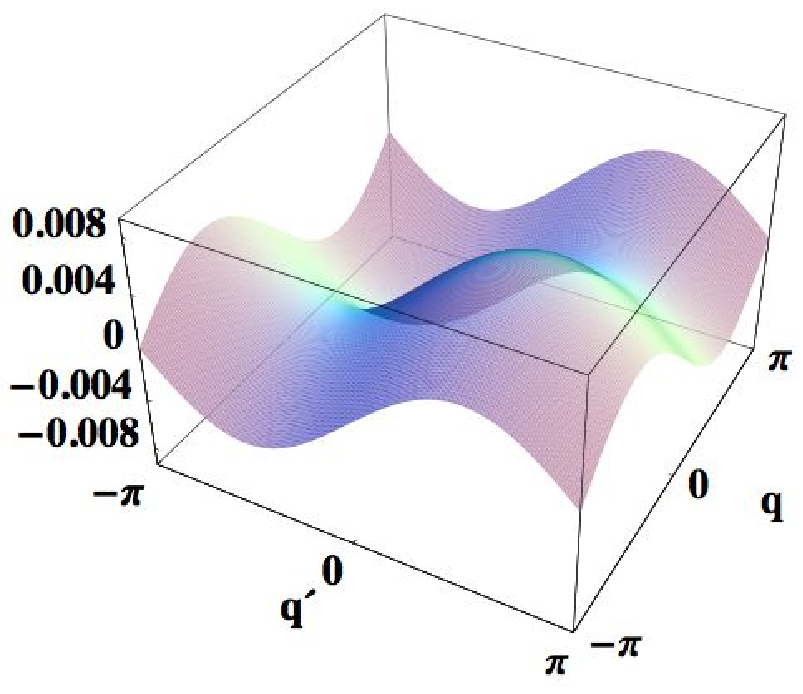}}
\subfigure{\includegraphics[width=7cm,height=7cm]{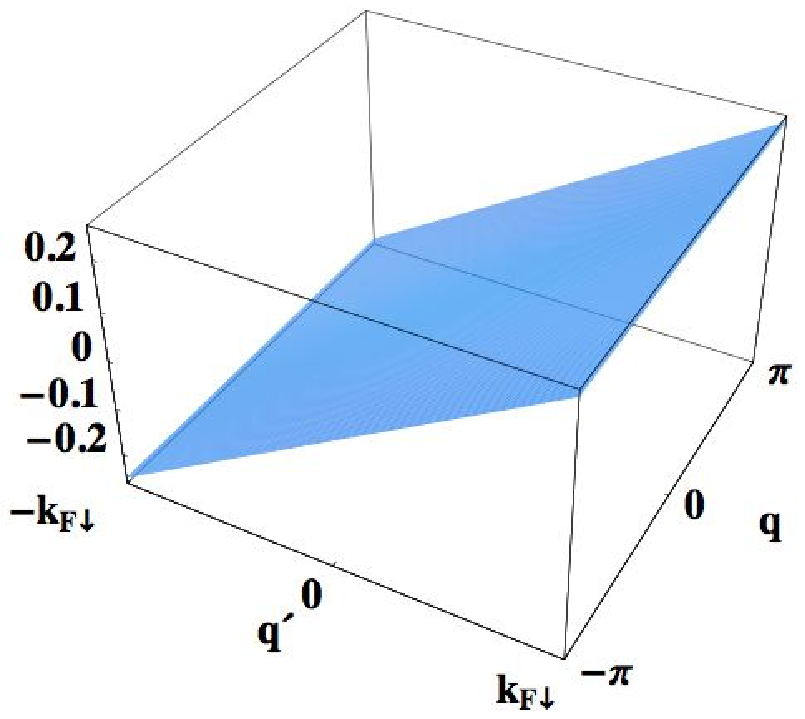}}
\subfigure{\includegraphics[width=7cm,height=7cm]{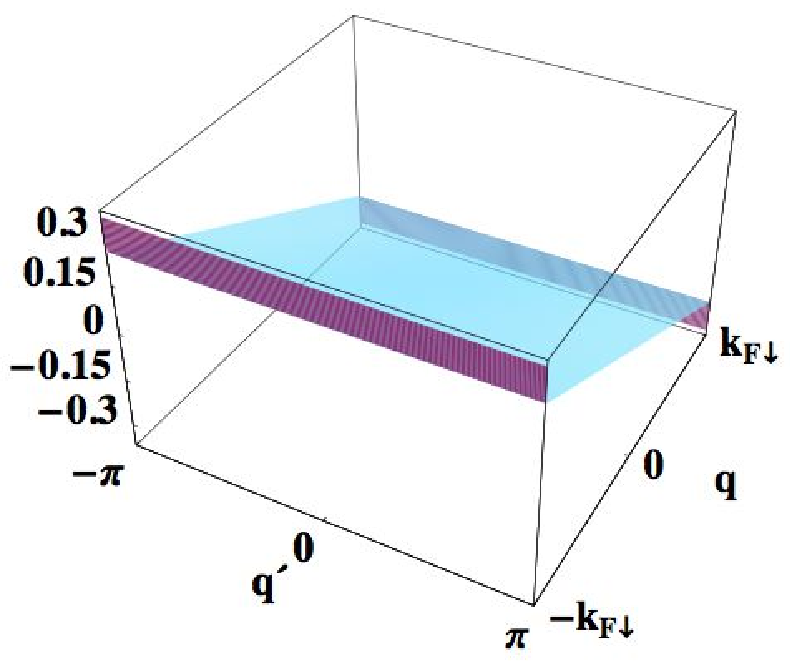}}
\subfigure{\includegraphics[width=7cm,height=7cm]{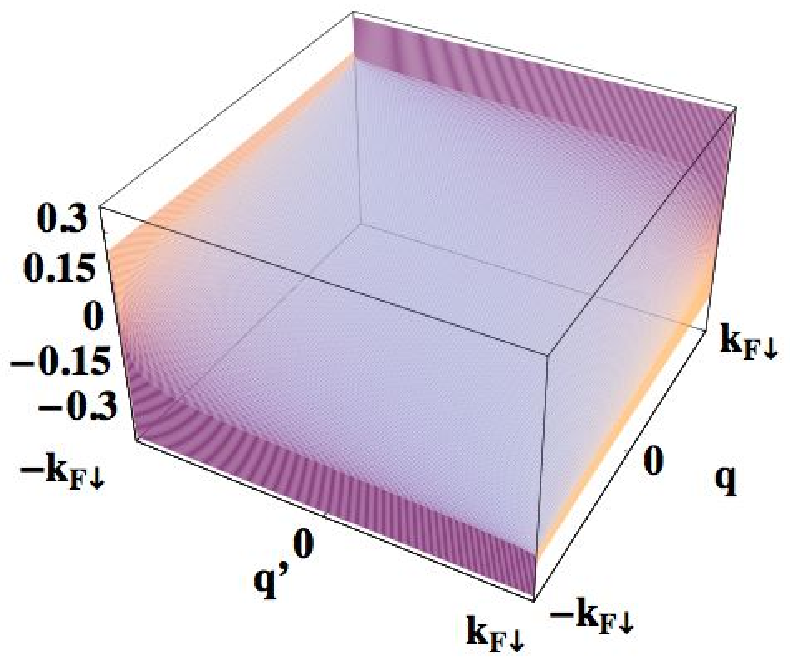}}
\caption{\label{PhiU100}The functions $\Phi_{\alpha\nu,\alpha'\nu'}(q,q')$ for $(U/t)=100$, $n=0.59$ and $m \rightarrow 0$, arranged according to: $\Phi_{c0,c0}(q,q')$ (upper left), $\Phi_{c0,s1}(q,q')$ (upper right), $\Phi_{s1,c0}(q,q')$ (lower left) and $\Phi_{s1,s1}(q,q')$ (lower right).}
\end{figure}

\begin{figure}
\subfigure{\includegraphics[width=7cm,height=7cm]{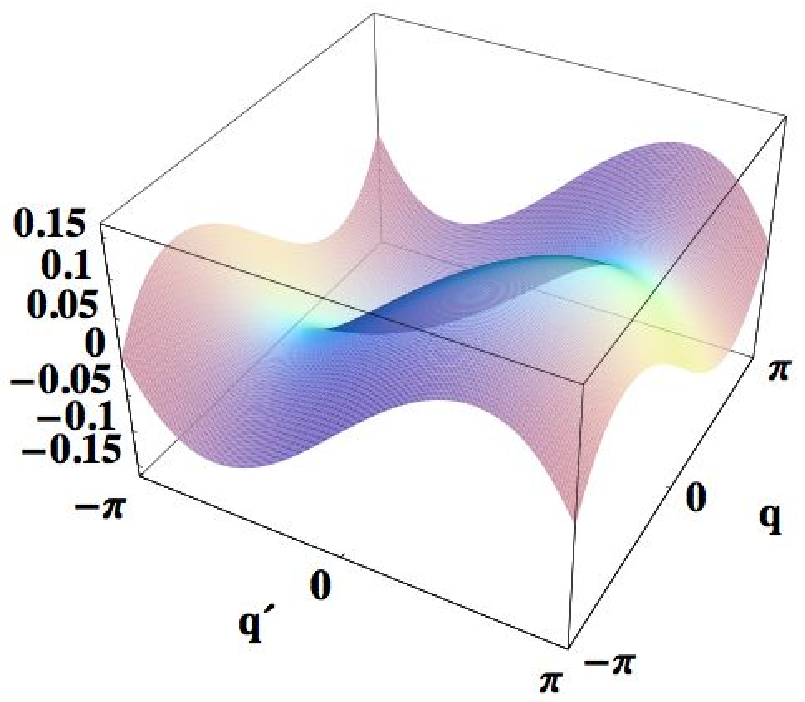}}
\subfigure{\includegraphics[width=7cm,height=7cm]{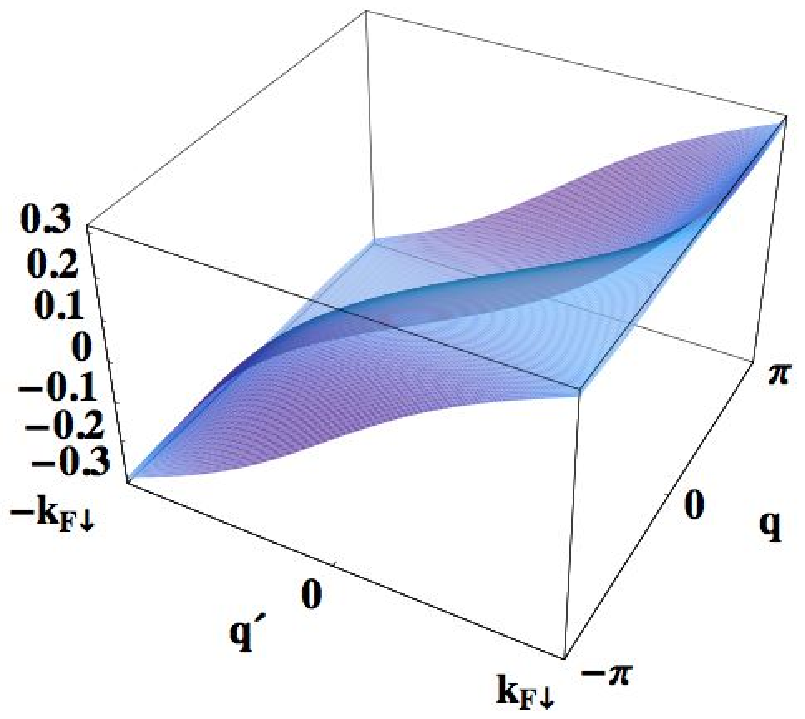}}
\subfigure{\includegraphics[width=7cm,height=7cm]{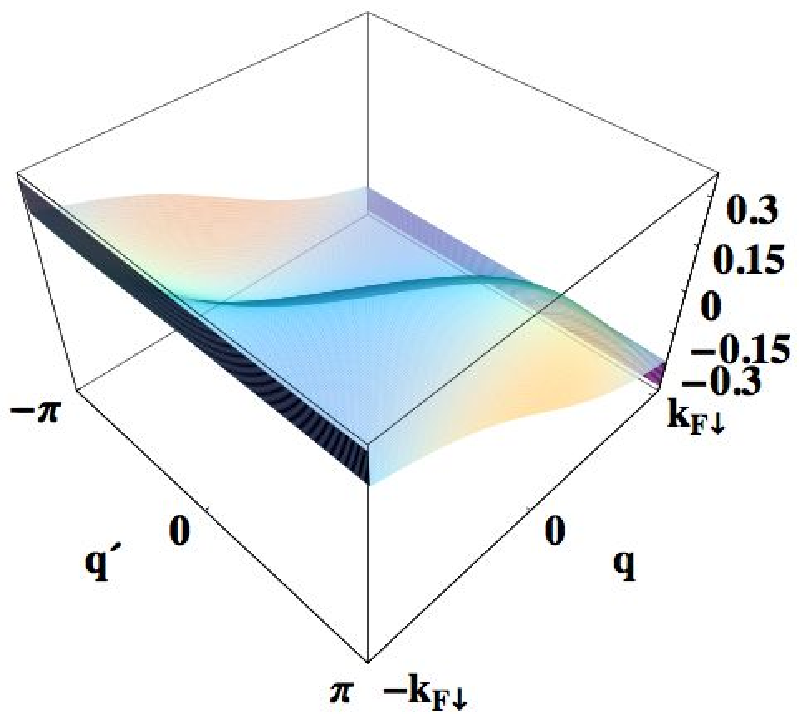}}
\subfigure{\includegraphics[width=7cm,height=7cm]{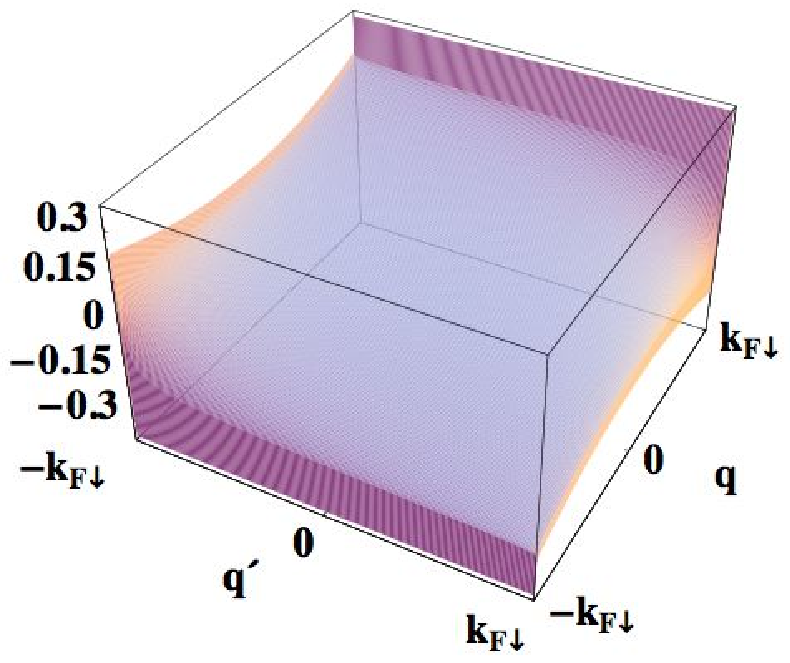}}
\caption{\label{PhiU4p9}The functions $\Phi_{\alpha\nu,\alpha'\nu'}(q,q')$ for $(U/t)=4.9$, $n=0.59$ and $m \rightarrow 0$, arranged according to: $\Phi_{c0,c0}(q,q')$ (upper left), $\Phi_{c0,s1}(q,q')$ (upper right), $\Phi_{s1,c0}(q,q')$ (lower left) and $\Phi_{s1,s1}(q,q')$ (lower right).}
\end{figure}

\begin{figure}
\subfigure{\includegraphics[width=7cm,height=7cm]{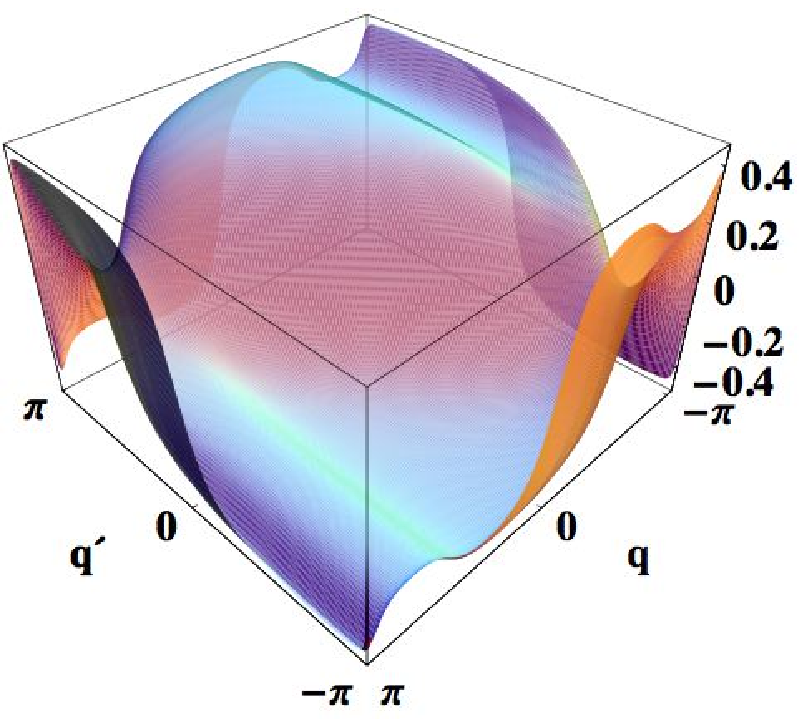}}
\subfigure{\includegraphics[width=7cm,height=7cm]{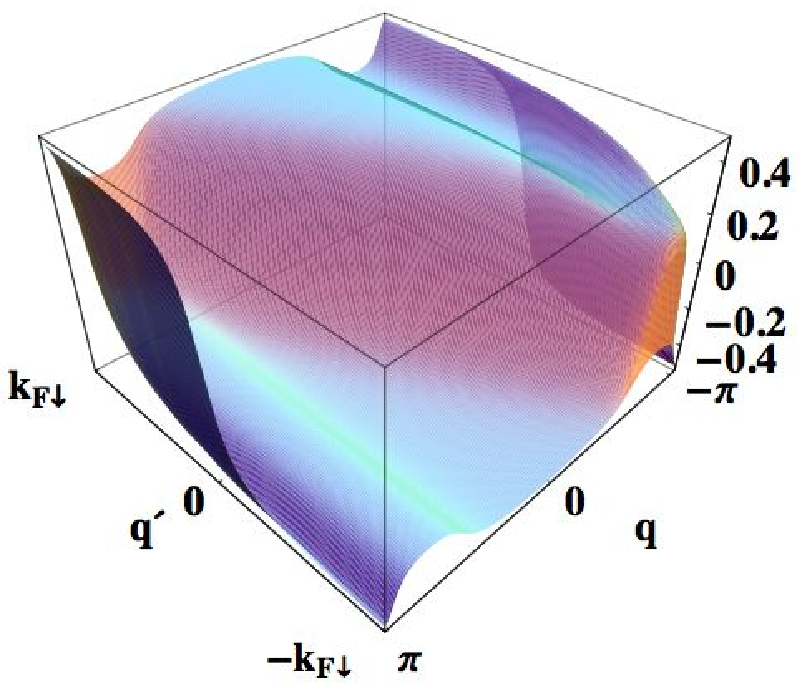}}
\subfigure{\includegraphics[width=7cm,height=7cm]{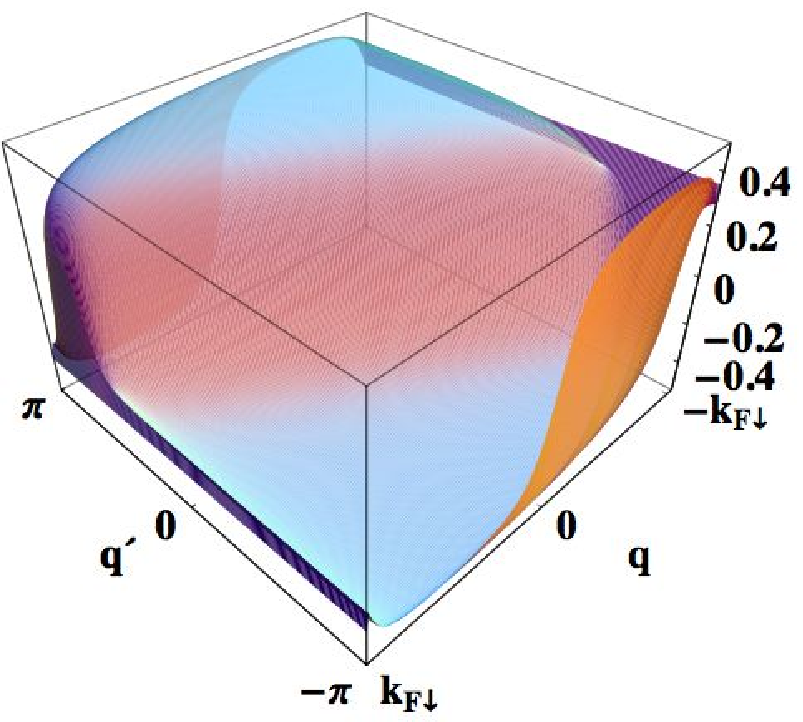}}
\subfigure{\includegraphics[width=7cm,height=7cm]{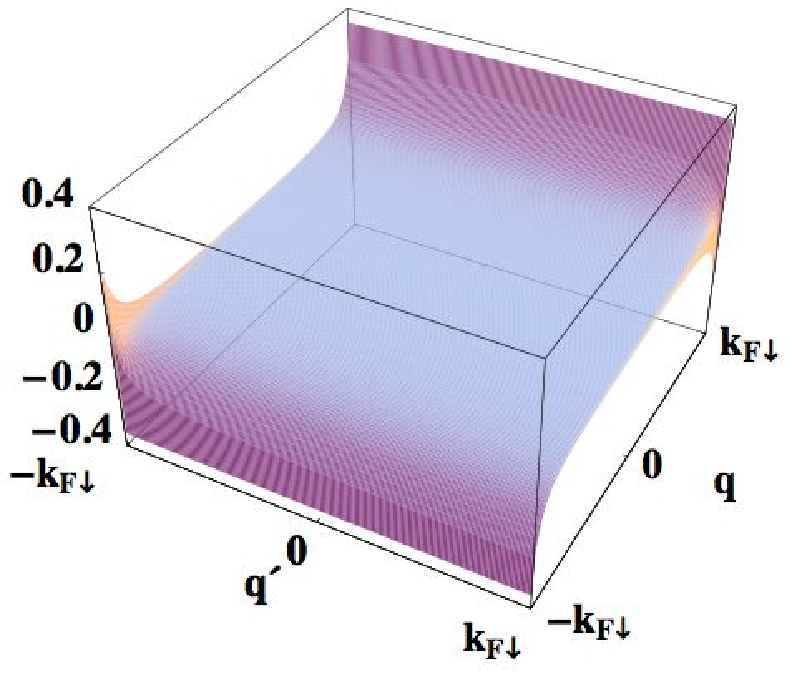}}
\caption{\label{PhiU0p3}The functions $\Phi_{\alpha\nu,\alpha'\nu'}(q,q')$ for $(U/t)=0.3$, $n=0.59$ and $m \rightarrow 0$, arranged according to: $\Phi_{c0,c0}(q,q')$ (upper left), $\Phi_{c0,s1}(q,q')$ (upper right), $\Phi_{s1,c0}(q,q')$ (lower left) and $\Phi_{s1,s1}(q,q')$ (lower right).}
\end{figure}

We want, according to convention, to fix the reference levels of these bands so that the $c0$ and the $s1$ bands gives zero at their respective Fermi points. The $c\nu$ ($\nu \geq 1$) and the $s\nu$ ($\nu \geq 2$) bands will then have their reference levels adjusted according to their $\nu=0$ ($\alpha=c$) and $\nu=1$ ($\alpha=s$) counterparts. This adjustment is a consequence of breaking one or both of the $SU(2)$ symmetries of the model. When this happens, the energies will depend on the chemical potential and the magnetic field strength, respectively. For $c\nu$ ($\nu \geq 1$) this energy difference is proportional to the number of doubly occupied rotated electron sites belonging to the $c\nu$ pseudoparticle, whilst for $s\nu$ ($\nu \geq 2$) it is proportional to the number of $\downarrow$-spin singly occupied rotated electron sites belonging to $s\nu$. Hence both of these contributions are equal to $\nu$. We define thus the following energy bands for the pseudoparticles:
\begin{eqnarray}
\epsilon_{c0}(q)&=&\epsilon_{c0}^0(q) - \epsilon_{c0}^0(2k_F) \nonumber \\
\epsilon_{s1}(q)&=&\epsilon_{s1}^0(q) - \epsilon_{s1}^0(k_{F\downarrow}) \nonumber \\
\epsilon_{c\nu}(q)&=&\epsilon_{c\nu}^0(q) + \mu_c \nu \label{rele} \\
\epsilon_{s\nu}(q)&=&\epsilon_{s\nu}^0(q) + \mu_s \nu \nonumber
\end{eqnarray}
where $\mu_c=2\mu$ and $\mu_s=2\mu_0 h$ were defined in section (\ref{so4sect}). The energy bands $\epsilon_{c0}(q)$, $\epsilon_{s1}(q)$ and $\epsilon_{\alpha\nu}^0(q)$ (for $\alpha\nu \neq c0,s1$) are even functions of $q$, and are such that:
\begin{equation}
\epsilon_{c0}(2k_F)=\epsilon_{s1}(k_{F\downarrow})=\epsilon^0_{c\nu}(\pi-2k_F)=\epsilon^0_{s\nu'}(k_{F\uparrow}-k_{F\downarrow})=0
\end{equation}
where $\nu \geq 1$ and $\nu' \geq 2$. The mathematical exercise of deriving the $\Phi_{\alpha \nu ,\alpha' \nu'}(q,q')$ functions payed off, judging by the beauty of the derived energy band expressions. We use Eq. (\ref{energybands}) to numerically obtain the dispersion relations shown in Figs. (\ref{figEc0}), (\ref{figEs1}) and (\ref{figEc1}), for a "very large", "intermediate" and "very small" value of ($U/t$). The filling dependence on these relations is discussed separately in section (\ref{densexp}) (for a further discussion on these energy bands, see Ref. \cite{CarmeloPedro}). In order to obtain these dispersions, we need (for example) the functions $\Phi_{c0,c0}$, $\Phi_{c0,s1}$, $\Phi_{s1,c0}$ and $\Phi_{s1,s1}$. These are plotted in Figs. (\ref{PhiU100}), (\ref{PhiU4p9}) and (\ref{PhiU0p3}), respectively, for three different values of ($U/t$), and are further discussed in Ref. \cite{CarmBoziPedro}.

This analysis can be carried to higher orders, where the terms of order $j$ includes $j$ factors of different $\Delta N_{\alpha \nu (q)}$'s. By keeping 2nd order terms, including products of type $\left[\Delta N_{\alpha \nu}(q) \Delta N_{\alpha' \nu'}(q')\right]$, we can derive the residual energy interaction term $E^{(2)}$ between the pseudoparticles. That this term is finite shows that the pseudoparticles have residual energy interactions. By following the same general scheme as for the first order (presented here above), we arrive to
\begin{eqnarray}
E^{(2)}&=&\frac 1 L \sum_{\alpha=c,s} \sum_{\nu=\delta_{\nu,s}}^{\infty} \sum_{j=1}^{N_{\alpha\nu}^0} v_{\alpha \nu} (q_j) \Delta Q_{\alpha \nu} (q_j) \Delta N_{\alpha \nu} (q_j) + \nonumber \\
&+&\frac L {4 \pi} \sum_{\alpha \nu = c0,s1} v_{ \alpha \nu} \sum_{j=\pm1} \left[ \Delta Q_{\alpha \nu} (j q_{F \alpha \nu})  \right]^2 \label{residualE}
\end{eqnarray}
after a considerable amount of algebra. Here we have introduced the pseudoparticle {\it group velocity} $v_{\alpha \nu}(q)$ and the pseudoparticle {\it Fermi velocity} $v_{\alpha \nu}$, defined by:
\begin{equation}
v_{\alpha \nu} (q) = \frac {d \epsilon^0_{\alpha \nu} (q)} {dq} \hspace{1.3cm} v_{ \alpha \nu}=v_{\alpha \nu} (q_{F\alpha \nu}) \label{Vvel}
\end{equation}
That $E^{(2)}$ only contains terms of order $\left[\Delta N_{\alpha \nu}(q) \Delta N_{\alpha' \nu'}(q')\right]$ can be seen by inserting Eq. (\ref{Qphase1}) into Eq. (\ref{residualE}). We note that the last term of $E^{(2)}$ is actually of order ($1/L$) due to the square of $\Delta Q_{\alpha\nu}(j q_{F \alpha \nu})$. 

\newpage
\subsection{Momentum deviations and inverse rapidities}

According to the two equivalent expressions for the value of the total momenta $P$ given in Eq. (\ref{energy}), we can in the same way as with the energies introduce the pseudoparticle number deviations into these two expressions. Since we have one expression for $P$ involving only the momenta ($P_{mom}$) and another involving only the rapidities ($P_{rap}$), we hope that by equalling the two expressions, it will be possible to deduce some interesting relationships between the pseudoparticle momentum and the ground state rapidities. In the same "normal ordered relative to the ground state" spirit as before, let now
\begin{eqnarray}
P=\sum_{j=0}^{\infty} P^{(j)} \hspace{1.0cm} \text{where} \ \ \ P^{(j)}=P_{mom}^{(j)}=P_{rap}^{(j)}
\end{eqnarray}
i.e. we will equal the terms order by order in the deviations. By reasons that will become obvious in section (\ref{pseudofermionI}), we will focus on the $j=0$ and the $j=1$ terms. By using Eqs. (\ref{defQ}) and (\ref{Qphase1}), we arrive to the following relationship:
\begin{eqnarray}
\int_{-2k_F}^{2k_F} dq \ \Delta k(q) &=& \int_{-q_{co}^0}^{q_{c0}^0} dq \ \Delta N_{c0} (q) \int_{-Q}^Q dk \  \tilde{\Phi}_{c0,c0} (k,k(q))  + \\
&+&\sum_{\alpha=c,s} \sum_{\nu=\delta_{\alpha,s}}^{\infty} \int_{-q_{\alpha\nu}^0}^{q_{\alpha\nu}^0} dq \ \Delta N_{\alpha \nu} (q) \int_{-Q}^Q dk \  \tilde{\Phi}_{c0,\alpha\nu} (k,\Lambda_{\alpha\nu}(q)) \nonumber
\end{eqnarray}

We find that by comparing the first order deviations with each other, the following relationships between the pseudoparticle momenta and their corresponding rapidities can be derived:
\begin{eqnarray}
q&=&k^0(q)+\int_{-Q}^Q dk \ \tilde{\Phi}_{c0,c0}(k,k^0(q)) \nonumber \\
q&=&k_{c\nu}^0(q)-\int_{-Q}^Q dk \ \tilde{\Phi}_{c0,c\nu}(k,\Lambda_{c\nu}^0(q)) \\
q&=&\int_{-Q}^Q dk \ \tilde{\Phi}_{c0,s\nu}(k,\Lambda_{s\nu}^0(q))
\end{eqnarray}
From these relationships it is possible, at least in principle, to obtain the explicit dependencies of the ground state rapidities on the momenta. However, there is more to be done than just to make an abstract mathematical claim. By defining the {\it inverse} of the ground state rapidities as functions of momenta according to
\begin{eqnarray}
q_{c0}(k)&=&k+\int_{-Q}^Q dk' \ \tilde{\Phi}_{c0,c0}(k',k) \nonumber \\
q_{s\nu}(\Lambda)&=&\int_{-Q}^Q dk' \ \tilde{\Phi}_{c0,s\nu}(k',\Lambda) \\
q_{c\nu}(\Lambda)&=& \frac 2 a \Re \left\{ \arcsin \left( \Lambda -i\nu u \right) \right\}-\int_{-Q}^Q dk' \ \tilde{\Phi}_{c0,c\nu}(k',\Lambda) \nonumber
\end{eqnarray}
we can by taking derivatives define the following {\it density functions}:
\begin{eqnarray}
2\pi\tilde{\rho}_{c0} (k)&=&\frac {dq_{c0}(k)} {dk}=1+\int_{-Q}^Q dk' \ \frac d {dk} \tilde{\Phi}_{c0,c0}(k',k) \nonumber \\
2\pi\tilde{\sigma}_{c\nu}(\Lambda)&=& \frac {dq_{s\nu}(\Lambda)} {d\Lambda}=\int_{-Q}^Q dk' \ \frac d {d\Lambda} \tilde{\Phi}_{c0,s\nu}(k',\Lambda) \label{invrap} \\
2\pi\tilde{\sigma}_{c\nu}(\Lambda)&=& \frac {dq_{c\nu}(\Lambda)} {d\Lambda}=\frac 2 a \text{Re} \left\{ \frac 1 {\sqrt{1-\left( \Lambda -i\nu u \right)^2}} \right\}-\int_{-Q}^Q dk' \ \frac d {d\Lambda} \tilde{\Phi}_{c0,c\nu}(k',\Lambda) \nonumber
\end{eqnarray}
Note that these functions can be written as
\begin{eqnarray}
\frac {dk^0(q)} {dq} &=& \frac 1 {2\pi \rho_{c0} (q)} \nonumber \\
\frac {d\Lambda_{\alpha\nu}^0(q)} {dq} &=& \frac 1 {2\pi \sigma_{\alpha\nu} (q)} \hspace{1.0cm} \alpha\nu=c\nu,s\nu \label{densityrap}
\end{eqnarray}
by inverting the equalities of Eq. (\ref{invrap}), where $2\pi \rho_{c0} (q)$ and $2\pi \sigma_{\alpha\nu} (q)$ are the corresponding pseudoparticle momentum dependent density functions. By inserting the former into the first derivative of the zeroth order "deviations" of the continuous Takahashi equations, we can obtain coupled integral equations in terms of these density functions. 

\newpage
\section{Pseudofermions}
\subsection{Introduction}
\label{pseudofermionI}

In this section we will introduce some new quantum objects, specify some of their properties and relate them to the pseudoparticles. The physical interpretation of some derived quantities in terms of relevant Hilbert spaces for our ground state $\rightarrow$ excited state transitions, and in terms of phase shifts and scatterers, will be presented in section (\ref{pseudofermionII}).

Eq. (\ref{residualE}) shows that the pseudoparticle residual energy interaction contains the quantities $\Delta Q_{\alpha \nu}(q)$, for $\alpha\nu=c0,c\nu,s\nu$. These quantities were, in turn, introduced to facilitate the mathematics, but they also carry an important physical meaning. In the following, we will incorporate these quantities in the definition of a quantum object related to the pseudoparticle, namely the {\it pseudofermion}. The pseudofermion will not have any residual energy interaction terms, and will allow us to construct a dynamical theory for the 1D Hubbard model. In the following, we will illustrate the "birth" of the pseudofermion picture by using the example of the $c0$ pseudoparticle, but is obviously valid for any pseudoparticle branch (by letting $k^0(q)\rightarrow \Lambda_{\alpha \nu}(q)$ whenever $\alpha \nu \neq c0$).

We have
\begin{equation}
k(q)=k^0(q)+\Delta k (q) = k^0(q)+ \frac {dk^0(q)} {dq} \Delta Q_{c0}(q)
\end{equation}
and equivalently for the other $\alpha \nu$ branches. However, a normal Taylor expansion of $k^0(q)$ yields
\begin{equation}
k^0(q+\delta(q))=k^0(q)+\frac {dk^0(q)} {dq} \delta(q)+\ldots
\end{equation}
where $\vert \delta(q) \vert$ is a small number. 

By defining $Q_{c0}^{\Phi}(q) = L \Delta Q_{co} (q)$, we see that the two expansions become equal if we define $\delta(q)$ as
\begin{equation}
\delta(q)=\Delta Q_{co} (q)=\frac {Q_{c0}^{\Phi}(q)} L
\end{equation}
That $\delta (q)$ is indeed of order ($1/L$) should be clear by observing that $\Delta Q_{c0} (q)$ carries a factor of ($2\pi/L$) in the discrete system. The summation over the effective Brillouin zone only contributes when the pseudoparticle deviation is nonzero, which only happens a finite number of times (for the cases that will become relevant when studying the one-electron spectral functions, we will see that this happens typically no more than two or three times, i.e. for two or three momentum values), which should leave no doubts concerning the order of magnitude of $\Delta Q_{c0} (q)$.

Up to first order, we can thus write the following relationship between the excited state rapidity and the ground state rapidity
\begin{eqnarray}
k(q)&=&k^0\left( q+\frac {Q_{c0}^{\Phi}(q)} L \right) \nonumber \\
\Lambda_{\alpha \nu}(q)&=&\Lambda_{\alpha \nu}^0 \left( q+\frac {Q_{\alpha \nu}^{\Phi}(q)} L \right) \hspace{0.75cm} \alpha=c,s \hspace{0.85cm} \nu=1,2,\ldots  
\end{eqnarray}

This is quite remarkable because it states that the excited state rapidity can be expressed by the ground state rapidity, if we shift the momenta by an amount $Q_{\alpha \nu}^{\Phi}(q)/L$. We see that all excited states that we are interested in can thus be expressed in terms of the ground state rapidities, if we are using a slightly shifted value for the discrete momenta. Also, all other properties of the pseudoparticles (for example their constitution in terms of rotated electrons, the fact that the $c\nu$ pseudoparticles are $\eta$-spin zero objects and that the $s\nu$ pseudoparticles are spin zero objects) remain intact. The "cruncher" of this new formulation is, however, that if we use these shifted discrete momentum values, the energy deviation expansion corresponding to Eq. (\ref{energyexp}) will yield zero or non physical expressions for all terms other than the first two:
\begin{equation}
E=\sum_{j=0}^{\infty} E^{(j)}=E^{(0)}+E^{(1)}=E_{GS}+\Delta E \hspace{0.5cm} \text{for} \hspace{0.5cm} q \rightarrow q+\frac {Q_{\alpha \nu}^{\Phi}(q)} L
\end{equation}
which is most easily understood by investigating the expressions for $E^{(2)}$, Eq. (\ref{residualE}), since this quantity is proportional to $Q_{\alpha \nu}^{\Phi}(q)$ itself. Since explicit calculation of the second order case is very lengthy with no contributions to the physical understanding, it seems more fruitful to present some simple reasons as to why this is true.

We remind ourselves that $Q_{\alpha \nu}^{\Phi}(q)$ is a measure of the discrete momentum shift, due to the ground state $\rightarrow$ excited state transition. By letting the original momentum values include this shift already "from the start", we have that there is no extra shift in the momenta to use in the rapidity expansions, since this shift is already recorded by the momentum values $q+Q_{\alpha \nu}^{\Phi}(q) / L$. Hence, we should put $Q_{\alpha \nu}^{\Phi}(q)=0$ in the expression for $E^{(2)}$, Eq. (\ref{residualE}), which renders $E^{(2)}=0$ exactly. Hence, the conclusion is that if we define a {\it new set of quantum objects}, with momentum values equal to the momenta plus this deviation, we find that these new objects undergo scattering events associated with the ground state $\rightarrow$ excited state transition with  {\it no energy exchange}. We will call these new objects {\it pseudofermions}. The pseudofermions have momenta $\bar{q}$, which we will call {\it canonical momenta}, due to the canonical pseudoparticle-pseudofermion transformation which will be defined in section (\ref{pseudoops}). This canonical momentum is defined as:
\begin{equation}
\bar{q}=q+\frac {Q_{\alpha \nu}^{\Phi}(q)} L \hspace{1.3cm} \alpha \nu = c0,c\nu,s\nu
\label{definitionbarmom}
\end{equation}

Note that for the ground state $Q_{\alpha\nu}^{\Phi}(q)=0$ since all the deviations in Eq. (\ref{Qphase1}) are zero. This means that for the ground state we have that $\bar{q}=q$. The absence of residual energy interactions between the pseudofermions will simplify our calculations of the one-electron spectral function tremendously. In fact, without this property of the pseudofermions, it would be pointless to introduce these new quantum objects. 

Similarily to the pseudoparticle, we will define creation and annihilation operators for the pseudofermions, which lead to a formal definition of the pseudofermions number operator $\mathcal{N}_{\alpha\nu}(\bar{q}_j)$, section (\ref{pseudoops}). However, by physical reasoning, there are some things that can be claimed without further due (see also Ref. \cite{wavefcnfact}). First off, since in the ground state the pseudoparticles and the pseudofermions are exactly the same objects, we have
\begin{equation}
\mathcal{N}_{\alpha\nu}^0(\bar{q}_j)=N_{\alpha\nu}^0(q_j)
\end{equation}
and moreover, when a pseudoparticle with momenta q is found in a configuration belonging to an excited energy eigenstate, we have that the corresponding pseudofermion has a canonical momentum value of $\bar{q}_j=\bar{q}_j(q_j)$ according to Eq. (\ref{definitionbarmom}). Since the ground state configuration of the two representations are equal, this means that
\begin{eqnarray}
\Delta \mathcal{N}_{\alpha\nu}(\bar{q}_j)&=& \Delta N_{\alpha\nu}(q_j) \nonumber \\
\Delta \mathcal{N}_{\alpha\nu}(\bar{q}_j)&=&\mathcal{N}_{\alpha\nu}(\bar{q}_j)-\mathcal{N}_{\alpha\nu}^0(\bar{q}_j)
\end{eqnarray}
which implies that $\mathcal{N}_{\alpha\nu}(\bar{q}_j)=N_{\alpha\nu}(q_j)$. However, this does not imply that in the continuous system $\mathcal{N}_{\alpha\nu}(\bar{q})=N_{\alpha\nu}(q)$, in fact this is in general {\it not true}. We remember that we reach the continuous system by letting ($2\pi/L) \rightarrow 0$ but that our pseudofermion theory carries physically relevant terms of order ($1/L$). 

Let us define the inverse of Eq. (\ref{definitionbarmom}) in the discrete system,
\begin{equation}
q_j=q_j(\bar{q}_j)=\bar{q}_j-\frac {2\pi} L \sum_{\alpha' \nu'} \sum_{j'=1}^{N_{\alpha' \nu'}^*} \Phi_{\alpha\nu,\alpha' \nu'} (\bar{q}_j,{\bar{q}'}_j) \Delta \mathcal{N}_{\alpha'\nu'}({\bar{q}'}_j)
\end{equation}
and then investigate the jacobian of the $q\rightarrow \bar{q}$ coordinate transformation:
\begin{eqnarray}
\sum_q F_{\alpha\nu}(q) = \frac L {2\pi} \int_q dq \  F_{\alpha\nu}(q) = \frac L {2\pi} \int_{\bar{q}} d\bar{q} \  \mathcal{F}_{\alpha\nu}(\bar{q}) \frac {dq(\bar{q})} {d\bar{q}}
\end{eqnarray}
where $F_{\alpha\nu}(q)$ and $\mathcal{F}_{\alpha\nu}(\bar{q})$ are some functions of the momenta and canonical momenta, respectively. The jacobian becomes
\begin{equation}
\frac {dq(\bar{q})} {d\bar{q}} = 1-\sum_{\alpha' \nu'} \int_{-q_{\alpha' \nu'}^0}^{q_{\alpha' \nu'}^0} d\bar{q} \ \frac {d\Phi_{\alpha\nu,\alpha' \nu'} (\bar{q},{\bar{q}'})} {d\bar{q}}  \Delta \mathcal{N}_{\alpha'\nu'}({\bar{q}'})
\end{equation}
which yields one only if $\mathcal{F}_{\alpha\nu}(\bar{q})$ is proportional to $\Delta \mathcal{N}_{\alpha'\nu'}({\bar{q}'})$, since we do not include second order terms in our theory. The canonical momenta spacing will be further discussed in section (\ref{pseudoops}). 

By similar reasoning, we see that the energy bands of Eq. (\ref{energybands}) transform according to
\begin{equation}
\epsilon_{\alpha\nu}(q(\bar{q}))=\epsilon_{\alpha\nu} (\bar{q})- v_{\alpha\nu}(\bar{q})\sum_{\alpha' \nu'} \int_{-q_{\alpha' \nu'}^0}^{q_{\alpha' \nu'}^0} d\bar{q} \  \Phi_{\alpha\nu,\alpha' \nu'} (\bar{q},{\bar{q}'}) \Delta \mathcal{N}_{\alpha'\nu'}({\bar{q}'})
\end{equation}
but since the energy bands always multiply the corresponding pseudoparticle or pseudofermion deviation, we find that the second term of the pseudofermion energy band is of second order in the pseudofermion number deviations, and hence falls outside the realm of our pseudofermion theory. It is therefore safe to use the same energy bands as previously derived. 

Since we do not change the total number of pseudoparticles, we only shift the momenta of them, the number operators as well as the corresponding number deviation operators, will have the same eigenvalues in the pseudoparticle basis as in the pseudofermion basis \cite{wavefcnfact}. This is due to the fact that the pseudoparticles, whose number operators commute with the hamiltonian, have the same composition in terms of rotated electrons as the pseudofermions. Therefore, the pseudofermion number operators should also commute with the hamiltonian. Thus, due to the equality of eigenvalues and the common eigenstates, it would be expected that by using a formal operator language, we find that the pseudoparticles and the pseudofermions are related to each other by a unitary transformation.

\newpage
\subsection{The PS subspace and quantum shake-up effects}
\label{pseudofermionII}

As already mentioned before (section \ref{pseudoparticle}), we are interested in formulating a normal ordered theory, relative to the ground state. This implies that all quantities needed to describe the dynamics of the model will be expressed in terms of deviations from the ground state configuration. Like this, we obtain different theories for different ground states. According to the pseudofermion picture, we will only retain terms of order ($1/L$). The fact that the pseudofermions do not have any physical properties of order $(1/L)^j$ for $j \geq 2$, makes it possible to formulate the normal ordered theory with only first order terms. Hence, we consider ground state $\rightarrow$ excited state transitions such that the number of pseudofermions change according to
\begin{equation}
\Delta N_{\alpha \nu} (q) = N_{\alpha \nu} (q) - N_{\alpha \nu}^0 (q) \hspace{1.0cm} -q_{\alpha \nu}^0 \leq q \leq q_{\alpha \nu}^0
\end{equation}
where we use the same symbol for pseudoparticle number deviation and the pseudofermion number deviation since they are always equal to each other. The deviations of $\alpha \nu$ pseudofermions, can equally be expressed in terms of deviations in the {\it electronic numbers}, according to:
\begin{eqnarray}
M_{c,-\frac 1 2} &=& \frac 1 2 (N-N_{c0}) \nonumber \\
M_{c,-\frac 1 2} &=& L_{c,- \frac 1 2}+\sum_{\nu=1}^{\infty} \nu N_{c\nu}
\end{eqnarray}
which by taking deviations leads to
\begin{equation}
\Delta N = \Delta N_{c0}+2\Delta L_{c,-\frac 1 2} +2\sum_{\nu=1}^{\infty} \nu \Delta N_{c\nu}
\label{deltaN}
\end{equation}
Secondly,
\begin{eqnarray}
M_{s,-\frac 1 2} &=& \frac 1 2 (N_{c0}-N_{\uparrow}+N_{\downarrow}) \nonumber \\
M_{s,-\frac 1 2} &=& L_{s,- \frac 1 2}+\sum_{\nu=1}^{\infty} \nu N_{s\nu}
\end{eqnarray}
implies that 
\begin{equation}
\Delta (N_{\uparrow}-N_{\downarrow}) = \Delta N_{c0}-2\Delta L_{s,-\frac 1 2} -2\Delta N_{s1}-2\sum_{\nu=2}^{\infty} \nu \Delta N_{s\nu}
\label{deltaNupNdown}
\end{equation}
Thirdly, due to Eq. (\ref{sumparticles}), we have that
\begin{eqnarray}
\Delta \left( M_{c,+\frac 1 2} + M_{c,-\frac 1 2} \right) &=& -\Delta N_{c0} \nonumber \\
\Delta \left( M_{s,+\frac 1 2} + M_{s,-\frac 1 2} \right) &=& \Delta N_{c0}
\end{eqnarray}
which, by taking deviations of Eq. (\ref{defLparts}) and equalling the resulting expressions with the ones above, eliminating the deviations for the $+\frac 1 2$ holons and spinons, respectively, leads to
\begin{eqnarray}
\Delta L_{c,+\frac 1 2}&=&-\Delta N_{c0} - 2\sum_{\nu=1}^{\infty} \nu N_{c\nu} - L_{c,-\frac 1 2} \nonumber \\
\Delta L_{s,+\frac 1 2}&=&\Delta N_{c0} - 2\Delta N_{s1} -2\sum_{\nu=2}^{\infty} \nu N_{s\nu} - L_{s,-\frac 1 2}
\end{eqnarray}
which implies that the numbers $\Delta L_{c,+\frac 1 2}$ and $\Delta L_{s,+\frac 1 2}$ are not independent and thus it suffices to specify only $L_{c,-\frac 1 2}$ and $L_{s,-\frac 1 2}$ when dealing with deviations of Yang holons and HL spinons. We note that for the LWS ground state, $N_{c\nu}=N_{s\nu}=L_{c,- \frac 1 2}=L_{s,- \frac 1 2}=0 \implies \Delta N_{c\nu}=N_{c\nu}$ (for $\nu \geq 1$), $\Delta N_{s\nu}=N_{s\nu}$ (for $\nu \geq 2$), $\Delta L_{c,- \frac 1 2}=L_{c,- \frac 1 2}$ and $\Delta L_{s,- \frac 1 2}=L_{s,- \frac 1 2}$. 

These relationships limit the number of pseudofermions created or annihilated whenever a finite number of electrons are created or annihilated. This is hence our first restriction: to only allow processes that create or annihilate a finite number of electrons. Ultimately, this will be seen in the number of electronic creation and annihilation operators present in the operators of any correlation function that we wish to calculate. Later, we will focus on the one-electron spectral problem, for which this issue is trivial (being only one electronic creation or annihilation operator). The finite electron creation and annihilation operator limitation, implies that the collection of excited states reachable by these operators, span a strict subspace of the entire Hilbert space of the model, a subspace that we will call the {\it pseudofermion subspace}, abbreviated {\it PS}. 

Having limited the number of electrons created or annihilated from the system, we can investigate how this change in the total number of electrons affect the number of pseudofermions and their lattice configurations. Within our theory, each $\alpha\nu$ pseudofermion is only existing inside the many-body quantum system, and doing so on a specific $\alpha\nu$ dependent lattice. This means that not only does the lattice constant and the total number of available canonical momentum values differ for each $\alpha\nu$ branch, but the latter number also changes whenever creating or annihilating electrons. This implies that the quantum numbers describing the occupancies of the specific $\alpha\nu$ branch under consideration, change from being integers or half-odd integers to being half-odd integers or integers, respectively. This effect is usually called {\it the quantum shake-up effect}. This effect takes place for all $\alpha\nu$ branches, even for the $c0$ branch in spite of the fact that $N_{c0}^*=N_a$ is constant, because for this branch it is the number apparent in Eq. (\ref{c0numbers}), and not $N_{c0}^*$, that decides whether or not the occupancies of the $c0$ branch are described by integers or half-odd integers. During a transition to an excited state, if the changes of the following numbers {\it are odd}, the quantum numbers describing the occupancies of the corresponding branch, change according to the quantum shake-up effect (derived using Eq. (\ref{pseudoholesII})):
\begin{eqnarray}
&c0:& \hspace{1.5cm} \Delta \left( \frac {N_a} 2 -\sum_{\alpha\nu=c\nu,s\nu} N_{\alpha\nu} \right) =\sum_{\alpha\nu=c\nu,s\nu} \Delta N_{\alpha\nu} \nonumber \\
&s\nu:& \hspace{1.5cm} \Delta N_{s\nu}^*=\Delta \left(N_{s\nu}+N_{c0}\right) -\sum_{\nu'=1}^{\infty} \left( \nu'+\nu - \vert \nu' - \nu \vert \right) \Delta N_{s\nu'} \hspace{1.0cm} \label{condishake} \\
&c\nu:& \hspace{1.5cm} \Delta N_{c\nu}^*=\Delta \left(N_{c\nu}-N_{c0}\right) -\sum_{\nu'=1}^{\infty} \left( \nu'+\nu - \vert \nu' - \nu \vert \right) \Delta N_{c\nu'} \hspace{1.0cm} \nonumber
\end{eqnarray}

The expressions of these number deviations will be simplified in section (\ref{scattering}). Note that the deviational numbers are purely expressed in terms of occupational numbers of pseudofermions. This is necessary since for the same {\it electronic} creation or annihilation process, the resulting quantum mechanical state may be a linear combination of several states, with different set-ups of pseudofermions. Thus, in order to properly account for the shake-up effect, we need to count the deviations of pseudofermions in the particular state that we are investigating. This is a consequence of the occupational numbers of electrons not being good quantum numbers: the same number of electrons, as well as the same numbers of $\uparrow$-spin and $\downarrow$-spin electrons, can be fitted with many different quantum mechanical states, whilst a certain specified set of pseudofermionic occupational deviation numbers specify {\it one and only one} quantum mechanical state.

\begin{figure}
\begin{center}
\includegraphics[width=13cm,height=10cm]{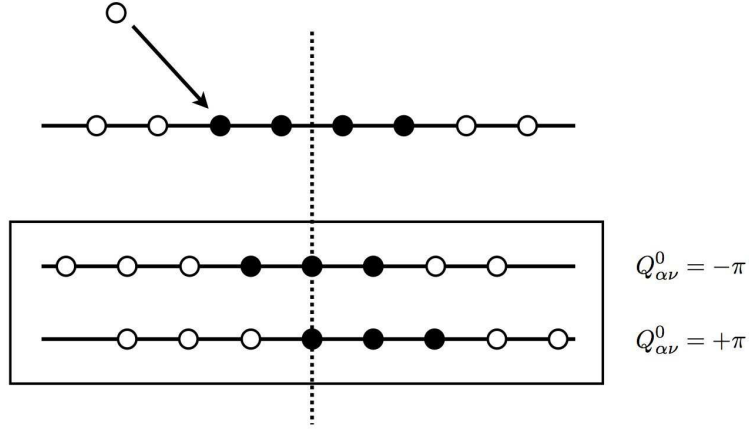}
\caption{\label{figshakeup} Schematical figure illustrating the shake-up effect on a toy lattice (where the filled circles depict pseudofermions and the empty circles pseudofermion holes). Let us, as a pedagogical example, suppose that we create a pseudofermion $\alpha\nu$ hole onto a ground state with a symmetrical distribution of pseudofermions around the zero momentum point (here indicated by a vertical dashed line), in such a way that we have a non zero shake up effect, according to Eq. (\ref{condishake}). We note that, in this example, the top configuration is energetically more favorable than the bottom configuration. The constant spacing between the lattice points is $2\pi /L$ and the size of the shake-up shift is $\pi / L$. \vspace{0.5cm}}
\end{center}
\end{figure}

Since these shake-up effects are measured relative to the ground state, we should study how they alter the momentum values in the ground state, which has $\bar{q}=q$ as already mentioned in section (\ref{pseudofermionI}). The "shifts" in the quantum numbers, which can be written as
\begin{equation}
I_j^{\alpha\nu} \rightarrow I_j^{\alpha\nu} + J_0 \hspace{1.0cm} J_0=-\frac 1 2\ ,\ 0\ ,\ \frac 1 2
\end{equation}
implies that the ground state momenta changes according to
\begin{equation}
\frac {2\pi} L \left( I_j^{\alpha\nu} + J_0 \right)=
\left\{
\begin{array} {c}
q_j-\frac {\pi} L\\
q_j \\
q_j+\frac {\pi} L
\end{array}
\right.
\end{equation}
so that we can write the shift in the ground state momenta as
\begin{equation}
q_j \rightarrow q_j+\frac {Q_{\alpha\nu}^0} L \hspace{1.0cm} Q_{\alpha\nu}^0=-\pi,0,\pi
\end{equation}
which defines the new quantity $Q_{\alpha\nu}^0$. This is not in contradiction with the definition of the canonical momentum, Eq. (\ref{definitionbarmom}). What the shake-up effect entails, is that the momentum values, the $q$'s in Eq. (\ref{definitionbarmom}), are shifted, but with the definition of the canonical momentum intact. One should note that if $\vert Q_{\alpha\nu}^0 \vert=\pi$, we have two resulting states that differ from each other in terms of the positions of the occupied quantum numbers, as shown in Fig. (\ref{figshakeup}). 

One could, and we certainly will, view the shake-up effect as producing a virtual excited state, which is a first step of any ground state $\rightarrow$ excited state transition, where the quantum objects have momenta $q+Q_{\alpha\nu}^0 / L$. The true final state would then be the one which is described by the canonical momenta, where the scattering events between the pseudofermions are governed by $Q_{\alpha\nu}^{\Phi}(q)$. In section (\ref{scattering}), where the scattering theory will be developed, we shall see that the usual quantum mechanical picture, "a shift in the momenta of the quantum particles produces a shift in the phase of the wave-function", will also apply here.

\subsection{Pseudofermion operators and momentum spacing}
\label{pseudoops}

By using the fact that pseudofermions, with canonical momenta $\bar{q}$, live on the same lattice as the pseudoparticles, with momenta $q$, and that furthermore there are as many pseudofermions as pseudoparticles for every $\alpha \nu$ branch, we can without further due define {\it pseudofermion creation} and  {\it pseudofermion annihilation} operators:
\begin{eqnarray}
f_{\bar{q},\alpha \nu}^{\dag}&=&\frac 1 {\sqrt{N_{\alpha \nu}^*}} \sum_{j=1}^{N_{\alpha \nu}^*} e^{i\bar{q}ja_{\alpha\nu}} f_{j,\alpha \nu}^{\dag} \nonumber \\
f_{\bar{q},\alpha \nu}&=&\frac 1 {\sqrt{N_{\alpha \nu}^*}} \sum_{j=1}^{N_{\alpha \nu}^*} e^{-i\bar{q}ja_{\alpha \nu}} f_{j,\alpha \nu} \label{feqf}
\end{eqnarray}
where $f_{j,\alpha \nu}^{\dag}$ creates a $\alpha \nu$ pseudofermion on a $\alpha \nu$ effective lattice site position $j$ (with space coordinate $x_j=ja_{\alpha \nu}$) and $f_{j,\alpha \nu}$ annihilates a $\alpha \nu$ pseudofermion on a $\alpha \nu$ effective lattice site position $j$. Note that there is only one $\bar{q}$ for each $q$. In other words, the function $\bar{q}=\bar{q} (q)$ is unique for each ground state $\rightarrow$ excited state transition, due to the uniqueness of the solution of Eq. (\ref{Phi}) .

The pseudofermion operators and the pseudoparticle operators are related through a unitary transformation $\hat{\mathcal{V}}_{\alpha \nu}$:
\begin{equation}
f_{\bar{q},\alpha \nu}^{\dag}=\hat{\mathcal{V}}_{\alpha \nu}^{\dag}b_{q,\alpha \nu}^{\dag}\hat{\mathcal{V}}_{\alpha \nu} \hspace{1.5cm} f_{\bar{q},\alpha \nu}=\hat{\mathcal{V}}_{\alpha \nu}^{\dag}b_{q,\alpha \nu}\hat{\mathcal{V}}_{\alpha \nu}
\end{equation}
where
\begin{equation}
\hat{\mathcal{V}}_{\alpha \nu}=\exp{\Big\{ \sum_{q} b_{q,\alpha \nu}^{\dag} \left[ b_{\bar{q},\alpha \nu}-b_{q,\alpha \nu} \right] \Big\}}
\end{equation}
These relationships, as well as the proof of $\hat{\mathcal{V}}_{\alpha \nu}$ being unitary, are given in Ref. \cite{wavefcnfact}. The pseudofermion picture thus allows us to write the energy as
\begin{equation}
E = E_{GS}+\Delta E = E_{GS} (\left \{ N_{\alpha \nu}^0 \right\}) + \Delta E (\left \{ \Delta N_{\alpha \nu} \right\})
\end{equation}
without any higher order terms in the deviations and
\begin{equation}
N_{\alpha \nu} = \Big\langle \sum_{q} b_{q,\alpha \nu}^{\dag}b_{q,\alpha \nu} \Big\rangle = \Big\langle \sum_{\bar{q}} f_{\bar{q},\alpha \nu}^{\dag}f_{\bar{q},\alpha \nu} \Big\rangle
\end{equation}

Before we move on, there are some important properties that need to be clarified. The name pseudofermion stems from the fact that the operators $f_{\bar{q},\alpha \nu}^{\dag}$ and $f_{\bar{q},\alpha \nu}$, just like the pseudoparticles, satisfy the fermionic anticommutation relations {\it almost}. To evaluate the anticommutator between these two operators, we use the fact that their {\it local} counterparts satisfy the fermionic anticommutation relations {\it exactly}:
\begin{equation}
\{ f_{\bar{q},\alpha \nu}^{\dag} , f_{{\bar{q}}',\alpha' \nu'} \} = \frac {\delta_{\alpha , \alpha'} \delta_{\nu , \nu'}} {N_{\alpha \nu}^*} \sum_{j=1}^{N_{\alpha \nu}^*} \sum_{j'=1}^{N_{\alpha \nu}^*} e^{ia_{\alpha \nu} \left( \bar{q}j-\bar{q}' j' \right)} \{ f_{j,\alpha \nu}^{\dag} , f_{j',\alpha \nu} \}
\end{equation}
By using the following equalities
\begin{eqnarray}
\{ f_{j,\alpha \nu}^{\dag} , f_{j',\alpha \nu} \}=\delta_{j,j'} \hspace{1.0cm} \{ f_{j,\alpha \nu} , f_{j',\alpha' \nu'} \} = 0 \nonumber \\
\sum_{j=1}^M e^{jz} = e^z \frac {e^{Mz}-1} {e^z-1} \hspace{1.75cm}
\end{eqnarray}
we arrive to the following anticommutators
\begin{eqnarray}
\{ f_{\bar{q},\alpha \nu}^{\dag} , f_{{\bar{q}}',\alpha' \nu'} \} &=& \frac {\delta_{\alpha , \alpha'} \delta_{\nu , \nu'}} {N_{\alpha \nu}^*} e^{i a_{\alpha \nu} (\bar{q}-\bar{q}') / 2} e^{i\left(Q_{\alpha \nu}(q)-Q_{\alpha \nu}(q') \right) / 2} \frac {\sin \left( \frac {Q_{\alpha \nu}(q)-Q_{\alpha \nu}(q')} 2 \right)} {\sin \left( \frac {a_{\alpha \nu} (\bar{q}-\bar{q}')} 2  \right)} \label{anticomm} \nonumber \\
\{ f_{\bar{q},\alpha \nu} , f_{{\bar{q}}',\alpha' \nu'} \} &=& 0
\end{eqnarray}
which will play a key role in the development of the theory. That the local pseudofermions satisfy the fermionic anticommutation relations exactly is a property due to the rotated electrons, whose number operators commute with the hamiltonian \cite{carmelocondmat}.

The second property of the pseudofermion description that we will mention here regards the discrete canonical momenta spacing $\Delta \bar{q}$. The discrete momenta spacing is, just like for free fermions, constant and equal to ($2\pi/L$):
\begin{equation}
\Delta q_j = q_{j+1}-q_j = \frac {2\pi} L
\end{equation}
The discrete canonical momenta, however, satisfies (for $\alpha \nu=c0,c\nu,s\nu$)
\begin{equation}
\Delta \bar{q}_j = \bar{q}_{j+1}-\bar{q}_j = \Delta q_j + \frac {Q_{\alpha \nu}^{\Phi}(q_{j+1}) - Q_{\alpha \nu}^{\Phi}(q_j)} L
\end{equation}
where the difference $\left(Q^{\Phi}(q_{j+1}) - Q^{\Phi}(q_j)\right)$, due to Eq. (\ref{Qphase1}), is
\begin{eqnarray}
&&\sum_{\alpha' \nu'} \int_{-q_{\alpha' \nu'}^0}^{q_{\alpha' \nu'}^0} dq' \  \left[ \Phi_{\alpha \nu,\alpha' \nu'}(q_{j+1},q')- \Phi_{\alpha \nu,\alpha' \nu'}(q_j,q') \right] \Delta N_{\alpha' \nu'} (q') = \nonumber \\
&&= \sum_{\alpha' \nu'} \int_{-q_{\alpha' \nu'}^0}^{q_{\alpha' \nu'}^0} dq' \ \Delta q_j \frac {d \Phi_{\alpha \nu,\alpha' \nu'} (q,q')} {dq} \Bigg\vert_{q=q_j} \Delta N_{\alpha' \nu'} (q')
\end{eqnarray}
which means that, since $\Delta q_j=2\pi/L$,
\begin{equation}
\Delta \bar{q}_j = \frac {2\pi} L + \mathcal{O} (1/L^2)
\end{equation}
or, in other words, since ($1/L^2$) terms do not contribute to the physics in the pseudofermion picture, we are erroneously led to conclude that $\Delta q_j = \Delta \bar{q_j}$, or that the discrete momenta and the discrete canonical momenta are equal to first order in $(1/L)$. These are erroneous conclusions based on the fact that for larger deviations, we have $q_{j'}-q_j=(2\pi/L) (j'-j)$ for the momenta, whilst in general $\bar{q}_{j'}-\bar{q}_j \neq (2\pi/L) (j'-j)$ for the canonical momenta. This is most easily seen when $j'$ and $j$ are very far apart, say $j'-j \sim (N_a/2)$. This means that we have 
\begin{equation}
\bar{q}_{j'}-\bar{q}_j \sim \frac {N_a} 2 \left[ \frac {2\pi} L +  \mathcal{O} (1/L^2) \right] \sim \frac {\pi} a +  \mathcal{O} (1/L)
\end{equation}
by averaging the values of the derivatives of $\Phi_{\alpha \nu, \alpha' \nu'} (q,q')$ which yields a factor proportional to $j'-j$. This difference between $\Delta q_j$ and $\Delta \bar{q_j}$ is of the order of ($1/L$), which is a difference {\it inside} the realm of the pseudofermion physics, i.e. a non-negligible difference. This difference implies that whenever we want to replace a sum by an integral, i.e. when $N_a \rightarrow \infty$ we are not allowed to use the "standard" replacement $\Delta q \rightarrow (2\pi/L)$ but instead we will need to use a state dependent jacobian (due to the state dependence of $\left[Q^{\Phi}(q_{j+1}) - Q^{\Phi}(q_j)\right]$).

Normal ordered operators are sometimes written as {\it :}$\hat{X}${\it :} with the definition $\text{{\it :}}\hat{X}\text{{\it:}} =\ \hat{X}-X^0$, for $X^0= \vert \langle GS \vert \hat{X} \vert GS \rangle \vert$ (where $\vert GS \rangle$ is the ground state). A typical eigenstate of such a normal ordered operator is thus the deviation between the eigenvalues of the excited state and the ground state respectively. Since our pseudofermionic operators are normal ordered by construction, it seems superfluous to use this notation for them, whilst the hamiltonian and the momentum operator becomes {\it :}$\hat{H}${\it:} and {\it :}$\hat{P}${\it:} respectively.

In the pseudofermionic basis, these operators are
\begin{eqnarray}
\text{{\it :}}\hat{H}\text{{\it:}} &=& \sum_{\alpha\nu} \sum_{j=1}^{N_{\alpha\nu}^*} \epsilon(\bar{q}_j) f_{\bar{q}_j,\alpha\nu}^{\dag} f_{\bar{q}_j,\alpha\nu} + \sum_{\alpha=c,s} \mu_{\alpha} \hat{L}_{\alpha,-\frac 1 2} \nonumber \\
\text{{\it :}}\hat{P}\text{{\it:}} &=&  \sum_{j=1}^{N_a} \bar{q}_j f_{\bar{q}_j,c0}^{\dag} f_{\bar{q}_j,c0} +\sum_{\nu=1}^{\infty} \sum_{j=1}^{N_{s\nu}^*} \bar{q}_j f_{\bar{q}_j,s\nu}^{\dag} f_{\bar{q}_j,s\nu}+ \label{normorderH} \\
&&+ \sum_{\nu=1}^{\infty} \sum_{j=1}^{N_{c\nu}^*} \left[ \left( 1+\nu \right) \frac {\pi} a - \bar{q}_j \right] f_{\bar{q}_j,c\nu}^{\dag} f_{\bar{q}_j,c\nu} + \frac {\pi} a \hat{L}_{c,-\frac 1 2} \nonumber
\end{eqnarray}

\subsection{Virtual states and pseudofermionic subspaces}
\label{scattering}

To describe the scatterers and the scattering centers of the theory, we need to describe the excited eigenstates  in terms of pseudofermions. We will follow the standard non-relativistic description of a quantum scattering theory \cite{scatteringbook}, in which the scattering S-matrix, which maps the "incoming" quantum state into the scattered "outgoing" quantum state, will play a central role. Since all our pseudofermions are either $\eta$-spin zero or spin zero objects, the scattering matrix will be of dimension $1\times1$, i.e. just a complex number, in contrast to the representation of Ref. \cite{holonspinon1}, where the scatterers are $\eta$-spin $\frac 1 2$ and spin $\frac 1 2$ objects. In this reference, the S-matrix has a larger dimension due to the the coupling of the $\eta$-spin and the spin channels. In the following, we will use the usual definition of a {\it phase shift} such that a shift in the momentum $\delta_l(q)$ of a quantum object with momentum $q$, produces a shift in its wave-function equal to $e^{2\delta_l(q)}$, where $l$ stands for a collection of quantum numbers used to fully describe the original (unscattered) incoming wave (also known as the {\it in asymptote}). The job of the S-matrix is then to supply the incoming wave with this phase shift, and transform it into the outgoing wave (also known as the {\it out asymptote}) \cite{scatteringbook}. Since the incoming and outgoing waves, according to the general quantum scattering theory, preserve the total momenta and the total energy, we see that we must introduce another scheme when describing the transitions. This is because the quantities that we have associated with a general ground state $\rightarrow$ excited state transition, namely the energy and momentum deviations according to
\begin{eqnarray}
E=E^{(0)}+E^{(1)}=E_{GS}+\Delta E \nonumber \\
P=P^{(0)}+P^{(1)}=P_{GS}+\Delta P \label{enemomdev}
\end{eqnarray}
do not preserve neither the total energy nor the total momentum. We therefore divide the entire transition into two steps: one scatteringless step and one in which all the scattering events occur. The scatteringless step yields a {\it virtual} state, or an intermediate state, which brings the system from having energy $E_{GS}$ and momentum $P_{GS}$, to having energy $E_{GS}+\Delta E$ and momentum $P_{GS}+\Delta P$ respectively \cite{CarmBoziPedro}. Under this ground state $\rightarrow$ virtual state transition, the quantum numbers describing the occupancies of the pseudofermions may "shake-up" as described in section (\ref{pseudofermionII}). Also, the energies and the momenta change according to Eq. (\ref{enemomdev}), i.e. the ground state $\rightarrow$ virtual state transition is a scatteringless finite energy and finite momentum transition. The virtual state is the in asymptote in the scattering theory. Thus, this virtual state is occupied by pseudofermions with momenta $q+Q^0_{\alpha\nu}/L$. Then, the virtual state will undergo scattering events governed by the quantities $Q_{\alpha\nu}^{\Phi}(q)$, for $\alpha\nu=c0,c\nu,s\nu$, that preserve total energy and total momentum. In other words, the additional state dependent shift $Q^{\Phi}_{\alpha\nu}(q)/L$ implies no extra energy nor momentum terms in the deviation expansions.

Before we move on, we need to specify the pseudofermion deviations that characterize a typical virtual state. Since electrons are the only quantum objects that can be created or annihilated, we have to classify the types of subspaces we obtain by fixing the deviations $\Delta N$ and $\Delta (N_{\uparrow}-N_{\downarrow})$. This has actually already been done in section (\ref{pseudofermionII}), where we saw that the electronic deviations can uniquely be expressed as
\begin{eqnarray}
\Delta N &=& \Delta N_{c0}+2\Delta L_{c,-\frac 1 2} +2\sum_{\nu=1}^{\infty} \nu \Delta N_{c\nu} \nonumber \\
\Delta (N_{\uparrow}-N_{\downarrow}) &=& \Delta N_{c0}-2\Delta L_{s,-\frac 1 2} -2\Delta N_{s1}-2\sum_{\nu=2}^{\infty} \nu \Delta N_{s\nu} \label{deltaNwspin}
\end{eqnarray}
which means that for each fixed set of numbers $\{ \Delta N_{\alpha\nu}\}_{\alpha\nu=c0,c\nu,s\nu}$ and $\{ \Delta L_{\alpha,-\frac 1 2} \}_{\alpha=c,s}$ we span one strict subspace of the entire Hilbert space of the model, which correspond to the actual electronic deviations at hand. One should then collect all possible sets of these pseudofermionic deviation numbers, to arrive to the complete set of virtual states that emerge due to nonzero deviations in Eq. (\ref{deltaNwspin}). 
 
The total energy and total momentum acquired during a ground state $\rightarrow$ virtual state transition, is easily obtained from Eq. (\ref{energy}):
\begin{eqnarray}
\Delta E &=& \frac L {2\pi} \sum_{\alpha=c,s} \sum_{\nu=\delta_{\alpha,s}}^{\infty} \int_{-q_{\alpha\nu}^0}^{q_{\alpha\nu}^0} dq \ \epsilon_{\alpha\nu}(q) \Delta N_{\alpha\nu}(q) + \omega_0 \nonumber \\
\Delta P &=&  \frac L {2\pi} \left[ \int_{-q_{c0}^0}^{q_{c0}^0} dq \ q\Delta N_{c0}(q) +  \sum_{\nu=1}^{\infty}\int_{-q_{s\nu}^0}^{q_{s\nu}^0} dq \ q\Delta N_{s\nu}(q)+\right. \label{EdevPdev} \\
&& + \left. \sum_{\nu=1}^{\infty} \int_{-q_{c\nu}^0}^{q_{c\nu}^0} dq \ \left( \frac {\pi} a-q \right) \Delta N_{c\nu}(q) \right] + \frac {\pi} a M_{c,-\frac 1 2} \nonumber
\end{eqnarray}
where we have defined the {\it minimum energy excitation} $\omega_0$ as
\begin{equation}
\omega_0=2\mu M_{c,-\frac 1 2} + 2\mu_0 h (M_{s,-\frac 1 2}-N_{s1})
\end{equation}
where $M_{c,-\frac 1 2}=M_{s,-\frac 1 2}-N_{s1}=0$ for the initial LWS ground state. As expected, this term is nonzero if we have broken the $SO(4)$ symmetry of the model. Hence, $\omega_0$ serves as a {\it gap parameter} that tells us whether or not our excitations live in a gapped or in a gapless system.

Note that for these expressions, we have $q_{j+1} - q_j = 2\pi/L$ due to the scatteringless property of the ground state $\rightarrow$ virtual state transition. However, the actual occupancy positions may shift {\it globally} according to the shake-up effect. The conditions on whether or not the quantum numbers for a particular $\alpha\nu$ branch are shaken up can be simplified as compared to the expression given in Eq. (\ref{condishake}). Since ($\nu+\nu' - \vert \nu' - \nu \vert$) is always an even number, we can exclude the $\nu$ summation in Eq. (\ref{condishake}) altogether  since we are only interested in terms that have a possibility to be odd. This is also the reason for why we neglect $-2\Delta N_{c0}$ in the $c\nu$ case, after having added and subtracted $\Delta N_{c0}$ on the right hand side of that equation. We thus arrive to the following statement: if the following deviations are odd, in connection with our ground state $\rightarrow$ virtual state transition, then the quantum numbers for the actual virtual state change from being integers (or half-odd integers) to being half-odd integers (or integers):
\begin{eqnarray}
&c0:& \hspace{1.5cm} \sum_{\alpha\nu=c\nu,s\nu} \Delta N_{\alpha\nu} \nonumber \\
&\alpha\nu:& \hspace{1.5cm} \Delta N_{\alpha\nu} + \Delta N_{c0} \label{shakeups}
\end{eqnarray}

\subsection{The S-matrix}
\label{humbadu}

According to the standard quantum mechanical scattering theory, the {\it S-matrix} is a unitary operator that maps the pre-scattered state $\vert \phi_{in} \rangle$ into the post-scattered state $\vert \phi_{out} \rangle$. As we have seen above, these states have many names, according to the rich history of scattering theory in general. An example of frequently used names are "incoming" waves and "outgoing" waves, due to the classical analogue of colliding billiard balls (thus incoming balls and outgoing balls). A more mathematical nomenclature includes "in asymptote" and "out asymptote", due to the mathematical formulation of modern scattering theory. In this case the pre-scattered state is regarded as the "untouched" state that existed at a time $t \rightarrow -\infty$ (i.e. as the time $t$ approaches this limit, the state approaches some asymptotic idealized form) and the post-scattered state is similarily regarded as the asymptotic state at time $t \rightarrow +\infty$. From now on, we will choose this latter nomenclature for the pre- and post-scattered states. Thus in general,
\begin{equation}
S \vert \phi_{in} \rangle=\vert \phi_{out} \rangle \hspace{1.0cm} S_{ll'}=e^{2i\delta_{ll'}}
\end{equation}
where $\delta_{ll'}$ is the total phase shift for the $l \rightarrow l'$ scattering process and $S_{ll'}$ is the corresponding matrix element of the S-matrix ($l$ and $l'$ are sets of quantum numbers that fully describe the scattering quantum objects). In our case, since the different $\alpha\nu$ branches do not mix with each other, we can define a $S_{\alpha\nu}(q)$ for each branch at momentum $q$. This quantity describes the scattering events of a $\alpha\nu$ pseudofermion of momentum $q$, with pseudofermions of all other $\alpha' \nu'$ branches created by the transition, as it travels around the lattice. We shall see that this operator can be described by products of quantities called $S_{\alpha\nu, \alpha' \nu'}(q,q')$, that gives the form for the individual scattering events between the $\alpha \nu$ and the $\alpha' \nu'$ pseudofermion. This should not be confused with the usual scattering notion of "mixing between different scattering channels". In our theory, there is no such mixing: we describe scattering events between $\eta$-spin and spin zero objects, which preserve the individual $\alpha \nu$ branches. Moreover, since the scattering itself does not change the energies nor the momenta of the {\it scatterers} nor of the {\it scattering centers} (upon which the scatterers scatter), the scattering events are of a trivial zero energy forward scattering type. We will clarify this claim in more detail later in this section (however, we refer to Ref. \cite{josescatt} for the main results and to Ref. \cite{CarmBoziPedro} for a detailed analysis).

Formally, in finding our expression for the total phase shift $\delta_{\alpha\nu}(q)$, we note that the phase of the in asymptote changes as our pseudofermion scatters with all the scattering centers of the system, i.e. as the scatterer travels around the lattice ring once, to arrive to its original starting position (remember that we adopted periodic boundary conditions for the original hamiltonian (\ref{H})). There are different choices of coordinates available for this picture, giving different defining expressions for $\delta_{\alpha\nu}(q)$, however always yielding the same S-matrix \cite{CarmBoziPedro}. We should note that all of the $\alpha\nu$ branches live on $\alpha\nu$ effective lattices with the same lattice length $L$. We will choose the following coordinates: let our pseudofermion depart from lattice position $x=-L/2$ and arrive at lattice position $x=L/2$, with a phase difference equal to $\delta_{\alpha\nu}(q)$.
We have
\begin{equation}
\left( q+ \frac {Q_{\alpha\nu}^0} L \right)x \rightarrow \bar{q}x=\left( q + \frac {Q_{\alpha\nu}^0+Q_{\alpha\nu}^{\Phi}(q)} L \right)x
\end{equation}
which yields
\begin{eqnarray}
-\frac {qL} 2 &\rightarrow& \frac {\bar{q}L} 2=\left( q + \frac {Q_{\alpha\nu}^0+Q_{\alpha\nu}^{\Phi}(q)} L \right) \frac L 2 = \frac {qL} 2 + \frac {Q_{\alpha\nu}^0+Q_{\alpha\nu}^{\Phi}(q)} 2 \nonumber \\
&\implies& \delta_{\alpha\nu}(q)=\frac {Q_{\alpha\nu}^0+Q_{\alpha\nu}^{\Phi}(q)} 2= \frac {Q_{\alpha\nu}(q)} 2 \label{phaseshiftdelta}
\end{eqnarray}
by comparing the momenta shifts between the ground state and the final state. Hence we can write the total phase shift of an $\alpha\nu$ pseudofermion as, Eq. (\ref{Qphase1}),
\begin{equation}
\frac {Q_{\alpha \nu}(q_j)} 2 = \frac {Q_{\alpha \nu}^0} 2 +\frac {Q_{\alpha \nu}^{\Phi} (q_j)} 2 = \frac {Q_{\alpha \nu}^0} 2 +\pi \sum_{\alpha' \nu'} \sum_{j'=1}^{N_{\alpha\nu}^*} \Phi_{\alpha \nu,\alpha' \nu'}(q_j,q_{j'}) \Delta N_{\alpha' \nu'} (q_{j'})
\end{equation}
and we obtain, by the formal definition of the S-matrix \cite{scatteringbook}, that
\begin{eqnarray}
S_{\alpha\nu}(q_j)&=&e^{2i\delta_{\alpha \nu}(q_j)}=e^{iQ_{\alpha \nu}^0} \prod_{\alpha\nu} \prod_{j'=1}^{N_{\alpha\nu}^*} S_{\alpha\nu, \alpha' \nu'} (q_j,q_{j'})= \nonumber \\
&=&e^{iQ_{\alpha \nu}^0} \prod_{\alpha\nu} \prod_{j'=1}^{N_{\alpha\nu}^*} e^{2\pi i \Phi_{\alpha \nu,\alpha' \nu'}(q_j,q_{j'}) \Delta N_{\alpha' \nu'} (q_{j'})} \label{Smatrix}
\end{eqnarray}
From this equation, we see that $\pi\Phi_{\alpha \nu,\alpha' \nu'}(q_j,q_{j'})$ measures the phase shift of the $\alpha\nu$ pseudofermion at momentum $q_j$ due to the individual scattering event with the $\alpha' \nu'$ pseudofermion at momentum $q_{j'}$. Note that it is the latter (primed) pseudofermions, that were not present in the original ground state, that make the S-matrix to differ from unity and hence it is these pseudofermions that are the {\it scattering centers} of the theory. The $\alpha\nu$ (unprimed) pseudofermions, on the other hand, are the {\it scatterers} of the theory. Note here that would we have chosen our pseudofermion to originate from $x=0$, and travel around the lattice ring until $x=L$, the resulting total phase shift would have been the same as above multiplied by $2$. However, the S-matrix would have remained the same by letting the phase shift of the individual scattering event be twice the expression given above, $2\pi\Phi_{\alpha \nu,\alpha' \nu'}(q_j,q_{j'})$. Moreover, there is nothing in this picture that distinguishes pseudofermions from pseudofermion holes, which means that an $\alpha\nu$ pseudofermion hole is also a scatterer on equal footing with the pseudofermion. In other words, whenever $\Delta N_{\alpha' \nu'}(q_{j'}) < 0$, we have that it is pseudofermion holes that act as scattering centers. Thus, the number of pseudofermions and pseudofermion holes for which the S-matrix has the form of Eq. (\ref{Smatrix}), equals the number of lattice sites for every non-empty branch (i.e. for every branch not consisting entirely out of holes).

Thus, the pseudofermion or pseudofermion hole S-matrix for the one dimensional Hubbard model is just a phase factor, given by Eq. (\ref{Smatrix}). This statement is consistent with the previous claims that the scattering events do not mix different $\eta$-spin or spin channels, and that the scatterers as well as the scattering centers are $\eta$-spin and spin zero objects. Moreover, it is not only the total energy and the total momentum that is conserved during these scattering events, but also the individual $\alpha\nu$ energies and canonical momenta components. This can most easily be seen on the form of the energy and momentum deviations, Eq. (\ref{EdevPdev}), that do not mix pseudofermion deviations from different $\alpha\nu$ branches after the substitution $q \rightarrow \bar{q}$.

Finally, we should note that the pseudofermion anticommutation relations of Eq. (\ref{anticomm}), can be solely expressed in terms of the pseudofermion S-matrix, according to

\begin{equation}
\{ f_{\bar{q},\alpha \nu}^{\dag} , f_{{\bar{q}}',\alpha' \nu'} \} = \frac {\delta_{\alpha , \alpha'} \delta_{\nu , \nu'}} {N_{\alpha \nu}^*} e^{i a_{\alpha \nu} (\bar{q}-\bar{q}') / 2} \Big[ S_{\alpha\nu} (q) \Big]^{1/2} \frac {\text{Im} \Big[ S_{\alpha\nu} (q) \Big]^{1/2}} {\sin [ a_{\alpha \nu} (\bar{q}-\bar{q}') / 2 ]} \label{acsmatr}
\end{equation}
with the important implication that the S-matrix introduced here fully controls the one electron spectral properties of the normal ordered 1D Hubbard hamiltonian, as will later become apparent by use of the anticommutation relations in the evaluation of matrix overlaps in chapter (\ref{onelecspec}).

\subsection{Properties of the pseudofermion scattering}
\label{prop}

Before closing the section on pseudofermion scattering theory, there are some properties that the theory implies which is worth mentioning, here numbered from (i) to (v).
\newline

(i) As noted by the explicit form of the S-matrix, we have reduced the many-body scattering events into two-body scattering events, as shown by the definition of $S_{\alpha\nu ,\alpha' \nu'}(q_j,q_{j'})$ in Eq. (\ref{Smatrix}). This means that the relative ordering between {\it any} pair of two-body scattering events is independent of the final expression for the S-matrix (mathematically due to the commutativity of complex numbers) \cite{CarmBoziPedro}. This is a stronger result that what an S-matrix satisfying the Yang-Baxter Equation \cite{YBE1}-\cite{YBE3} could claim \cite{holonspinon1} \cite{wrongscattering}. In these references, a representation different from the one of our scattering theory regarding the active scattering centers is made. Indeed, the scatterers and scattering centers of that representation are "spinons" and "holons" with $\eta$-spin and spin projection equal to $\pm \frac 1 2$, whilst in our theory the scattering centers are $\eta$-spinless and spinless. This explains the differing dimension of the S-matrix in those references, as compared to our representation.

(ii) One should note that the $\pm \frac 1 2$ Yang holons and the $\pm \frac 1 2$ HL spinons have not played any role in the pseudofermion scattering theory. On the contrary, they have constant momentum values during the ground state $\rightarrow$ virtual state $\rightarrow$ final state transitions, only one of them nonzero (the $-\frac 1 2$ Yang holon with momentum equal to $\pi$). This means that the S-matrix for these quantum objects equals unity, due to the absence of phase shifts. We remind ourselves that these objects are exactly the same in the original electronic frame as in the rotated electronic frame, and thus have no quantity corresponding to the $\Delta Q(q)$, Eq. (\ref{Qphase1}), previously defined for the quantum objects described by the Takahashi string hypothesis, Eq. (\ref{takacont}). We thus conclude that the $\pm \frac 1 2$ Yang holons and the $\pm \frac 1 2$ HL spinons are neither scatterers nor scattering centers.

(iii) There is an elegant theorem, called the {Levinson's theorem} \cite{levinson}, which states that in the infinite wavelength limit, i.e. when the momenta of the scatterer tends to zero in the reference frame of the scattering center, the phase shift becomes an integer multiple of $\pi$, where this integer is nothing but the number of bound states $N_b$. In our notation, this means that the momenta of the scattering center $q'$ should be replaced by $0$ and that the momenta of the scatterer $q$ should be replaced by $q-q'$ and we should thus have
\begin{equation}
\lim_{q-q' \rightarrow 0} \frac {Q_{\alpha\nu}^{\Phi} (q-q')} {\pi} = N_b
\label{levinsoneq}
\end{equation}
according to the theorem, since we in our case have that the scatterer feels the effect of the scattering centers during the virtual state $\rightarrow$ final state transition only. For our theory to comply with this theorem, we should have that the above limit is equal to zero, since we by construction have no bound states in the theory (this can also be seen mathematically: our S-matrix has no poles). In section (\ref{sectPhi}) it was found that $\Phi_{\alpha\nu ,\alpha' \nu'}(q,q')=-\Phi_{\alpha\nu ,\alpha' \nu'}(-q,-q')$ which in the alternative reference frame translates into 
\begin{equation}
\Phi_{\alpha\nu ,\alpha' \nu'}(q-q',0)=-\Phi_{\alpha\nu ,\alpha' \nu'}(-(q-q'),0)
\implies \lim_{q-q' \rightarrow 0} \Phi_{\alpha\nu ,\alpha' \nu'}(q-q',0) = 0
\end{equation}
which means that Eq. (\ref{levinsoneq}) is fulfilled for our scattering theory.

(iv) There is another exact result for which we can check our derived results, known as the {\it Fumi theorem} \cite{Mahan}, \cite{Fumi}. This theorem, originally formulated for electrons in a metal, states that the total energy $E_i$ due to the existence of an impurity upon which otherwise free electrons scatter, can be written as an integral over all the phase shifts caused by this impurity:
\begin{equation}
E_i=-\int_0^{E_F} dk \  \frac {dE(k)} {dk} \sum_l \frac {\delta_l (k)} {\pi} \label{Fumieq}
\end{equation}
where $E(k) \sim k^2$ for free electrons, and $l$ represents a set of relevant quantum numbers (for three dimensional scattering events of electrons, $l$ usually denotes the angular momentum components). 

In addition to the pseudoparticle and the pseudofermion representation, one can introduce a third related description in terms of quantum objects that carry rapidity momentum $k_j$. It will be shown elsewhere \cite{fumiprep} that the phase shift of such quantum objects obey the Fumi theorem.  The energy $E_i$ above will then correspond to that part of the energy deviation $E^{(1)}$, which is of scattering origin.

The occupied rapidity momentum $k_j=k^0(q_j)$ obeys for the ground state $-Q<k_j<Q$, i.e. the Fermi points of these quantum objects are defined by the value $Q$. Moreover, their dispersion relation goes as $E(k) \sim \cos k$, and their phase shift is $\delta_{c0}(k)=\tilde{Q}_{c0}(k) / 2$, where $\tilde{Q}_{c0}(k)$ is the quantity equivalent to $Q_{c0}(q)$ in this representation.

(v) The following final properties will only be briefly mentioned here, since they correspond to quite exotic cases of the pseudofermion theory, and are only valid for the $c\nu$ ($\nu \geq 1$) and the $s\nu$ ($\nu \geq 2$) branches and thus will not be considered in the dynamical theory. The interested reader should go to Ref. \cite{CarmBoziPedro} for a complete analysis. In section (\ref{sectPhi}) we saw that the energy bands in the case of $\alpha\nu=c\nu,s\nu$ equal zero for momenta equal to the limiting value of the $\alpha\nu$ effective Brillouin zone. Thus at these momenta points, the energies of the $c\nu$ and the $s\nu$ pseudofermions becomes the sum of the energies of the individual quantum objects of which they are constituted. One can understand this by the "handwaving" analogue that the "binding energies" between the $\pm \frac 1 2$ holons ($\alpha=c$) and $\pm \frac 1 2$ spinons ($\alpha=s$) vanish, so that there is nothing to hold the pseudofermions together. Interestingly enough, one can show that for these momentum values, and for $0<na<1$ and $0<ma<n$, we have the following equality
\begin{equation}
\pm \Delta \bar{q}=\pm \Delta q + \frac {Q_{\alpha\nu}^{\Phi}(\pm q_{\alpha\nu}^0)} L = 0
\end{equation}
which implies that the canonical momentum spacing vanishes, as opposed to canonical momenta spacings at other points in the $\alpha\nu$ canonical momentum Fourier space. We should recall that the corresponding {\it pseudoparticle} shift is nonzero whenever $\Delta N_{\alpha\nu}^* \neq 0$, i.e. whenever we have a finite shake-up. This means that not only does the $c\nu$ and $s\nu$ pseudofermions break up at the boundaries of the effective Brillouin zone, but their canonical momentum values at these boundaries are the same for the ground state as for the final state, as there are no nonzero momenta spacings that allows a shift in the canonical momentum values, thus becoming non dynamical. 

This phenomena is ultimately demonstrated by the fact that at these canonical momentum values, the $c\nu$ and the $s\nu$ pseudofermions become invariant to the unitary operator transforming electrons into rotated electrons. There is however, in the many-body system, some "memory" of these pseudofermions left, since it is possible to express $\Phi_{\alpha\nu,\alpha'\nu'}(q_{F\alpha\nu}, \iota q_{\alpha'\nu'}^0)$ (where $\alpha'\nu'=c\nu'$ or $s\nu'$ and $\iota=\pm$) solely in terms of $\Phi_{\alpha\nu,c0}(q_{F\alpha\nu},\pm q_{Fc0})$ and $\Phi_{\alpha\nu,s1}(q_{F\alpha\nu},\pm q_{Fs1})$, \cite{CarmBoziPedro}. This means that even though the $c\nu$ and $s\nu$ pseudofermions fall apart into their constituents, their momentum values are "carried over" in the system, in such a way that the $c0$ and the $s1$ pseudofermions feel the usual two-body scattering events with these pseudofermions {\it as if they were $c0$ and $s1$ pseudofermion scattering centers, respectively, at their corresponding Fermi points}. 

By letting $na\rightarrow 1$ and/or $ma\rightarrow 0$, we have that the $c\nu$ and/or the $s\nu$ bands shrink until they finally disappear in the half filled case ($c\nu$) or in the case with zero magnetization ($s\nu$). This is easily seen by the fact that the limiting momentum values for the two bands become zero, i.e. for $c\nu$ we have that $(\pi / a)-2k_F \rightarrow 0$ whilst for $s\nu$ we have that $k_{F\uparrow}-k_{F\downarrow} \rightarrow 0$. 
\setcounter{chapter}{2}
\setcounter{section}{2}

\chapter{Pseudofermion Dynamical Theory}
\label{dynamics}
\section{Excited energy eigenstates}

\subsection{Introduction (spectral function)}
\label{class}

The goal of this chapter is to derive a pseudofermion dynamical theory which will enable us to calculate the one electron spectral function for the Hubbard hamiltonian (\ref{H}). By "one electron spectral function" we mean the spectral function for one electron removal and one electron addition, respectively. The removal function gives rise to spectral weight in the so called {\it Removal Hubbard Band} (RHB) whilst the addition function gives rise to the {\it Lower Hubbard Band} (LHB) and the {\it Upper Hubbard Band} (UHB), respectively. For $(U/t) \rightarrow \infty$, these latter two bands are separated by an energy proportional to the effective Coloumb interaction strength $U$, since the LHB gives the spectral weight for an electron added at an empty site whilst the UHB gives the spectral weight for an electron added at a singly occupied site. There is only one band in the removal case due to the fact that our ground state, upon which we act with suitable electronic creation or annihilation operators, is a LWS that is void of doubly occupied sites. In this thesis report, the spectral function for the RHB and the LHB will be calculated, even though the method used here is general and can perfectly well be applied to the UHB as well as to correlation functions involving creation or annihilation of several electrons \cite{spectral1} \cite{Carmspec1}. We will use arbitrary values for the parameters ($U/t$) and $n$, but we will keep a small magnetization $ma>0$ and later let $ma \rightarrow 0$ (to confine $ma$ to zero at the start of the calculations has been seen to be quite pathological, for reasons that we will give later). Formally at zero temperature, a spectral function is defined as the imaginary part of the time ordered Green's function $G(k,\omega)$ at electron momentum $k$ and electron energy $\omega$, multiplied by a constant for which there is no conventional fixed value, but that is uniquely defined by applying suitable sumrules. These sumrules stem from the fact that the spectral function is interpreted as a probability function, and that thus the integral over the domain of this function must equal a certain positive value. The definition of the one electron removal spectral function $B^-(k,\omega)$ and the one electron addition spectral function $B^+(k,\omega)$ is
\begin{eqnarray}
B^-(k,\omega)&=&\sum_{\sigma=\uparrow,\downarrow} \sum_{f_-} \big\vert \langle f_- \vert c_{k\sigma} \vert GS \rangle \big\vert^2 \delta \left( \omega - \Delta E_- \right) \hspace{1.0cm} (RHB) \nonumber \\
B^+(k,\omega)&=&\sum_{\sigma=\uparrow,\downarrow} \sum_{f_+} \big\vert \langle f_+ \vert c_{k\sigma}^{\dag} \vert GS \rangle \big\vert^2 \delta \left( \omega - \Delta E_+ \right) \hspace{1.0cm} (LHB) \label{defspecfs}
\end{eqnarray}
where the summation over $\sigma$ will yield nothing but a factor of $2$ in the zero magnetization limit since creating or annihilating a $\uparrow$-spin electron will give exactly the same spectral function as creating or annihilating a $\downarrow$-spin electron. These spectral functions are then directly proportional to the probability of finding the added electron or the added electron hole at momentum $k$ and energy $\omega$, respectively. Many times we will summarize these two functions by using $l=\pm$ (which we will treat equivalently to $l=\pm 1$):
\begin{equation}
B^l(k,\omega)=\sum_{f_l} \big\vert \langle f_l \vert c_{k\sigma}^l \vert GS \rangle \big\vert^2 \delta \left( \omega - \Delta E_l \right) \hspace{1.0cm} l=\pm
\end{equation}
where $c_{k\sigma}=c_{k\sigma}^-$ and $c_{k\sigma}^{\dag}=c_{k\sigma}^+$. $\vert f_l \rangle$ denotes a final state, i.e. the energy eigenstate of the $N+l$ electron system, where $N$ is the number of electrons in the ground state, here denoted by $\vert GS \rangle$. Due to the nonzero phase shift of the pseudofermions, the states $\vert f_l \rangle$ and $\vert GS \rangle$ have different boundary conditions for each final state and $\alpha \nu$ pseudofermion branch. This implies that the evaluation of the matrix overlaps leads to the orthogonal catastrophe, originally due to the canonical momentum shifts of order ($1/L$) of the theory \cite{ortognl1} \cite{ortognl2}. Since the scattering phase shift is state dependent, we would expect a different contribution due to the orthogonal catastrophe for each ground state $\rightarrow$ final state transition. $\Delta E_l$ is the energy difference between the ground state and the final state, defined according to
\begin{equation}
\Delta E_l = l \left( E_{f_l} - E_{GS} \right)
\end{equation}
This definition of $\Delta E_l$ measures the energies relative to the chemical potential, that hence never enters the calculations explicitly. The relation to the zero temperature Green's functions is
\begin{eqnarray}
B^-(k,\omega)&=& - \frac 1 {\pi} \ \text{Im} \left\{ G(k,\omega) \right\} \hspace{1.0cm} \omega < 0 \nonumber \\
B^+(k,\omega)&=& \frac 1 {\pi} \ \text{Im} \left\{ G(k,\omega) \right\} \hspace{1.35cm} \omega > 0
\end{eqnarray}
where $\text{Im}$ is the imaginary part. The Kramer-Kronig relations give us the inversion of these relationships, expressing $G(k,\omega)$ in terms of $B^l(k,\omega)$:
\begin{equation}
G(k,\omega)= \int_0^{\infty} d\omega' \ \frac {B^+ (k,\omega')} {\omega-\omega'+i\delta_0} + \int_{-\infty}^0 d\omega' \ \frac {B^- (k,\omega')} {\omega-\omega'-i\delta_0}
\end{equation}
where $\delta_0$ is a positive infinitesimal quantity. This Green's function can, at least formally, be used to obtain the expectation value of {\it any} one electron correlation function \cite{Mahan} \cite{fetterwalecka}. The momentum distribution function $n_k$ is just the $\omega$ integral over $B^-(k,\omega)$. Integrating this function over $k$ gives then the density of electrons $na$. The remaining spectral weight, from the LHB and the UHB spectral functions respectively, must then by construction have a $k$ and $\omega$ integrated value of $\ 2-na$. These integral values constitute the sum rules for the spectral functions
\begin{eqnarray}
\int_{-\infty}^{\infty} \frac {dk} {2\pi} \ \int_{-\infty}^0 d\omega \ B^-(k,\omega) &=& na \nonumber \\
\int_{-\infty}^{\infty} \frac {dk} {2\pi} \ \int_0^{\infty} d\omega \ B^+(k,\omega) &\approx& 2(1-na) \label{sumrule}
\end{eqnarray}
where the approximative sign in the last equality stands for the (very) weak ($U/t$) dependence in the sum rule \cite{UHBweight}. This dependence is due to the allocated weight in the UHB, which varies slightly as ($U/t$) varies, and is approximatively equal to $na$. We see that at half filling, $na \rightarrow 1$, the LHB weight vanishes, as all the one electron addition spectral weight is transferred to the UHB.

One could argue that since the exact wave function of the model is known \cite{explicitwavefcn}, the calculation of $B^l(k,\omega)$ is just a matter of explicit brute force calculation. Unfortunately, however, the complex form of this wave function and the fact that it has remained unknown how to express the generators of the excited energy eigenstates in terms of electronic creation and annihilation operators, it has so far been practically impossible to calculate $B^l(k,\omega)$ by brute force. This means that, from a theoretical standpoint, we have to choose an approach between either finding alternative methods, using the exact solution in some limit that simplifies the expressions, or discovering the missing link between the electrons and the quantum objects that diagonalize the normal ordered hamiltonian. The pseudofermion theory allows to do the latter, but the main contributions in the literature on the subject has primarily been focused on the former. 

In Ref. \cite{spectralattempt1} a lattice dividing technique is used to calculate the spectral weight for small system sizes, with $(U/t)=4$, $na=1$ and $na=\frac 5 6$. The purpose of dividing the lattice into small "clusters" is that the Green's function can be obtained by exact diagonalization, when the system size is very small. The "intercluster" hopping integral is then treated perturbatively to obtain the full Green's function. In the perturbation theory, the exactly solvable hamiltonian is taken to be the entire Hubbard hamiltonian for one cluster (using cluster size of 12 lattice sites). Like this, a spectral weight for the cases of RHB and LHB with somewhat distinguishable spin and charge dispersions is obtained, even though the shape of the two dimensional surface in the ($k,\omega$) plane could be improved.

Another method was used in Ref. \cite{spectralattempt2}, using parameter values suitable for comparison with experimental results on the charge transfer salt TTF-TCNQ, namely filling $na=0.59 \approx 0.6$ and effective Coloumb repulsion $(U/t)=4.9$. For this the "dynamical density matrix renormalization group method" (DDMRG) was employed. With open boundary conditions and system sizes up to 90 lattice sites, the spectral weight was calculated by using the eigenstates of the particle-in-a-box problem in the DDMRG routine. A more reliable association of the spectral weight with different quantum object dispersions is made, as well as some estimates for the exponents with which the spectral function diverges along the dispersive lines. Unfortunately, the DDMRG routine becomes non applicable as the system size approaches the thermodynamic limit. 

In Refs. \cite{spectralhalffill1} and \cite{spectralhalffill2},  the one electron spectral function is investigated using Green's functions, conformal field theory (briefly described below) and the Bethe ansatz solution in the half filled Mott-Hubbard insulating phase, for finite values of ($U/t$). Using a holon and spinon picture, some lines along which these holons and spinons disperse are identified to display singular features of the spectral function, with momentum line shape dependent exponents (however, these holons and spinons are different than the quantum objects dubbed "holons" and "spinons" in this thesis report).
 
In Refs. \cite{Karlo1}-\cite{Karlo3}, the RHB and the LHB spectral functions were calculated in the $(U/t)\rightarrow \infty$ limit, by using the exact solution and exploiting the wave function factorization in that limit. The quantum objects that describe the occupancies of the two resulting parts of the factorized wave function are called "spinless fermions" and "spinons", respectively, and account for the spin-charge separation in the $(U/t) \rightarrow \infty$ limit. The momentum of the spin wave, obtained by mapping the spin part of the Hubbard hamiltonian to the 1D Heisenberg spin hamiltonian, imposes a twisted boundary condition on the otherwise periodic lattice of the spinless fermions. This is however, the only remnant of the coupled Takahashi equations (\ref{Takahashi}). Moreover, the spin spectrum collapses as demonstrated by the fact that the group velocity of the propagation of the spin wave goes to zero in this limit. The "spinless fermions" introduced in these references are nothing but the $c0$ pseudofermions as $(U/t) \rightarrow \infty$. Thus they are not "real" fermions, but rather quantum objects obeying the Haldane statistics \cite{haldane}, introduced in section (\ref{pseudoops}). However, these quantum objects account for all the excited states available since at infinite repulsion there can exist no $c\nu$ pseudofermions at finite energy, and at zero magnetization there can exist no $s\nu$ pseudofermions, due to the fact that the $s\nu$ band is non existent for the ground state at zero magnetization. The technique of calculating the matrix overlap of the charge part, expressing the full matrix overlap in the spectral function as a determinant of the anticommutators of the spinless fermion operators, will in this work be generalized to finite values of ($U/t$).

Some of the main difficulties of this model lay in the fact that the Hubbard hamiltonian cannot be treated by perturbative methods due to the non pertubative character of the electronic interactions. Indeed, in 1D the Coloumb interaction parameter $U$, however weak, qualitatively changes the correlations between the electrons, as compared to the free system. By restricting the Hilbert space to low energy eigenstates only, we can apply various methods that ultimately depend on the linearization of the elementary excitation energy bands. The assumption one makes is that all relevant low lying excitations can be constructed by taking into account states with momenta close to $\pm k_F$ only. The technique of {\it bosonization} separates the hamiltonian into two bosonic hamiltonians, one describing the charge part and another the spin part. In this way, the problem is reformulated into two massless bosonic theories describing the charge ($c$) and spin ($s$) degrees of freedom, respectively, with dispersions $\omega_{c,s}(k)=v_{c,s} \left( k \pm k_{c,s} \right)$, where $v_{c,s}$ and $k_{c,s}$ are the Fermi velocity and Fermi momentum of the $c$ (charge) or $s$ (spin) branches, respectively. The charge-spin separation of the electronic degrees of freedom is thus explicit for the bosonization technique \cite{boson1}-\cite{boson3}. In the case of weak coupling, it is then possible to compute the critical exponents of the correlation functions for the Hubbard model \cite{crit1}-\cite{crit4}. 

Other powerful techniques used to calculate correlation functions in the same low energy excitation regime are {\it conformal field theories} and {\it finite size scaling}. The basic ideas of these techniques are simple scaling arguments, due to the fact that at large distances the behavior of correlation functions does not depend on the microscopic hamiltonian. Moreover, correlation functions for systems without any internal scale have to decay algebraically, for example as simple power laws, due to the universality class of the Hubbard hamiltonian. The exponents of these power laws of the conformal theory are then used to obtain finite size corrections of the energy and momentum \cite{crit5}-\cite{crit7}. In this way, the low lying excitations can be obtained as "towers" of states by adding ($\pm 2\pi v_{c,s}/L$) to the energy and ($\pm 2\pi/L$) to the momentum \cite{conf1}. The spectral function and its exponents have been described near the Fermi points of the elementary excitations (usually referred to as charge and spin excitations respectively) \cite{spectrallutt1}-\cite{spectrallutt3}. In these references, the spectral function of the related Tomonaga-Luttinger model was examined, and yielded the characteristic Luttinger-type power-law behavior in the vicinity of the elementary excitation Fermi energies: $\left[ \omega_{c,s}-v_{c,s} \left( k \pm k_{c,s} \right) \right]^{\gamma_1}$, where the exponent $\gamma_1$ is given in these references.

In the following, we will attempt to calculate this spectral function for all energy scales, i.e. for the entire ($k,\omega$) domain. Obviously, we will exploit some of the ideas briefly discussed here, for example, we will see that for each pseudofermion branch, the spectral weight close to the dispersive lines in the ($k,\omega$) plane obeys simple power law behaviors whose exponents are related to the tower of states close to these lines. In section (\ref{classstates}) we will classify the processes leading to the final states of the model, as well as the various subspaces that these final states span. However, we will simplify many expressions of the general theory (as presented in Refs. \cite{Carmspec1} and \cite{Carmspec2}). These simplifications are heavily dependent on the findings of Ref. \cite{spectral1}, in which the partial sum rules, i.e. contributions to the total sum rule from different excited state subspaces, are measured. Thus the use of the theory developed here will involve approximations in terms of compliance with the sum rules, however {\it all relevant features} of the one electron spectral function will be accounted for. The theory presented in the following is formally developed in Refs. \cite{Carmspec1} and \cite{Carmspec2}.

\subsection{Fourier transform and rotated electrons}

Since the final states are energy eigenstates of the hamiltonian, which in turn is diagonalized in the pseudofermionic basis, it would be suitable to describe all quantities in terms of pseudofermions. Thus, apart from the description of the final states in terms of occupancies of pseudofermions, we need to express the electronic creation and annihilation operators in terms of their pseudofermionic counterparts. The goal is to allow for a unique description of the generators of all relevant eigenstates in terms of pseudofermionic creation operators acting onto the vacuum. Our first step however, is a little bit more modest. Since the unitary transformation that maps electrons onto rotated electrons is defined with local operators, we need to Fourier transform the $c_{k\sigma}^l$ operators appearing in the defining expressions for the spectral functions, into operators creating or annihilating local electrons. Using Eq. (\ref{Fourierelectr}), we obtain a sum over the lattice sites $j$ inside the spectral function which, due to the translational invariance of the system, reduces to $N_a$ times one typical term of the sum, say the term with $j=0$. Also, since the spectral function continues to be a function of the momentum $k$, the above mentioned translational invariance introduces a Kr\"{o}necker $\delta$-function:
\begin{equation}
B^l(k,\omega)= N_a \sum_{f_l} \big\vert \langle f_l \vert c_{0\sigma}^l \vert GS \rangle \big\vert^2 \delta \left( \omega - \Delta E_l \right) \delta_{k,\Delta P_l} \label{Belectr}
\end{equation}
The next step is to express the electronic operator in terms of rotated electronic operators. Even though a closed form expression relating the former in terms of the latter is unknown, there are some things that can be done to shed some light on the procedure. By using the $(U/t) \gg 1$ expansion of $\hat{V}(U/t)$ presented in section (\ref{sectrotE}), we have for example to first order, that
\begin{equation}
c_{i\sigma} = e^{\hat{Y^{(1)}}} \tilde{c}_{i\sigma} e^{-\hat{Y^{(1)}}} = \tilde{c}_{i\sigma} + \left[ \hat{Y^{(1)}} , \tilde{c}_{i\sigma} \right] + \ldots = \tilde{c}_{i\sigma} + \frac 1 U \left[ \tilde{T}_U-\tilde{T}_{-U} , \tilde{c}_{i\sigma}  \right] + \ldots 
\end{equation}
which after introducing the explicit expressions for $\tilde{T}_U$ and $\tilde{T}_{-U}$ and evaluating the commutators yields
\begin{equation}
c_{i\sigma}= \tilde{c}_{i\sigma} - \frac t U \sum_{\delta=\pm 1} \left[ \tilde{c}_{i+\delta,\sigma} \left( \tilde{n}_{i+\delta,\bar{\sigma}} - \tilde{n}_{i\bar{\sigma}} \right) - \tilde{c}_{i\bar{\sigma}}^{\dag}\tilde{c}_{i+\delta,\bar{\sigma}}\tilde{c}_{i\sigma}  + \tilde{c}_{i+\delta,\bar{\sigma}}^{\dag} \tilde{c}_{i\bar{\sigma}} \tilde{c}_{i\sigma} \right] +\mathcal{O} \left( \frac {t^2} {U^2} \right) \label{crotexp}
\end{equation}
where $\bar{\sigma}=-\sigma$. Note that as expected,
\begin{equation}
\tilde{c}_{i\sigma} \rightarrow c_{i\sigma} \hspace{2.0cm} \frac t U \rightarrow 0
\end{equation}
which expresses the fact that the electron - rotated electron unitary transformation becomes the identity transformation in this limit. This shows an example of how to replace the electronic operator with the rotated electronic operator. However, we will not be depending on the large-($U/t$) expansion of the electron - rotated electron unitary transformation from now on, but instead use physical reasoning when introducing the rotated electrons into the problem. Our basic consideration is based upon the results of Refs. \cite{spectral1}, \cite{Carmspec1} and \cite{Carmspec2}. In these references it is shown that for the one electron spectral weight, the substitution $c_{i\sigma}^l \rightarrow \tilde{c}_{i\sigma}^l$ accounts for over $99\%$ of the total spectral weight, as measured by the sum rules (\ref{sumrule}). This does not mean however, that we let $(U/t)=\infty$ by the erroneous assumption that we only keep the first term of the expansion of $\hat{V}(U/t)$, Eq. (\ref{crotexp}). On the contrary, all quantities will be evaluated for the actual value of ($U/t$) that the original problem refers to. For example, the phase shifts $Q_{\alpha\nu}^{\Phi}(q)$ are strongly dependent on ($U/t$) via the functions $\Phi_{\alpha\nu,\alpha'\nu'} (q,q')$, and there will be no limiting procedure in evaluating these functions, which means that all exponents will inherit this ($U/t$) dependence as well. 

The substitution $c_{i\sigma}^l \rightarrow \tilde{c}_{i\sigma}^l$ is just a statement of the fact that the subsequent terms only contribute marginally to the total spectral weight, with the $i=1$ term contributing not more than $1\%$ to the sum rules. However, one could argue that since we are measuring deviations from the exact sum rule there could still be some singular behavior that is left unaccounted for. This could be the case with a very strong (i.e. narrow) singularity with a dominant contribution as compared to other singular features but with a small contribution to the total sum rule. That this is not the case is easily concluded from the type of terms in the expansion of Eq. (\ref{crotexp}): the higher order terms are generated by particle-hole processes of rotated electrons and do not bring about any new {\it types} of excitations that could lead to some sort of critical behavior that the first term does not bring about (an example of a new type of excitation would be, for example, a net creation or annihilation of $2$ rotated electrons in the ($t/U$) term). Eq. (\ref{Belectr}) is then rewritten as
\begin{equation}
B^l(k,\omega)= N_a \sum_{f_l} \big\vert \langle f_l \vert \tilde{c}_{0\sigma}^l \vert GS \rangle \big\vert^2 \delta \left( \omega - \Delta E_l \right) \delta_{k,\Delta P_l}
\end{equation}
where the final states are described by occupations of rotated electrons, which we wish to reformulate in terms of pseudofermions. 

\subsection{Classification of the eigenstates of the model}
\label{classstates}

The purpose of this section is to "sketch" the decomposition of the state summation appearing in the expression for the spectral function, into summations over subspaces defined by pseudofermion deviational numbers and occupational configurations. The mathematical details necessary for an exact computation of the spectral function will then be presented in subsequent sections. For reasons apparent in section (\ref{YangHL}), the following theory will not include any finite numbers of $-\frac 1 2$ Yang holons or HL spinons, to the contrary of the theory developed in Refs. \cite{Carmspec1} and \cite{Carmspec2}. This will simplify the definitions of the relevant processes and subspaces introduced in this section.

In section (\ref{pseudofermionII}) the PS subspace was introduced. This subspace is spanned by the ground state and all states generated from it by a finite number of electronic processes, i.e. by finite deviations $\Delta N$ and $\Delta (N_{\uparrow}-N_{\downarrow})$. Now, due to Eqs. (\ref{deltaN}) and (\ref{deltaNupNdown}), these quantities are uniquely expressible in terms of deviations of pseudofermions, $- \frac 1 2$ Yang holons, and $- \frac 1 2$ HL spinons, respectively:
\begin{eqnarray}
\Delta N &=& \Delta N_{c0}+2 L_{c,-\frac 1 2} +2\sum_{\nu=1}^{\infty} \nu N_{c\nu} \nonumber \\
\Delta (N_{\uparrow}-N_{\downarrow}) &=& \Delta N_{c0}-2 L_{s,-\frac 1 2} -2\Delta N_{s1}-2\sum_{\nu=2}^{\infty} \nu N_{s\nu} \label{devnums}
\end{eqnarray}
which means that we can equivalently say that the PS is spanned by the ground state and all excited energy eigenstates with finite deviations of $c0$ and $s1$ pseudofermions, with finite (or zero) numbers of $c\nu$ ($\nu \geq 1$) and $s\nu$ ($\nu \geq 2$) pseudofermions and with finite (or zero) numbers of $- \frac 1 2$ Yang holons and $- \frac 1 2$ HL spinons, respectively. One should note that 1) the ground state, labeled $\vert GS \rangle$, is void of $c\nu$ ($\nu \geq 1$) and $s\nu$ ($\nu \geq 2$) pseudofermions as well as of $- \frac 1 2$ Yang holons and $- \frac 1 2$ HL spinons, respectively, 2) apart from pure creation and annihilation of pseudofermions, the generation of the excited states which span the PS also involves a finite number of particle-hole processes in the $\alpha\nu=c0,s1$ bands, 3) the $\alpha\nu \neq c0,s1$ branches have no ground state pseudofermion occupancy and thus do not have any Fermi points. However in these cases, the limiting canonical momentum values for the $\alpha\nu$ effective Brillouin zone play the role of the Fermi points. To summarize from section (\ref{occpp}):
\begin{eqnarray}
N_{c0}^0 (q)&=&\theta (q_{Fc0}-\vert q \vert)= \theta (2k_F-\vert q \vert)\hspace{1.0cm} \vert q \vert \leq q_{c0}^0=\frac {\pi} a \nonumber \\
N_{s1}^0 (q)&=&\theta (q_{Fs1}-\vert q \vert)= \theta (k_{F\downarrow}-\vert q \vert)\hspace{1.05cm} \vert q \vert \leq q_{s1}^0=k_{F\uparrow} \\
N_{\alpha \nu}^0 (q)&=&0 \hspace{5.7cm} \vert q \vert \leq q_{\alpha \nu}^0 \nonumber
\end{eqnarray}
where
\begin{eqnarray}
q_{s \nu}^0&=&k_{F\uparrow}-k_{F\downarrow} \hspace{2.95cm} \nu=2,3,\ldots \nonumber \\
q_{c \nu}^0&=&\frac {\pi} a-2k_F=\pi (\frac 1 a-n) \hspace{1.0cm} \nu=1,2,\ldots
\end{eqnarray}

In the following, let the index $\iota=\pm$ denote the left Fermi point and/or shake-up discrete momentum shift of pseudofermions towards smaller canonical momentum values ($\iota=-$) and the right Fermi point and/or shake-up discrete momentum shift of pseudofermions towards larger canonical momentum values ($\iota=+$). Let us now count the number of pseudofermions created and annihilated at the Fermi points ($\alpha\nu=c0,s1$) and at the limiting canonical momentum values of the effective Brillouin zone ($\alpha\nu \neq c0,s1$) on the one hand, and the number of pseudofermions created and annihilated away from these points on the other. The principal reason for this division is due to the fact that in the continuum momentum limit, the Fermi seas become compact since the momentum spacing $(2\pi /L) \rightarrow 0$. Hence  a non zero phase shift $Q_{\alpha\nu}^{\Phi}(q)$ inside the Fermi sea can not be detected due to the uniform occupation of pseudofermions. 

However, the situation is different with the {\it outmost} canonical momentum values $\iota q_{F\alpha\nu}$ that become shifted to a value for which there are no occupancies on the positive ($\iota=+$) and the negative ($\iota=-$) side of that canonical momentum value. This means that the value of $Q_{\alpha\nu}^{\Phi}(\iota q_{F\alpha\nu})$ defines the new {\it canonical Fermi points}, which are unique for each value of ($U/t$), $n$ and the transition in consideration. Thus, creating or annihilating pseudofermions at their Fermi points should contribute much more to the dynamics than the corresponding actions on pseudofermions inside the Fermi sea. 

Let the number of $\alpha\nu=c0,s1$ pseudofermions created or annihilated at the positive ($\iota=+$) and negative ($\iota=-$) Fermi points respectively, be denoted by $\Delta N_{\alpha\nu,\iota}^{0,F}$. For $\alpha\nu \neq c0,s1$ the definition is the same but with "Fermi point" replaced by "limiting effective Brillouin zone canonical momentum value" (however, we will stick to "Fermi point" even when $\alpha\nu \neq c0,s1$, keeping in mind that we actually refer to the limiting effective Brillouin zone canonical momentum values). If we add the extra contribution from the shake-up effect, we obtain a number $\Delta N_{\alpha\nu,\iota}^F= \Delta N_{\alpha\nu,\iota}^{0,F} + \iota Q_{\alpha\nu}^0 / 2\pi$ of $\alpha\nu$ pseudofermions at the positive ($\iota=+$) and negative ($\iota=-$) Fermi points respectively. This "half particle addition" reflects the shift from integers or half-odd integers to half-odd integers or integers, respectively, of the quantum numbers introduced by the Takahashi string hypothesis. Thus, the total number deviation of $\alpha\nu$ pseudofermions at the Fermi points is $\Delta N_{\alpha\nu}^F=\Delta N_{\alpha\nu,+}^F+\Delta N_{\alpha\nu,-}^F$. Similarily, we define the $\alpha\nu$ {\it pseudofermion current deviation} $\Delta J_{\alpha\nu}^F$ logically as the difference between the number of $\alpha\nu$ pseudofermions created or annihilated at the right and the left Fermi points respectively, i.e. $2\Delta J_{\alpha\nu}^F= \Delta N_{\alpha\nu,+}^F-\Delta N_{\alpha\nu,-}^F$. We thus obtain the following
\begin{equation}
\left.
\begin{array} {c}
\Delta N_{\alpha\nu}^F=\Delta N_{\alpha\nu,+}^F+\Delta N_{\alpha\nu,-}^F \\
2\Delta J_{\alpha\nu}^F= \Delta N_{\alpha\nu,+}^F-\Delta N_{\alpha\nu,-}^F
\end{array}
\right\} \implies \Delta N_{\alpha\nu,\iota}^F = \iota \Delta J_{\alpha\nu}^F + \frac {\Delta N_{\alpha\nu}^F} 2
\end{equation}
The corresponding number deviation of $\alpha\nu$ pseudofermions created or annihilated away from the right and the left Fermi points respectively, is denoted $\Delta N_{\alpha\nu}^{NF}$.

These different types of deviational numbers correspond to different types of ground state $\rightarrow$ final state processes. We will classify these processes as $\mathbb{A}$, $\mathbb{B}$ and $\mathbb{C}$ respectively, according to:

\begin{itemize}
  \item $\mathbb{A}$: Creation or annihilation of $\alpha\nu$ pseudofermions {\it away from} the Fermi points ($\alpha\nu=c0,s1$) or the limiting canonical momentum values for the effective Brillouin zone ($\alpha\nu \neq c0,s1$). This is a finite energy and finite momentum process, affecting the number $\Delta N_{\alpha\nu}^{NF}$.
  \item $\mathbb{B}$: Creation or annihilation of $\alpha\nu$ pseudofermions {\it at} the Fermi points ($\alpha\nu=c0,s1$) or the limiting canonical momentum values for the effective Brillouin zone ($\alpha\nu \neq c0,s1$). For the $\alpha\nu=c0,s1$ branches, this is a zero energy and finite momentum process affecting the number $\Delta N_{\alpha\nu}^F$. It transforms the $c0$ and the $s1$ densely packed ground state configurations into {\it excited state} densely packed configurations.
  \item $\mathbb{C}$: Small momentum and low energy particle-hole processes near the $\alpha\nu = c0,s1$ left and right Fermi points respectively, relative to the densely packed configurations obtained through processes $\mathbb{B}$. For these processes, we will assume that the $c0$ and the $s1$ pseudofermions disperse linearly.
\end{itemize}

At a later stage, we will use this classification of the pseudofermion processes when expressing the (rotated) electronic operators in terms of pseudofermionic operators. In Ref. \cite{Carmspec2} there is a similar classification of the different types of processes, however also including the numbers $L_{\alpha,-\frac 1 2}$. For reasons apparent in section (\ref{YangHL}), we do not need to include these numbers here. Also, we will not include particle-hole processes which are not in the vicinity of the Fermi points, due to the very small effect these excitations have on the one electron spectral weight \cite{spectral1}.

As the ground state is well defined in terms of occupational numbers of $\alpha\nu$ pseudofermions, the excited energy eigenstates are well defined in terms of deviational numbers of these pseudofermions. Indeed, each excited state is characterized by a number $N_{\alpha\nu}^0+\Delta N_{\alpha\nu}$ of pseudofermions, as well as by a number $L_{\alpha,-\frac 1 2}$ of Yang holons ($\alpha=c$) and HL spinons ($\alpha=s$). For each combination of these numbers, there exists then a subspace of many states all with the same deviational numbers, but with different canonical momentum dependent occupancy configurations. This is particularly evident when considering the $\alpha\nu=c0,s1$ particle-hole processes $\mathbb{C}$, which for each set of numbers $\left\{ \Delta N_{\alpha\nu} \right\}_{\alpha\nu=c0,s1}$ correspond to many different eigenstates of the model with each of these having a different configuration of $c0$ and $s1$ pseudofermions, as given by $\Delta N_{\alpha\nu} (q)$ where $\sum_q \Delta N_{\alpha\nu} (q) = \Delta N_{\alpha\nu}$. 

The subspace of states with fixed deviational numbers $\Delta N_{\alpha\nu}$ for all $\alpha\nu$ branches is called "c0 pseudofermion, holon and spinon ensemble subspace", abbreviated "CPHS ensemble subspace", in Refs. \cite{Carmspec1} and \cite{Carmspec2}, if we also include fixed (possibly nonzero) numbers $L_{\alpha,-\frac 1 2}$ of Yang holons ($\alpha=c$) and HL spinons ($\alpha=s$). 

Note that there can be many states characterized by the same values of the deviational numbers $\Delta N_{\alpha\nu}$, but with different values of the deviational numbers  $\Delta N_{\alpha\nu,\iota}^F$ (for $\alpha\nu=c0,s1$ and $\iota=\pm$). Moreover, we note that for each state with a fixed number of the following deviations and numbers: $\Delta N_{\alpha\nu}$, $\Delta N_{\alpha\nu,\iota}^F$ and $L_{\alpha,-\frac 1 2}$ (for $\alpha\nu=c0,s1$ and $\iota=\pm$), there exists a subspace  with a total number of $N_{\alpha\nu}^{ph}$ particle-hole pairs for $\alpha\nu=c0,s1$, due to the $\mathbb{C}$ processes. A typical element of this subspace contains one specific particle-hole configuration of $N_{c0}^{ph}$ number of $c0$ pseudofermion pairs and $N_{s1}^{ph}$ number of $s1$ pseudofermion pairs, with energy and momentum as specified by the numbers $m_{\alpha\nu,\iota}$ of Eq. (\ref{enemoms}). Therefore, we will see that the summation over the particle-hole towers of states will reduce to the summation over the integer numbers $m_{\alpha\nu,\iota}$.

The total number of subspaces here considered is less than what is being considered in Refs. \cite{Carmspec1} and \cite{Carmspec2}. However, since we only study one electron spectral functions, to the contrary of the studies of these references, we can afford to simplify the subspace descriptions and minimize the total number of subspaces needed. 

\subsection{Energy, momentum and number deviations}
\label{tjontiflex}

As before, we will not bother with finite occupancies of $L_{\alpha,-\frac 1 2}$, due to the findings of section (\ref{YangHL}), which will simplify the expressions for the energy and momentum deviations. 

The energy deviations $\Delta E_{\alpha\nu} = \Delta E_{\alpha\nu} (\mathbb{A}) +\Delta E_{\alpha\nu} (\mathbb{C})$ for $\alpha\nu = c0,s1$ give the total energy deviation from the ground state due to the $\mathbb{A}$ and $\mathbb{C}$ processes, respectively, and similarily $\Delta E_{\alpha\nu} = \Delta E_{\alpha\nu} (\mathbb{A})$ for $\alpha\nu \neq c0,s1$ branches. We remind ourselves that the $\mathbb{B}$ processes are zero energy processes. After defining the Fourier transform of the electronic operator inside the spectral function, we will obtain another delta function for the momenta, of the form $\delta_{k,l\Delta P}$, where $\Delta P = \sum_{\alpha\nu} \Delta P_{\alpha\nu}$. Here we have $\Delta P_{\alpha\nu} = \Delta P_{\alpha\nu} (\mathbb{A})+ \Delta P_{\alpha\nu} (\mathbb{B})+ \Delta P_{\alpha\nu} (\mathbb{C})$ for $\alpha\nu = c0,s1$ and $\Delta P_{\alpha\nu} = \Delta P_{\alpha\nu} (\mathbb{A})+ \Delta P_{\alpha\nu} (\mathbb{B})$ for $\alpha\nu \neq c0,s1$ (however note that  $\Delta P_{s\nu} (\mathbb{B})=0$, for $\nu \geq 2$). The above energy spectra are given in terms of the energy bands in Eq. (\ref{rele}) together with the pseudofermion number deviations under consideration. We thus obtain a connection between the variables of the spectral function and the energies and momentum occupancies of our pseudofermions:
\begin{eqnarray}
\Delta E_{\alpha\nu} \left( \mathbb{A} \right) &=& \text{sgn}\left( \Delta N_{\alpha\nu}^{NF} \right) \sum_{i=1}^{\vert \Delta N_{\alpha\nu}^{NF} \vert} \epsilon_{\alpha\nu} (q_i) \nonumber \\
\Delta E_{\alpha\nu} \left( \mathbb{C} \right) &=&  \frac {2\pi v_{\alpha\nu}} L \sum_{\iota=\pm} m_{\alpha\nu,\iota} \hspace{2.1cm} \left( \alpha\nu=c0,s1\right) \nonumber \\
\Delta P_{\alpha\nu} \left( \mathbb{A} \right) &=& \text{sgn}\left( \Delta N_{\alpha\nu}^{NF} \right) \sum_{i=1}^{\vert \Delta N_{\alpha\nu}^{NF} \vert} q_i \hspace{1.0cm} \left( \alpha\nu=c0,s1\right) \nonumber \\
\Delta P_{c\nu} \left( \mathbb{A} \right) &=&\left( 1+ \nu \right) \pi \Delta N_{c\nu}^{NF} - \sum_{i=1}^{\Delta N_{c\nu}^{NF}} q_i \label{enemoms} \\
\Delta P_{s\nu} \left( \mathbb{A} \right) &=& \sum_{i=1}^{\Delta N_{s\nu}^{NF}} q_i \nonumber \\
\Delta P_{c0} \left( \mathbb{B} \right) &=& 4k_F  \left( \Delta J_{c0}^F + \sum_{\nu=1}^{\infty} \Delta J_{c\nu}^F + \sum_{\nu=2}^{\infty} \Delta J_{s\nu}^F \right) \hspace{1.0cm} \left( na \neq 1 \right) \nonumber \\
\Delta P_{s1} \left( \mathbb{B} \right) &=& 2k_{F\downarrow}  \left( \Delta J_{s1}^F -2 \sum_{\nu=2}^{\infty} \Delta J_{s\nu}^F \right) \hspace{2.6cm} \left( ma \neq 0 \right) \nonumber
\end{eqnarray}
\begin{eqnarray}
\Delta P_{c\nu} \left( \mathbb{B} \right) &=& \pi \sum_{\nu=1}^{\infty} \nu N_{c\nu}^F \nonumber \\
\Delta P_{\alpha\nu} \left( \mathbb{C} \right) &=&  \frac {2\pi} L \sum_{\iota=\pm} \iota m_{\alpha\nu,\iota} \hspace{2.6cm} \left( \alpha\nu=c0,s1\right) \nonumber
\end{eqnarray}
where the expressions for $\Delta P_{c0} \left( \mathbb{B} \right)$ and $\Delta P_{s1} \left( \mathbb{B} \right)$ stem from the effect accounted for in section (\ref{prop}), namely that the $\alpha\nu \neq c0,s1$ pseudofermions at the limiting canonical momentum values are felt by the $\alpha\nu = c0,s1$ pseudofermions {\it as if} they were scattering centers at the $c0$ and $s1$ Fermi points, respectively (for an exact derivation, see Ref. \cite{Carmspec1}). Assuming a given momentum and energy reached by the $\mathbb{A}$ and $\mathbb{B}$ processes, $m_{\alpha\nu,\iota}$ measures the number of momentum steps in units of ($2\pi/L$), to the left ($\iota=-$) or to the right ($\iota=+$), of such a momentum value. This simulates the particle-hole $\mathbb{C}$ processes in the vicinity of the $\alpha\nu=c0,s1$ Fermi points. There are many particle-hole processes contributing to one number $m_{\alpha\nu,\iota}$ (this description of the particle-hole excitations coincides with that of Ref. \cite{Karlo3}). 

The number deviations are expressed as $\delta$-functions at the corresponding canonical momentum values. However, in order to achieve the right dimension of our expressions, we need to remember that the quantities that have to equal each other in order for the $\delta$-function to contribute, are the quantum numbers of the Takahashi equations, say $I$ and $I'$. When going to the continuous system, we then have that:
\begin{equation}
\delta_{I,I'}=\delta_{\frac L {2\pi} q , \frac L {2\pi} q'} \rightarrow \delta \left( \frac L {2\pi} q - \frac L {2\pi} q' \right) =  \frac {2\pi} L \delta \left( q - q' \right) \hspace{1.0cm} N_a \gg 1 \label{krondirac}
\end{equation}
where $L=aN_a$.

In this fashion, we have for the various number deviations introduced in the sections above, that
\begin{eqnarray}
\Delta N_{\alpha\nu} &=& \Delta N_{\alpha\nu}^{NF} + \Delta N_{\alpha\nu}^F \nonumber \\
\Delta N_{\alpha\nu} (q) &=& \Delta N_{\alpha\nu}^{NF} (q) + \Delta N_{\alpha\nu}^F (q) + \Delta N_{\alpha\nu}^{ph} (q) \nonumber \\
\Delta N_{\alpha\nu}^{NF} (q) &=& \frac {2\pi} L \text{sgn} \left( \Delta N_{\alpha\nu}^{NF} \right) \sum_{i=1}^{\vert \Delta N_{\alpha\nu}^{NF} \vert} \delta \left( q - q_i \right) \label{devnumsstate} \\
\Delta N_{\alpha\nu}^F (q) &=& \frac {2\pi} L \Delta N_{\alpha\nu,\iota}^F \delta_{\iota,\text{sgn}(q)} \delta \left( \vert q \vert - q_{F\alpha\nu} \right) \nonumber \\
\Delta N_{\alpha\nu}^{ph} (q) &=& \frac {2\pi} L \sum_{\iota=\pm} \sum_{i_{\iota}=1}^{N_{\alpha\nu,\iota}^{ph}} \left[ \delta \left( q - q_{p,i_{\iota}} \right) - \delta \left( q - q_{h,i_{\iota}} \right) \right] \nonumber
\end{eqnarray}
where $N_{\alpha\nu}^{ph}=N_{\alpha\nu,+}^{ph}+N_{\alpha\nu,-}^{ph}$. It is these deviations that enter expressions like Eq. (\ref{energydev}) and Eq. (\ref{Qphase1}) in the evaluation of the energy deviations and phase shifts, respectively.

\subsection{Yang holons and HL spinons}
\label{YangHL}

Having defined the relevant pseudofermionic processes and subspaces, we need to clarify how the rotated electronic creation/annihilation operator inside the spectral function relates to the pseudofermions. The subsequent analysis is further explained in Refs. \cite{spectral1} \cite{Carmspec1} \cite{Carmspec2}, even though here we can profit from the simpler case of having only one electronic creation or annihilation operator. This fact will simplify our studies and we will not need to derive selection rules connecting the operators to the total number of $L_{\alpha,-\frac 1 2}$ Yang holons ($\alpha=c$) and HL spinons ($\alpha=s$) that can be created (these selection rules become trivial in the one electron spectral function case).

To start with, let us note that the operator $\tilde{c}_{j\sigma}^l$ stands for $4$ different operators: $\tilde{c}_{j\uparrow}^{\dag}$, $\tilde{c}_{j\downarrow}^{\dag}$, $\tilde{c}_{j\uparrow}$ and $\tilde{c}_{j\downarrow}$ respectively, where $j=0,1,2,\ldots,N_a-1$. However, considering our specific ground state and that we do not consider the UHB, we introduce the projection operators $1- (\tilde{c}_{j\sigma}^{\dag}\tilde{c}_{j\sigma})=1-\tilde{n}_{j\sigma}$ according to:
\begin{eqnarray}
&\text{RHB} \hspace{2.0cm} &\tilde{c}_{j\sigma} \rightarrow \tilde{c}_{j\sigma} \left( 1-\tilde{n}_{j\bar{\sigma}} \right) \nonumber \\
&\text{LHB} \hspace{2.0cm} &\tilde{c}_{j\sigma}^{\dag} \rightarrow \tilde{c}_{j\sigma}^{\dag} \left( 1-\tilde{n}_{j\bar{\sigma}} \right)
\end{eqnarray}
where $\bar{\sigma}=-\sigma$ and $j=0$ in the spectral expressions of this chapter and of chapter (\ref{onelecspec}).

We have many times claimed that we do not need to consider final states with finite occupancies of $-\frac 1 2$ Yang holons and $-\frac 1 2$ HL spinons. In the following, we will motivate this claim. 

Due to not considering the UHB, we note instantly that we can never create any $L_{c,-\frac 1 2}$ Yang holons nor any $c\nu$ ($\nu \geq 1$) pseudofermions, $L_{c,-\frac 1 2}=N_{c\nu}=0$ always. Moreover, the total number of $- \frac 1 2$ HL spinons can never exceed one. This is easily seen by the following two considerations: First, it is impossible to form a $- \frac 1 2$ HL spinon either when annihilating a $\downarrow$-spin rotated electron (the only possibility is that the $\downarrow$-spin rotated electron came from a $s1$ pseudofermion, thus leaving an unpaired $+ \frac 1 2$ HL spinon) or when creating a $\uparrow$-spin rotated electron (of the simple reason that there is no combination of quantum objects that would allow a formation of a $- \frac 1 2$ HL spinon in this case), respectively. Second, we can create either zero or one $- \frac 1 2$ HL spinon(s) when annihilating a $\uparrow$-spin rotated electron or when creating a $\downarrow$-spin rotated electron, respectively. In the first case, we may (or may not) annihilate the $\uparrow$-spin rotated electron from a $s1$ pseudofermion, leading to a single unpaired $- \frac 1 2$ HL spinon left in the system. On the other hand, if the spin degrees of freedom of the $\uparrow$-spin rotated electron was a $+ \frac 1 2$ HL spinon, the resulting system will continue to be void of $- \frac 1 2$ HL spinons. In the second case, the spin part of the created $\downarrow$-spin rotated electron can either couple with a $s1$ pseudofermion (thus decreasing the number of $+\frac 1 2$ HL spinons by one), or it can remain uncoupled in the system, giving rise to one $- \frac 1 2$ HL spinon. 

We must remind ourselves here that the Yang holons and HL spinons are quantum objects that are invariant under the electron - rotated electron unitary operator $\hat{V}$(U/t).  Moreover, their energy and momentum values remain constant during any ground state $\rightarrow$ final state transition. As pointed out in section (\ref{prop}), the Yang holons and HL spinons are neither scatterers nor scattering centers. This can easily be seen since they do not suffer any phase shifts under an arbitrary ground state $\rightarrow$ final state transition. It follows that these quantum objects do not affect the dynamics of the model.

In the case of the HL spinons, there are some straightforward estimates one can do comparing the spectral weight between final states with $L_{s,-\frac 1 2}=1$ to final states with $L_{s,-\frac 1 2}=0$, respectively. Let us first form the candidate final state:
\begin{equation}
\vert f \rangle = \hat{S}_{\alpha}^{\dag} \vert LWS \rangle
\end{equation}
where $\hat{S}_{\alpha}^{\dag}$ brings the LWS state up one notch on the LWS $\rightarrow$ HWS ladder. By direct evaluation of the norm, we obtain
\begin{eqnarray}
\langle f \vert f \rangle = \langle LWS \vert \hat{S}_{\alpha} \hat{S}_{\alpha}^{\dag} \vert LWS \rangle =\langle LWS \vert \left[ \hat{S}_{\alpha} , \hat{S}_{\alpha}^{\dag} \right] \vert LWS \rangle - \nonumber \\ 
- \langle LWS \vert \hat{S}_{\alpha}^{\dag} \hat{S}_{\alpha} \vert LWS \rangle = -2 \langle LWS \vert \hat{S}_{\alpha}^z \vert LWS \rangle = L_{\alpha} \hspace{0.15cm}
\end{eqnarray}
by using the commutation relations of the $SU(2)$ algebras and by noting that $\hat{S}_{\alpha} \vert LWS \rangle=0$. This means that a proper normalized state which is not a LWS but with one $L_{\alpha,-\frac 1 2}$ occupancy, is given by
\begin{equation}
\vert f \rangle = \frac 1 {\sqrt{L_{\alpha}}}  \hat{S}_{\alpha}^{\dag} \vert LWS \rangle
\end{equation}
Now, by using the explicit form for $\hat{S}_s$, Eq. (\ref{gensu2}), together with the usual fermionic anticommutation relations of the rotated electrons, we see after some algebra that
\begin{equation}
\hat{S}_s \tilde{c}_{j\sigma}^l - \tilde{c}_{j\sigma}^l \hat{S}_s = -\sigma \ \tilde{c}_{j,-\sigma}^l \ \big( \delta_{l,+} \delta_{\sigma,\downarrow} +  \delta_{l,-} \delta_{\sigma,\uparrow} \big) = -\sigma \ \tilde{\mathcal{G}}_s
\end{equation}
defining the operator $\tilde{\mathcal{G}}_s$. Remember that according to our notation convention, the symbol $l$ stands for creation/annihilation for the operator $\tilde{c}_{j\sigma}^l$, as well as the numerical values $\pm1$, as well as the signs $\pm$.

For a final state with $L_{s,-\frac 1 2}=1$, we now have for a typical matrix overlap that
\begin{eqnarray}
\big\vert \langle f \vert \tilde{c}_{0\sigma}^l \vert GS \rangle \big\vert &=& \frac 1 {\sqrt{L_s}} \big\vert \langle LWS \vert \hat{S}_s \tilde{c}_{0\sigma}^l \vert GS \rangle \big\vert = \nonumber \\
&=& \frac 1 {\sqrt{L_s}} \big\vert \langle LWS \vert  \tilde{c}_{0\sigma}^l \hat{S}_s \vert GS \rangle - \sigma\langle LWS \vert \tilde{\mathcal{G}}_s \vert GS \rangle \big\vert
\end{eqnarray}
where the first term of the last line is always zero since the ground state is a LWS (note that we follow previous considerations of letting $j=0$ whenever inside a matrix overlap expression). Since we will study systems where the magnetization $ma \rightarrow 0$, it does not matter whether we choose to study matrix elements with $\sigma=\uparrow=+$ or $\sigma=\downarrow=-$. 

This means that by choosing $\sigma = \downarrow$ for the RHB and $\sigma=\uparrow$ for the LHB, we obtain final states which carry no spectral weight if $L_{s,-\frac 1 2} =1$.

We hence conclude that by choosing the following rotated electronic operators for the RHB and the LHB, respectively:
\begin{eqnarray}
&\text{RHB} \hspace{2.0cm} & \tilde{c}_{j,\downarrow} \left( 1-\tilde{n}_{j,\uparrow} \right) \nonumber \\
&\text{LHB} \hspace{2.0cm}  & \tilde{c}_{j,\uparrow}^{\dag} \left( 1-\tilde{n}_{j,\downarrow} \right)
\end{eqnarray}
we obtain final states completely void of any occupancies of the $-\frac 1 2$ Yang holons and $-\frac 1 2$ HL spinons, respectively. Thus, from now on, these quantum objects will not enter in the following analysis of the one electron spectral function.

\subsection{Restricted subspace approximation}
\label{specdevnum}
 
By restricting ourselves to excited state subspaces such that $L_{\alpha,-\frac 1 2}=N_{c\nu}=0$ ($\nu \geq 1$) the deviational expressions for $\Delta N$ and $\Delta (N_{\uparrow}-N_{\downarrow})$ given in Eqs. (\ref{devnums}) can be simplified. Moreover, due to the studies of Ref. \cite{spectral1}, final states with finite occupancies of $N_{s\nu}$ ($\nu \geq 3$) contribute very marginally to the spectral function. In this reference, the following was found for the thermodynamic limit at zero magnetization, and for values of the filling $0 < na < 1$ and arbitrary values of ($U/t$): Final states with $N_{s1}$ number of $s1$ pseudofermions and with $N_{s\nu}=0$ for $\nu \geq 2$, generate approximatively $94\%$ of the total LHB spectral weight and $98\%$ of the total RHB spectral weight, respectively. Moreover, final states with $N_{s1}$ number of $s1$ pseudofermions and $N_{s2} > 0$ number of $s2$ pseudofermions, but with $N_{s\nu}=0$ for $\nu \geq 3$, generate at most approximatively $6 \%$ of the total LHB spectral weight, and at most approximatively $2 \%$ of the total RHB spectral weight, respectively. Thus, with this motivation, Eqs. (\ref{devnums}) become:
\begin{eqnarray}
\Delta N &=& \Delta N_{c0} \nonumber \\
\Delta (N_{\uparrow}-N_{\downarrow}) &=& \Delta N_{c0}-2\Delta N_{s1}-4 \Delta N_{s2}
\end{eqnarray}

However, these are not the only relevant deviations to study. To the contrary of systems with free fermions, or even systems of spinless fermions in the $(U/t)\rightarrow \infty$ Hubbard model, the total number of discrete momentum values $N_{\alpha\nu}^*$ is in general not constant, but deviates from their respective ground state values according to the specific transition in consideration. We have:
\begin{eqnarray}
N_{\alpha\nu}^* &=& N_{\alpha\nu} + N_{\alpha\nu}^h \nonumber \\
N_{\alpha\nu}^* &=& N_{\alpha\nu}^{*,0} + \Delta N_{\alpha\nu}^* \\
\Delta N_{\alpha\nu}^* &=& \Delta N_{\alpha\nu} + \Delta N_{\alpha\nu}^h \nonumber
\end{eqnarray}
where contra-intuitively $\Delta N_{\alpha\nu} \neq -\Delta N_{\alpha\nu}^h$ in general (however, for $c0$ we will indeed always have that $\Delta N_{c0}^*=0$). From section (\ref{occpp}) we have that $N_{\alpha\nu}^{*,0}$ is the corresponding ground state number, given for the different branches by
\begin{eqnarray}
N_{c0}^{*,0}&=&N_a \nonumber \\
N_{s1}^{*,0}&=&N_{\uparrow} \label{pseudolattice} \\
N_{s2}^{*,0}&=&N_{\uparrow}-N_{\downarrow} \nonumber
\end{eqnarray}
as well as the number of holes for each branch
\begin{eqnarray}
N_{c0}^h&=&N_a-N_{c0} \nonumber \\
N_{s1}^h&=&N_{c0}-2N_{s1}-2N_{s2} \\
N_{s2}^h&=&N_{c0}-2N_{s1}-4N_{s2} \nonumber
\end{eqnarray}
from which it is easily deduced that
\begin{eqnarray}
\Delta N_{c0}^* &=& 0 \nonumber \\
\Delta N_{s1}^* &=& \Delta N_{c0} - \Delta N_{s1} -2 \Delta N_{s2} \\
\Delta N_{s2}^* &=& \Delta N_{c0} - 2\Delta N_{s1} -3 \Delta N_{s2} \nonumber
\end{eqnarray}
Note that the last deviation also can be expressed as
\begin{equation}
\Delta N_{s2}^* = \Delta N_{s2} + \Delta (N_{\uparrow}-N_{\downarrow}) 
\end{equation}
which expresses the fact that in the zero magnetization limit, the entire band shrinks as $ma$ vanishes. In the zero magnetization limit, the entire $s2$ band is nonexistent for the ground state since $N_{s2}^*=0$ (and similarly for $s\nu$ bands with $\nu \geq 3$). However, if we create one $s2$ pseudofermion, there appears a canonical momentum band with a single discrete value, in order to accommodate for this quantum object. Thus in this case, we will have one static $s2$ pseudofermion, with zero energy and with zero canonical momentum.

The last deviational numbers we have to consider regards the eventual contribution from the shake-up effect, as given by Eqs. (\ref{shakeups}). In our case, the shake-up effect will contribute for the specific branch, if the following corresponding deviations {\it are odd}:
\begin{eqnarray}
&c0:& \hspace{1.5cm} \Delta N_{s1} + \Delta N_{s2} \nonumber \\
&s1:& \hspace{1.5cm} \Delta N_{s1} + \Delta N_{c0} \\
&s2:& \hspace{1.5cm} \Delta N_{s2} + \Delta N_{c0} \nonumber
\end{eqnarray}

Due to this consideration and the fact that states with many $s2$ pseudofermions are highly unlikely (i.e. produce negligible spectral weight), we will confine our final states to having maximum one $s2$ pseudofermion. Due to the findings of Ref. \cite{spectral1}, we will label the final states as "{\it Basic}" or "{\it Exotic}" in accordance with how much the corresponding final state contributes to the sum rule. In the following, if no $s2$ pseudofermions are created, the corresponding $s2$ deviations will not be accounted for. Also, the approximative percentage with which the transition contribute to the total sum rule is given after the $\rightarrow$ symbol in the heading of the each table.
\newpage

RHB \ \ \ \ \ \ \ \ \ \ {\it Basic} \ \ \ \ \ \ \ \ \ \ $\rightarrow \  98 \%$
\newline
\begin{tabular} {llll}
\hline
\ & $\Delta N = -1$ & $\Delta (N_{\uparrow}-N_{\downarrow}) = +1$ & \  \\
\ & $\Delta N_{c0} = -1$ & $\Delta N_{s1} = -1$ & \ \\
\ & $\Delta N_{s1}^h = +1$ & \  & \ \\
\ & $\Delta N_{s1} = -1$ & \ & $\implies \vert Q_{c0}^0 \vert = \pi$ \\
\ & $\Delta N_{s1} + \Delta N_{c0} =-2$ & \ & $\implies \vert Q_{s1}^0 \vert = 0 $
\end{tabular}
\newline
\newline

RHB \ \ \ \ \ \ \ \ \ \ {\it Exotic}  \ \ \ \ \ \ \ \ \ \ $\rightarrow \  2 \%$
\newline
\begin{tabular} {llll}
\hline
\ & $\Delta N = -1$ & $\Delta (N_{\uparrow}-N_{\downarrow}) = +1$ & \  \\
\ & $\Delta N_{c0} = -1$ & $\Delta N_{s1} = -3$ & $\Delta N_{s2} = +1$ \\
\ & $\Delta N_{s1}^h = +3$ & \  & \ \\
\ & $\Delta N_{s1} +\Delta N_{s2} =-2$ & \ & $\implies \vert Q_{c0}^0 \vert = 0$ \\
\ & $\Delta N_{s1} + \Delta N_{c0} =-4$ & \ & $\implies \vert Q_{s1}^0 \vert = 0 $ \\
\ & $\Delta N_{s2} + \Delta N_{c0} =0$ & \ & $\implies \vert Q_{s2}^0 \vert = 0 $
\end{tabular}
\newline
\newline

LHB \ \ \ \ \ \ \ \ \ \ {\it Basic}  \ \ \ \ \ \ \ \ \ \ $\rightarrow \  94 \%$
\newline
\begin{tabular} {llll}
\hline
\ & $\Delta N = +1$ & $\Delta (N_{\uparrow}-N_{\downarrow}) = +1$ & \  \\
\ & $\Delta N_{c0} = +1$ & $\Delta N_{s1} = 0$ & \ \\
\ & $\Delta N_{s1}^h = +1$ & \  & \ \\
\ & $\Delta N_{s1} = 0$ & \ & $\implies \vert Q_{c0}^0 \vert = 0$ \\
\ & $\Delta N_{s1} + \Delta N_{c0} =+1$ & \ & $\implies \vert Q_{s1}^0 \vert = \pi $
\end{tabular}
\newline
\newline

LHB \ \ \ \ \ \ \ \ \ \ {\it Exotic}  \ \ \ \ \ \ \ \ \ \ $\rightarrow \  6 \%$
\newline
\begin{tabular} {llll}
\hline
\ & $\Delta N = +1$ & $\Delta (N_{\uparrow}-N_{\downarrow}) = +1$ & \  \\
\ & $\Delta N_{c0} = +1$ & $\Delta N_{s1} = -2$ & $\Delta N_{s2} = +1$ \\
\ & $\Delta N_{s1}^h = +3$ & \  & \ \\
\ & $\Delta N_{s1}+\Delta N_{s2} = -1$ & \ & $\implies \vert Q_{c0}^0 \vert = \pi$ \\
\ & $\Delta N_{s1} + \Delta N_{c0} =-1$ & \ & $\implies \vert Q_{s1}^0 \vert = \pi $ \\
\ & $\Delta N_{s2} + \Delta N_{c0} =+2$ & \ & $\implies \vert Q_{s2}^0 \vert = 0 $ \\
\end{tabular}
\newline
\newline

In similar fashion, for example, can the most relevant UHB transitions be classified, by keeping a non zero number $N_{c1}$ in the expressions for the deviations. Note that $\Delta N_{s1} \neq - \Delta N_{s1}^h$ for both of the LHB transitions. Here, this means that the transition induces one extra $s1$ pseudofermionic hole in the $s1$ band.

In the following, if not stated otherwise, we will only consider the "Basic" transitions, which only involves finite pseudofermion deviations in the $c0$ and the $s1$ bands. In other words, {\it all other branches} will be assumed to be completely void of pseudofermions.

\section{State dependent dynamics}

\subsection{Scattering phase shifts:  particle-hole processes ($\mathbb{C}$)}
\label{constantQ}

Using the deviational numbers defined in section (\ref{specdevnum}) together with Eq. (\ref{devnumsstate}), we can calculate the scattering phase shifts $Q_{c0}^{\Phi} (q)$ and $Q_{s1}^{\Phi} (q)$ by use of the quantity defined in Eq. (\ref{Qphase1}) divided by $L$, which in the continuous momentum representation becomes:
\begin{equation}
Q_{\alpha \nu}^{\Phi} (q) = L \sum_{\alpha' \nu'} \int_{-q_{\alpha' \nu'}^0}^{q_{\alpha' \nu'}^0} dq' \ \Phi_{\alpha \nu,\alpha' \nu'}(q,q') \Delta N_{\alpha' \nu'} (q')
\end{equation}

We can now specify the scattering phase shifts for each transition and for each process. Schematically, we would then have $Q_{\alpha \nu}^{\Phi} (q) = Q_{\alpha \nu}^{\Phi} (q,\mathbb{A})+Q_{\alpha \nu}^{\Phi} (q,\mathbb{B})+Q_{\alpha \nu}^{\Phi} (q,\mathbb{C})$ for each transition. However, as we will see by the following analysis, we will always have that $Q_{\alpha \nu}^{\Phi} (q,\mathbb{C})=0$ independently of $\alpha\nu$ and the transition under consideration, due to a pairwise cancellation of the phase shift of each "particle" and "hole" pair. To show this, we must recall that the $\mathbb{C}$ processes are defined in the vicinity of the Fermi points only, where the linearization of the dispersion relations remains a valid approximation. This means that the momentum values for the particle ($q_p$) and the hole ($q_h$) are only a distance $2\pi \mathcal{J} /L$ apart, where $\mathcal{J}$ is a {\it finite} number. We thus have for these processes a canonical momentum shift given by
\begin{eqnarray}
&& L \int_{-q_{\alpha' \nu'}^0}^{q_{\alpha' \nu'}^0} dq' \ \Phi_{\alpha \nu,\alpha' \nu'}(q,q') \Delta N_{\alpha' \nu'}^{ph} (q') =  \nonumber \\
&&= 2 \pi  \int_{-q_{\alpha' \nu'}^0}^{q_{\alpha' \nu'}^0} dq' \ \Phi_{\alpha \nu,\alpha' \nu'}(q,q') \bigg( \delta \left( q'-q_p \right) - \delta \left( q' -q_h \right) \bigg) = \nonumber \\
&&= 2 \pi \bigg( \Phi_{\alpha \nu,\alpha' \nu'}(q,q_p) - \Phi_{\alpha \nu,\alpha' \nu'}(q,q_h) \bigg)= \\
&&= 2 \pi \bigg( \Phi_{\alpha \nu,\alpha' \nu'}(q,q_h) +\frac {2\pi \mathcal{J}} L \frac {d \Phi_{\alpha\nu. \alpha' \nu'}(q,q')} {dq'} \Bigg\vert_{q'=q_h} - \ \Phi_{\alpha \nu,\alpha' \nu'}(q,q_h) \bigg) = \nonumber \\
&&= 2 \pi \frac {2\pi \mathcal{J}} L \frac {d \Phi_{\alpha\nu. \alpha' \nu'}(q,q')} {dq'} \Bigg\vert_{q'=q_h} \hspace{1.5cm} \implies \frac {Q_{\alpha \nu}^{\Phi} (q,\mathbb{C})} L = \mathcal{O}(1/L^2) \nonumber
\end{eqnarray}

We remember that in our pseudofermion theory, discrete momentum contributions of order $(1/L)^j$ do not have any physical relevance for $j \geq 2$. Thus, such a canonical momentum shift is to be treated as equalling zero exactly. It is by this consideration that we claim that all the particle-hole processes belonging to the {\it same} tower of states all share the same phase shift. 

\subsection{Relative spectral weights: Tower of states}
\label{relwetower}

The property of the scattering phase shifts derived in the previous section allows us to treat the spectral function in two "steps": the first step due to processes $\mathbb{A}$ and $\mathbb{B}$, respectively, producing a spectral weight at ($k_0,\omega_0$), given by a lowest peak weight $A_{\alpha\nu}^{(0,0)}$. The second step, due to processes $\mathbb{C}$, produces a spectral weight with energy given by $\omega_{c0}+\omega_{s1}$, where $\omega_{\alpha\nu}=(2\pi/L)\sum_{\iota=\pm} v_{\alpha\nu}m_{\alpha\nu,\iota}$ and with momentum given by $k_{c0}+k_{s1}$, where $k_{\alpha\nu}=(2\pi/L) \sum_{\iota=\pm} \iota m_{\alpha\nu,\iota}$, respectively. Thus, such $\mathbb{A}$, $\mathbb{B}$ and $\mathbb{C}$ processes contribute to the weight at the point ($k_0+k_{c0}+k_{s1},\omega_0+\omega_{c0}+\omega_{s1}$).

The final weight is a convolution of the weights for $\alpha\nu=c0$ and $\alpha\nu=s1$, respectively, where $m_{\alpha\nu,\iota}$ is the number of momenta steps, measured in units of ($2\pi/L$), to the left ($\iota=-$) or the right ($\iota=+$) of the ($k_0,\omega_0$) point. The superscript $(0,0)$ refers to $(m_{\alpha\nu,-}=0,m_{\alpha\nu,+}=0)$. This procedure is inspired by that of Refs. \cite{Karlo1}-\cite{Karlo3}, where the same problem is studied in the $(U/t) \rightarrow \infty$ limit. The spinless fermions used to describe the model in these references correspond to the $c0$ pseudofermions in the arbitrary ($U/t$) model. The "spinons" on the other hand are carried over from the 1D Heisenberg model, which cannot be done in the arbitrary ($U/t$) case. Indeed, the $s1$ pseudofermion should not be compared to the notion of "spinons" that is used in Refs. \cite{Karlo1}-\cite{Karlo3}. In these references, the spinons have a spin projected value of $\pm \frac 1 2$, whilst in our pseudofermion theory for arbitrary ($U/t$), the corresponding $s1$ pseudofermion is a two-spinon object with zero spin projection.

As distinguished by the numbers ($m_{\alpha\nu,-},m_{\alpha\nu,+}$), we then obtain a tower of states where $m_{\alpha\nu,+}+m_{\alpha\nu,-}$ is proportional to the energy of the particular particle-hole process and where $m_{\alpha\nu,+}-m_{\alpha\nu,-}$ is proportional to the momentum of the same process. The number of particle-hole processes contributing to the ($m_{\alpha\nu,-},m_{\alpha\nu,+}$) point grows exponentially as we build the tower. Due to convention, we go to successively more negative energies whenever in the RHB and to successively more positive energies whenever in the LHB, as we build the tower of particle-hole states emerging from momentum $k$ and energy $\omega$.

As in, for example, Ref. \cite{Karlo3}, we now define the {\it relative spectral weight} $a_{\alpha\nu}^{(m_{\alpha\nu,-},m_{\alpha\nu,+})}$ according to
\begin{equation}
A_{\alpha\nu}^{(m_{\alpha\nu,-},m_{\alpha\nu,+})}=a_{\alpha\nu}(m_{\alpha\nu,-},m_{\alpha\nu,+}) A_{\alpha\nu}^{(0,0)} \label{defweight}
\end{equation}

In the cited reference only one one spinless fermion and one spinon is created or annihilated, during the one electron addition or removal process. Thus the explicit form for $a_{\alpha\nu}(m_{\alpha\nu,-},m_{\alpha\nu,+})$ is simpler than what we will need here, where in general a number of $\Delta N_{\alpha\nu}$ pseudofermions is created or annihilated for each $\alpha\nu$ branch. This means that wherever in Ref. \cite{Karlo3}, the excited energy eigenstate has $N+l$ spinless fermions or spinons (for $l=\pm1$), we in our case have $N_{\alpha\nu}^0+\Delta N_{\alpha\nu}$ $\alpha\nu$ pseudofermions. In the following analysis, we will explicitly calculate $a_{\alpha\nu}(1,0)$ and $a_{\alpha\nu}(0,1)$ which then easily combines to form $a_{\alpha\nu}(1,1)$. For reasons of clarity, we will consider excited energy eigenstates resulting from processes $\mathbb{B}$ only, such that $\Delta N_{\alpha\nu}^{NF}=0$, i.e. such that there are no $\alpha\nu$ pseudofermions away from the densely packed excited Fermi sea. This means that we will consider $\Delta N_{\alpha\nu}=\Delta N_{\alpha\nu}^F=\Delta N_{\alpha\nu,+}^F + \Delta N_{\alpha\nu,-}^F$. Furthermore, we will consider only one pseudofermion taking part in the most basic particle-hole excitation process: the one in which the hole momentum is at the shifted (excited state) $\iota$ Fermi point and the particle momentum is just one step away from this point, i.e. $\bar{q}_p=\bar{q}_h+\iota (2\pi /L)$. Adapted to our notation the generalized expression for the relative weights, for one particle-hole pair, is then
\begin{equation}
a_{\alpha\nu}(\bar{q}_h,\bar{q}_p)=\frac {\underset{j=1}{\overset{N_{\alpha\nu}^0+\Delta N_{\alpha\nu}^F} \prod} \sin^2 \left( \frac {\bar{q}_j-\bar{q}_p} 2 \right) \underset{j=1}{\overset{N_{\alpha\nu}^0} \prod} \sin^2 \left( \frac {q_j-\bar{q}_h} 2 \right)} {\sin^2 \left( \frac {\bar{q}_p-\bar{q}_h} 2 \right) \underset{\substack{j=1 \\ j \neq h}}{\overset{N_{\alpha\nu}^0+\Delta N_{\alpha\nu}^F} \prod} \sin^2 \left( \frac {\bar{q}_j-\bar{q}_h} 2 \right) \underset{j=1}{\overset{N_{\alpha\nu}^0} \prod} \sin^2 \left( \frac {q_j-\bar{q}_p} 2 \right)} \label{relweight}
\end{equation}
Thus, this relative weight is only valid for one particle-hole pair. Moreover, the notation $a_{\alpha\nu}(\bar{q}_h,\bar{q}_p)$ is allowed here due to the simplicity of the particle-hole process in consideration. For any other particle-hole process, we would have to write $a_{\alpha\nu}(m_{\alpha\nu,-},m_{\alpha\nu,+})$ since the canonical momentum values of each particle-hole pair uniquely define two integers ($m_{\alpha\nu,-},m_{\alpha\nu,+}$) whilst the converse is not true. As it stands, this expression is only valid in the finite system, the continuous momentum limit will be taken at a later stage. The equivalent expressions for higher numbers of particle-hole pairs can be found for $(U/t) \rightarrow \infty$ in Ref. \cite{Karlo3} and for arbitrary ($U/t$) in Ref. \cite{Carmspec2}. 

Since we are considering densely packed excited state pseudofermion occupation configurations, most of the factors above will cancel with each other, which can be easily seen by explicit calculation of $a_{\alpha\nu}(\bar{q}_h,\bar{q}_p)$ using the following expressions for the relevant canonical momenta:
\begin{eqnarray}
\bar{q}_h&=&q_h + \frac {Q_{\alpha\nu}^{\Phi}(q_h)} L \hspace{1.5cm} q_h=\frac {2\pi} L I_h = \frac {2\pi} L \left( I_{F\iota}^{\alpha\nu} + \iota \Delta N_{\alpha\nu,\iota}^F \right) \nonumber \\
\bar{q}_p&=&\bar{q}_h + \iota \frac {2\pi} L \implies \hspace{1.15cm} I_p = I_h + \iota \label{excstatenum}
\end{eqnarray}
However, this fact is only due to some quite obvious approximations, based on the difference
\begin{equation}
\bar{q}_j-\bar{q}_i = \frac {2\pi} L \left( I_j - I_i  + \frac {Q_{\alpha\nu}^{\Phi}(q_j)-Q_{\alpha\nu}^{\Phi}(q_i)} {2\pi} \right)
\end{equation}

We note that in the second term above, the difference of the scattering phase shifts, is always bounded and small due to the boundedness of the $\Phi_{\alpha\nu,\alpha' \nu'} (q,q')$ functions, as compared to the difference $\vert I_j - I_i \vert = 1,2,\ldots$ For example, considering the "Basic" excitations of section (\ref{specdevnum}), we have that $\left( Q_{\alpha\nu}^{\Phi}(q_j)-Q_{\alpha\nu}^{\Phi}(q_i) \right) /2\pi$ attains a typical maximum value between $1$ and $2$ when $q_j$ and $q_i$ are on opposite sides of the $c0$ or $s1$ Fermi sea \cite{CarmBoziPedro}, and is always smaller than this value for any other pair of $q_j$ and $q_i$. Moreover, since the scattering phase shifts are in general continuous functions of the momentum, when the two canonical momenta are close to each other, the difference of their scattering phase shifts becomes negligible. This means that the difference between the two scattering phase shifts can always be neglected. Moreover, since the particle and the hole canonical momenta both are in the vicinity of the same Fermi point, we have that $Q_{\alpha\nu}^{\Phi}(q_h) \approx Q_{\alpha\nu}^{\Phi}(q_p) \approx Q_{\alpha\nu}^{\Phi}(\iota q_{F\alpha\nu})$. By following this scheme, we obtain for the relative weights that:
\begin{equation}
a_{\alpha\nu}(\bar{q}_h,\bar{q}_p) \approx \prod_{j=1}^{N_{\alpha\nu}^0} \frac {\sin^2 \left( \frac {\pi} L \left( I_j - I_h \right) - \frac {Q_{\alpha\nu}^{\Phi}(\iota q_{F\alpha\nu})} {2L} \right)} {\sin^2 \left( \frac {\pi} L \left( I_j - I_p \right) - \frac {Q_{\alpha\nu}^{\Phi}(\iota q_{F\alpha\nu})} {2L} \right)} \prod_{\substack{j=1 \\ j \neq h}}^{N_{\alpha\nu}^0 + \Delta N_{\alpha\nu}^F} \frac {\sin^2 \left( \frac {\pi} L \left( I_j - I_p \right) \right)} {\sin^2 \left( \frac {\pi} L \left( I_j - I_h \right) \right)} \label{relweight2}
\end{equation}
which by explicit investigation of the factors allows for the above mentioned cancellation. After this cancellation, what remains is 
\begin{equation}
a_{\alpha\nu}(\bar{q}_h,\bar{q}_p) \approx \frac {\sin^2 \left( \frac {\iota \pi \Delta N_{\alpha\nu,\iota}^F} L + \frac {Q_{\alpha\nu}^{\Phi}(\iota q_{F\alpha\nu})} {2L} \right) \sin^2 \left( \pi n + \frac {\pi \Delta N_{\alpha\nu}^F } L \right)} {\sin^2 \left( \pi n + \frac {\pi \Delta N_{\alpha\nu,\iota}^F} L + \frac { \iota Q_{\alpha\nu}^{\Phi}(\iota q_{F\alpha\nu})} {2L} \right) \sin^2 \left( \frac {\pi} L \right)}
\end{equation}
which in the large system limit goes as
\begin{eqnarray}
a_{\alpha\nu}(\bar{q}_h,\bar{q}_p) & \approx &\left( \iota \Delta N_{\alpha\nu,\iota}^F + \frac {Q_{\alpha\nu}^{\Phi}(\iota q_{F\alpha\nu})} {2\pi} \right)^2 \cdot \\
&\cdot& \left[1 + \frac {2\pi} L \left( \Delta N_{\alpha\nu,-\iota}^F - \frac {\iota Q_{\alpha\nu}^{\Phi}(\iota q_{F\alpha\nu})} {2\pi} \right) \cot \pi n + \mathcal{O}\left(1/L^2 \right) \right] \nonumber
\end{eqnarray}
Even though strictly speaking this formula for the relative weight is only valid for the lowest particle hole excitations $a_{\alpha\nu}(1,0)$ and $a_{\alpha\nu}(0,1)$, we notice that there is a leading order term which does not depend on the system size. This leading order term will remain untouched as we go to higher particle-hole processes, as relative weights for successively higher particle-hole excitations only contribute to the ($1/L$) term, which is verifiable by suitable modification of Eq. (\ref{relweight2}). 

\subsection{Relative spectral weights: closed form expressions}
\label{relwetower2}

Generally, we have particle-hole processes at both sides of the Fermi sea at the same time. This means that in the expressions for the relative weights of the preceding section, we will have to consider many particle-hole pairs ($\bar{q}_{h1},\bar{q}_{p1}$), ($\bar{q}_{h2},\bar{q}_{p2}$), $\ldots$ , ($\bar{q}_{hN_{\alpha\nu}^{ph}},\bar{q}_{pN_{\alpha\nu}^{ph}}$). The expression for the relative weights for more than one particle-hole pair are given in Refs. \cite{Karlo3} ($U/t \rightarrow \infty$) and \cite{Carmspec2} (arbitrary $U/t$). For example, to continue our study for $m_{\alpha\nu,\pm}=1$ of the previous section, it is easily shown that the relative spectral weight of two particle-hole processes can be written as (adapted to our notation):
\begin{equation}
a_{\alpha\nu}(1,1) = a_{\alpha\nu}(0,1) \ a_{\alpha\nu}(1,0) \  \frac {\sin^2 \left( \frac {\bar{q}_{h1}-\bar{q}_{h2}} 2 \right) \sin^2 \left( \frac {\bar{q}_{p1}-\bar{q}_{p2}} 2 \right)}    {\sin^2 \left( \frac {\bar{q}_{p1}-\bar{q}_{h2}} 2 \right) \sin^2 \left( \frac {\bar{q}_{h1}-\bar{q}_{p2}} 2 \right) }
\end{equation}
which by introduction of the quantum numbers in the same way as before becomes
\begin{equation}
a_{\alpha\nu}(1,1) = a_{\alpha\nu}(0,1) a_{\alpha\nu}(1,0) \cdot \bigg[ 1 + \mathcal{O} (1/L^2) \bigg]
\end{equation}
where the exact cancellation of the ($1/L$) terms is a result of the $m_{\alpha\nu,\iota} = 1$ special case. Exactly the same procedure can be repeated for successively higher particle-hole excitations, i.e. for numbers $m_{\alpha\nu,\pm} = 1,2,\ldots$, with successive contributions to the first order correction term. In general as $m_{\alpha\nu,\pm}$ increases we will have more and more factors that do not cancel in the expression for the relative weight. The number of factors that do not cancel grows exponentially with $m_{\alpha\nu,\pm}$, with an additional exponential increase of the factors that mixes the canonical momenta for the holes and the particles of the left and the right Fermi points, respectively. As shown above, this effect is not present in our $m_{\alpha\nu,\pm}=1$ example. The expressions involving many particle-hole pairs are very involved and an exact derivation of the various cases for increasing $m_{\alpha\nu,\pm}$ will not be given here since it is more confusing than enlightening. However the method used is exactly the same as in the example shown above. Due to the number of non-cancelling factors present, we find that we have to relax the $\mathcal{O}(1/L)$ correction term slightly, according to
\begin{equation}
a_{\alpha\nu}(m_{\alpha\nu,-},m_{\alpha\nu,+}) =a_{\alpha\nu,-}(m_{\alpha\nu,-}) a_{\alpha\nu,+}(m_{\alpha\nu,+}) \cdot \bigg[ 1 + \mathcal{O} (\ln L/L) \bigg]
\end{equation}
which nevertheless allows for a complete description of the spectral weights of the processes $\mathbb{C}$ in the thermodynamic limit.

In section (\ref{classstates}) it was mentioned that the dynamics of the system should, in the continuum limit, depend heavily on the scattering phase shifts at the Fermi points. Consider now a ground state $\rightarrow$ final state transition. The pseudofermions that are inside the Fermi sea become shifted from one canonical momentum value to another that is also inside the Fermi sea, and that thus was also occupied in the original ground state. Thus, the corresponding phase shift becomes like a grain of salt in a bowl of water that technically is visible during the transition itself, but that in the final state becomes "dissolved" in the filled densely packed excited energy eigenstate. Note however that this is only true for the continuous momentum limit. The shifted Fermi points on the other hand, {\it define} the boundaries for the new Fermi sea. Because of this, the amount that the canonical momenta at the Fermi points have shifted, $\Delta \bar{q}_{F\alpha\nu,\iota}$, is an important quantity for the description of the dynamics of the model. We can see for example in Eq. (\ref{excstatenum}) that
\begin{equation}
\Delta \bar{q}_{F\alpha\nu,\iota} = \frac {2\pi} L \left( \iota \Delta N_{\alpha\nu,\iota}^F + \frac {Q_{\alpha\nu}^{\Phi}(\iota q_{F\alpha\nu})} {2\pi} \right)
\end{equation}
which means that by defining
\begin{equation}
2 \Delta_{\alpha\nu}^{\iota}=\left( \frac {\Delta \bar{q}_{F\alpha\nu,\iota}} {\left[ 2\pi / L \right]} \right)^2 \label{def2D}
\end{equation}
we have indeed
\begin{eqnarray}
&& a_{\alpha\nu} (1,0) = a_{\alpha\nu,-} (1) = 2 \Delta_{\alpha\nu}^- \nonumber \\
&& a_{\alpha\nu} (0,1) = a_{\alpha\nu,+} (1) = 2 \Delta_{\alpha\nu}^+ \\
&& a_{\alpha\nu} (1,1) = a_{\alpha\nu,-} (1) \ a_{\alpha\nu,+} (1) = \left( 2 \Delta_{\alpha\nu}^- \right) \left( 2 \Delta_{\alpha\nu}^+ \right) \nonumber
\end{eqnarray}
for $L \rightarrow \infty$. The quantities $2 \Delta_{\alpha\nu}^{\iota}$ will be very important in our subsequent study of this problem. As already mentioned, this analysis can be carried to higher orders, which give for example,
\begin{eqnarray}
&& a_{\alpha\nu} (2,0) = a_{\alpha\nu,-} (2) = \frac {2 \Delta_{\alpha\nu}^- \left( 2\Delta_{\alpha\nu}^- +1\right)} 2 \nonumber \\
&& a_{\alpha\nu} (0,2) = a_{\alpha\nu,+} (2) = \frac {2 \Delta_{\alpha\nu}^+ \left( 2\Delta_{\alpha\nu}^+ +1 \right)} 2
\end{eqnarray}
and so forth, reaching the general result
\begin{equation}
a_{\alpha\nu,\iota} (m_{\alpha\nu,\iota}) = \frac {2 \Delta_{\alpha\nu}^{\iota} \left( 2\Delta_{\alpha\nu}^{\iota} +1\right) \ldots \left( 2\Delta_{\alpha\nu}^{\iota} + m_{\alpha\nu,\iota} -1 \right)} {\left(m_{\alpha\nu,\iota}\right)!}
\end{equation}
Algebraically, this can be rewritten as
\begin{equation}
a_{\alpha\nu,\iota} (m_{\alpha\nu,\iota}) = \frac {\Gamma \left( m_{\alpha\nu,\iota} + 2 \Delta_{\alpha\nu}^{\iota} \right)} {\Gamma \left( m_{\alpha\nu,\iota} + 1 \right) \Gamma \left( 2 \Delta_{\alpha\nu}^{\iota} \right)} \approx \left[ \Gamma \left( 2 \Delta_{\alpha\nu}^{\iota} \right) \right]^{-1} \left( m_{\alpha\nu,\iota} \right)^{2 \Delta_{\alpha\nu}^{\iota}-1} \label{relpower}
\end{equation}
where the approximation is almost exact except for the first few ($1/L$) positions in the tower of particle-hole states, provided that $2 \Delta_{\alpha\nu}^{\iota} \neq 0$. For $2 \Delta_{\alpha\nu}^{\iota} = 0$ there is no change in the canonical momentum value of the $\iota$ Fermi point and hence there is no dynamical change between the ground state configuration and the excited energy eigenstate configurations of $\alpha\nu$ pseudofermions at that Fermi point.

Eq. (\ref{relpower}) shows the power law type behavior of the spectral weight generated by the particle-hole processes, with exponent equal to ($2 \Delta_{\alpha\nu}^{\iota} - 1$). This exponent obviously changes for each position in the ($k,\omega$) plane that the processes $\mathbb{A}$ and $\mathbb{B}$ brings the excitation to. Furthermore, the total spectral weight at a certain position in the tower of states will then become a summation of contributions from different particle-hole processes originating from neighboring points in the ($k,\omega$) plane due to processes $\mathbb{A}$ and $\mathbb{B}$. Therefore, the final exponent will be different from the exponent given here, as we will take into account contributions from many different overlapping towers of states.

\subsection{Scattering phase shifts: $\mathbb{A}$ and $\mathbb{B}$ processes}
\label{AandB}

In the previous section, we derived a closed form expression for the relative spectral weight for the particle-hole processes following a given $\mathbb{A}$ and $\mathbb{B}$ process. The spectral weight generated by the particle-hole processes is crucially dependent on the phase shift that the processes $\mathbb{A}$ and $\mathbb{B}$ produces. 
Since $\Delta N_{\alpha\nu}(q)=\Delta N_{\alpha\nu}^{NF}(q)+\Delta N_{\alpha\nu}^F(q)$ for the $\mathbb{A}$ and $\mathbb{B}$ processes, we similarily have by definition $Q_{\alpha\nu}^{\Phi}(\iota q_{F\alpha\nu})= Q_{\alpha\nu}^{\Phi (NF)}(\iota q_{F\alpha\nu})+Q_{\alpha\nu}^{\Phi (F)}(\iota q_{F\alpha\nu})$. We have

\begin{eqnarray}
\frac {Q_{\alpha\nu}^{\Phi}(\iota q_{F\alpha\nu})} {2\pi} &=& \sum_{\alpha' \nu'} \sum_{q'} \Phi_{\alpha\nu, \alpha'\nu'} \left( \iota q_{F \alpha\nu} , q' \right) \Delta N_{\alpha' \nu'} (q') = \nonumber \\
&=& \sum_{\alpha' \nu'} \sum_{q'} \Phi_{\alpha\nu, \alpha'\nu'} \left( \iota q_{F \alpha\nu} , q' \right) \bigg[ \Delta N_{\alpha' \nu'}^F (q') +  \Delta N_{\alpha' \nu'}^{NF} (q') \bigg]
\end{eqnarray}
where by the use of Eq. (\ref{devnumsstate}) we see that

\begin{eqnarray}
\frac {Q_{\alpha\nu}^{\Phi (F)}(\iota q_{F\alpha\nu})} {2\pi} &=& \sum_{\alpha' \nu'} \int dq' \ \Phi_{\alpha\nu, \alpha'\nu'} \left( \iota q_{F \alpha\nu} , q' \right) \Delta N_{\alpha' \nu',\iota}^F \delta_{\iota,\text{sgn}(q')} \delta \left( \vert q' \vert - q_{F,\alpha' \nu'} \right) = \nonumber \\ 
& = &\sum_{\alpha' \nu'} \sum_{\iota'} \Phi_{\alpha\nu, \alpha'\nu'} \left( \iota q_{F \alpha\nu} , \iota' q_{F \alpha' \nu'} \right) \Delta N_{\alpha' \nu' ,\iota'}^F \label{Qfermi}
\end{eqnarray}

To this end, we will define some quantities $\xi_{\alpha\nu,\alpha'\nu'}^j$ where $j=0,1$:

\begin{equation}
\xi_{\alpha\nu,\alpha'\nu'}^j = \delta_{\alpha\nu, \alpha'\nu'} + \Phi_{\alpha\nu, \alpha'\nu'} \left( \iota q_{F \alpha\nu} , q_{F \alpha' \nu'} \right) + (-1)^j \Phi_{\alpha\nu, \alpha'\nu'} \left( \iota q_{F \alpha\nu} , - q_{F \alpha' \nu'} \right)
\end{equation}
which implies that $\Phi_{\alpha\nu, \alpha'\nu'} \left( \iota q_{F \alpha\nu} , \iota' q_{F \alpha' \nu'} \right)$ can be expressed as

\begin{equation}
\Phi_{\alpha\nu, \alpha'\nu'} \left( \iota q_{F \alpha\nu} , \iota' q_{F \alpha' \nu'} \right) = \frac {\iota} 2 \xi_{\alpha\nu,\alpha'\nu'}^0 + \frac {\iota'} 2 \xi_{\alpha\nu,\alpha'\nu'}^1 - \iota \delta_{\iota,\iota'} \delta_{\alpha\nu, \alpha'\nu'} \label{phixi}
\end{equation}

The introduction of these quantities simplify the expressions of the scattering phase shifts as we will see below. Together with the identity $\Delta N_{\alpha' \nu' ,\iota'}^F = \iota' \Delta J_{\alpha' \nu'}^F + \Delta N_{\alpha' \nu'}^F/2$ we can substitute Eq. (\ref{phixi}) into Eq. (\ref{Qfermi}) to obtain

\begin{equation}
\frac {Q_{\alpha\nu}^{\Phi (F)}(\iota q_{F\alpha\nu})} {2\pi} = -\Delta J_{\alpha\nu}^F -\iota \frac {\Delta N_{\alpha\nu}^F} 2 + \sum_{\alpha' \nu'} \bigg(  \xi_{\alpha\nu,\alpha'\nu'}^1 \Delta J_{\alpha' \nu'}^F + \iota \xi_{\alpha\nu,\alpha'\nu'}^0 \frac {\Delta N_{\alpha'\nu'}^F} 2 \bigg)
\end{equation}
which obviously implies that

\begin{equation}
\frac {Q_{\alpha\nu}^{\Phi (F)}(\iota q_{F\alpha\nu})} {2\pi} + \Delta N_{\alpha\nu ,\iota}^F = \sum_{\alpha' \nu'} \bigg(  \xi_{\alpha\nu,\alpha'\nu'}^1 \Delta J_{\alpha' \nu'}^F + \iota \xi_{\alpha\nu,\alpha'\nu'}^0 \frac {\Delta N_{\alpha'\nu'}^F} 2 \bigg)
\end{equation}
so that we finally obtain, by performing the $\alpha' \nu'=c0,s1$ summation,

\begin{equation}
2\Delta_{\alpha\nu}^{\iota} = \bigg( \xi_{\alpha\nu,c0}^1 \Delta J_{c0}^F + \xi_{\alpha\nu,s1}^1 \Delta J_{s1}^F + \iota \xi_{\alpha\nu,c0}^0 \frac {\Delta N_{c0}^F} 2 + \iota \xi_{\alpha\nu,s1}^0 \frac {\Delta N_{s1}^F} 2 + \frac {Q_{\alpha\nu}^{\Phi (NF)}(\iota q_{F\alpha\nu}) } {2\pi} \bigg)^2 \label{twodelta}
\end{equation}
where generally

\begin{eqnarray}
\frac {Q_{\alpha\nu}^{\Phi (NF)}(\iota q_{F\alpha\nu}) } {2\pi} = \sum_{\alpha' \nu'} \text{sgn}\left( \Delta N_{\alpha'\nu'}^{NF} \right)  \bigg[ &\Phi_{\alpha\nu,\alpha' \nu'}(\iota q_{F\alpha\nu},q'_1) + \Phi_{\alpha\nu,\alpha' \nu'}(\iota q_{F\alpha\nu},q'_2) + \ldots + \nonumber \\
& + \: \Phi_{\alpha\nu,\alpha' \nu'}(\iota q_{F\alpha\nu},q'_{\vert \Delta N_{\alpha'\nu'}^{NF} \vert }) \bigg] 
\end{eqnarray}
is the scattering phase shift at the $\iota$ Fermi point due to the created or annihilated $\alpha'\nu'$ pseudofermions with canonical momenta $q'_1$, $q'_2$, $\ldots$, $q'_{\vert \Delta N_{\alpha'\nu'}^{NF} \vert }$, for $\alpha'\nu'=c0,s1$. Note that for the "Basic" transitions of section (\ref{specdevnum}), we will always have that $\vert \Delta N_{\alpha'\nu'}^{NF} \vert = 0,1$. The quantities $2\Delta_{\alpha\nu}^{\iota}$ ultimately control the behavior of the spectral function in the ($k,\omega$) plane. We will see later that different linear combinations of $2\Delta_{c0}^{\pm}$ and $2\Delta_{s1}^{\pm}$ form the exponents of the one electron spectral function for the excitations for the "Basic" transitions, which lead to a power-law behavior of the spectral function. This mixing of the various $2\Delta_{\alpha\nu}^{\iota}$'s is a result of the fact that several different processes contribute to the same vicinity of a typical point in the ($k,\omega$) plane, as is evident in the case of overlapping towers of states.

\subsection{Excited state characterization}
\label{character}

In section (\ref{specdevnum}), we found that nearly all of the total spectral weight was dominated by two types of transitions for both the RHB and the LHB. These transitions were called "Basic" and "Exotic", respectively, due to their pseudofermionic content (as shown by the occupancy number deviations) and their contribution to the total sum rule. As mentioned in that section, and as implicitly assumed in the subsequent sections, we only consider the "Basic" transitions. These transitions are such that the only bands with finite occupancies are the $c0$ and the $s1$ bands. They contribute to about $98\%$ (RHB) and $94\%$ (LHB) of the total sum rule, respectively. In this section we will express the momentum $k$ and energy $\omega$, which are the variables of the spectral function, in terms of the pseudofermionic or the pseudofermionic hole canonical momenta. The procedure is actually quite straightforward, considering that we have already calculated all the relevant quantities needed. All that remains is to specify exactly how the specific transitions here considered affect these quantities. From section (\ref{class}) we have that the variables of the spectral function obey

\begin{eqnarray}
E_{f_l} - E_{GS} &=& \sum_{\alpha\nu=c0,s1} \bigg[ \Delta E_{\alpha\nu} \left( \mathbb{A} \right) +\ \Delta E_{\alpha\nu} \left( \mathbb{C} \right) \bigg] \nonumber \\
P_{f_l} - P_{GS} &=& \sum_{\alpha\nu=c0,s1} \bigg[ \Delta P_{\alpha\nu} \left( \mathbb{A} \right) + \Delta P_{\alpha\nu} \left( \mathbb{B} \right) + \Delta P_{\alpha\nu} \left( \mathbb{C} \right) \bigg]
\end{eqnarray}
Since we will only consider the "Basic" transitions we can simplify the expressions of Eq. (\ref{enemoms}) according to:

\begin{eqnarray}
l \Delta E &=& l \left[ \text{sgn}\left( \Delta N_{c0}^{NF} \right) \epsilon_{c0} (q_{c0})  + \text{sgn}\left( \Delta N_{s1}^{NF} \right) \epsilon_{s1} (q_{s1}) \right] \label{dEdPspectral} \\
l \Delta P &=& l \left[ \text{sgn}\left( \Delta N_{c0}^{NF} \right) q_{c0} + \text{sgn}\left( \Delta N_{s1}^{NF} \right) q_{s1} + 4k_F  \Delta J_{c0}^F + 2k_{F\downarrow} \Delta J_{s1}^F \right] \nonumber
\end{eqnarray}
where we have used the fact that we will never have more than one $c0$ or $s1$ pseudofermion created or annihilated. $q_{c0}$ and $q_{s1}$ are the corresponding momentum values for the created or annihilated scattering centers, which should not be confused with the Fermi momentum $q_{F\alpha\nu}$, nor with the limiting momentum value for the $\alpha\nu$ effective Brillouin zone $q_{\alpha\nu}^0$. Judging from Eq. (\ref{twodelta}), we see that we will have qualitatively different expressions for the quantities $2\Delta_{\alpha\nu}^{\iota}$ when pseudofermions or pseudofermion holes are created at any of the Fermi points on the one hand, and when they are not, on the other. This leads us to consider four different cases (where in the following {\it "P"} stands for those {\it P}seudofermions created or annihilated away from the Fermi points):
\begin{enumerate}
\item {\it 2P contribution:} Neither the $c0$ nor the $s1$ pseudofermion or pseudofermion hole are created at any of the Fermi points. This contribution will lead to the overall "background" of the weight distribution of the spectral function, as both pseudofermions or pseudofermion holes are dispersive, leading to contributions over nearly the whole range of allowed $k$ and $\omega$ values.
\item {\it s-branch (1P):} The $c0$ pseudofermion or pseudofermion hole is created at the left or the right $c0$ Fermi point and the $s1$ pseudofermion hole is created away from any of the $s1$ Fermi points. This will lead to a line in the ($k,\omega$) plane, following the dispersion of the $s1$ pseudofermion hole. 
\item {\it c-branch (1P):} The $s1$ pseudofermion hole is created at the left or the right $s1$ Fermi point and the $c0$ pseudofermion hole is created away from any of the $c0$ Fermi points. This will lead to a line in the ($k,\omega$) plane, following the dispersion of the $c0$ pseudofermion or pseudofermion hole.
\item {\it Fermi contribution (0P):} Both pseudofermions or pseudofermion holes are created at their left or right Fermi points, respectively. This contribution leads to a spectral weight distribution in the vicinity of certain points in the ($k,\omega$) plane.
\end{enumerate}

This distinction is merely due to the number of different ways that we can calculate spectral weight in ($k,\omega$) domains topologically different from each other, as well as due to the observation that the numbers $\Delta N_{\alpha\nu}^F$ and $\Delta J_{\alpha\nu}^F$ are different in each of the cases above. Before we move on, however, there are two special cases which will be important in the following. They are classified as
\begin{enumerate}
\item {\it 2P contribution - Border Lines:} These line shapes share the definition of the general {\it 2P contribution} described above, but with the additional requirement that the velocity of the non Fermi $c0$ pseudofermion is equal to the velocity of the non Fermi $s1$ pseudofermion hole. This extra requirement confines the 2P spectral weight to certain lines in the ($k,\omega$) plane. We will see that the general expressions for the 2P contribution are singular as the two velocities become equal, giving rise to a divergent feature of the overall spectral function.
\item {\it s-branch and c-branch - Luttinger contribution (1P):} This "partial line shape" shares the definitions of the {\it s-branch} and the {\it c-branch} described above, with the difference that the dispersive $s1$ (s-branch) or $c0$ (c-branch) pseudofermion has a finite but small energy. These excitations, which are close to the Fermi energy level, belong to the subspace of excitations usually described by the Luttinger liquid theory. We will see that excitations belonging to the branch lines but in the region of very small energies will have different final expressions of the full spectral function, as compared to the expressions of the finite energy s-branch and the finite energy c-branch, respectively. The values of the critical exponents at low energies are obtained in this regime, and have been shown to reproduce known results obtained by conformal field theory \cite{equival} \cite{crit6} \cite{crit61}.
\end{enumerate}

To make the following analysis easier, we note that 

\begin{eqnarray}
\Delta N_{\alpha\nu}^F &=& \Delta N_{\alpha\nu,+}^{0,F} + \Delta N_{\alpha\nu,-}^{0,F} \nonumber \\
2 \Delta J_{\alpha\nu}^F &=& \Delta N_{\alpha\nu,+}^{0,F} - \Delta N_{\alpha\nu,-}^{0,F} + \lambda_{\alpha\nu}
\end{eqnarray}
by using $\Delta N_{\alpha\nu,\iota}^F = \Delta N_{\alpha\nu,\iota}^{0,F} + \iota Q_{\alpha\nu}^0 / 2\pi$ and where $\lambda_{\alpha\nu}=Q_{\alpha\nu}^0/\pi=-1,0,1$.

Lastly, we will express the RHB and the LHB annihilation and creation rotated electronic operators in terms of pseudofermionic operators. The latter operators are defined in Eq. (\ref{feqf}), and denoted $f_{\bar{q},\alpha\nu}^{\dag}$ and $f_{\bar{q},\alpha\nu}$ in the canonical momentum representation, and $f_{j,\alpha\nu}^{\dag}$ and $f_{j,\alpha\nu}$ in the effective $\alpha\nu$ lattice representation (where $j$ is denotes the effective lattice site coordinate). The rotated electronic operators occurring in the expression for the spectral function are local operators and can be uniquely expressed in terms of local pseudofermionic operators. Without further due, the conversion between the two representations results in the following leading order expressions for the $\alpha\nu=c0,s1$ pseudofermions:

\begin{eqnarray}
&\text{RHB} \hspace{2.0cm} & \tilde{c}_{j,\downarrow} \left( 1-\tilde{n}_{j,\uparrow} \right) \approx f_{j,c0}f_{j',s1} \nonumber \\
&\text{LHB} \hspace{2.0cm}  & \tilde{c}_{j,\uparrow}^{\dag} \left( 1-\tilde{n}_{j,\downarrow} \right) \approx f_{j,c0}^{\dag} \label{kukhuvud}
\end{eqnarray}
where the $c0$ pseudofermion is annihilated (RHB) or created (LHB) at position $x_j = a_{c0} j$ and the $s1$ pseudofermion hole is created (RHB) at position $x_{j'} = a_{s1} j'$, such that $x_j \approx x_{j'}$ \cite{Carmspec1}. In the latter case, we also produce a $s1$ pseudofermion hole, however, this hole results from the emergence of one extra canonical momentum value in the $s1$ momentum space, and not due to the destruction of a $s1$ pseudofermion, as confirmed by the analysis in section (\ref{specdevnum}). The LHB transition is further described in section (\ref{LHBtrans}).

In Eq. (\ref{kukhuvud}), we use the approximation sign for the following reason: There should be a pre-factor on the right side of the equalities above, which in Ref. \cite{Carmspec1} is denoted $1/C_J$. For example, the strict operator equality for the dominant contribution of the LHB transition, is given by Eq. (47) of that reference. However, for our cases, this pre-factor reduces to a simple phase factor, since the quantities denoted $G_C$ and $G_J$ in this reference are both equal to one, as can be seen in Eq. (57) of that reference. Moreover, the argument of this phase factor is given by the discussion and the equalities on page 17 of that reference. Since this phase factor will not be important for the subsequent analysis, it is omitted here. 

A note on the effective $\alpha\nu$ lattice: The lattice site index $j$ is not arbitrary. Rather, a rotated electron being created or annihilated on a lattice site position $j$, corresponds in the cases considered here to creation or annihilation of $\alpha\nu=c0,s1$ pseudofermions on strictly defined $\alpha\nu$ effective lattice site coordinates. These coordinates can be found in the discussion on page 12 of Ref. \cite{Carmspec1}. We find that the $c0$ effective lattice site is the same as the rotated electronic lattice site, whilst the $s1$ effective lattice site $j'$ equals the closest integer number to $jn_{\uparrow} \rightarrow jn / 2$ as $m \rightarrow 0$. Note that $x_j\approx x_{j'}$, i.e. that the two pseudofermions are (approximatively) created or annihilated at the same spatial coordinate.

\newpage
\subsection{The RHB "Basic" transition}
\label{RHBtrans}

The RHB "Basic" transition involves creation of one $c0$ pseudofermion hole and one $s1$ pseudofermion hole. It is characterized by $\vert Q_{c0}^0 \vert = \pi$ and $\vert Q_{s1}^0 \vert =0$ and thus $\lambda_{c0}=\text{sgn}\left( Q_{c0}^0 \right)$ and  $\lambda_{s1}=0$ respectively. In the case of the "s-branch" and the "c-branch", we will need to specify at which Fermi point the $c0$ and the $s1$ pseudofermion hole, respectively, is created. For this, we will define the quantity $\iota_{\alpha\nu}=\text{sgn}(q_{F\alpha\nu})$.

\begin{enumerate}
\item {\it RHB 2P contribution:} $\hspace{1.0cm} -2k_F < q_{c0} < 2 k_F \hspace{0.5cm} -k_{F\downarrow} < q_{s1} < k_{F\downarrow} \vspace{0.5cm}$ 

$
\begin{array}{lll}
\Delta N_{c0}^{NF}=-1 \ \ \ \  & \Delta N_{c0}^F=0 \ \ \ \  & \Delta J_{c0}^F=\lambda_{c0} / 2 \vspace{0.5cm} \\
\Delta N_{s1}^{NF}=-1 \ \ \ \  & \Delta N_{s1}^F=0 \ \ \ \  & \Delta J_{s1}^F=0 \vspace{0.5cm}
\end{array}
$

$
\begin{array}{l}
Q_{c0}^{\Phi (F)} \left( \iota 2k_F \right) / 2\pi = \lambda_{c0} \left( - 1 + \xi_{c0,c0}^1\right) / 2 \vspace{0.5cm} \\
Q_{s1}^{\Phi (F)} \left( \iota k_{F\downarrow} \right) / 2\pi = \lambda_{c0} \  \xi_{s1,c0}^1 /2 \vspace{0.5cm} \\
Q_{c0}^{\Phi (NF)} \left( \iota 2k_F \right) / 2\pi = - \bigg( \Phi_{c0,c0} (\iota 2 k_F , q_{c0}) + \Phi_{c0,s1} (\iota 2 k_F , q_{s1}) \bigg) \vspace{0.5cm} \\
Q_{s1}^{\Phi (NF)} \left( \iota k_{F\downarrow} \right) / 2\pi = - \bigg( \Phi_{s1,c0} (\iota k_{F\downarrow} , q_{c0}) + \Phi_{s1,s1} (\iota k_{F\downarrow} , q_{s1}) \bigg) \vspace{0.5cm} \\
2 \Delta_{c0}^{\iota} =\displaystyle{ \left( \frac {\lambda_{c0} \xi_{c0,c0}^1} 2 + \frac {Q_{c0}^{\Phi (NF)} \left( \iota 2k_F \right)} {2\pi} \right)^2 } \vspace{0.5cm} \\
2 \Delta_{s1}^{\iota} = \displaystyle{ \left( \frac {\lambda_{c0} \xi_{s1,c0}^1} 2 + \frac {Q_{s1}^{\Phi (NF)} \left( \iota k_{F\downarrow} \right)} {2\pi} \right)^2 } \vspace{0.5cm} \\
\omega = \epsilon_{c0}(q_{c0}) + \epsilon_{s1}(q_{s1}) \hspace{1.5cm} k=q_{c0}+q_{s1}-\lambda_{c0}2k_F \vspace{0.5cm}
\end{array}
$

\newpage
\item {\it RHB s-branch:} $\hspace{1.0cm} q_{c0} = \iota_{c0} 2 k_F \hspace{0.5cm} -k_{F\downarrow} < q_{s1} < k_{F\downarrow} \vspace{0.5cm}$ 

$
\begin{array}{lll}
\Delta N_{c0}^{NF}=0  \ \ \ \  & \Delta N_{c0}^F=-1 \ \ \ \  & \Delta J_{c0}^F=(\lambda_{c0}-\iota_{c0}) / 2 \vspace{0.5cm} \\
\Delta N_{s1}^{NF}=-1 \ \ \ \  & \Delta N_{s1}^F=0 \ \ \ \  & \Delta J_{s1}^F=0 \vspace{0.5cm}
\end{array}
$

$
\begin{array}{l}
Q_{c0}^{\Phi (F)} \left( \iota 2k_F \right) / 2\pi = \iota (1-\xi_{c0,c0}^0) / 2 + (\lambda_{c0}-\iota_{c0})(\xi_{c0,c0}^1 -1) / 2 \vspace{0.5cm} \\
Q_{s1}^{\Phi (F)} \left( \iota k_{F\downarrow} \right) / 2\pi = -\iota \xi_{s1,c0}^0 / 2 + (\lambda_{c0}-\iota_{c0}) \xi_{s1,c0}^1 / 2 \vspace{0.5cm} \\
Q_{c0}^{\Phi (NF)} \left( \iota 2k_F \right) / 2\pi = -\Phi_{c0,s1} (\iota 2 k_F , q_{s1}) \vspace{0.5cm} \\
Q_{s1}^{\Phi (NF)} \left( \iota k_{F\downarrow} \right) / 2\pi = - \Phi_{s1,s1} (\iota k_{F\downarrow} , q_{s1}) \vspace{0.5cm} \\
2 \Delta_{c0}^{\iota} =\displaystyle{ \left( - \frac {\iota \xi_{c0,c0}^0} 2 + \frac {(\lambda_{c0}-\iota_{c0}) \xi_{c0,c0}^1} 2 + \frac {Q_{c0}^{\Phi (NF)} \left( \iota 2k_F \right)} {2\pi} \right)^2 } \vspace{0.5cm} \\
2 \Delta_{s1}^{\iota} = \displaystyle{ \left( - \frac {\iota \xi_{s1,c0}^0} 2 + \frac {(\lambda_{c0}-\iota_{c0}) \xi_{s1,c0}^1} 2 + \frac {Q_{s1}^{\Phi (NF)} \left( \iota k_{F\downarrow} \right)} {2\pi} \right)^2 } \vspace{0.5cm} \\ 
\omega = \epsilon_{s1}(q_{s1}) \hspace{1.5cm} k=q_{s1}-(\lambda_{c0}-\iota_{c0})2k_F \vspace{0.5cm}
\end{array}
$

\newpage
\item {\it RHB c-branch:} $\hspace{1.0cm} -2k_F < q_{c0} < 2 k_F \hspace{0.5cm} q_{s1} = \iota_{s1} k_{F\downarrow} \vspace{0.5cm}$ 

$
\begin{array}{lll}
\Delta N_{c0}^{NF}=-1  \ \ \ \  & \Delta N_{c0}^F=0 \ \ \ \  & \Delta J_{c0}^F=\lambda_{c0} / 2 \vspace{0.5cm} \\
\Delta N_{s1}^{NF}=0 \ \ \ \  & \Delta N_{s1}^F=-1 \ \ \ \  & \Delta J_{s1}^F=- \iota_{s1} / 2 \vspace{0.5cm}
\end{array}
$

$
\begin{array}{l}
Q_{c0}^{\Phi (F)} \left( \iota 2k_F \right) / 2\pi = - \iota \xi_{c0,s1}^0 / 2 + \lambda_{c0}(\xi_{c0,c0}^1 -1) / 2 - \iota_{s1} \xi_{c0,s1}^1 / 2 \vspace{0.5cm} \\
Q_{s1}^{\Phi (F)} \left( \iota k_{F\downarrow} \right) / 2\pi = \iota (1-\xi_{s1,s1}^0) / 2 + \lambda_{c0} \xi_{s1,c0}^1/ 2 + \iota_{s1} (1-\xi_{s1,s1}^1) / 2 \vspace{0.5cm} \\
Q_{c0}^{\Phi (NF)} \left( \iota 2k_F \right) / 2\pi = -\Phi_{c0,c0} (\iota 2 k_F , q_{c0}) \vspace{0.5cm} \\
Q_{s1}^{\Phi (NF)} \left( \iota k_{F\downarrow} \right) / 2\pi = - \Phi_{s1,c0} (\iota k_{F\downarrow} , q_{c0}) \vspace{0.5cm} \\
2 \Delta_{c0}^{\iota} =\displaystyle{ \left( - \frac {\iota \xi_{c0,s1}^0} 2 + \frac {\lambda_{c0} \xi_{c0,c0}^1} 2 - \frac {\iota_{s1} \xi_{c0,s1}^1} 2 + \frac {Q_{c0}^{\Phi (NF)} \left( \iota 2k_F \right)} {2\pi} \right)^2 } \vspace{0.5cm} \\
2 \Delta_{s1}^{\iota} = \displaystyle{ \left( - \frac {\iota \xi_{s1,s1}^0} 2 + \frac {\lambda_{c0} \xi_{s1,c0}^1} 2 - \frac {\iota_{s1} \xi_{s1,s1}^1} 2 + \frac {Q_{s1}^{\Phi (NF)} \left( \iota k_{F\downarrow} \right)} {2\pi} \right)^2 } \vspace{0.5cm} \\
\omega = \epsilon_{c0}(q_{c0}) \hspace{1.5cm} k=q_{c0}-\lambda_{c0}2k_F + \iota_{s1}k_{F\downarrow} \vspace{0.5cm}
\end{array}
$

\newpage
\item {\it RHB Fermi contribution:} $\hspace{1.0cm} q_{c0} = \iota_{c0} 2 k_F \hspace{0.5cm} q_{s1} = \iota_{s1} k_{F\downarrow} \vspace{0.5cm}$ 

$
\begin{array}{lll}
\Delta N_{c0}^{NF}=0 \ \ \ \  & \Delta N_{c0}^F=-1 \ \ \ \  & \Delta J_{c0}^F=(\lambda_{c0}-\iota_{c0}) / 2 \vspace{0.5cm} \\
\Delta N_{s1}^{NF}=0 \ \ \ \  & \Delta N_{s1}^F=-1 \ \ \ \  & \Delta J_{s1}^F=- \iota_{s1} / 2 \vspace{0.5cm}
\end{array}
$

$
\begin{array}{l}
Q_{c0}^{\Phi (F)} \left( \iota 2k_F \right) / 2\pi = \iota (1 - \xi_{c0,c0}^0 - \xi_{c0,s1}^0) / 2 + (\lambda_{c0}-\iota_{c0}) (\xi_{c0,c0}^1 - 1) / 2 - \iota_{s1} \xi_{c0,s1}^1 / 2 \vspace{0.5cm} \\
Q_{s1}^{\Phi (F)} \left( \iota k_{F\downarrow} \right) / 2\pi = \iota (1 - \xi_{s1,c0}^0 - \xi_{s1,s1}^0) / 2 + (\lambda_{c0}-\iota_{c0}) \xi_{s1,c0}^1 / 2 + \iota_{s1} (1 - \xi_{s1,s1}^1) / 2 \vspace{0.5cm} \\
Q_{c0}^{\Phi (NF)} \left( \iota 2k_F \right) / 2\pi = 0 \vspace{0.5cm} \\
Q_{s1}^{\Phi (NF)} \left( \iota k_{F\downarrow} \right) / 2\pi = 0 \vspace{0.5cm} \\
2 \Delta_{c0}^{\iota} =\displaystyle{ \left( \frac {\iota (\xi_{c0,c0}^0 + \xi_{c0,s1}^0)} 2 - \frac {(\lambda_{c0}-\iota_{c0}) \xi_{c0,c0}^1} 2 + \frac {\iota_{s1} \xi_{c0,s1}^1} 2 \right)^2} \vspace{0.5cm} \\
2 \Delta_{s1}^{\iota} = \displaystyle{ \left(\frac {\iota (\xi_{s1,c0}^0 + \xi_{s1,s1}^0)} 2 - \frac {(\lambda_{c0}-\iota_{c0}) \xi_{s1,c0}^1} 2 + \frac {\iota_{s1} \xi_{s1,s1}^1} 2 \right)^2} \vspace{0.5cm} \\
\omega = 0 \hspace{1.5cm} k= \iota_{s1} k_{F\downarrow} - (\lambda_{c0}-\iota_{c0}) 2k_F \vspace{0.5cm}
\end{array}
$
\end{enumerate}

\newpage
\subsection{The LHB "Basic" transition}
\label{LHBtrans}

The LHB "Basic" transition has a peculiarity that is absent in the RHB case, namely that even though no $s1$ pseudofermions are annihilated from the system, one $s1$ pseudofermion hole is being created. The created $\uparrow$-spin rotated electron opens up the canonical momentum space for the $s1$ pseudofermions, allowing one more discrete canonical momentum value in the $s1$ band, without changing the number of $s1$ pseudofermions present. Thus we have that $\Delta N_{s1}=0$ and $\Delta N_{s1}^h=+1$, which induces a shake-up in the $s1$ band, i.e. $\vert Q_{c0}^0 \vert = 0$ and $\vert Q_{s1}^0 \vert =\pi$ and thus $\lambda_{c0}=0$ and  $\lambda_{s1}=\text{sgn}\left( Q_{s1}^0 \right)$ respectively. The canonical momentum value of this $s1$ pseudofermion hole controls the dynamics of the $s1$ band, and we will thus obtain different dynamical descriptions of the spectral weight depending on whether this hole is created at any of the Fermi points or away from the Fermi points. Actually, the $s1$ pseudofermionic current will depend on where the $s1$ pseudofermion hole is created, which is also why the number $\lambda_{s1}$ will not appear in any of the expressions below. The fact that $\vert Q_{s1}^0 \vert =\pi$ does not mean that we first shake-up the $s1$ band and then create a $s1$ pseudofermion hole in it, rather, it is the emergence of one extra canonical momentum value, i.e. the increase in the total number of allowed canonical momentum points, which induces the shake-up effect. 

Thus, in spite of the creation of one $s1$ pseudofermion hole, the expression of the rotated electron creation operator in terms of pseudofermionic operators does not include a $s1$ pseudofermionic annihilation operator. However, the $s1$ pseudofermion hole that emerges, gives rise to a finite current if it appears on any of the $s1$ Fermi points (which for $ma \rightarrow 0$ coincide with the effective Brillouin zone limits). 

The way to see this is that if the hole emerges {\it inside} the $s1$ Fermi sea, it "pushes" all the $s1$ pseudofermions on its left side a half step towards the left and all the $s1$ pseudofermions on its right side a half step towards the right. This means that we will have a "half-particle addition" at each Fermi point, implying that $\Delta N_{s1}^F= \frac 1 2 + \frac 1 2 = 1$ even though $\Delta J_{s1}^F=0$. If the hole emerges {\it at} the $\iota$ Fermi point then it will only have $s1$ pseudofermions on the right ($\iota=-$) or on the left ($\iota=+$) side of it, and the above mentioned "pushing" will result in a global relocation of {\it all} $s1$ pseudofermions present in the band. In this case we have that $\Delta N_{s1}^F= 0$ (since the emergence of the hole at one of the Fermi points is cancelled by the appearance of another "pushed" $s1$ pseudofermion at the other Fermi point) and that $2\Delta J_{s1}^F=-\iota_{s1}=-\text{sgn}(q_{s1})$, where $q_{s1}$ is the momentum of the emerging $s1$ pseudofermion hole. With this peculiar effect in mind, we obtain the following characterization of the LHB "Basic" transition:

\newpage
\begin{enumerate}
\item {\it LHB 2P contribution:} $\hspace{1.0cm} 2k_F< \vert q_{c0} \vert < \pi \hspace{0.5cm} -k_{F\downarrow} < q_{s1} < k_{F\downarrow} \vspace{0.5cm}$ 

$
\begin{array}{lll}
\Delta N_{c0}^{NF}=+1 \ \ \ \  & \Delta N_{c0}^F=0 \ \ \ \  & \Delta J_{c0}^F=0 \vspace{0.5cm} \\
\Delta N_{s1}^{NF}=-1 \ \ \ \  & \Delta N_{s1}^F=1 \ \ \ \  & \Delta J_{s1}^F=0 \vspace{0.5cm}
\end{array}
$

$
\begin{array}{l}
Q_{c0}^{\Phi (F)} \left( \iota 2k_F \right) / 2\pi = \iota \xi_{c0,s1}^0 / 2 \vspace{0.5cm} \\
Q_{s1}^{\Phi (F)} \left( \iota k_{F\downarrow} \right) / 2\pi = \iota (\xi_{s1,s1}^0 -1 ) / 2 \vspace{0.5cm} \\
Q_{c0}^{\Phi (NF)} \left( \iota 2k_F \right) / 2\pi = \bigg( \Phi_{c0,c0} (\iota 2 k_F , q_{c0}) - \Phi_{c0,s1} (\iota 2 k_F , q_{s1}) \bigg) \vspace{0.5cm} \\
Q_{s1}^{\Phi (NF)} \left( \iota k_{F\downarrow} \right) / 2\pi = \bigg( \Phi_{s1,c0} (\iota k_{F\downarrow} , q_{c0}) - \Phi_{s1,s1} (\iota k_{F\downarrow} , q_{s1}) \bigg) \vspace{0.5cm} \\
2 \Delta_{c0}^{\iota} =\displaystyle{ \left(  \frac {\iota \xi_{c0,s1}^0} 2 + \frac {Q_{c0}^{\Phi (NF)} \left( \iota 2k_F \right)} {2\pi} \right)^2 } \vspace{0.5cm} \\
2 \Delta_{s1}^{\iota} = \displaystyle{ \left( \frac {\iota \xi_{s1,s1}^0} 2 + \frac {Q_{s1}^{\Phi (NF)} \left( \iota k_{F\downarrow} \right)} {2\pi} \right)^2 } \vspace{0.5cm} \\
\omega = \epsilon_{c0}(q_{c0}) - \epsilon_{s1}(q_{s1}) \hspace{1.5cm} k=q_{c0}-q_{s1} \vspace{0.5cm}
\end{array}
$

\newpage
\item {\it LHB s-branch:} $\hspace{1.0cm} q_{c0} = \iota_{c0} 2 k_F \hspace{0.5cm} -k_{F\downarrow} < q_{s1} < k_{F\downarrow} \vspace{0.5cm}$ 

$
\begin{array}{lll}
\Delta N_{c0}^{NF}=0  \ \ \ \  & \Delta N_{c0}^F=+1 \ \ \ \  & \Delta J_{c0}^F=\iota_{c0} / 2 \vspace{0.5cm} \\
\Delta N_{s1}^{NF}=-1 \ \ \ \  & \Delta N_{s1}^F=+1 \ \ \ \  & \Delta J_{s1}^F=0 \vspace{0.5cm}
\end{array}
$

$
\begin{array}{l}
Q_{c0}^{\Phi (F)} \left( \iota 2k_F \right) / 2\pi = \iota (\xi_{c0,c0}^0 + \xi_{c0,s1}^0 -1) / 2 + \iota_{c0} (\xi_{c0,c0}^1 -1) /2 \vspace{0.5cm} \\
Q_{s1}^{\Phi (F)} \left( \iota k_{F\downarrow} \right) / 2\pi = \iota (\xi_{s1,c0}^0 + \xi_{s1,s1}^0 -1) / 2 + \iota_{c0} \xi_{s1,c0}^1 /2 \vspace{0.5cm} \\
Q_{c0}^{\Phi (NF)} \left( \iota 2k_F \right) / 2\pi = -\Phi_{c0,s1} (\iota 2 k_F , q_{s1}) \vspace{0.5cm} \\
Q_{s1}^{\Phi (NF)} \left( \iota k_{F\downarrow} \right) / 2\pi = - \Phi_{s1,s1} (\iota k_{F\downarrow} , q_{s1}) \vspace{0.5cm} \\
2 \Delta_{c0}^{\iota} =\displaystyle{ \left( \frac {\iota \left( \xi_{c0,c0}^0 + \xi_{c0,s1}^0 \right) } 2 + \frac {\iota_{c0} \xi_{c0,c0}^1} 2 + \frac {Q_{c0}^{\Phi (NF)} \left( \iota 2k_F \right)} {2\pi} \right)^2 } \vspace{0.5cm} \\
2 \Delta_{s1}^{\iota} = \displaystyle{ \left( \frac {\iota \left( \xi_{s1,c0}^0 + \xi_{s1,s1}^0 \right) } 2 + \frac {\iota_{c0} \xi_{s1,c0}^1} 2 + \frac {Q_{s1}^{\Phi (NF)} \left( \iota k_{F\downarrow} \right)} {2\pi} \right)^2 } \vspace{0.5cm} \\ 
\omega = - \epsilon_{s1}(q_{s1}) \hspace{1.5cm} k=\iota_{c0}2k_F - q_{s1} \vspace{0.5cm}
\end{array}
$

\newpage
\item {\it LHB c-branch:} $\hspace{1.0cm} 2k_F< \vert q_{c0} \vert < \pi \hspace{0.5cm} q_{s1} = \iota_{s1} k_{F\downarrow} \vspace{0.5cm}$ 

$
\begin{array}{lll}
\Delta N_{c0}^{NF}=+1  \ \ \ \  & \Delta N_{c0}^F=0 \ \ \ \  & \Delta J_{c0}^F=0 \vspace{0.5cm} \\
\Delta N_{s1}^{NF}=0 \ \ \ \  & \Delta N_{s1}^F=0 \ \ \ \  & \Delta J_{s1}^F=- \iota_{s1} / 2 \vspace{0.5cm}
\end{array}
$

$
\begin{array}{l}
Q_{c0}^{\Phi (F)} \left( \iota 2k_F \right) / 2\pi = -\iota_{s1} \xi_{c0,s1}^1 / 2 \vspace{0.5cm} \\
Q_{s1}^{\Phi (F)} \left( \iota k_{F\downarrow} \right) / 2\pi = \iota_{s1} \left( 1-\xi_{s1,s1}^1 \right) / 2 \vspace{0.5cm} \\
Q_{c0}^{\Phi (NF)} \left( \iota 2k_F \right) / 2\pi = \Phi_{c0,c0} (\iota 2 k_F , q_{c0}) \vspace{0.5cm} \\
Q_{s1}^{\Phi (NF)} \left( \iota k_{F\downarrow} \right) / 2\pi = \Phi_{s1,c0} (\iota k_{F\downarrow} , q_{c0}) \vspace{0.5cm} \\
2 \Delta_{c0}^{\iota} =\displaystyle{ \left( -\frac {\iota_{s1} \xi_{c0,s1}^1} 2 + \frac {Q_{c0}^{\Phi (NF)} \left( \iota 2k_F \right)} {2\pi} \right)^2 } \vspace{0.5cm} \\
2 \Delta_{s1}^{\iota} = \displaystyle{ \left( -\frac {\iota_{s1} \xi_{s1,s1}^1} 2 + \frac {Q_{s1}^{\Phi (NF)} \left( \iota k_{F\downarrow} \right)} {2\pi} \right)^2 } \vspace{0.5cm} \\
\omega = \epsilon_{c0}(q_{c0}) \hspace{1.5cm} k=q_{c0} - \iota_{s1}k_{F\downarrow} \vspace{0.5cm}
\end{array}
$

\newpage
\item {\it LHB Fermi contribution:} $\hspace{1.0cm} q_{c0} = \iota_{c0} 2 k_F \hspace{0.5cm} q_{s1} = \iota_{s1} k_{F\downarrow} \vspace{0.5cm}$ 

$
\begin{array}{lll}
\Delta N_{c0}^{NF}=0 \ \ \ \  & \Delta N_{c0}^F=+1 \ \ \ \  & \Delta J_{c0}^F=\iota_{c0} / 2 \vspace{0.5cm} \\
\Delta N_{s1}^{NF}=0 \ \ \ \  & \Delta N_{s1}^F=0 \ \ \ \  & \Delta J_{s1}^F=- \iota_{s1} / 2 \vspace{0.5cm}
\end{array}
$

$
\begin{array}{l}
Q_{c0}^{\Phi (F)} \left( \iota 2k_F \right) / 2\pi = \iota (\xi_{c0,c0}^0 - 1) / 2 + \iota_{c0} (\xi_{c0,c0}^1 - 1) / 2 -\iota_{s1} \xi_{c0,s1}^1 / 2 \vspace{0.5cm} \\
Q_{s1}^{\Phi (F)} \left( \iota k_{F\downarrow} \right) / 2\pi = \iota \xi_{s1,c0}^0 / 2 + \iota_{c0} \xi_{s1,c0}^1 / 2 + \iota_{s1} (1 - \xi_{s1,s1}^1 ) / 2 \vspace{0.5cm} \\
Q_{c0}^{\Phi (NF)} \left( \iota 2k_F \right) / 2\pi = 0 \vspace{0.5cm} \\
Q_{s1}^{\Phi (NF)} \left( \iota k_{F\downarrow} \right) / 2\pi = 0 \vspace{0.5cm} \\
2 \Delta_{c0}^{\iota} =\displaystyle{ \left( \frac {\iota \xi_{c0,c0}^0} 2 + \frac {\iota_{c0} \xi_{c0,c0}^1} 2 - \frac {\iota_{s1} \xi_{c0,s1}^1} 2 \right)^2} \vspace{0.5cm} \\
2 \Delta_{s1}^{\iota} = \displaystyle{ \left(\frac {\iota \xi_{s1,c0}^0} 2 + \frac {\iota_{c0} \xi_{s1,c0}^1} 2 - \frac {\iota_{s1} \xi_{s1,s1}^1} 2 \right)^2} \vspace{0.5cm} \\
\omega = 0 \hspace{1.5cm} k= \iota_{c0} 2k_F - \iota_{s1} k_{F\downarrow} \vspace{0.5cm}
\end{array}
$
\end{enumerate}
\setcounter{chapter}{3}
\setcounter{section}{3}

\chapter{The One Electron Spectral Function}
\label{onelecspec}

\section{Basic Derivation}

\subsection{Matrix elements and pseudofermion operators}
\label{opfact}

In this chapter, we express the state generators and the operators of the matrix elements of the spectral function in terms of pseudofermionic creation and annihilation operators. We remind ourselves that the operator expressions presented here are special cases of the ones calculated in Ref. \cite{Carmspec1}. In the following, we will only focus on the necessary ingredients of the one electron spectral function. Thus, for example, Fourier transforms of a product of an arbitrary number of pseudofermionic operators will not be formally performed here, but the procedure will be described in general terms and the final results presented.
Moreover, in this section, we will temporarily revert to the momentum representation (which is also used in Ref. \cite{Carmspec1}) and thus write $\tilde{c}_{k,\sigma}^l$ for the rotated electron creation ($l=+$) or annihilation ($l=-$) operator. 

In section (\ref{occpp}), we defined $q_{F\alpha\nu}$ as the positive ground state $\alpha\nu=c0,s1$ Fermi momentum in the thermodynamic limit, $q_{Fc0} = 2k_F$ and $q_{Fs1} = k_{F\downarrow}$. These quantities are dubbed $q_{F\alpha\nu}^0$ in Ref. \cite{Carmspec2}. First off, we will define the following quantities:
\begin{eqnarray}
q_{F\alpha\nu,\iota}&=&\iota q_{F\alpha\nu}+\Delta q_{F\alpha\nu,\iota} \hspace{1.0cm} \Delta q_{F\alpha\nu,\iota} = \iota \frac {2\pi} L \Delta N_{\alpha\nu,\iota}^{F,0} \nonumber \\
\bar{q}_{F\alpha\nu,\iota}&=&\iota q_{F\alpha\nu}+\Delta \bar{q}_{F\alpha\nu,\iota} \hspace{1.0cm} \Delta \bar{q}_{F\alpha\nu,\iota} = \iota \frac {2\pi} L \Delta N_{\alpha\nu,\iota}^F + \frac {Q_{\alpha\nu}^{\Phi}(\iota q_{F\alpha\nu})} L
\end{eqnarray}
which are nothing but the deviations of the Fermi point momenta and canonical momenta, respectively, in the excited state configuration relative to that of the ground state. We notice that since the ground state consists of two densely packed minimum energy Fermi seas (i.e. with all the canonical momenta symmetrically distributed around zero), one for $c0$ and one for $s1$, it can be quite easily expressed in terms of $c0$ and $s1$ pseudofermionic creation operators
\begin{equation}
\vert GS \rangle = \prod_{\alpha\nu=c0,s1} \prod_{i=1}^{N_{\alpha\nu}^0} f_{q_i, \alpha\nu}^{\dag} \vert 0 \rangle \label{gs}
\end{equation}
where $\vert 0 \rangle$ is the pseudofermion vacuum state. 

Now, a typical matrix element occurring in the definition of the spectral function of Eq. (\ref{defspecfs}), is of the form $\langle f_l \vert \tilde{c}_{k\sigma}^l \vert GS \rangle$. Here $\vert f_l \rangle$ denotes an excited energy eigenstate and $\vert GS \rangle$ is the ground state of Eq. (\ref{gs}). Due to the findings of chapter (\ref{dynamics}), we can rewrite the rotated electron creation ($l=+$) or annihilation ($l=-$) operator, in terms of operators that creates and/or annihilates $\alpha\nu=c0,s1$ pseudofermions according to the $\mathbb{A}$ and the $\mathbb{B}$ processes, respectively. Thus, let us write $\tilde{c}_{k\sigma}^l = \tilde{c}_{\mathbb{A},k\sigma}^l \tilde{c}_{\mathbb{B}\sigma}^l$, where $\tilde{c}_{\mathbb{A},k\sigma}^l$ is associated with the $\mathbb{A}$ processes, and $\tilde{c}_{\mathbb{B}\sigma}^l$ with the $\mathbb{B}$ processes. The latter operator creates a number $\sum_{\alpha\nu=c0,s1} \Delta N_{\alpha\nu}^F$ of pseudofermions at the $\alpha\nu=c0,s1$ Fermi points, producing a densely packed state denoted $\vert \widetilde{GS} \rangle = \tilde{c}_{\mathbb{B}\sigma}^l \vert GS \rangle$. This state can be expressed solely in terms of pseudofermion creation operators as:

\begin{equation}
\vert \widetilde{GS} \rangle = \prod_{\alpha\nu=c0,s1} \prod_{i=1}^{N_{\alpha\nu}^0+\Delta N_{\alpha\nu}^F} f_{q_i,\alpha\nu}^{\dag} \vert 0 \rangle \label{cgs}
\end{equation}
where $q_1=-q_{F\alpha\nu}+\Delta q_{F\alpha\nu,-}$ and $q_{N_{\alpha\nu}^0+\Delta N_{\alpha\nu}^F}=q_{F\alpha\nu}+\Delta q_{F\alpha\nu,+}$. The matrix element $\langle f_l \vert \tilde{c}_{k\sigma}^l \vert GS \rangle$ can then be written as $\langle f_l \vert \tilde{c}_{\mathbb{A},k\sigma}^l \vert \widetilde{GS} \rangle$.

Consider now the state $\langle \widetilde{f_l} \vert = \langle f_l \vert \tilde{c}_{\mathbb{A},k\sigma}^l$. Upon acting onto $\langle f_l \vert$, the operator $\tilde{c}_{\mathbb{A},k\sigma}^l$ removes the finite energy pseudofermions or pseudofermion holes created under the $\vert GS \rangle \rightarrow \vert f_l \rangle$ transition, due to the $\mathbb{A}$ processes, and thus $\langle \widetilde{f_l} \vert = \langle f_l \vert \tilde{c}_{\mathbb{A},k\sigma}^l$ is also a densely packed state. The state $\vert \widetilde{f_l} \rangle$ can be generated from the pseudofermion vacuum, by acting with pseudofermion creation operators carrying canonical momentum values $\bar{q}_i=q_i+Q_{\alpha\nu}^{\Phi}(q_i) / L$:

\begin{equation}
\vert \widetilde{f_l} \rangle = \prod_{\alpha\nu=c0,s1} \prod_{i=1}^{N_{\alpha\nu}^0+\Delta N_{\alpha\nu}^F} f_{\bar{q}_i,\alpha\nu}^{\dag} \vert 0 \rangle \label{finalst}
\end{equation}

Here, $Q_{\alpha\nu}^{\Phi}(q_i) / L$ is the scattering part of the canonical momentum shift of the excited energy eigenstate $\vert f_l \rangle$, relative to the corresponding ground state discrete momentum $q_i$. This means that after taking into account the $\mathbb{A}$ and the $\mathbb{B}$ processes, the matrix element $\langle f_l \vert \tilde{c}_{\mathbb{A},k\sigma}^l \vert \widetilde{GS} \rangle$ can be written as $\langle \widetilde{f_l} \vert \widetilde{GS} \rangle$. 

These processes take the excitation to a certain energy and momentum in the ($k,\omega$) plane, upon which the spectral weight is calculated. On top of this we build a tower of particle-hole states, generated by the application of the operator $F_{p-h,\alpha\nu}^{\dag}$ onto the state $\vert f_l \rangle$. This operator reads
\begin{equation}
F_{p-h,\alpha\nu}^{\dag} = \prod_{i=1}^{N_{\alpha\nu}^{ph}} f_{\bar{q}_i,\alpha\nu}^{\dag} f_{\bar{q}'_i,\alpha\nu} \label{Fph}
\end{equation}
for a total number of $N_{\alpha\nu}^{ph}$ particle-hole pairs in the vicinity of any of the two $\alpha\nu$ Fermi points.

In order to proceed, we focus now on the matrix element $\langle \widetilde{f_l} \vert F_{p-h,\alpha\nu} \vert \widetilde{GS} \rangle$. This quantity originates from the matrix element $\langle f_l \vert F_{p-h,\alpha\nu} \tilde{c}_{k\sigma}^l \vert GS \rangle$, that involves the final state with $N_{\alpha\nu}^{ph}$ particle-hole pairs, namely $F_{p-h,\alpha\nu}^{\dag} \vert f_l \rangle$. This state obeys the following equality:
\begin{equation}
\langle f_l \vert F_{p-h,\alpha\nu} \tilde{c}_{k\sigma}^l \vert GS \rangle = \langle f_l \vert F_{p-h,\alpha\nu} \tilde{c}_{\mathbb{A},k\sigma}^l \vert \widetilde{GS} \rangle \label{almosttheere}
\end{equation}

Importantly, we note that the excited states $\vert f_l \rangle$ and $F_{p-h,\alpha\nu}^{\dag}\vert f_l \rangle$ have the same canonical momentum shift $Q_{\alpha\nu}(q_i) / L = Q_{\alpha\nu}^0 / L + Q_{\alpha\nu}^{\Phi}(q_i) / L$, for every $q_i$, due to the findings of section (\ref{constantQ}).

Since $F_{p-h,\alpha\nu}$ involve pseudofermion operators with canonical momenta {\it in the vicinity} of the $\alpha\nu=c0,s1$ Fermi points, whilst $\tilde{c}_{\mathbb{A},k\sigma}^l$ can be expressed in terms of pseudofermion operators with canonical momenta {\it away from} the $\alpha\nu=c0,s1$ Fermi points, we have that

\begin{equation}
\Big[ F_{p-h,\alpha\nu} , \tilde{c}_{\mathbb{A},k\sigma}^l \Big] = 0
\end{equation}
and therefore, $\langle f_l \vert F_{p-h,\alpha\nu} \tilde{c}_{\mathbb{A},k\sigma}^l \vert \widetilde{GS} \rangle = \langle \widetilde{f_l} \vert F_{p-h,\alpha\nu} \vert \widetilde{GS} \rangle$. Since the momentum values $q_i$ of Eq. (\ref{cgs}) and the canonical momentum values $\bar{q}_i$ of Eq. (\ref{finalst}) are slightly different, the general matrix elements of the form $\langle \widetilde{f_l} \vert F_{p-h,\alpha\nu} \vert \widetilde{GS} \rangle$ are finite, even for $F_{p-h,\alpha\nu}^{\dag} \neq \bm{1}$.

In conclusion, after taking into account the occupancy configuration transformations produced by the $\mathbb{A}$ and $\mathbb{B}$ processes, the typical matrix elements  $\langle f_l \vert F_{p-h,\alpha\nu} \tilde{c}_{k\sigma}^l \vert GS \rangle$ can be rewritten as $\langle \widetilde{f_l} \vert F_{p-h,\alpha\nu} \vert \widetilde{GS} \rangle$. The states $\vert \widetilde{GS} \rangle$ and $\vert \widetilde{f_l} \rangle$ correspond to the densely packed occupancy configurations given in Eqs. (\ref{cgs}) and (\ref{finalst}), respectively, and the hermitian conjugate of the operator $F_{p-h,\alpha\nu}$ is given in Eq. (\ref{Fph}). This overlap involves an excited state whose particle-hole occupancy configuration includes $N_{\alpha\nu}^{ph}$ pseudofermion particle-hole pairs in the $\alpha\nu=c0,s1$ bands. In the subsequent section, we will then sum over all possible particle-hole occupancy configurations corresponding to the same tower of states. We recall that the value of the phase shift $Q_{\alpha\nu}^{\Phi}(\iota q_{F\alpha\nu})$ is constant for each particle-hole tower of states. 

The only remainder of the finite energy pseudofermion (or pseudofermion hole), is then a Fourier canonical momentum summation, introduced by the Fourier transform of the corresponding pseudofermion operator defined on the effective $\alpha\nu$ lattice. This Fourier transform is formally treated in Ref. \cite{Carmspec1}. The finite energy canonical momentum will in the following be denoted $\tilde{q}_{\alpha\nu}$, for $\alpha\nu=c0,s1$. In section (\ref{closedform}), this summation will be crucial to the derivation of closed form expressions for the full spectral function. We note that the energy $l \Delta E$ and the momentum $l \Delta P$ are functions of these canonical momentum variables.

\subsection{Convolutions}
\label{convo}

Having dealt with the contribution from the operators creating or annihilating pseudofermions away from any of the Fermi points, it still remains to treat the problem involving the matrix overlap with the operator $F_{p-h,\alpha\nu}^{\dag}$, Eq. (\ref{almosttheere}). In the following, we define the subspace $\mathfrak{C}_{\alpha\nu}$ as {\it equivalent} to the subspace of Ref. \cite{Carmspec2} called "J-CPHS-$\alpha\nu$-(C)" occurring for example in Eq. (45) of that reference. Thus, the states of this subspace are described by pseudofermion occupancy configurations generated by the $\mathbb{C}$ particle-hole processes, with fixed values of the deviation numbers $\{ \Delta N_{\alpha\nu} \}$ and $\{ \Delta N_{\alpha\nu,\iota}^F \}$ (for $\alpha\nu=c0,s1$ and $\iota=\pm$). Let a typical element of this subspace describe a particle-hole occupancy configuration with energy $\omega_{\alpha\nu} = \Delta E_{\alpha\nu} (\mathbb{C})$ and with momentum $k_{\alpha\nu}= \Delta P_{\alpha\nu} (\mathbb{C})$, according to Eq. (\ref{enemoms}). We define the total particle-hole energy and momentum as the $\alpha\nu$ sum of the energy and momentum of the individual branches: $\Delta \omega = \omega_{c0}+\omega_{s1}$ and $\Delta \omega / v = k_{c0}+k_{s1}$, respectively. Note that the last definition also defines the particle-hole velocity $v$.

In order to evaluate the relevant matrix elements, we have already seen in section (\ref{opfact}) that we can describe a general excited energy eigenstate in terms of pseudofermion creation and annihilation operators in canonical momentum space, acting upon a vacuum state. As a consequence, the spectral weight associated with the amplitude $\mathcal{M}_{Q_{\alpha\nu}}^l = \big\vert \langle \widetilde{f_l} \vert F_{p-h,\alpha\nu} \vert \widetilde{GS} \rangle \big\vert^2$ can be written as a convolution between the $c0$ and the $s1$ branches. This involves the definition of an auxiliary function that reads

\begin{eqnarray}
&&\breve{B}^l (v, \Delta \omega) = \sum_{\mathfrak{C}_{c0}} \sum_{\mathfrak{C}_{s1}} \int d\omega' \ \ \delta (\omega' -\omega_{c0}) \delta (\Delta \omega - \omega' -\omega_{s1}) \cdot \nonumber \\ 
&\cdot& \frac 1 {N_a} \sum_{k'}  \delta_{k',k_{c0}}  \delta_{\Delta \omega / v,k'+k_{s1}} \  \mathcal{M}_{Q_{c0}}^l \cdot \mathcal{M}_{Q_{s1}}^l = \label{convolute} \\
&=&  \frac {\text{sgn}(v)} {2\pi} \left( \frac {2\pi} L \right)^2 \sum_{\mathfrak{C}_{c0}} \sum_{\mathfrak{C}_{s1}} \int_0^{\Delta \omega} d\omega' \  \int_{-\text{sgn}(v)\Delta \omega / v_{c0}}^{\text{sgn}(v) \Delta \omega / v_{c0}} dk' \ \delta (\omega' - \omega_{c0}) \delta (k' - k_{c0}) \cdot  \nonumber \\
&&\cdot  \ \delta (\Delta \omega-\omega' - \omega_{s1}) \delta (\Delta \omega / v-k' - k_{s1}) \ \mathcal{M}_{Q_{c0}}^l \cdot \mathcal{M}_{Q_{s1}}^l \nonumber
\end{eqnarray}
where $\Delta \omega$ and $\Delta \omega / v$ are the particle-hole energy and momentum, respectively, The expression of the amplitude $\mathcal{M}_{Q_{\alpha\nu}}^l$ in terms of pseudofermion creation and annihilation operators acting on the pseudofermion vacuum reads:

\begin{eqnarray}
&& \mathcal{M}_{Q_{\alpha\nu}}^l =\big\vert \langle 0 \vert  \prod_{\bar{q}=-q_{F\alpha\nu}+\Delta \bar{q}_{F\alpha\nu,-}}^{q_{F\alpha\nu}+\Delta \bar{q}_{F\alpha\nu,+}}\! \! \! \! \! \! \! \! \! f_{\bar{q},\alpha\nu} \ \  F_{p-h,\alpha\nu} \! \! \! \! \! \! \! \! \! \prod_{q=-q_{F\alpha\nu}+\Delta q_{F\alpha\nu,-}}^{q_{F\alpha\nu}+\Delta q_{F\alpha\nu,+}} \! \! \! \! \! \! \! \! \! f_{q,\alpha\nu}^{\dag} \ \vert 0 \rangle \big\vert ^2
\end{eqnarray}

In Eq. (\ref{convolute}), the factor ($1/N_a$) in the second line stems from the introduction of the discrete $k'$ sum. The jacobian that arises when making the momentum a continuous variable cancels this factor and the pre-factor of $(2\pi / L)^2$ arises due to turning the Kr\"onecker $\delta$-functions into Dirac $\delta$-functions, according to Eq. (\ref{krondirac}).

The introduction of $\text{sgn}(v)$ is necessary in order to keep the integration limits of the $k'$ integration in proper order: for $l=-$ we have that the particle-hole excitations grow in the negative $\omega$ direction whilst for $l=+$ we have the opposite situation. In other words, in the tower of particle-hole states, positive velocities for $l=+$ is equivalent to negative velocities for $l=-$ and vice versa. This ensures that the lower integration limit is always smaller than the upper integration limit.

The energy $\Delta \omega$ runs by definition over a small range of particle-hole energies from zero to a small negative number (RHB) or from zero to a small positive number (LHB), whilst the momentum runs over a small symmetrical interval around zero. 

The "Luttinger contribution" specified in section (\ref{character}) is not described by the function given by Eq. (\ref{convolute}). In the region of small excitation energy, the branch line group velocity equals one of the velocities of the particle-hole excitations. In this case, the pseudofermion or pseudofermion hole is created at the Fermi points by the $\mathbb{B}$ processes. We can then view the build-up of the $\alpha$-branch line, as moving this pseudofermion along its dispersion starting at one of its Fermi points, and in this way tracing out the branch line in the ($k,\omega$) plane. However, for the first low lying energies that the $\alpha\nu$ pseudofermion assumes after having left the Fermi point, the dispersion is in its linear regime. This means that for this intermediate region, the $\mathbb{C}$ processes replace the $\mathbb{A}$ processes in generating the branch line. This special case will be treated separately in section (\ref{luttinger}). In the following, we will always assume that we are outside of the region belonging to the Luttinger regime.

We note that the final expression of Eq. (\ref{convolute}) is equivalent to Eq. (45) of Ref. \cite{Carmspec2}, by identifying the spectral function $B_{Q_{\alpha\nu}}^l$ occurring in that reference with the expressions found here, according to:

\begin{eqnarray}
\frac {2\pi} L \sum_{\mathfrak{C}_{\alpha\nu}}  \mathcal{M}_{Q_{\alpha\nu}}^l \delta (\omega' - \omega_{\alpha\nu}) \delta (k' - k_{\alpha\nu}) &=& \nonumber \\ 
\sum_{\mathfrak{C}_{\alpha\nu}}  \mathcal{M}_{Q_{\alpha\nu}}^l \delta (\omega' - \omega_{\alpha\nu}) \delta_{k' , k_{\alpha\nu}} &=& B_{Q_{\alpha\nu}}^l (k',\omega') \label{fuckingjose}
\end{eqnarray}

\subsection{The energy cutoff ${\bf \Omega}$}

In this section, we will focus on the highest possible excitation energy value that a tower of particle-hole states can achieve. This energy is then ultimately a measure of how good an approximation the linearization regime will be, since, by forcing this value to be very small, we obtain a low tower of states albeit with good accuracy. Letting this value grow, we indeed obtain a higher tower of particle-hole states, allowing us to take into account more and more particle-hole processes and thereby aspire to account for almost the entire spectral weight of the problem (as measured by the sum rules). However, if we let this energy to be too large, we will start to consider non physical processes as the linear approximation of the dispersion relations become less and less accurate.

This quantity hereby introduced will be called the {\it energy cutoff}, denoted by $\Omega$. There is in principle an exact value for $\Omega$, which depends on the width of the linear regime of the pseudofermion band and on the four $2\Delta_{\alpha\nu}^{\iota}$'s, given in Eq. (\ref{twodelta}). However, since the dependence of $\Omega$ on these quantities is an open problem, we will replace the exact value by an effective value. In the following, we will assume a constant cutoff for the entire ($k,\omega$) plane for all types of contributions. We can then deduce an average value of such an effective $\Omega$ by imposing the sum rules. 

As a prologue to the introduction of the cutoff into our expressions, we should first mention that the pseudofermion velocities
\begin{eqnarray}
v_{c0}(q_{c0}) &=& \frac {d\epsilon_{c0}(q_{c0})} {dq} \hspace{1.0cm} \vert v_{c0}(q_{c0}) \vert \leq v_{c0}\ \ \ ,\ \vert q_{c0} \vert \leq q_{Fc0} \nonumber \\
v_{s1}(q_{s1}) &=& \frac {d\epsilon_{s1}(q_{s1})} {dq} \hspace{1.0cm} \vert v_{s1}(q_{s1}) \vert \leq v_{s1}\ \ \ ,\ \vert q_{s1} \vert \leq q_{Fs1}
\end{eqnarray}
obey the following relations (for $m \rightarrow 0$):
\begin{eqnarray}
\begin{array}{lll}
v_{s1} \leq v_{c0}  & \ \ \ \ \ \ \  & 0 < n < 1, \ {\displaystyle \frac U t } < \infty  \vspace{0.3cm} \\
v_{s1} \rightarrow 0  & \ \ \ \ \ \ \  & {\displaystyle \frac U t } \rightarrow \infty \vspace{0.3cm} \\
v_{s1} \rightarrow v_{c0}  & \ \ \ \ \ \ \  & {\displaystyle \frac U t } \rightarrow 0
\end{array}
\label{relvels}
\end{eqnarray}
which are easily deduced by suitable expansions of Eq. (\ref{Vvel}).

From Eq. (\ref{convolute}), we see that the domain of $\Delta \omega$ is defined by the value of $\Omega$. Indeed, we have that $0 < l \Delta \omega < \Omega$ by the definition of the cutoff. For the energy $\omega$ due to the $\mathbb{A}$ and $\mathbb{C}$ processes, we have that $\omega =  \Delta \omega + l \Delta E$. In other words, the criteria induced by the cutoff is that the spectral function treated in Eq. (\ref{convolute}), should be multiplied by two $\theta$-functions: $\theta \left(l \left[ \omega - l\Delta E \right] \right) \theta \left( \Omega - l \left[ \omega - l\Delta E \right] \right)$, in order for the energy $\Delta \omega$ of that equation to stay in the valid regime for the particle-hole tower of states.

For the momentum part, we will use the fact that between the two pseudofermion branches available for particle-hole processes, it is the $s1$ branch that always has the lowest Fermi velocity between the two, as stated in Eq. (\ref{relvels}). In other words, the velocity $v$ of Eq. (\ref{convolute}) must be such that $\vert v \vert > v_{s1}$. This can be implemented by introducing $\theta (\vert v \vert - v_{s1})=\theta( l \left[ \omega - l\Delta E \right] - v_{s1} \vert k - l\Delta P \vert)$

Thus, we arrive to the following expression for the full spectral function:

\begin{eqnarray}
B^l (k,\omega) &=& \sum_{ \{ \Delta N_{\alpha\nu} \} } \sum_{\{ \Delta N_{\alpha\nu,\iota}^F \}} \sideset {} {_{}^{\bf{\prime}}} \sum_{\tilde{q}_{\alpha\nu}} \theta \left(l \left[ \omega - l\Delta E \right] \right) \theta \left( \Omega - l \left[ \omega - l\Delta E \right] \right) \cdot  \nonumber \\
&\cdot& \theta \Big( l \left[ \omega - l\Delta E \right] - v_{s1} \vert k - l\Delta P \vert \Big) \breve{B}^l \left( \frac {\omega- l\Delta E} {k-l\Delta P} \ ,\  \omega- l\Delta E \right) \label{fullsofar} \\
\breve{B}^l (v, \Delta \omega) &=& \frac {\text{sgn}(v)} {2\pi} \int_0^{\Delta \omega} d\omega' \  \int_{-\text{sgn}(v) \Delta \omega / v_{c0}}^{\text{sgn}(v) \Delta \omega / v_{c0}} dk' \  B_{Q_{c0}} (k',\omega') B_{Q_{s1}} (\Delta \omega / v - k', \Delta \omega - \omega') \nonumber
\end{eqnarray}
where the primed summation in the first line stands for $\prod_{\alpha\nu=c0,s1} \frac 1 {N_{\alpha\nu}^*} \sum_{\tilde{q}_{\alpha\nu}}$ in the case of the 2P contribution, $\frac 1 {N_{\alpha\nu}^*} \sum_{\tilde{q}_{\alpha\nu}}$ for the $\alpha\nu=c0$ or $\alpha\nu=s1$ 1P contribution or just $1$ (no summation) in the case of the 0P contribution. Here, the variable $\tilde{q}_{\alpha\nu}$ stands for the canonical momentum of the created or annihilated finite energy $\alpha\nu=c0,s1$ pseudofermion.

Note that when creating this finite energy $\alpha\nu$ pseudofermion, the range of the $\tilde{q}_{\alpha\nu}$ canonical momentum summation goes from $-q_{\alpha\nu}^0$ to $-q_{F\alpha\nu}$ and from $q_{F\alpha\nu}$ to $q_{\alpha\nu}^0$. When annihilating the finite energy $\alpha\nu$ pseudofermion, this range goes from $-q_{F\alpha\nu}$ to $q_{F\alpha\nu}$. 

\subsection{The lowest peak weight $A^{(0,0)}$}

In order to calculate the lowest peak weight, from which the particle-hole excitations part, we need to consider matrix overlaps with $F_{p-h,\alpha\nu}^{\dag}=\bm{1}$. This means that we do not allow any particle-hole excitations for the lowest peak weight, defining $A^{(0,0)}$ as the spectral weight associated with the $\mathbb{A}$ and $\mathbb{B}$ processes. To evaluate the matrix overlap we will use the following general result, which is easily proved by induction:
\begin{equation}
\big \vert \langle 0 \vert \prod_{j=1}^M f_{\bar{q}_j} \prod_{j=1}^M f_{\bar{q}'_j}^{\dag} \vert 0 \rangle \big\vert^2 = \Bigg\vert \left\vert
\begin{array}{cccc}
\left\{ f_{\bar{q}'_1}^{\dag} , f_{\bar{q}_1} \right\}  & \left\{ f_{\bar{q}'_1}^{\dag} , f_{\bar{q}_2} \right\}  & \ldots & \left\{ f_{\bar{q}'_1}^{\dag} , f_{\bar{q}_M} \right\} \vspace{0.5cm} \\
\left\{ f_{\bar{q}'_2}^{\dag} , f_{\bar{q}_1} \right\}  & \left\{ f_{\bar{q}'_2}^{\dag} , f_{\bar{q}_2} \right\}  & \ldots & \left\{ f_{\bar{q}'_2}^{\dag} , f_{\bar{q}_M} \right\}  \\
\vdots  & \  & \ddots & \  \\
\left\{ f_{\bar{q}'_M}^{\dag} , f_{\bar{q}_1} \right\}  & \left\{ f_{\bar{q}'_M}^{\dag} , f_{\bar{q}_2} \right\}  & \ldots & \left\{ f_{\bar{q}'_M}^{\dag} , f_{\bar{q}_M} \right\}  
\end{array}
\right\vert \Bigg\vert^2
\end{equation}

In other words, the evaluation of the spectral function $B_{Q_{\alpha\nu}}^l (k',\omega')$ with $F_{p-h,\alpha\nu}^{\dag}=\bm{1}$ reduces to evaluate a $(N_{\alpha\nu}^0 + \Delta N_{\alpha\nu}^F) \times (N_{\alpha\nu}^0 + \Delta N_{\alpha\nu}^F)$ determinant.  The anticommutator is given by Eq. (\ref{anticomm}), but that expression simplifies here since all of the creation operators refer to the ground state with $Q_{\alpha\nu}^{\Phi}(q)=0$ for all $q$ in the entire Fermi sea. This means that Eq. (\ref{anticomm}) becomes:

\begin{equation}
\{ f_{q',\alpha \nu}^{\dag} , f_{\bar{q},\alpha\nu} \} = \frac 1 {N_{\alpha \nu}^*} e^{i a_{\alpha \nu} (q'-\bar{q}) / 2}  e^{-iQ_{\alpha \nu}(q)/ 2} \frac {\sin \left( \frac {Q_{\alpha \nu}(q)} 2 \right)} {\sin \left( \frac {a_{\alpha \nu} (\bar{q}-q')} 2  \right)} \label{anticomm2}
\end{equation}
which can be substituted into the expression for the determinant. In the evaluation, we will use the following exact result for the so called "Cauchy determinants":

\begin{equation}
\text{det} \left( \frac 1 {a_i - b_j} \right) = (-1)^{M(M-1)/2} \cdot  \frac {\underset{i<j}\prod (a_i - a_j)(b_i - b_j)} {\underset{i,j}\prod (a_i - b_j)}
\end{equation}

After some algebra, we find that by introducing the expression for the anticommutator into the determinant, we can express each entry as the difference $1/[\cot (q'/2) - \cot (\bar{q}/2)]$, which allows for a direct comparison with the Cauchy formula. We thus arrive to:

\begin{eqnarray}
&&A^{(0,0)}= \prod_{\alpha\nu=c0,s1} A_{\alpha\nu}^{(0,0)} \label{karloexpr} \\
&&A_{\alpha\nu}^{(0,0)} = \frac 1 {\left[N_{\alpha\nu}^* \right]^{2(N_{\alpha\nu}^0 + \Delta N_{\alpha\nu}^F)}} \underset{i=1}{\overset{N_{\alpha\nu}^0 + \Delta N_{\alpha\nu}^F}\prod} \sin^2 \left( \frac {Q_{\alpha\nu}(q_i)} 2 \right) \cdot  \frac {\underset{j=1}{\overset{i-1}\prod} \sin^2 \left( \frac {\bar{q}_i - \bar{q}_j} 2 \right) \sin^2 \left( \frac {q_i - q_j} 2 \right)} {\underset{j=1}{\overset{N_{\alpha\nu}^0 + \Delta N_{\alpha\nu}^F}\prod} \sin^2 \left( \frac {q_i - \bar{q}_j} 2  \right)} \nonumber
\end{eqnarray}
where one should not mistake canonical momenta and momenta with the same index $i$ as being equal: it is just a way to enumerate the momentum values, hence $q_i \neq \bar{q}_i$ in general. This formula is exact, however, it poses some problems from a numerical point of view. This is due to the fact that in this expression, we have as many scattering phase shifts as pseudofermions. This makes the problem of finding the dependence of $A^{(0,0)}$ on the system size and the filling a very tedious problem. In other words, we are not able to deduce a closed form expression for the productorials. The same expression is used in Ref. \cite{Karlo3} for the limit $(U/t) \rightarrow \infty$, but for a constant scattering phase shift $Q$. This simplifies the problem tremendously, since we have a similar cancellation of the factors in the above expression, as in section (\ref{relwetower}) for the relative weights. In that section, the calculations depended on the fact that for each tower of states, the scattering phase shift is constant. Here, on the other hand, we cannot depend on such a result. However, inspired by the solution of the problem in the $(U/t) \rightarrow \infty$ limit, there are some asymptotic behaviors that can be deduced.

Since the scattering phase shifts away from the left and the right Fermi points do not contribute to the leading order terms of the dynamical quantities, we will use a trial approximation of letting {\it all} the scattering phase shifts in the above expression be equal to $Q_{\alpha\nu}(\iota q_{F\alpha\nu})$, for $\iota=\pm$. This would allow us to evaluate the lowest peak weight for each $\iota$, leading to a weight that we can call $A_{\alpha\nu,\iota}^{(0,0)}$. We then form the product
\begin{equation}
A_{\alpha\nu}^{(0,0)} \approx \prod_{\iota=\pm} A_{\alpha\nu,\iota}^{(0,0)}
\end{equation}
which should be a reasonable approximation in the thermodynamic limit. Note that this approximation must coincide with the expressions of the known limits, in this case the limit $(U/t) \rightarrow \infty$. Inspired by the results of Ref. \cite{Karlo1}-\cite{Karlo3}, we here propose
\begin{equation}
A_{\alpha\nu,\iota}^{(0,0)} = \frac {\sqrt{f(Q_{\alpha\nu}(\iota q_{F\alpha\nu})-\pi)}} {(L \: \mathcal{S}_{\alpha\nu})^{2\Delta_{\alpha\nu}^{\iota}-1/2}} \label{lowestweight}
\end{equation}
where the even function $f(x)$ is the same function as that of Ref. \cite{Karlo3}, and $\mathcal{S}_{\alpha\nu}$ is an unknown quantity, depending on the density, the magnetization and ($U/t$). By comparison with the corresponding expressions of Ref. \cite{Karlo3}, we have $\mathcal{S}_{c0}\mathcal{S}_{s1} \rightarrow 1$, and at the same time $\mathcal{S}_{c0} \rightarrow \sin \pi na$, in the limit $(U/t) \rightarrow \infty$. Thus, we must have $\mathcal{S}_{s1} \rightarrow 1/\sin \pi na$ in that limit (for $m \rightarrow 0$ which is always our case). Note that (1) the square root of the function $f(x)$ will guarantee the correct behavior of the full lowest weight at $(U/t) \rightarrow \infty$ and that (2) we will in the following assume the quantity $\mathcal{S}_{\alpha\nu}$ to be equal to its value in this limit. One should note that the argument of the function $f(x)$ is here shifted by $\pi$ due to the fact that in Ref. \cite{Karlo3}, the ground state phase shift is taken from the 1D Heisenberg model for the spinons, and is equal to $\pi$ exactly, whilst in our case due to the normal ordered formulation of the problem we have by definition that $Q_{\alpha\nu}(\iota q_{F\alpha\nu})=0$ always for the ground state. Another example of this is also related to the scattering phase shift, namely the exponent in the denominator. We note that in our case, we can create or annihilate any number of pseudofermions at the Fermi points, whilst in Ref. \cite{Karlo3}, there is always exactly one spinless fermion and one spinon being created or annihilated, respectively. This is manifested by the absence of the numbers $\Delta N_{\alpha\nu,\iota}$ in that reference, which ultimately leads to a different expression for the exponents. Thus, the generalization from Ref. \cite{Karlo3} is two-folded: from infinite repulsion to arbitrary repulsion and from one quantum object creation / annihilation to many quantum object creation / annihilation. 

The form of Eq. (\ref{lowestweight}) follows the form of the equivalent quantity of Ref. \cite{Karlo3}. This form for the lowest peak weight is due to the equivalence of Eq. (\ref{karloexpr}) to the corresponding expression of that reference. Due to this likeness, the thermodynamic limit expression of the lowest peak weight must by necessity be of the form presented in Eq. (\ref{lowestweight}). 

\subsection{Merging the lowest weight and the relative weights}
\label{merge}

Having dealt with the relative weights in sections (\ref{relwetower}) and (\ref{relwetower2}), and the lowest peak weight in the preceding section, we can finally put some real meaning to Eq. (\ref{defweight}), by noting that the weight $A_{\alpha\nu}^{(m_{\alpha\nu,-},m_{\alpha\nu,+})}$ of that equation is nothing but the spectral weight we would obtain using the derived expressions for the spectral function, with $F_{p-h,\alpha\nu}^{\dag}$ given by Eq. (\ref{Fph}). This spectral weight is given by $B_{Q_{\alpha\nu}}^l(k',\omega')$ of Eq. (\ref{fuckingjose}), which means that we can replace the explicit "absolute value of the squared overlap" with the following summation over the allowed positions in the ($k,\omega$) plane, of the particle-hole excitations
\begin{eqnarray}
B_{Q_{\alpha\nu}}^l(k',\omega') &=& A_{\alpha\nu}^{(0,0)} \sum_{\iota=\pm}\sum_{m_{\alpha\nu,\iota}} a_{\alpha\nu} (m_{\alpha\nu,+} , m_{\alpha\nu,-}) \cdot \\
&\cdot& \delta \bigg( \omega' - l \frac {2\pi v_{\alpha\nu}} L  \left[ m_{\alpha\nu,+} + m_{\alpha\nu,-} \right] \bigg) \delta \bigg( k' -  l \frac {2\pi} L \left[ m_{\alpha\nu,+} - m_{\alpha\nu,-} \right] \bigg) \nonumber
\end{eqnarray}
which leads to

\begin{equation}
B_{Q_{\alpha\nu}}^l(k',\omega')= \frac L {8\pi v_{\alpha\nu}} A_{\alpha\nu}^{(0,0)} (k,\omega) \prod_{\iota=\pm} a_{\alpha\nu,\iota} \left( \frac {l[\omega' + \iota v_{\alpha\nu} k']} {4\pi v_{\alpha\nu} / L} \right) \label{Bq}
\end{equation}
where we have used the following identity for the $\delta$-functions:
\begin{equation}
\delta(a+x+y) \delta(b+x-y) = \frac 1 4 \delta\left(\frac {b+a} 2 -x \right) \delta \left( \frac {b-a} 2 -y \right)
\end{equation}
and in the thermodynamic limit $a_{\alpha\nu}(m_{\alpha\nu,+} , m_{\alpha\nu,-})=a_{\alpha\nu,-}(m_{\alpha\nu,-}) a_{\alpha\nu,+}(m_{\alpha\nu,+})$. Here and in the following we use the following expressions for the energy and momentum in terms of the numbers $m_{\alpha\nu,\iota}$:
\begin{equation}
m_{\alpha\nu,\iota} = \frac {l[\omega' + \iota v_{\alpha\nu} k']} {4\pi v_{\alpha\nu} / L} \geq 0 \label{defm}
\end{equation}
From Eq. (\ref{relpower}) we have that
\begin{equation}
a_{\alpha\nu,\iota} (m_{\alpha\nu,\iota}) = \theta(m_{\alpha\nu,\iota}) \left[ \Gamma \left( 2 \Delta_{\alpha\nu}^{\iota} \right) \right]^{-1} \left( m_{\alpha\nu,\iota} \right)^{2 \Delta_{\alpha\nu}^{\iota}-1} \label{relpower2}
\end{equation}
where the $\theta$-function seems unnecessary due to Eq. (\ref{defm}), however it will be handy in the continuous limit as we shall see later. Substituting Eq. (\ref{defm}) into Eq. (\ref{relpower2}), and substituting the resulting expression for $a_{\alpha\nu,\iota}$ into Eq. (\ref{Bq}), together with the expression of Eq. (\ref{lowestweight}) for the lowest peak weight, we arrive to the following expression:
\begin{eqnarray}
B_{Q_{\alpha\nu}}^l(k',\omega') &=& \frac {\mathcal{S}_{\alpha\nu}^{1-2\Delta_{\alpha\nu}}} {8\pi v_{\alpha\nu}} \prod_{\iota=\pm} \frac {\sqrt{f(Q_{\alpha\nu}(\iota q_{F\alpha\nu})-\pi)}} {\Gamma(2\Delta_{\alpha\nu}^{\iota})} \cdot \label{corespec} \\
&\cdot&  \theta \Big( l[\omega' + \iota v_{\alpha\nu} k'] \Big) \left( \frac {l[\omega' + \iota v_{\alpha\nu} k']} {4\pi v_{\alpha\nu}} \right)^{2 \Delta_{\alpha\nu}^{\iota}-1}  \nonumber
\end{eqnarray}
where we have defined the convenient sum
\begin{equation}
2 \Delta_{\alpha\nu} = 2 \Delta_{\alpha\nu}^+ +2 \Delta_{\alpha\nu}^- \label{sumdelta}
\end{equation}
To recapitulate, we have now that
\begin{eqnarray}
\breve{B}^l (v,\Delta \omega) &=& \frac {\text{sgn}(v)} {2\pi} \int_0^{\Delta \omega} d\omega' \  \int_{-\text{sgn}(v)\Delta \omega / v_{c0}}^{\text{sgn}(v) \Delta \omega / v_{c0}} dk' \cdot \label{conph} \\
&&\cdot B_{Q_{c0}}^l(k',\omega') B_{Q_{s1}}^l(\Delta \omega / v - k', \Delta \omega - \omega') \nonumber 
\end{eqnarray}
Introduce now the following change of integration variables:
\begin{equation}
\left.
\begin{array}{l}
x=\omega' / \Delta \omega \\
y=\text{sgn}(v) v_{c0} \: k' / \Delta \omega  
\end{array}
\right\}
\implies \frac {\text{sgn}(v)} {2\pi} d\omega' dk' = \frac {(\Delta \omega)^2} {2\pi v_{c0}} dx dy
\label{jabbe}
\end{equation}
which means that $0<x<1$ and $-1<y<1$ and also that
\begin{equation}
l [ \omega' + \iota v_{\alpha\nu} k' ] = \left\{
\begin{array}{lc}
l\Delta \omega \left[ x +  \iota \text{sgn}(v) y \right] & \ \ \  \alpha\nu=c0 \vspace{0.5cm} \\
l\Delta \omega \left[ 1-x + \iota v_{s1} \text{sgn}(v) {\displaystyle \left( \frac 1 {\vert v \vert} - \frac y {v_{c0}} \right)} \right] & \ \ \  \alpha\nu=s1
\end{array}
\right.
\end{equation}
Thus the quantity to integrate becomes:

\begin{eqnarray}
&&\frac {(\Delta \omega)^2} {2\pi v_{c0}} \Bigg[\prod_{\alpha\nu=c0,s1} \frac {\mathcal{S}_{\alpha\nu}^{1-2\Delta_{\alpha\nu}}} {8\pi v_{\alpha\nu}} \prod_{\iota=\pm} \frac {\sqrt{f(Q_{\alpha\nu}(\iota q_{F\alpha\nu})-\pi)}} {\Gamma(2\Delta_{\alpha\nu}^{\iota})} \left( \frac {l\Delta \omega} {4\pi v_{\alpha\nu}} \right)^{2\Delta_{\alpha\nu}^{\iota}-1} \Bigg] \cdot \nonumber \\
&&\cdot \prod_{\iota=\pm} \theta \Big( x +  \iota \text{sgn}(v) y \Big) \  \theta \Big( 1-x + \iota v_{s1} \text{sgn}(v) {\displaystyle \Big[ \frac 1 {\vert v \vert} - \frac y {v_{c0}} \Big]} \Big) \cdot \\
&& \cdot \ \Big( x +  \iota \text{sgn}(v) y \Big)^{2\Delta_{c0}^{\iota}-1} \  \Big( 1-x + \iota v_{s1} \text{sgn}(v) {\displaystyle \Big[ \frac 1 {\vert v \vert} - \frac y {v_{c0}} \Big]} \Big)^{2\Delta_{s1}^{\iota}-1} \nonumber
\end{eqnarray}
where the first line of this expression is constant in the integration.

Let us define the function $\mathcal{F}(1/v)$ as
\begin{eqnarray}
\mathcal{F}(1/v) &=& \frac {D_0} 2 \int_0^1 dx \ \int_{-1}^1 dy \ \prod_{\iota=\pm} \frac {\theta \Big( x +  \iota \text{sgn}(v) y \Big)} {\Gamma (2\Delta_{c0}^{\iota})} \  \frac {\theta \Big( 1-x + \iota v_{s1} \text{sgn}(v) {\displaystyle \Big[ \frac 1 {\vert v \vert} - \frac y {v_{c0}} \Big]} \Big)} {\Gamma (2\Delta_{s1}^{\iota})} \cdot  \nonumber \\
&& \cdot \ \Big( x +  \iota \text{sgn}(v) y \Big)^{2\Delta_{c0}^{\iota}-1} \  \Big( 1-x + \iota v_{s1} \text{sgn}(v) {\displaystyle \Big[ \frac 1 {\vert v \vert} - \frac y {v_{c0}} \Big]} \Big)^{2\Delta_{s1}^{\iota}-1} \label{mathcalF}
\end{eqnarray}
This function plays the role of the function called $F_0$ in Ref. \cite{Carmspec2}, and differs from this function only due to the various factors of $v_{c0}$ and $v_{s1}$ occurring in the definition of $F_0$. In this thesis report, we have chosen to collect all of these factors, as can be seen in Eq. (\ref{FAAN}). We get, after some algebra simplifying the constant factors, that 
\begin{eqnarray}
\breve{B}^l (v, \Delta \omega) &=& \frac {v_{c0}^{-2\Delta_{c0}} \ v_{s1}^{1-2\Delta_{s1}}} {4\pi} \left( \frac {l\Delta \omega} {4 \pi} \right)^{\zeta_0 -2} \cdot \mathcal{F}(1/v) \nonumber \\
D_0 &=& \prod_{\alpha\nu} \mathcal{S}_{\alpha\nu}^{1-2\Delta_{\alpha\nu}} \prod_{\iota=\pm} \sqrt{f(Q_{\alpha\nu}(\iota q_{F\alpha\nu})-\pi)} \label{FAAN}
\end{eqnarray}
where we have defined $\zeta_0 = \sum_{\alpha\nu} 2\Delta_{\alpha\nu}$.

Note that this expression is equivalent with that of Eq. (61) in Ref. \cite{Carmspec2}, even though the definitions of the different quantities differ from each other. It is a simple algebraic task to extract all the different square roots and powers of $v_{c0}$ and $v_{s1}$ from $F_0$ of Eq. (61) and (62) of that reference, to show that they are equal to the ones presented here. The difference in powers of these two velocities stem from choosing the $c0$ branch to be the convoluting branch. This introduces a factor ($1/v_{c0}$) in the jacobian of Eq. (\ref{jabbe}).

\section{Closed form expressions}
\label{closedform}

\subsection{The final step: canonical momentum integrations}

The expression for the spectral function derived in section (\ref{merge}), Eq. (\ref{FAAN}), is the base for all the expressions given in the remainder of this thesis report, except for the expressions for the Luttinger contribution, which will be presented separately. The function $\mathcal{F}(1/v)$, where $v$ is the velocity of the particle-hole excitations, is crucial for the evaluation of the state dependent pre-factors of the spectral function. The particle-hole energy $\Delta \omega$, and the exponent $\zeta_0 -2$ of Eq. (\ref{FAAN}), will become crucial for the power law type behavior of the spectral function. The double integral occurring in the definition of $\mathcal{F}(1/v)$ can not be expressed in a closed form and has to be treated numerically. 

What remains now is to perform the finite energy pseudofermion summations, with summation variables $\tilde{q}_{c0}$ and $\tilde{q}_{s1}$ in case of the 2P contribution, $\tilde{q}_{s1}$ in case of the s-branch, $\tilde{q}_{c0}$ in case of the c-branch, and with no summation at all in case of the 0P Fermi point contribution. The canonical momentum $\tilde{q}_{\alpha\nu}$ was introduced at the end of section (\ref{opfact}), and entered the derivation of the spectral function in Eq. (\ref{fullsofar}). Moreover, since we only consider one type of transition, called the "Basic" transition defined in sections (\ref{RHBtrans}) and (\ref{LHBtrans}), we can drop the summations over the deviation numbers $\{ \Delta N_{\alpha\nu} \}$ and $\{ \Delta N_{\alpha\nu,\iota}^F \}$. 

We have that
\begin{eqnarray}
&&B_{2P}^l (k,\omega) = \frac 1 {(2\pi)^2} \int d\tilde{q}_{c0} \  \int d\tilde{q}_{s1} \  \mathcal{D}_{ph}(v,\omega) \: \breve{B}^l (v,\omega - l\Delta E) \nonumber \\
&&B_{\alpha}^l (k,\omega) = \frac 1 {2\pi} \int d\tilde{q}_{\alpha\nu} \  \mathcal{D}_{ph} (v,\omega) \: \breve{B}^l (v,\omega - l\Delta E) \nonumber \\
&&B_{0P}^l (k,\omega) = \mathcal{D}_{ph} (v,\omega) \: \breve{B}^l (v,\omega - l\Delta E) \label{syftet} \\
&& \mathcal{D}_{ph} (v,\omega)= \theta (l[\omega - l\Delta E])\theta (\Omega-l[\omega - l\Delta E])\theta (\vert v \vert -v_{s1}) \nonumber
\end{eqnarray}
where $\alpha=c$ or $s$ for the c-branch line and the s-branch line, respectively, and the two special cases $B_{Border}^l (k,\omega)$ and $B_{Lutt}^l (k,\omega)$ will be treated separately. 

The domain defining function $\mathcal{D}_{ph}(v,\omega)$ illustrates that we are always integrating over such canonical momentum values, which all contribute to that certain point in the ($k,\omega$) plane, to which the $\mathbb{A}$, $\mathbb{B}$ {\it and} the $\mathbb{C}$ processes bring the excitation to. Note that the integration domain, as controlled by $\mathcal{D}_{ph}$, is very small as compared to the whole ($k,\omega$) plane. It covers a region in this plane, which comprises all particle-hole excitations that can reach the momentum $k$ and the energy $\omega$. Note that the spectral weight at this point has contributions from particle-hole processes originating from many different surrounding points in the ($k,\omega$) plane. In other words, there are many $\mathbb{A}$ and $\mathbb{B}$ processes whose excitations brings us to a finite momentum and a finite energy in the vicinity of ($k,\omega$), and that by the $\mathbb{C}$ processes actually reaches the point ($k,\omega$). This demonstrates the overlap of the different particle-hole tower of states, originating from the {\it vicinity} of ($k,\omega$), and contributing to the spectral weight {\it at} ($k,\omega$).

\subsection{2P contribution}
\label{2D}

We will perform the 2P contribution integrations explicitly, to obtain the final expression for $B_{2P}^l (k,\omega)$. The calculations here are depending on the fact that the 2P contribution covers a compact subspace of the ($k,\omega$) plane. Thus a typical point in this compact subspace is both reachable by a $\mathbb{A}$ and $\mathbb{B}$ process, and by a similar process for a slightly different set of canonical momenta values and then adding particle-hole excitations, due to process $\mathbb{C}$. The limiting lines of this complete 2D subspace of the ($k,\omega$) plane are the lines which limit the effective Brillouin zone, and the so called border lines described in the subsequent section. 

Consider an energy $\omega$ and a momentum $k$, reached by the $\mathbb{A}$, $\mathbb{B}$ and the $\mathbb{C}$ processes. Due to the compactness of the subspace of the ($k,\omega$) plane, we can define a canonical momentum value $\tilde{q}_{\alpha\nu}^{\: 0}$ for each $\alpha\nu$ branch, such that
\begin{eqnarray}
\omega = l\Delta E (\tilde{q}_{\alpha\nu}^{\: 0}) \nonumber \\
k = l\Delta P (\tilde{q}_{\alpha\nu}^{\: 0})
\end{eqnarray}
where $l \Delta E (q)$ and $l \Delta P (q)$ are given in Eq. (\ref{dEdPspectral}). Note the adopted shorthand notation: technically we should have written 
$"l\Delta E (\tilde{q}_{c0}^{\: 0},\tilde{q}_{s1}^{\: 0})"$ (and similarily for $l\Delta P \: $) in the equation above.

Let now $\tilde{q}_{\alpha\nu}^{\: 0} + \tilde{q}_{\alpha\nu}$ denote a typical canonical momentum value, that due to the $\mathbb{A}$ and $\mathbb{B}$ processes have energy and momentum $l \Delta E (\tilde{q}_{\alpha\nu}^{\: 0} + \tilde{q}_{\alpha\nu})$ and $l \Delta P (\tilde{q}_{\alpha\nu}^{\: 0} + \tilde{q}_{\alpha\nu})$, respectively. We are looking to integrate over canonical momentum values $\tilde{q}_{\alpha\nu}$, such that the differences $\big[ \omega - l\Delta E (\tilde{q}_{\alpha\nu}^{\: 0} + \tilde{q}_{\alpha\nu}) \big]$ and $\big[k - l\Delta P (\tilde{q}_{\alpha\nu}^{\: 0} + \tilde{q}_{\alpha\nu})\big]$ are inside the allowed region for the particle-hole towers of states, i.e. such that $0 < l[\omega - l\Delta E] < \Omega$, and $\vert v \vert > v_{s1}$.

This means that
\begin{eqnarray}
\omega - l\Delta E (\tilde{q}_{\alpha\nu}^{\: 0} + \tilde{q}_{\alpha\nu}) &=& -l \sum_{\alpha\nu} \text{sgn}\left( \Delta N_{\alpha\nu}^{NF} \right) v_{\alpha\nu} (\tilde{q}_{\alpha\nu}) \tilde{q}_{\alpha\nu} \nonumber \\
k - l\Delta P (\tilde{q}_{\alpha\nu}^{\: 0} + \tilde{q}_{\alpha\nu}) &=& -l \sum_{\alpha\nu} \text{sgn}\left( \Delta N_{\alpha\nu}^{NF} \right) \tilde{q}_{\alpha\nu} \label{2Domk}
\end{eqnarray}
by expanding in powers of $\tilde{q}_{\alpha\nu}$ and retaining the linear terms only. Defining the energy variable $\omega' = \omega - l\Delta E$ and the reciprocal velocity variable $z=1/v=(k - l\Delta P) / (\omega - l\Delta E)$, we can solve the relationships of Eq. (\ref{2Domk}) for $\tilde{q}_{\alpha\nu}$, to obtain
\begin{eqnarray}
&& \tilde{q}_{\alpha\nu} = -l\text{sgn}\left( \Delta N_{\alpha\nu}^{NF} \right) \omega'  \frac {1-v_{\alpha' \nu'}(\tilde{q}_{\alpha' \nu'}) z} {v_{\alpha \nu}(\tilde{q}_{\alpha\nu})-v_{\alpha' \nu'}(\tilde{q}_{\alpha' \nu'})} \nonumber \\
&& \tilde{q}_{\alpha' \nu'} = l\text{sgn}\left( \Delta N_{\alpha'\nu'}^{NF} \right) \omega'  \frac {1-v_{\alpha\nu}(\tilde{q}_{\alpha\nu}) z} {v_{\alpha \nu}(\tilde{q}_{\alpha\nu})-v_{\alpha' \nu'}(\tilde{q}_{\alpha' \nu'})} 
\end{eqnarray}

At this stage, we remind ourselves that we are integrating the variable $\tilde{q}_{\alpha\nu}$, as we are scanning for all possible $\mathbb{A}$ and $\mathbb{B}$ processes contributing to the energy and momentum given by $\omega$ and $k$ above. However, the domain of integration is governed by the quantity $\mathcal{D}_{ph}$ defined in Eq. (\ref{syftet}), which is expressed in terms of $\omega'$ and $z$. The jacobian of this transformation is equal to $\omega' / \vert v_{\alpha \nu}(\tilde{q}_{\alpha\nu}^{\: 0})-v_{\alpha' \nu'}(\tilde{q}_{\alpha' \nu'}^{\: 0}) \vert$. We thus obtain
\begin{eqnarray}
B_{2P}^l (k,\omega) &\sim& \frac 1 {(2\pi)^2} \int d\tilde{q}_{c0} \  \int d\tilde{q}_{s1} \  \theta (l\omega')\theta (\Omega-l\omega') \theta \left( \frac 1 {v_{s1}} - \vert z \vert \right) \left( \frac {l \omega'} {4 \pi} \right)^{\zeta_0 -2} \: \mathcal{F}(z) \nonumber \\
&=& \frac l {(2\pi)^2} \frac {4\pi} {\vert v_{c0}(\tilde{q}_{c0}^{\: 0})-v_{s1}(\tilde{q}_{s1}^{\: 0}) \vert} \int_0^{l\Omega} d\omega' \ \left( \frac {l \omega'} {4 \pi} \right)^{\zeta_0 -1} \int_{-1/v_{s1}}^{1/v_{s1}} dz \ \mathcal{F}(z) \nonumber \\
&=& \frac 4 {\zeta_0 \vert v_{c0}(\tilde{q}_{c0}^{\: 0})-v_{s1}(\tilde{q}_{s1}^{\: 0}) \vert} \left( \frac {l \Omega} {4 \pi} \right)^{\zeta_0}  \int_{-1/v_{s1}}^{1/v_{s1}} dz \ \mathcal{F}(z) \label{ontheway}
\end{eqnarray}
so that the final expression for the full spectral function in the case of the 2P contribution becomes:
\begin{equation}
B_{2P}^l (k,\omega) = \frac {v_{c0}^{-2\Delta_{c0}} \ v_{s1}^{1-2\Delta_{s1}}} {\pi \zeta_0 \vert v_{c0}(\tilde{q}_{c0}^{\: 0})-v_{s1}(\tilde{q}_{s1}^{\: 0}) \vert} \left( \frac {l \Omega} {4 \pi} \right)^{\zeta_0}  \int_{-1/v_{s1}}^{1/v_{s1}} dz \ \mathcal{F}(z) \label{2P}
\end{equation}
where we have $\omega=\omega(\tilde{q}_{c0}^{\: 0} \: ,\tilde{q}_{s1}^{\: 0})$, $k=k(\tilde{q}_{c0}^{\: 0} \: ,\tilde{q}_{s1}^{\: 0})$ as well as $\zeta_0=\zeta_0 (\tilde{q}_{c0}^{\: 0} \: ,\tilde{q}_{s1}^{\: 0})$, as defined in sections (\ref{RHBtrans}) (RHB) and (\ref{LHBtrans}) (LHB) (in these sections, the canonical momentum $\tilde{q}_{\alpha\nu}^{\: 0}$ is denoted $q_{\alpha\nu}$).

We see that this expression becomes singular as $\vert v_{c0}(\tilde{q}_{c0}^{\: 0})-v_{s1}(\tilde{q}_{s1}^{\: 0}) \vert \rightarrow 0$. Indeed, the transformation from the $\tilde{q}_{\alpha\nu}$ variables to $\omega'$ and $z$ is not well defined as the two velocities approach each other, which is clearly seen from the jacobian of the transformation: it shoots off to infinity. The mathematical condition $v_{c0}(\tilde{q}_{c0}^{\: 0})=v_{s1}(\tilde{q}_{s1}^{\: 0})$ traces out the border lines in the ($k,\omega$) plane. 

\newpage
\subsection{Border lines}
\label{Border}

The border lines are truly the "borders" of the spectral weight due to the $\mathbb{A}$ and the $\mathbb{B}$ processes. For example, consider the RHB "Basic" transition.  The point in the ($k,\omega$) plane, reachable by the 2P contribution, corresponding to pseudofermion canonical momenta both equal to zero, belongs trivially to the border line, since $v_{c0}(0)=v_{s1}(0)=0$. But this point also has the smallest energy of all one electron removal excitations, i.e. $\epsilon_{c0}(0)+\epsilon_{s1}(0) < \epsilon_{c0}(q_{c0})+\epsilon_{s1}(q_{s1})$ for all $q_{c0} \neq 0$ and $q_{s1} \neq 0$. This means that there are no finite energy and finite momentum processes available to put spectral weight {\it below} this point. The same reasoning can be applied to the one electron addition case as well. Note however, that we can still reach areas outside the border lines by particle-hole excitations originating from a point sufficiently close to the border line itself. This effect forces us to consider two separate cases: one in which we consider areas of the ($k,\omega$) plane available to {\it either} $\mathbb{A}$ and the $\mathbb{B}$ processes on the one hand {\it or} $\mathbb{C}$ processes on the other, and one in which the ($k,\omega$) plane is reachable by the $\mathbb{C}$ processes {\it only}. We will thus have two different contributions, here dubbed $B_{Border}^{l,<} (k,\omega)$ for the former case and $B_{Border}^{l,>} (k,\omega)$ for the latter.

Let us focus on $B_{Border}^{l,<} (k,\omega)$. The parametric equations of the border lines are given by the energy $\omega_{BL}$ and the momentum $k_{BL}$, according to 

\begin{eqnarray}
\omega_{BL}&=& l  \delta_{\displaystyle{v_{c0} (\tilde{q}_{c0}^{\: 0}) , v_{s1}(\tilde{q}_{s1}^{\: 0})}} \sum_{\alpha\nu} \text{sgn} (\Delta N_{\alpha\nu}^{NF}) \epsilon (\tilde{q}_{\alpha\nu}^{\: 0}) \\
k_{BL} &=& l  \delta_{\displaystyle{v_{c0} (\tilde{q}_{c0}^{\: 0}) , v_{s1}(\tilde{q}_{s1}^{\: 0})}} \sum_{\alpha\nu} \text{sgn} (\Delta N_{\alpha\nu}^{NF}) \tilde{q}_{\alpha\nu}^{\: 0} \nonumber
\end{eqnarray}

These are then the energy and the momenta of the border lines themselves. As before, we now fix a point in the vicinity of this line, with energy and momentum $\omega$ and $k$, such that this point is reachable from a small but finite region of the border line by particle-hole processes. We introduce the small variation $\tilde{q}_{\alpha\nu}^{\: 0} + \tilde{q}_{\alpha\nu}$ ($\alpha\nu=c0,s1$), where $\tilde{q}_{\alpha\nu}^{\: 0}$ is a canonical momenta bringing the excitation to the line. Since this line is truly a "border" of the spectral weight, this means that in this first step, we are scanning for particle-hole processes slightly below ($l=+$) or slightly above ($l=-$) the line itself, which motivates the introduction of $\tilde{q}_{\alpha\nu}$. In these regions, we have points which are reachable by the $\mathbb{A}$ and $\mathbb{B}$ processes as well, whilst on the other side of these lines, the ($k,\omega$) plane is completely void of any spectral weight from the $\mathbb{A}$ and $\mathbb{B}$ processes, so that the only spectral weight stems from particle-hole excitations originating from regions slightly below ($l=+$) or slightly above ($l=-$) the line itself.

We have now the following slight displacement in the energy and the momentum:

\begin{eqnarray}
\Delta E(\tilde{q}_{\alpha\nu}^{\: 0} + \tilde{q}_{\alpha\nu}) &=& l \omega_{BL} + v_{\alpha\nu}(\tilde{q}_{\alpha\nu}^{\: 0}) \sum_{\alpha\nu} \text{sgn}(\Delta N_{\alpha\nu}^{NF}) \tilde{q}_{\alpha\nu} \nonumber \\
\Delta P(\tilde{q}_{\alpha\nu}^{\: 0} + \tilde{q}_{\alpha\nu}) &=& l k_{BL} + \sum_{\alpha\nu} \text{sgn}(\Delta N_{\alpha\nu}^{NF}) \tilde{q}_{\alpha\nu} 
\end{eqnarray}

The particle-hole velocity $v$ is thus

\begin{eqnarray}
v = \frac {\omega - l\Delta E (\tilde{q}_{\alpha\nu}^{\: 0} + \tilde{q}_{\alpha\nu}) } {k -l\Delta P (\tilde{q}_{\alpha\nu}^{\: 0} + \tilde{q}_{\alpha\nu}) } &=& \frac {\omega - \omega_{BL} - l v_{\alpha\nu}(\tilde{q}_{\alpha\nu}^{\: 0}) \sum_{\alpha\nu} \text{sgn}(\Delta N_{\alpha\nu}^{NF})\tilde{q}_{\alpha\nu} } {l \sum_{\alpha\nu} \text{sgn}(\Delta N_{\alpha\nu}^{NF})\tilde{q}_{\alpha\nu}} = \nonumber \\ 
&=& v_{\alpha\nu} (\tilde{q}_{\alpha\nu}^{\: 0}) - \frac {l [\omega - \omega_{BL}] } {\sum_{\alpha\nu} \text{sgn}(\Delta N_{\alpha\nu}^{NF})\tilde{q}_{\alpha\nu}}
\end{eqnarray}
which is equivalent to

\begin{equation}
\sum_{\alpha\nu} \text{sgn}(\Delta N_{\alpha\nu}^{NF})\tilde{q}_{\alpha\nu} = - \frac {l [\omega - \omega_{BL}]} {v-v_{\alpha\nu}(\tilde{q}_{\alpha\nu}^{\: 0})} \label{gg0}
\end{equation}
where $\alpha\nu=c0,s1$ as usual and $v_{\alpha\nu}(\tilde{q}_{\alpha\nu}^{\: 0})$ could stand for either pseudofermion velocity, since they are equal to each other. Now, this equation permits us to extract the relationship between the two different deviations $\tilde{q}_{\alpha\nu}$, that the border line demands. This constraint comes from the condition that the two pseudofermion velocities must be equal, and hence, a small deviation in the canonical momentum of one of the branches induces a small change in the canonical momentum of the other branch, in order to keep the velocities equal. We can simplify matters a lot, by restricting ourselves to infinite particle-hole velocity, which would make Eq. (\ref{gg0}) equal zero. Like this, we are only considering points in a straight vertical line originating from the point ($k_{BL},\omega_{BL}$). This means that we only consider such particle-hole excitations which are straight over ($l=-$) or straight under ($l=+$) the border line (which is something we can do without loss of generality since we will integrate over all relevant momentum and energy values). We obtain:

\begin{eqnarray}
&&\sum_{\alpha\nu} \text{sgn}(\Delta N_{\alpha\nu}^{NF})\tilde{q}_{\alpha\nu} = 0 \implies \tilde{q}_{\alpha\nu} = -g_{\alpha\nu,\alpha'\nu'}  \tilde{q}_{\alpha'\nu'} \nonumber \\
&& g_{\alpha\nu,\alpha'\nu'} = \text{sgn}(\Delta N_{\alpha\nu}^{NF}) \text{sgn}(\Delta N_{\alpha'\nu'}^{NF}) = \pm 1\label{gg1}
\end{eqnarray}

This means, for example,

\begin{eqnarray}
\Delta E(\tilde{q}_{\alpha\nu}^{\: 0} + \tilde{q}_{\alpha\nu}) &=& \text{sgn}(\Delta N_{c0}^{NF}) \epsilon_{c0} (\tilde{q}_{c0}^{\: 0} + \tilde{q}_{c0}) + \text{sgn}(\Delta N_{s1}^{NF}) \epsilon_{s1} (\tilde{q}_{s1}^{\: 0} -g_{c0,s1} \tilde{q}_{c0}) = \nonumber \\
&=& l \omega_{BL} + \frac {\tilde{q}_{c0}^{\: 2}} 2 \Big[ \text{sgn}(\Delta N_{c0}^{NF}) a_{c0} (\tilde{q}_{c0}^{\: 0}) + \text{sgn}(\Delta N_{s1}^{NF}) a_{s1} (\tilde{q}_{s1}^{\: 0}) \Big] \nonumber \\
\label{desperate}
\end{eqnarray}
where the energy is expanded up to the second order in $\tilde{q}_{c0}$, and we have defined the pseudofermion {\it accelerations} $a_{\alpha\nu}(q) = d v_{\alpha\nu} (q) / dq$. Now, we need a way to treat the difference in the pseudofermion velocities. We obtain:

\begin{eqnarray}
&&v_{\alpha\nu} (\tilde{q}_{\alpha\nu}^{\: 0} + \tilde{q}_{\alpha\nu}) - v_{\alpha'\nu'} (\tilde{q}_{\alpha'\nu'}^{\: 0} -g_{\alpha\nu,\alpha'\nu'} \tilde{q}_{\alpha\nu}) = \nonumber \\
&&=\text{sgn} (\Delta N_{\alpha\nu}^{NF}) \tilde{q}_{\alpha\nu} \sum_{\alpha'\nu'} \text{sgn}(\Delta N_{\alpha'\nu'}^{NF}) a_{\alpha'\nu'} (\tilde{q}_{\alpha'\nu'}^{\: 0})
\end{eqnarray}
and thus

\begin{eqnarray}
\tilde{q}_{\alpha\nu} = \text{sgn} (\Delta N_{\alpha\nu}^{NF}) \frac {v_{\alpha\nu} (\tilde{q}_{\alpha\nu}^{\: 0} + \tilde{q}_{\alpha\nu}) - v_{\alpha'\nu'} (\tilde{q}_{\alpha'\nu'}^{\: 0} -g_{\alpha\nu,\alpha'\nu'} \tilde{q}_{\alpha\nu})} {\sum_{\alpha'\nu'} \text{sgn}(\Delta N_{\alpha'\nu'}^{NF}) a_{\alpha'\nu'} (\tilde{q}_{\alpha'\nu'}^{\: 0})}
\end{eqnarray}
which is the quantity that, when squared, can be introduced into Eq. (\ref{desperate}), to produce
 
\begin{eqnarray}
\frac 1 {\vert v_{\alpha\nu} (\tilde{q}_{\alpha\nu}^{\: 0} + \tilde{q}_{\alpha\nu}) - v_{\alpha'\nu'} (\tilde{q}_{\alpha'\nu'}^{\: 0} -g_{\alpha\nu,\alpha'\nu'} \tilde{q}_{\alpha\nu}) \vert} =
 \frac {\Big(2l [ \omega - \omega_{BL} ] \Big)^{-\frac 1 2}} {\sqrt{\sum_{\alpha\nu} \vert a_{\alpha\nu} (\tilde{q}_{\alpha\nu}^{\: 0}) \vert }} \nonumber \\
\end{eqnarray}
which finally yields the expression for the fully integrated spectral function, for the spectral weight in the vicinity of the border line, just above ($l=-$) or below ($l=+$) this line:

\begin{eqnarray}
B_{Border}^{l,<} (k,\omega) &=& \frac {2\theta \Big( \Omega - l[\omega_{BL} - \omega] \Big) v_{c0}^{-2\Delta_{c0}} \ v_{s1}^{1-2\Delta_{s1}}} {\pi \zeta_0 \sqrt{\vert a_{c0}(\tilde{q}_{c0}^{\: 0}) \vert + \vert a_{s1}(\tilde{q}_{s1}^{\: 0}) \vert }} \cdot \nonumber \\
&\cdot& \left( \frac {l \Omega} {4 \pi} \right)^{\zeta_0} \Big(2l [ \omega - \omega_{BL} ] \Big)^{-\frac 1 2} \int_{-1/v_{s1}}^{1/v_{s1}} dz \ \mathcal{F}(z) \label{borde1}
\end{eqnarray}
where $\omega-\omega_{BL}$ is a small energy, sufficiently small to be reached from the border line by some particle-hole processes, as demonstrated by the $\theta$-function. Note that as before, the energy $\omega$ and the momentum $k$, is connected to the energy and the momenta of the dispersive quantum objects through $\omega=\omega(\tilde{q}_{c0}^{\: 0} \: ,\tilde{q}_{s1}^{\: 0})$, $k=k(\tilde{q}_{c0}^{\: 0} \: ,\tilde{q}_{s1}^{\: 0})$ as well as $\zeta_0=\zeta_0(\tilde{q}_{c0}^{\: 0} \: ,\tilde{q}_{s1}^{\: 0})$ and $\omega_{BL}=\omega_{BL}(\tilde{q}_{c0}^{\: 0} \: ,\tilde{q}_{s1}^{\: 0})$. Finally, the factor of $2$ arises from the consideration that there are always two values ($\tilde{q}_{c0}^{\: 0}$ , $\tilde{q}_{s1}^{\: 0}$) contributing to the spectral weight at the same point ($k,\omega$).

\begin{figure}
\begin{center}
\includegraphics[width=11cm,height=6cm]{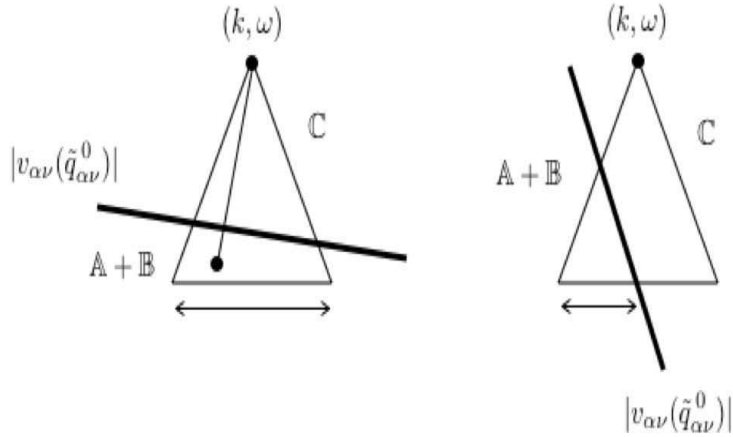}
\caption{\label{figborder} Schematic figures of the geometrical considerations needed for the integrations in the $z=1/v$ variable, for spectral weight contribution to $B^{l,>}_{Border}(k.\omega)$, for $l=+$. In the left figure, we see a border line with a "smaller" velocity $\vert v_{\alpha \nu}(\tilde{q}_{\alpha\nu}^{\: 0}) \vert$. The border line divides the depicted region into two parts: below the line, where the $\mathbb{A}$ and the $\mathbb{B}$ processes allocate spectral weight, and above the line which is only reachable via particle-hole excitations from energy and momentum points below the line. The line with dotted endpoints originates from a point under the line, and through the $\mathbb{C}$ processes reaches the point ($k,\omega$). Note that the arrowheaded horizontal line covers the entire particle-hole range, i.e. $-1/v_{s1} < z < 1/v_{s1}$, in contrast to the right figure where the "larger" value of $\vert v_{\alpha \nu}(\tilde{q}_{\alpha\nu}^{\: 0}) \vert$ causes a cut of the base of the triangle. On the right side of this line, we have only spectral weight due to the $\mathbb{C}$ processes. The integration range is limited by the velocity of the border line, but also by the value of $\omega - \Delta E$. Indeed, if we consider the left figure with a larger value of $\omega$, the border line will cut through the base of the left triangle as well. Note that for the spectral function $B^{l,<}_{Border}(k.\omega)$, the ($k,\omega$) point is below the border line. \vspace{0.5cm}}
\end{center}
\end{figure}

The other border line expression is very similar to the one derived above. The differences stem from the fact that above ($l=+$) or below ($l=-$) the border line, there is no spectral weight due to the $\mathbb{A}$ and $\mathbb{B}$ processes alone. The spectral weight of that region is generated by the $\mathbb{C}$ processes, on towers of states originating from the region below ($l=+$) or above ($l=-$) the border lines. The first difference is due to the $\omega'$ integration of Eq. (\ref{ontheway}). In this case, this integration can not run from $\omega'=0$ since we do not have any finite energy and finite momentum processes {\it at} the ($k,\omega$) point under consideration. In fact, we must move away a minimum distance of $\omega-\omega_{BL}$ from this point, in order to reach the "allowed" region for the $\mathbb{A}$ and the $\mathbb{B}$ processes. Thus, the integration can only start at this energy value. Furthermore Eq. (\ref{gg0}) introduces a factor $1 /[1-z v_{\alpha\nu}(\tilde{q}_{\alpha\nu}^{\: 0}) ] $, since we can no longer scan all particle-hole energy values from $0$ to $\Omega$ at infinite particle-hole velocity, as above. Moreover, we are integrating over a two dimensional region that is {\it tilted} with a slope proportional to $v_{\alpha\nu}(\tilde{q}_{\alpha\nu}^{\: 0})$. The velocity of the border line measures at what angle it cuts through the region available for the particle-hole excitations. Due to the inclination of the line, there is thus a non negligible region for which the particle-hole processes can not enter the domain of the $\mathbb{A}$ and $\mathbb{B}$ processes. However, the size of the region which {\it is} available depends also on the value of the particle-hole energy (and yet for some particle-hole velocities, we will never reach the region allowed for the $\mathbb{A}$ and $\mathbb{B}$ processes). In other words, the interplay between these quantities influences both the region for which we will have a finite spectral weight as well as the limits of integration in the variables depending on the particle-hole excitations, in our case in the variable $z=1/v$. These considerations are further explained in Fig. (\ref{figborder}). We will skip the mathematical details of this analysis and merely present the result. In the expression below, the first $\theta$-function refers to a border line cutting through the particle-hole region in such a way that the "base" of this region is completely {\it inside} the allowed region for the $\mathbb{A}$ and $\mathbb{B}$ processes. This is why the accompanied integration can run over the entire particle-hole velocity range, from $-v_{s1}$ to $v_{s1}$. The second $\theta$-function corresponds to a border line cutting through this "base" and the accompanied integration limits are thus modified to only integrate in the allowed region for the $\mathbb{A}$ and $\mathbb{B}$ processes. Note that the integrand is always the same, with the term that was zero in the ordinary 2P case now replaced with a term proportional to $l[\omega - \omega_{BL}]^{\zeta_0}$. The quantities with index $\alpha\nu$ remain unspecified since these quantities are equal for both pseudofermion branches.

\begin{eqnarray}
&&B_{Border}^{l,>} (k,\omega) = \frac {2\theta \Big( \Omega - l[\omega_{BL} - \omega] \Big) v_{c0}^{-2\Delta_{c0}} \ v_{s1}^{1-2\Delta_{s1}}} {\pi \zeta_0 \sqrt{\vert a_{c0}(\tilde{q}_{c0}^{\: 0}) \vert + \vert a_{s1}(\tilde{q}_{s1}^{\: 0}) \vert }} \Big(2l [ \omega - \omega_{BL} ] \Big)^{-\frac 1 2} \cdot \nonumber \\
&&\cdot \: \Bigg\{ \theta \bigg( v_{s1} \Big[ 1 - \frac {l [\omega - \omega_{BL}]} {\Omega} \Big] - \vert v_{\alpha\nu} (\tilde{q}_{\alpha\nu}^{\: 0}) \vert \bigg) \int_{-\frac 1 {v_{s1}}}^{\frac 1 {v_{s1}}} dz \ + \nonumber \\ 
&& +\: \text{sgn}(\tilde{q}_{\alpha\nu}^{\: 0}) \theta \bigg(\vert v_{\alpha\nu} (\tilde{q}_{\alpha\nu}^{\: 0}) \vert - v_{s1} \Big[ 1 - \frac {l [\omega - \omega_{BL}]} {\Omega} \Big] \bigg) \int_{-\frac {\text{sgn}(\tilde{q}_{\alpha\nu}^{\: 0})} {v_{s1}}}^{\left(1 - \frac {l [\omega - \omega_{BL}]} {\Omega}\right)/v_{\alpha\nu} (\tilde{q}_{\alpha\nu}^{\: 0})} dz \ \Bigg\} \cdot \nonumber \\
&&\cdot \Bigg[ \left( \frac {l \Omega} {4 \pi} \right)^{\zeta_0} - \left( \frac {l [\omega - \omega_{BL}]} {4 \pi [1-z v_{\alpha\nu}(\tilde{q}_{\alpha\nu}^{\: 0})]} \right)^{\zeta_0} \Bigg] \mathcal{F}(z) \label{borde2}
\end{eqnarray}

\subsection{The $\alpha$-branch lines}
\label{branch}

The $\alpha$-branch lines ($\alpha=c,s$) are lines which contribute significantly to the overall shape of the full spectral function. These lines are formed by the $\alpha\nu$ pseudofermion or pseudofermion hole assuming values along the entire range of its dispersion, whilst the other pseudofermion or pseudofermion hole is created or annihilated at one of its Fermi points. The only reservation we will have in this section is that when the dispersive pseudofermion is sufficiently close to either one of its Fermi points, the mathematical treatment of the problem will become different, and will be dealt with in the subsequent section. 

We saw in sections (\ref{RHBtrans}) and (\ref{LHBtrans}), that the scattering phase shifts have different expressions in the 2P case as compared to the branch line cases. Indeed, in the former case we have two scattering centers, which disperse in the many body system. Here, however, we have only one dispersive scattering center, as the other pseudofermion or pseudofermion hole is confined to one of its Fermi points. Thus, the displacement of the integrating pseudofermions in sections (\ref{2D}) and (\ref{Border}), is in this section confined to a line. This means that we fix a point in the ($k,\omega$) plane in the vicinity of the branch line, and then integrate over canonical momentum values {\it on the line only}, in order to account for the spectral weight due to the processes that brings us from the line to the point in the ($k,\omega$) plane under consideration. In other words, if the integrating pseudofermion would leave the branch line, we would be considering spectral weights described by another set of the quantities $2\Delta_{\alpha\nu}^{\iota}$, than that of the branch line itself.

\begin{figure}
\begin{center}
\includegraphics[width=7cm,height=5cm]{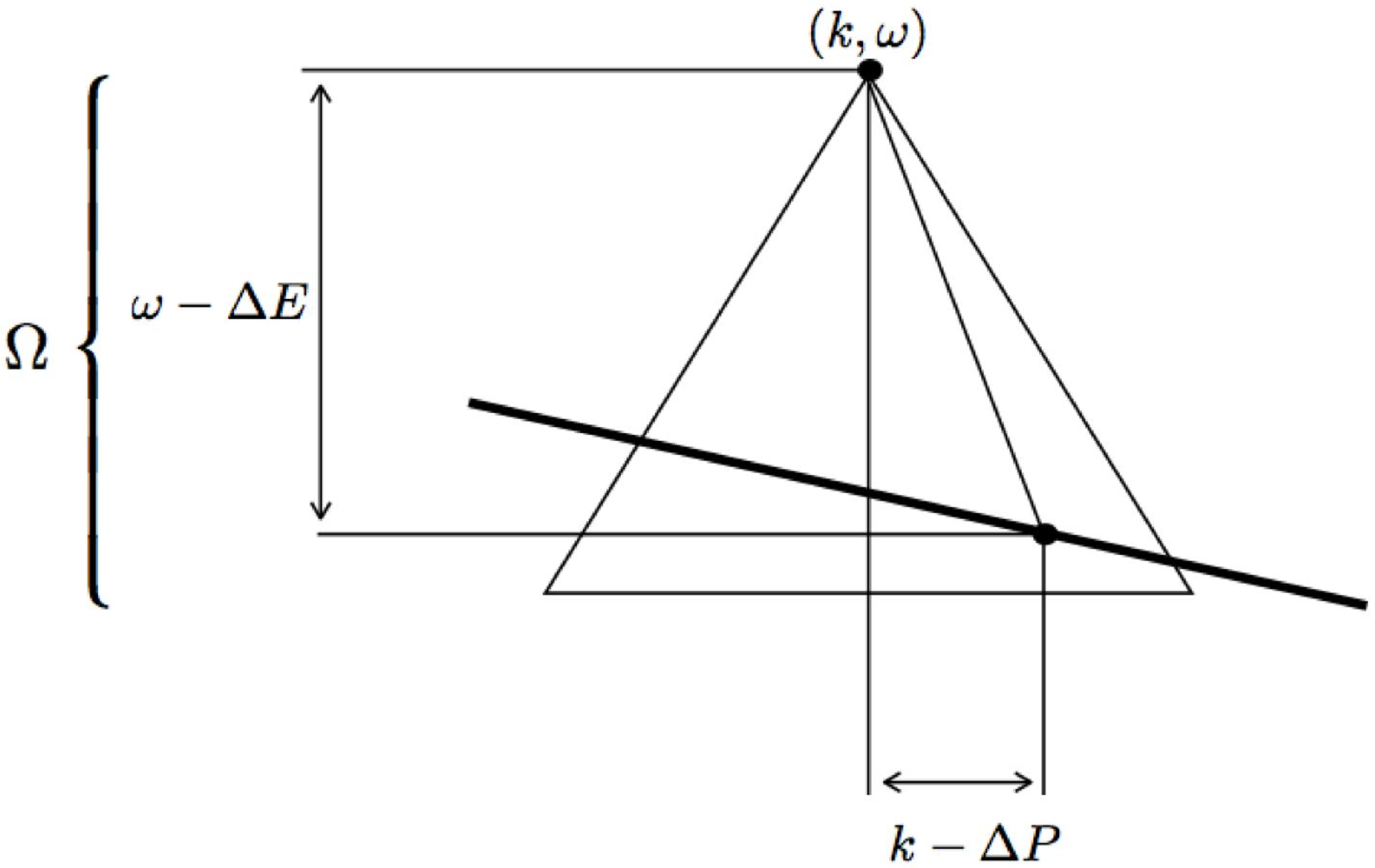}
\caption{\label{figbranch} Schematic figure for a typical branch line integration, for $l=+$. The line with dotted endpoints has one end fixed at ($k,\omega$), and one end varying on the branch line itself. The latter point must not be further away in energy as what is dictated by $\Omega$. Note that, depending on the branch line velocity and the value of $\omega-\Delta E$, we will have a similar situation as already discussed for the border line case, accounted for in Fig. (\ref{figborder}). This means that the branch line may cut through the base of the triangle in this picture, rendering a smaller integration interval of $z$ than what is depicted here. \vspace{0.5cm}}
\end{center}
\end{figure}

We now have that the small canonical momentum denoted $\tilde{q}_{\alpha\nu}$, only varies inside a small domain on the branch line such that the ($k,\omega$) point can be reached by particle-hole processes with energy less than or equal to $\Omega$. Since $\tilde{q}_{\alpha\nu}$ will vary on a tilted line, with the amount of inclination proportional to the velocity of the branch line, we have to consider a situation topologically equivalent to the one considered for the spectral function $B_{Border}^{l,>} (k,\omega)$, i.e. the situation where the slope of the branch line might be such that for some values of the particle-hole velocity $v$, a non negligible region of the tower of states can not be reached from the branch line, assuming that we only allow particle-hole energies between $0$ and $\Omega$. This topological effect is depicted in Fig. (\ref{figbranch}) and further discussed in Ref. \cite{Carmspec2}. The effective impact that these considerations have, is the introduction of the integration limits and the $\theta$-functions in the expression given below.

Motivated by the discussion above, we now introduce a small canonical momentum value $\tilde{q}_{\alpha\nu}$, which varies {\it on the branch line} around the canonical momentum value $\tilde{q}_{\alpha\nu}^{\: 0}$. Thus the energy and momentum of this slightly displaced canonical momentum will be

\begin{eqnarray}
\Delta E(\tilde{q}_{\alpha\nu}^{\: 0} + \tilde{q}_{\alpha\nu}) &=& \text{sgn}(\Delta N_{\alpha\nu}^{NF}) \Big[ \epsilon_{\alpha\nu} (\tilde{q}_{\alpha\nu}^{\: 0}) + \tilde{q}_{\alpha\nu} \: v_{\alpha\nu}(\tilde{q}_{\alpha\nu}^{\: 0}) \Big] \nonumber \\
\Delta P(\tilde{q}_{\alpha\nu}^{\: 0} + \tilde{q}_{\alpha\nu}) &=& \text{sgn}(\Delta N_{\alpha\nu}^{NF}) \Big[ \tilde{q}_{\alpha\nu}^{\: 0} + \tilde{q}_{\alpha\nu} \Big] 
\end{eqnarray}

To integrate over the particle-hole contributions, we will need the particle-hole velocity $v$, which is readily found to be:
\begin{equation}
v = \frac {\omega - l\Delta E(\tilde{q}_{\alpha\nu}^{\: 0} + \tilde{q}_{\alpha\nu})} {k-l\Delta P(\tilde{q}_{\alpha\nu}^{\: 0} + \tilde{q}_{\alpha\nu})} = v_{\alpha\nu}(\tilde{q}_{\alpha\nu}^{\: 0}) - \text{sgn}(\Delta N_{\alpha\nu}^{NF}) \frac {l \big[ \omega - l \text{sgn}(\Delta N_{\alpha\nu}^{NF}) \epsilon_{\alpha\nu} (\tilde{q}_{\alpha\nu}^{\: 0}) \big]} {\tilde{q}_{\alpha\nu} }
\end{equation}

We thus obtain

\begin{equation}
\tilde{q}_{\alpha\nu} = - \text{sgn}(\Delta N_{\alpha\nu}^{NF}) \frac {l [\omega - l \omega_{\alpha}]} {v-v_{\alpha\nu}(\tilde{q}_{\alpha\nu}^{\: 0})} \label{ggbranch}
\end{equation}
where $\omega_{\alpha} = \text{sgn}(\Delta N_{\alpha\nu}^{NF}) \epsilon (\tilde{q}_{\alpha\nu}^{\: 0})$ is the energy of the $\alpha$-branch line, where $\alpha=c,s$. The jacobian becomes:
\begin{equation}
d \tilde{q}_{\alpha\nu} = \frac 1 {v^2} \frac {l [\omega - l \omega_{\alpha}]} {\Big( 1- \displaystyle{\frac {v_{\alpha\nu}(\tilde{q}_{\alpha\nu}^{\: 0})} v} \Big)^2} dv = \frac {l [\omega - l \omega_{\alpha}]} {\big( 1- z v_{\alpha\nu}(\tilde{q}_{\alpha\nu}^{\: 0}) \big)^2} dz
\end{equation}
where the neglected sign is later taken care of when defining the integration limits. This jacobian is presented here because it will change the behavior of the spectral function in the energy. The particle-hole energy occurring in the argument of the spectral function $\breve{B}^l$ can now be expressed as $l [\omega - l \omega_{\alpha}] \cdot v / \big(v-v_{\alpha\nu}(\tilde{q}_{\alpha\nu}^{\: 0})\big)$. By expressing the energy in this form, we show that we are scanning the branch line in such an interval, where it can reach the ($k,\omega$) point under consideration. Note for example that directly under this particle-hole point, i.e. for $v=\pm \infty$, we have that this energy expression is equal to $l [\omega - l \omega_{\alpha}]$. We are then, by introducing a small quantity $\tilde{q}_{\alpha\nu}$, and allowing it to be both positive and negative, scanning the branch line in an interval that covers both sides of this energy point. Thus, by defining the energy in this way, we are automatically accounting for the region of the branch line of interest. We have that

\begin{eqnarray}
&& \int d\tilde{q}_{\alpha\nu} \ \breve{B}^l \Big( v,l [\omega -l \omega_{\alpha}] \cdot v / \big(v-v_{\alpha\nu}(\tilde{q}_{\alpha\nu}^{\: 0})\big) \Big) = \nonumber \\
&& = \int dz \ \left( \frac {l [\omega - l \omega_{\alpha}]} {4 \pi} \right)^{\zeta_0 -2} \bigg( \frac 1 {1-z v_{\alpha\nu}(\tilde{q}_{\alpha\nu}^{\: 0})}\bigg)^{\zeta_0 -2} \frac {l [\omega -l \omega_{\alpha}]} {\big( 1- z v_{\alpha\nu}(\tilde{q}_{\alpha\nu}^{\: 0}) \big)^2} \cdot \mathcal{F}(z) \nonumber \\
&& = 4 \pi \ \left( \frac {l [\omega - l \omega_{\alpha}]} {4 \pi} \right)^{\zeta_0 -1} \int dz \ \frac {\mathcal{F}(z)} {\big[ 1- z v_{\alpha\nu}(\tilde{q}_{\alpha\nu}^{\: 0}) \big]^{\zeta_0}}
\end{eqnarray}

So that the full spectral function in the vicinity of the branch lines becomes:

\begin{eqnarray}
&&B_{\alpha}^l (k,\omega) = \frac {v_{c0}^{-2\Delta_{c0}} v_{s1}^{1-2\Delta_{s1}}} {2\pi} \theta (\Omega - l [\omega - l\omega_{\alpha}]) \theta (l [\omega - l\omega_{\alpha}])\left( \frac {l [\omega -l \omega_{\alpha}]} {4 \pi} \right)^{\zeta_0 -1} \nonumber \\
&&\cdot \: \text{sgn}(\tilde{q}_{\alpha\nu}^{\: 0}) \Bigg\{ \theta \bigg( v_{s1} \Big[ 1 - \frac {l [\omega - l\omega_{\alpha}]} {\Omega} \Big] - \vert v_{\alpha\nu} (\tilde{q}_{\alpha\nu}^{\: 0}) \vert \bigg) \int_{-\frac {\text{sgn}(\tilde{q}_{\alpha\nu}^{\: 0})} {v_{s1}}}^{\frac {\text{sgn}(\tilde{q}_{\alpha\nu}^{\: 0})} {v_{s1}}} dz \ + \nonumber \\ 
&& +\ \theta \bigg(\vert v_{\alpha\nu} (\tilde{q}_{\alpha\nu}^{\: 0}) \vert - v_{s1} \Big[ 1 - \frac {l [\omega - l\omega_{\alpha}]} {\Omega} \Big] \bigg) \int_{-\frac {\text{sgn}(\tilde{q}_{\alpha\nu}^{\: 0})} {v_{s1}}}^{\left(1 - \frac {l [\omega - l\omega_{\alpha}]} {\Omega}\right)/v_{\alpha\nu} (\tilde{q}_{\alpha\nu}^{\: 0})} dz \  \Bigg\}\cdot \nonumber \\
&&\hspace{3.0cm} \cdot \frac {\mathcal{F}(z)} {\big[ 1- z v_{\alpha\nu}(\tilde{q}_{\alpha\nu}^{\: 0}) \big]^{\zeta_0}} \label{alphabr}
\end{eqnarray}
where the factor $\text{sgn}(\tilde{q}_{\alpha\nu}^{\: 0})$ is introduced together with the integration limits, in order to always produce a positive number from the $z$ integral. Note that this expression becomes singular whenever we approach the branch line for states such that $\zeta_0 -1 < 0$. This expression will be responsible for the characteristic line shapes of the spectral function, following the dispersion of the dispersive pseudofermion or pseudofermion hole. However, when it enters the linear region of its dispersion, this expression ceases to be valid. 

\subsection{The Luttinger contribution}
\label{luttinger}

The "Luttinger contribution" is a special case of the $\alpha$-branch line, defined in section (\ref{character}), where the dispersive pseudofermion is very close to one of its Fermi points so that the dispersion relation is in its linear region. This case needs to be treated separately from the general $\alpha$-branch case since the formulas applied in that case are not valid as the dispersive pseudofermion enters the linear region. We remind ourselves that in this region, the dispersive pseudofermion is in the same region as some of the particle-hole excitations. In other words, the created pseudofermion or pseudofermion hole, and the particle-hole excitations, share the same velocity. In this way, the "Luttinger contribution" case arises from a "velocity resonance effect".

For this reason, we have to take a step back in our analysis, all the way to Eq. (\ref{conph}). In this equation, we must change the domain of the $k'$ integration, due to the limited range of momenta available for the linear regime. We will introduce a small quantity denoted $\Delta q$ which measures the width of momentum over which we will integrate, for each value of $\omega'$. This more careful procedure of integrating the particle-hole processes accounts for the linear regime, as the integration runs successively along the dispersive line, according to the integration limits of the $k'$ integral. We will thus let
\begin{equation}
\int_0^{\Delta \omega} d\omega' \  \int_{-\text{sgn}(v)\Delta \omega / v_{c0}}^{\text{sgn}(v) \Delta \omega / v_{c0}} dk' \ \longrightarrow \int_0^{\Delta \omega} d\omega' \  \int_{\iota ( \omega' / v_{\alpha\nu} - \Delta q)}^{\iota \omega' / v_{\alpha\nu}} dk' 
\end{equation}
We then define $\Delta q = 2\pi y / L$ where $y$ is a number between $0$ and $1$, and will be further specified later. Since the Fermi velocities differ from each other for the different branches and for the two different Fermi points of the same branch, we will change the notation of the spectral function $\breve{B}^l$, to $\breve{B}^{l,\iota}$. In the following we let $\overline{c0}=s1$ and $\overline{s1}=c0$:

\begin{eqnarray}
&& \breve{B}^{l,\iota} = \frac 1 {\pi} \int_0^{\Delta \omega} d\omega' \int_{\iota ( \omega' / v_{\alpha\nu} - \Delta q)}^{\iota \omega' / v_{\alpha\nu}} dk' \ B_{Q_{\alpha\nu}}^l (k' , \omega') B_{Q_{\overline{\alpha\nu}}}^l (\Delta \omega / v - k' , \Delta \omega - \omega') \approx \nonumber \\
&& \approx \frac {\Delta q} {\pi} \int_0^{\Delta \omega} d\omega' \ B_{Q_{\alpha\nu}}^l (\iota \omega' / v_{\alpha\nu} - \iota \Delta q / 2 \:  , \omega') B_{Q_{\overline{\alpha\nu}}}^l (\Delta \omega / v - \iota \omega' / v_{\alpha\nu} + \iota \Delta q / 2 \: , \Delta \omega - \omega')  \nonumber
\end{eqnarray}

This intermediate step can now be continued by using Eq. (\ref{corespec}), and by changing integration variable from $\omega'$ to $x=\omega' / \Delta \omega$. There will be some constant factors in the following expressions, which we for now bundle up into one overall constant, denoted by $\mathcal{C}$. In this way, we will arrive to the following expression after some straightforward algebra:
\begin{eqnarray}
&&\mathcal{C} \: \frac {\Delta q} {2 \pi} (l \Delta \omega)^{\zeta_0 -3} (4 \pi v_{\alpha\nu})^{2-2\Delta_{\alpha\nu}} (4 \pi v_{\overline{\alpha\nu}})^{2-2\Delta_{\overline{\alpha\nu}}} \int_0^1 dx \ \cdot \\
&&\cdot \prod_{\iota'=\pm} \theta\left( 1-x+\iota' v_{\overline{\alpha\nu}} \left\{ \frac 1 v - \frac {\iota x} {v_{\alpha\nu}} + \frac {\iota \Delta q} 2  \right\} \right) \theta\left( x+\iota' v_{\alpha\nu} \left\{ \frac {\iota x} {v_{\alpha\nu}} - \frac {\iota \Delta q} 2  \right\} \right) \cdot \nonumber \\
&& \cdot \left( 1-x+\iota' v_{\overline{\alpha\nu}} \left\{ \frac 1 v - \frac {\iota x} {v_{\alpha\nu}} + \frac {\iota \Delta q} 2  \right\} \right)^{2\Delta_{\overline{\alpha\nu}}^{\iota'}-1} \left( x+\iota' v_{\alpha\nu} \left\{ \frac {\iota x} {v_{\alpha\nu}} - \frac {\iota \Delta q} 2  \right\} \right)^{2\Delta_{\alpha\nu}^{\iota'}-1} \nonumber
\end{eqnarray}

We note now that for the $\alpha\nu$ branch, the argument of the $\theta$-function is always larger than $0$, which makes it convenient to perform the $\iota'$ product explicitly for this branch, which leads to
\begin{eqnarray}
&& \prod_{\iota'=\pm} \theta\left( x+\iota' v_{\alpha\nu} \left\{ \frac {\iota x} {v_{\alpha\nu}} - \frac {\iota \Delta q} 2  \right\} \right) \left(x+\iota' v_{\alpha\nu} \left\{ \frac {\iota x} {v_{\alpha\nu}} - \frac {\iota \Delta q} 2  \right\} \right)^{2\Delta_{\alpha\nu}^{\iota'}-1} = \nonumber \\
&& \hspace{4.0cm} = (2x)^{2\Delta_{\alpha\nu}^{\iota}-1} \left( \frac {l \Delta q v_{\alpha\nu}} 2 \right)^{2\Delta_{\alpha\nu}^{-\iota}-1} (l \Delta \omega)^{1-2\Delta_{\alpha\nu}^{-\iota}}
\end{eqnarray}

By extracting all the different exponents of the Fermi velocities, of the energy $\Delta \omega$ and of the momentum $\Delta q$, as well as of numerical factors and factors of $\pi$, we find that by defining the following function:
\begin{eqnarray}
&&\mathcal{F}_{\alpha\nu,\iota} (1/v) = D_0 \bigg[ \prod_{\alpha'\nu'=c0,s1} \prod_{\iota'=\pm} \big[ \Gamma (2\Delta_{\alpha'\nu'}^{\iota'})\big]^{-1} \bigg]  \int_0^1 dx \ (2x)^{2\Delta_{\alpha\nu}^{\iota}-1}  \cdot \\
&& \cdot \prod_{\iota'=\pm} \theta\left( 1-x+ \frac {\iota' v_{\overline{\alpha\nu}}} v - \frac {\iota \iota' v_{\overline{\alpha\nu}}} {v_{\alpha\nu}} x \right) \left( 1-x+\frac {\iota' v_{\overline{\alpha\nu}}} v - \frac {\iota \iota' v_{\overline{\alpha\nu}}} {v_{\alpha\nu}} x \right)^{2\Delta_{\overline{\alpha\nu}}^{\iota'}-1} \nonumber
\end{eqnarray}

we can reach the following expression for the spectral function: 

\begin{eqnarray}
&& \breve{B}^{l,\iota} (v,\Delta \omega)= \frac {4^{-2\Delta_{\alpha\nu}^{-\iota}}} {2\pi N_a}   \left( \frac {l\Delta \omega} {4\pi} \right)^{\zeta_{\alpha\nu,\iota}-2} v_{\alpha\nu}^{-2\Delta_{\alpha\nu}^{\iota}} v_{\overline{\alpha\nu}}^{1-2\Delta_{\overline{\alpha\nu}}} \cdot \mathcal{F}_{\alpha\nu,\iota} (1/v) \nonumber \\
\end{eqnarray}
where we have defined $\zeta_{\alpha\nu,\iota}=\zeta_0 -2\Delta_{\alpha\nu}^{-\iota}$ and the factor ($1/N_a$) comes from the definition of the quantity $y$ occurring in the definition of $\Delta q$: $y=N_a^{1- 1/ 2\Delta_{\alpha\nu}^{\iota}}$. This $y$ is chosen so that $y \rightarrow 0$ when $2\Delta_{\alpha\nu}^{\iota} \rightarrow 0$ and so that $y \rightarrow 1$ when $2\Delta_{\alpha\nu}^{\iota} \rightarrow 1$, where it is assumed that for the Luttinger case $0< 2\Delta_{\alpha\nu}^{\iota} <1$.

The remaining procedure is now exactly equivalent to that of the $\alpha$-branch case, with the same considerations as already dealt with. This is the consequence of the Luttinger contribution being a "special case" of the $\alpha$-branch line: what remains is to disperse our pseudofermion, albeit confined to the linear regime of the dispersion relation, and fix a point in the ($k,\omega$) plane which is in the vicinity of this dispersive line. Where in the $\alpha$-branch case we dealt with the energy difference $\omega-\omega_{\alpha}$, here we deal with a similar energy difference $\omega - \omega_{Lutt}$. The exponent of this energy difference in the former case was $\zeta_0 - 1$, here this exponent is $\zeta_{\alpha\nu,\iota} - 1$. The arguments of the $\theta$-functions will consist of Fermi velocities, and not the general momentum dependent velocity $\vert v_{\alpha\nu} (q) \vert$, simulating the confinement to the vicinity of the Fermi points. Moreover, the jacobian of the integration will be identical to the $\alpha$-branch case, but with a velocity equal to $\iota v_{\alpha\nu}$ instead of a velocity $v_{\alpha\nu} (\tilde{q}_{\alpha\nu}^{\: 0})$. Thus, with an analysis identical to the one of the $\alpha$-branch line, we obtain the full spectral function for the Luttinger contribution:

\begin{eqnarray}
B_{Lutt}^l (k,\omega) &=& \frac {4^{-2\Delta_{\alpha\nu}^{-\iota}} v_{\alpha\nu}^{-2\Delta_{\alpha\nu}^{\iota}} v_{\overline{\alpha\nu}}^{1-2\Delta_{\overline{\alpha\nu}}}} {2\pi} \theta (\Omega - l [\omega - \omega_{Lutt}]) \theta (l [\omega - \omega_{Lutt}])\left( \frac {l [\omega - \omega_{Lutt}]} {4 \pi} \right)^{\zeta_{\alpha\nu,\iota} -1} \nonumber \\
&&\cdot \: \iota \Bigg\{ \theta \bigg(v_{\alpha\nu} - v_{\overline{\alpha\nu}} \Big[ 1 - \frac {l [\omega - \omega_{Lutt}]} {\Omega} \Big] \bigg)\int_{-\frac {\iota} {v_{\overline{\alpha\nu}}}}^{\iota \left(1 - \frac {l [\omega - \omega_{Lutt}]} {\Omega}\right)/v_{\alpha\nu}} dz \ + \nonumber \\ 
&& +\  \theta \bigg( v_{\overline{\alpha\nu}} \Big[ 1 - \frac {l [\omega - \omega_{Lutt}]} {\Omega} \Big] - v_{\alpha\nu} \bigg) \int_{-\frac {\iota} {v_{\overline{\alpha\nu}}}}^{\frac {\iota} {v_{\overline{\alpha\nu}}}} dz \  \Bigg\}\cdot  \frac {\mathcal{F}_{\alpha\nu,\iota}(z)} {\big[ 1- \iota z v_{\alpha\nu} \big]^{\zeta_{\alpha\nu,\iota}}}  \nonumber \\ \label{speclutt}
\end{eqnarray}

\subsection{Fermi point contribution}

The last case to consider is the 0P case, where both pseudofermions are confined to one of their $\iota$ Fermi points. We will have 4 different points in the ($k,\omega$) plane of this kind. The spectral function expression will then be valid in the vicinity of these 4 points only, as the particle-hole contributions allow the spectral weight from this case to extend a maximum energy of $\Omega$ from the Fermi level. There is actually not much work to be done to account for the spectral function for this case, since there is no finite energy pseudofermion to integrate over: both pseudofermions or pseudofermion holes are confined to one of their Fermi points. The resulting expression for the spectral function in this case can thus easily be read from Eq. (\ref{FAAN}):

\begin{eqnarray}
B_{0P}^l (k,\omega) &=& \frac {v_{c0}^{-2\Delta_{c0}} \ v_{s1}^{1-2\Delta_{s1}}} {4\pi} \theta (\Omega - l\omega) \theta(l\omega) \left( \frac {l \omega} {4 \pi} \right)^{\zeta_0 -2} \cdot \mathcal{F}(1/v) \label{0P}
\end{eqnarray}
where the introduced $\theta$-functions restrict us to particle-hole energies between $0$ and $\Omega$. We note that this spectral function contributes in the vicinity of the specific zero energy momentum points specified in sections (\ref{RHBtrans}) (RHB) and (\ref{LHBtrans}) (LHB), respectively. However, of all these points it is only in the vicinity of ($k_F,0$), that the spectral function has a singular behavior. Indeed, for the other points, the exponent $\zeta_0 -2$ is positive and hence the spectral weight vanishes as the zero energy level is approached. Moreover, in the vicinity of ($k_F,0$), this spectral function has the smallest value for the exponent of all exponents derived.
\setcounter{chapter}{4}
\setcounter{section}{4}

\chapter{One Electron Spectral Weight}
\label{theorweight}

\section{General ($U/t$) and $n$ dependence}
\label{genun}

The expressions for the spectral weight distributions derived in chapter (\ref{onelecspec}) depend on the value of the ratio of the effective Coloumb interaction strength $U$ and the transfer integral $t$, as well as on the electronic density of the system $n$ (note that we always have the magnetization $m \rightarrow 0$). This follows from the ($U/t$) and $n$ dependence of most quantities involved, as for example the phase shifts, the dispersion relations $\epsilon_{\alpha\nu}(q)$ and their corresponding group velocities $v_{\alpha\nu}(q)$ (these quantities are defined by Eqs. (\ref{energydev}), (\ref{energybands}), (\ref{rele}) and (\ref{Vvel}), respectively). The dispersion relations determine the shape of the branch lines in the ($k,\omega$) plane. Since one of the pseudofermions or pseudofermion holes is created at one of its Fermi points, we can associate these characteristic lines with distinctive charge type ($c0$) or spin type ($s1$) excitations (where the "type" of the excitation stands for the charge or spin content of the dispersive pseudofermion or pseudofermion hole). These spectral features are only singular, however, for negative exponents which produce divergent expressions as we approach the branch lines. In the following, whenever referring to electrons, we will use units such that the lattice constant $a=1$.

We remind ourselves that all of the $2\Delta_{\alpha\nu}^{\iota}$ quantities are larger than zero and thus $\zeta_0 > 0$ as well, where $\zeta_0$ is defined in the text under Eq. (\ref{FAAN}). This means that the general 2P contribution does not exhibit any singular behavior, with the sole exception of the border line case. As we approach these lines, the exponent becomes negative, equal to ($-1/2$), and thus we would expect a singular "rim" along the line described by the condition $v_{c0}(\tilde{q}_{c0}^{\: 0})=v_{s1}(\tilde{q}_{s1}^{\: 0})$, where $k$ and $\omega$ depend on $\tilde{q}_{c0}^{\: 0}$ and $\tilde{q}_{s1}^{\: 0}$ through the relationships given in sections (\ref{RHBtrans}) and (\ref{LHBtrans}). It is interesting to note that the "velocity resonance effect" of having the two elementary excitations propagating at the same group velocity, produces significant spectral features far away from the zero energy Fermi level.

From now on, we will change back to the original notation by letting $\tilde{q}_{\alpha\nu} \rightarrow q_{\alpha\nu}$ denote the canonical momentum of the created or annihilated pseudofermion, independently if this is a finite energy pseudofermion or not. Coming back to the branch lines, these are controlled by an exponent $\zeta_{\alpha}=\zeta_0 - 1$, which may or may not be greater than zero ($\zeta_c$ denotes the c-branch line exponent and $\zeta_s$ the s-branch line exponent). If we are aspiring to compare our theoretical results with experiments, then these singular features must be visible in a ($k,\omega$) photo emission or photo absorption scan of the spectral weight of the material in question. 

In the notation of sections (\ref{RHBtrans}) and (\ref{LHBtrans}), we will find in the subsequent sections that the lines exhibiting singular behavior, correspond to the numbers $\lambda_{c0}=\iota_{c0}$ (RHB s-branch line), $\lambda_{c0}=\iota_{s1}$ (RHB c-branch line), $\iota_{c0}=+$ (LHB s-branch line), and $\iota_{s1}=+$ (LHB c-branch line) respectively. For these cases, the momentum dependence of the branch line exponent is plotted for various values of ($U/t$).

One can envision the $\alpha$-branch line contribution as a dispersive $\alpha\nu$ pseudofermion or pseudofermion hole, "moving" along the line dictated by its dispersion relation. As it disperses, the particle-hole towers of states gives rise to spectral features in the vicinity of the branch line. However, as this pseudofermion or pseudofermion hole reaches one of the end points of its dispersive line shape, it enters an intermediate regime where the valid expression for the spectral weight is not that of the $\alpha$-branch line, but rather that of the Luttinger contribution. This special case can be likened to the border line case, since it too arises from a "velocity resonance effect": when the dispersive pseudofermion or pseudofermion hole enters the regime where its dispersion relation becomes linear, it has a velocity equal to one of the velocities of the particle-hole excitations. In this case, the spectral features are described by another exponent than that of the $\alpha$-branch line case. 

The exponents obtained for the Luttinger contribution is equivalent to the exponent obtained by low elementary excitation energy methods, such as conformal field theory \cite{equival}Ê\cite{crit6} \cite{crit61}.

As the above mentioned $\alpha\nu$ pseudofermion or pseudofermion hole approaches the very end  of its dispersive line shape, it enters yet another regime within the Luttinger-liquid behavior, here dubbed the 0P regime. In this regime, notable singular spectral features can be found, as it corresponds to the most divergent exponent. We will expect some diverging peaks at the zero energy Fermi points, i.e. for $k=k_F$. However, for other integer multiples of $k_F$ (at the zero energy level), the exponents are positive and does not give rise to any singular features. Lastly, since $k_F=\pi n / 2$, we have that the distinctive Fermi point peaks, as well as the shape of the branch lines, change proportionally to $n$. As we approach half filling, $n \rightarrow 1$, more and more spectral weight is transferred from the LHB to the UHB. Indeed, for $n=1$, the LHB is completely empty. We will briefly discuss the filling dependence of the exponents in section (\ref{densexp}).

Thus, the two ingredients one needs in order to deduce the general spectral weight behavior are the dispersion relations and the values of the exponents, respectively. However, the greatest numerical challenge is to compute the pre-factors of the spectral function expressions, which were found to be proportional to an integration in the $z=1/v$ variable of the function $\mathcal{F}(z)$. The results for the theoretical spectral features reported in the remainder of this thesis work were obtained by employing the programming language {\it Fortran} as well as the {\it Mathematica} software.

{\bf Numerical considerations:}

Since we assume that the magnetization is vanishing, the $s1$ pseudofermion band is almost completely filled. However, at strict zero magnetization, we have that $2 \Delta_{s1}^{\pm}=0$ by definition, since the $s1$ pseudofermions become non dynamical. This leads to an ill defined expression for the function $\mathcal{F}(z)$, Eq. (\ref{mathcalF}). Also, many quantities show discontinuities in this limit, i.e. that $\lim_{m \rightarrow 0} f(m) \neq f(0)$. See for example Figs. (\ref{PhiU100}) - (\ref{PhiU0p3}), where $\Phi_{s1,\alpha\nu}(q,q')$ has a sudden jump at the boundary $q=\pm k_{F\downarrow}=\pm k_F$. Therefore, we do allow a {\it very small} yet finite magnetization in our calculations. In this way, we avoid problems with having coinciding values for the Fermi momenta and the limiting momenta for the Brillouin zone. Typically in our calculations, we have $k_{F\uparrow}-k_{F\downarrow} \lesssim 0.001$.

To arrive to a suitable value for our cutoff $\Omega$, we assume for the following discussion that we are in the vicinity of a spectral feature described by a negative exponent. The particle-hole tower of states will then produce a decaying tail, as the particle-hole energy increases. The value of the cutoff $\Omega$ must be chosen in such a way as to properly account for this decaying tail. If chosen too small, there will be an unphysical (abrupt) end to the tower of states, producing a step-like feature at the cutoff. However, if chosen too large, we will take into account unphysical processes as we approach the cutoff energy. Moreover, the cutoff has to be chosen so that the sum rules are satisfied. Under the approximation that the cutoff is only weakly state dependent, we find that an average value of $\Omega \approx 0.2t $ produces a spectral weight that fulfills these criteria. 

\newpage
\section{The RHB spectral weight}

In this section we present the full one electron spectral functions for the RHB, as obtained by use of Eqs. 
(\ref{2P}), (\ref{borde1}), (\ref{borde2}), (\ref{alphabr}), (\ref{speclutt}), and (\ref{0P}), respectively. Furthermore, we present the momentum and ($U/t$) dependence of the $\alpha$-branch line exponents, given by $\zeta_0 -1= -1+\sum_{\alpha\nu} 2\Delta_{\alpha\nu}$, where $2\Delta_{\alpha\nu}=\sum_{\iota=\pm} 2 \Delta_{\alpha\nu}^{\iota}$ is defined by Eq. (\ref{sumdelta}), and $2 \Delta_{\alpha\nu}^{\iota}$ is given by the expressions found in section (\ref{RHBtrans}). We also plot the regions in the ($k,\omega$) plane where the contributions to the one electron spectral function generates a finite spectral weight. These regions were obtained by the defining equations for $k$ and $\omega$, respectively, presented in section (\ref{RHBtrans}).

From section (\ref{RHBtrans}), we see that the sign of the shake-up phase shift, equal to $\lambda_{c0}=\pm$, can be combined with the sign of the Fermi point of that pseudofermion hole which is confined to such a point, when listing all possible branch lines. We remember that the spectral function is an even function of its momentum variable and that hence for simplicity we are only interested in positive momentum values. We now have two distinct c-branch lines, both involving a $s1$ pseudofermion hole being created at its positive Fermi point. 

\begin{figure}
\subfigure{\includegraphics[width=7cm,height=5cm]{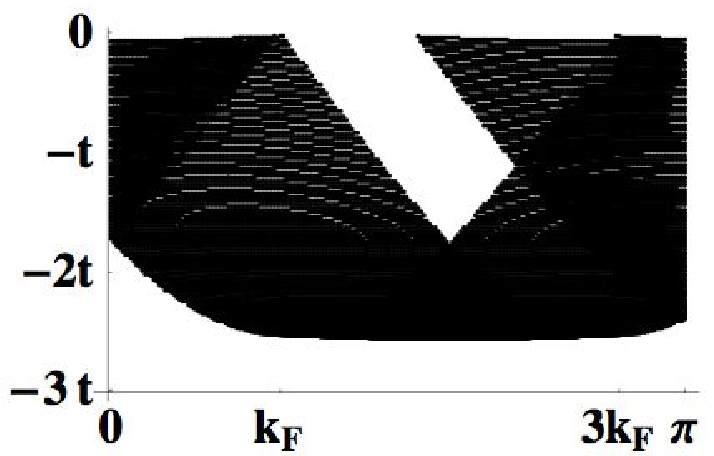}}
\subfigure{\includegraphics[width=7cm,height=5cm]{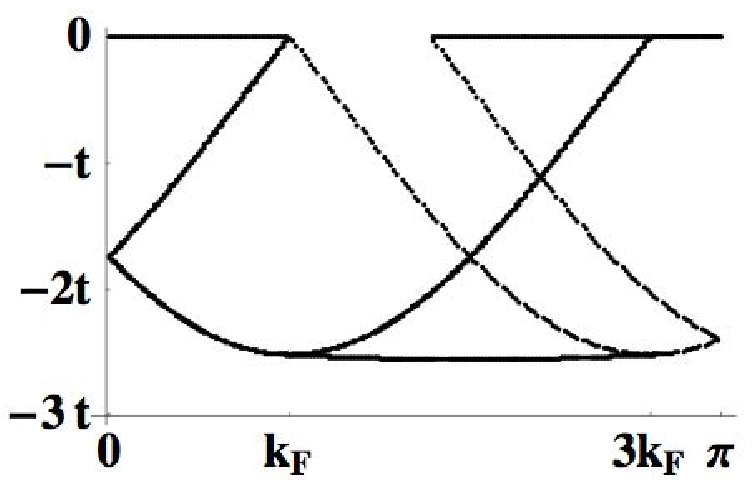}}
\caption{\label{fig2DRHBU100} The region of the ($k,\omega$) plane with a finite spectral weight from the 2P contribution (left) and the branch lines and the border line (right), respectively, for the one electron removal band (RHB) with $(U/t)=100$, $n=0.59$ and $m \rightarrow 0$. In the right figure, there are two c-branch lines emerging from the point ($k,\omega$)=($k_F,0$). The one extending towards smaller momenta is characterized by $\lambda_{c0} = \iota_{s1} =-1$, the other one by $\lambda_{c0} = -\iota_{s1} = -1$. The former line obeys $k=q_{c0}+k_F$ and the latter $k=q_{c0}+3k_F$, in the $m \rightarrow 0$ limit. Both of these lines are folded back into the positive momentum section of the first Brillouin zone. Note the almost completely flat $s1$ dispersion. The border line can be seen connecting   the minimum energy points of the two c-branch lines, having $v_{c0}(q_{c0})=v_{s1}(q_{s1}) \approx 0$.\vspace{0.5cm}}
\end{figure}

\begin{figure}
\subfigure{\includegraphics[width=7cm,height=5cm]{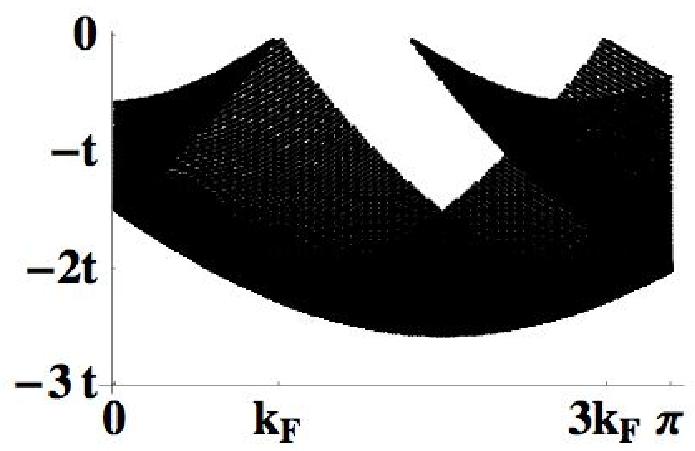}}
\subfigure{\includegraphics[width=7cm,height=5cm]{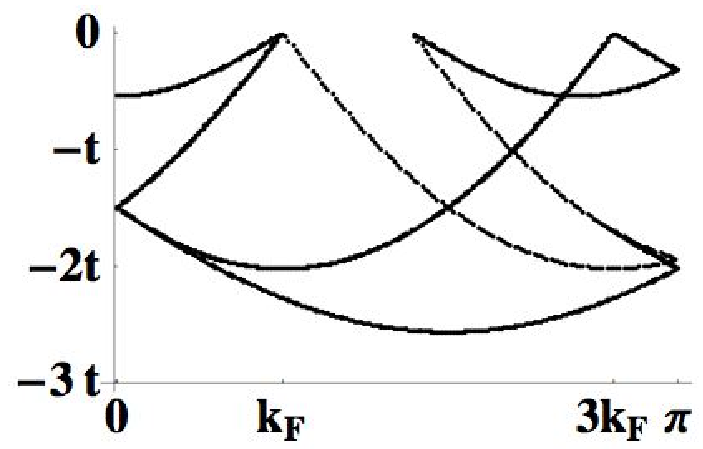}}
\caption{\label{fig2DRHBU4p9} The region of the ($k,\omega$) plane with a finite spectral weight from the 2P contribution (left) and the branch lines and the border line (right), respectively, for the one electron removal band (RHB) with $(U/t)=4.9$, $n=0.59$ and $m \rightarrow 0$. The main difference as compared with the $(U/t)=100$ case, is that the $s1$ s-branch lines have now a non negligible energy width. The s-branch line between $-k_F$ and $k_F$ is characterized by $\lambda_{c0} = \iota_{c0}=\pm 1$ (the two choices of the sign produces two superposing line shapes), and the s-branch line between $3k_F$ and $5k_F$ is characterized by $\lambda_{c0} = - \iota_{c0}=-1$ ($5k_F$ is folded back to $2\pi - 5k_F$). The border line velocity $v_{BL}$ assumes all values in the domain $v_{BL}=v_{c0}(q_{c0})=v_{s1}(q_{s1})$ and brings the region of finite spectral weight to smaller energies than that of the $(U/t)=100$ case due to the larger value of $v_{s1}$. \vspace{0.5cm}}
\end{figure}

\begin{figure}
\subfigure{\includegraphics[width=7cm,height=5cm]{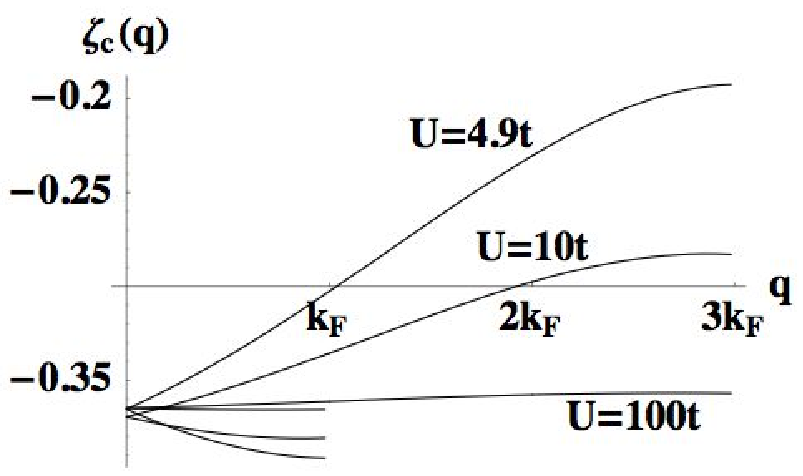}} 
\subfigure{\includegraphics[width=7cm,height=5cm]{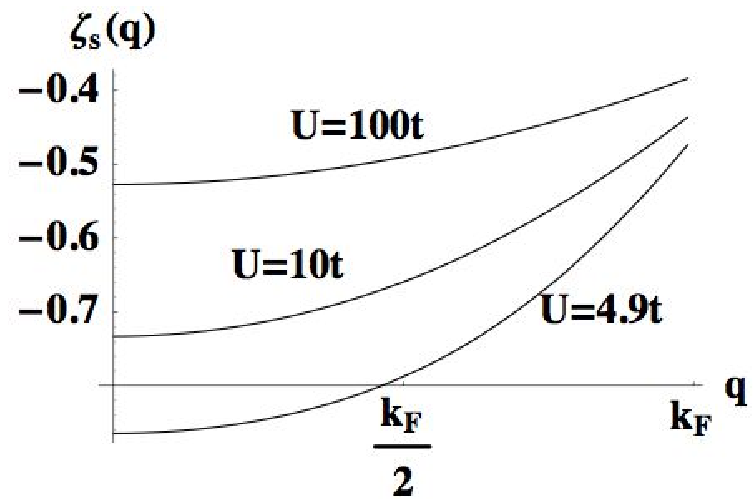}}
\caption{\label{figexpoRHB} The value of the exponents for the c-branch lines (left) and the s-branch lines (right), for various values of ($U/t$), $n=0.59$ and $m \rightarrow 0$. For the s-branch line exponent, the tick mark at $k_F / 2$ is inserted to aid the eye. Note that the c-branch line segment between $-k_F$ and $0$ is folded over into the positive momentum region. The value of the c-branch exponent for $(U/t)=100$ is almost constant. For the folded momenta values, the values of the c-branch line exponent for intermediate values of ($U/t$), is smaller than the corresponding exponent for $(U/t)=100$. For this subbranch, the exponent is smaller for smaller $(U/t)$, however for all other values of $q$, we have the opposite situation.  The s-branch line exponent is always smaller for smaller $(U/t)$, however with a decreasing difference as we approach the Fermi momentum $k_F$. \vspace{0.5cm}}
\end{figure}

In Figs. (\ref{fig2DRHBU100}), (\ref{fig2DRHBU4p9}) and (\ref{figexpoRHB}), the lines originating from excitations with momentum values outside the first Brillouin zone (i.e. such that $k > \pi$) are folded back into this zone. Moreover, for $0 < k < k_F$ we have two c-branch lines joined at $k=0$. The one lowest in $\vert \omega \vert$ can equivalently be described in terms of the other one, but with negative momentum values. This c-branch line segment is then folded over into the positive momentum region. The momentum dependent branch line exponents are plotted in (\ref{figexpoRHB}), for the branches with negative exponents, i.e. for the singular c-branch line between $-k_F$ and $3k_F$ and for the s-branch line between $-k_F$ and $k_F$. All other branch lines have positive exponents and thus their weight decreases as we approach the branch lines. For the Fermi point contributions, the only negative exponent occurs at $k=k_F$. For $(U/t)=100$ this exponent is $\zeta_F \approx -0.867$ whilst for $(U/t)=4.9$ it is $\zeta_F \approx-0.951$.

\begin{figure}
\subfigure{\includegraphics[width=7cm,height=7cm]{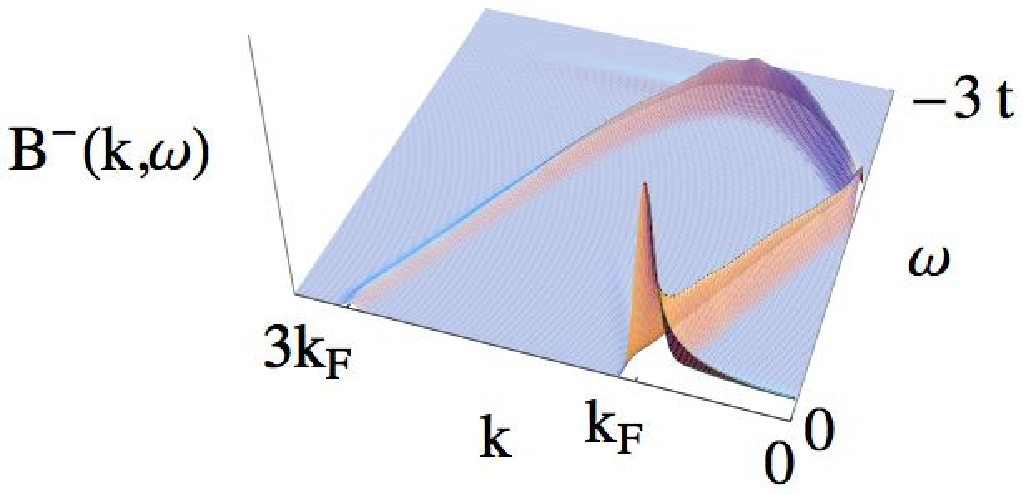}}
\hspace{1.0cm}
\subfigure{\includegraphics[width=7cm,height=7cm]{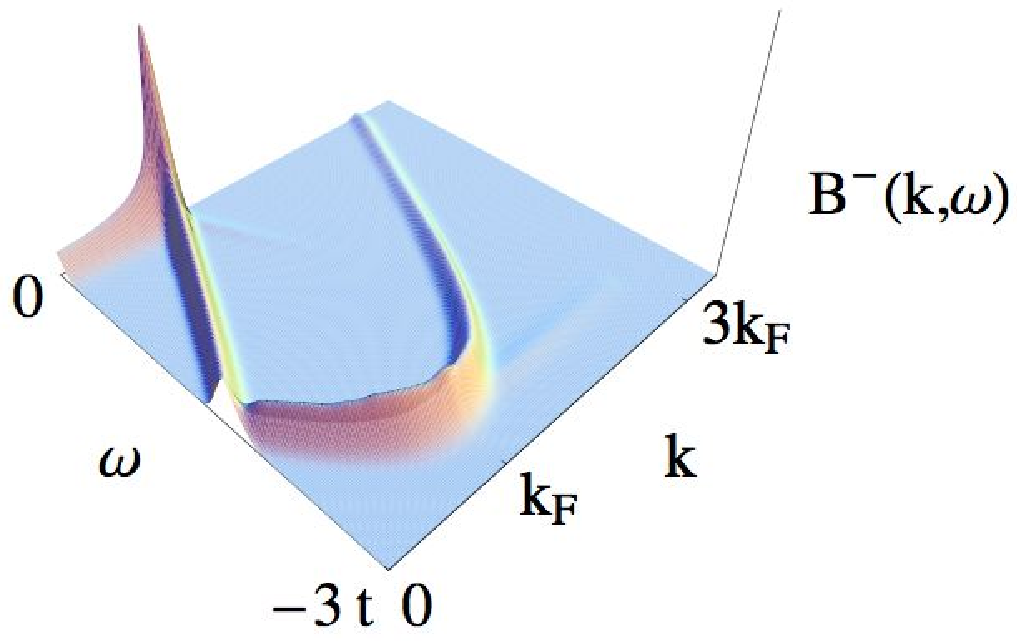}}
\caption{\label{figspecU100} The one-electron removal (RHB) full spectral function, for $(U/t)=100$, $n=0.59$ and $m \rightarrow 0$, as viewed from two different angles. The most divergent peak is to be found at the zero energy $k=k_F$ point. From this point, one c-branch and one s-branch emerge. For both of these branches, we have the intermediate "Luttinger contribution" which brings the spectral weight down as compared to the value at $(k,\omega)=(k_F,0)$. For the c-branch, the total weight does not change significantly for negative velocities. However, as this line passes the zero velocity point the weight starts to vanish as we approach $k=3k_F$. The border line contribution can be seen to produce very little weight, however visible in the figure. The s-branch weight decreases continuously as we depart from $k=k_F$ and approach $k=0$. The entire s-branch is concentrated at excitation energies close to zero, due to the very weakly dispersing $s1$ pseudofermion holes for large values of ($U/t$). \vspace{0.5cm}}
\end{figure}
\begin{figure}
\subfigure{\includegraphics[width=7cm,height=7cm]{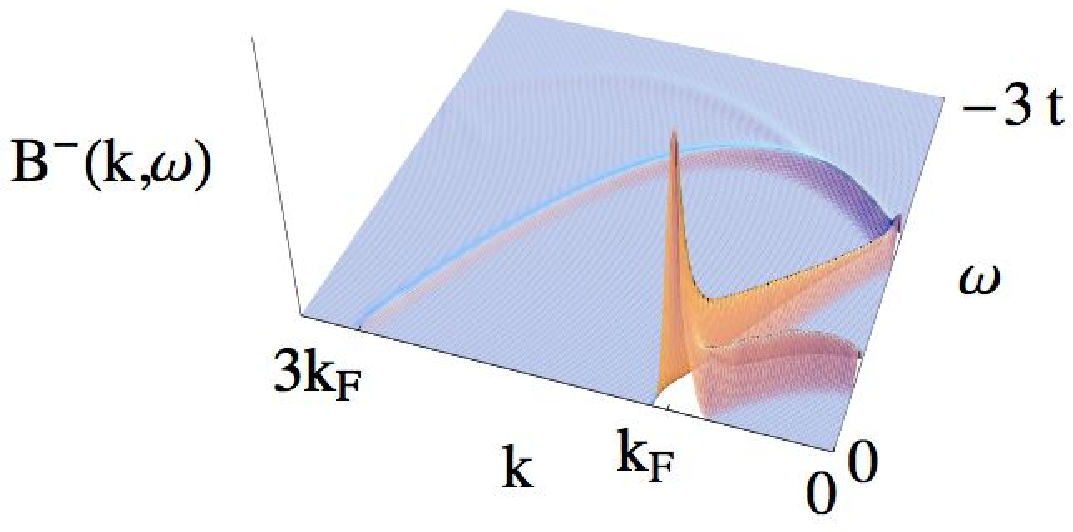}}
\hspace{1.0cm}
\subfigure{\includegraphics[width=7cm,height=7cm]{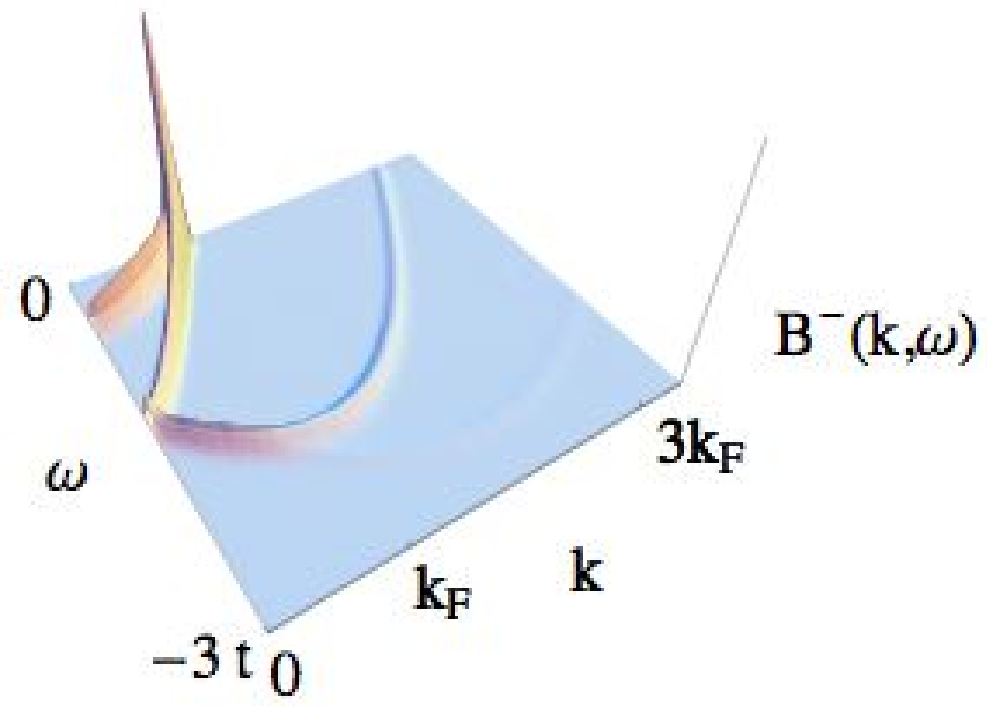}}
\caption{\label{figspecU4p9} The one-electron removal (RHB) full spectral function, for $(U/t)=4.9$, $n=0.59$ and $m \rightarrow 0$, as viewed from two different angles. Some of the features are similar to the ones of the $(U/t)=100$ case, for example the strong divergence of the spectral weight at the point ($k,\omega$)=($k_F,0$). The s-branch is however much more dispersive, as can be seen on the s-branch line feature between $k=0$ and $k=k_F$. Note that the border line is no longer flat, as compared to the $(U/t)=100$ case, mainly due to the larger $s1$ group velocity. \vspace{0.5cm}}
\end{figure}

\begin{figure}
\begin{center}
\includegraphics[width=7cm,height=7cm]{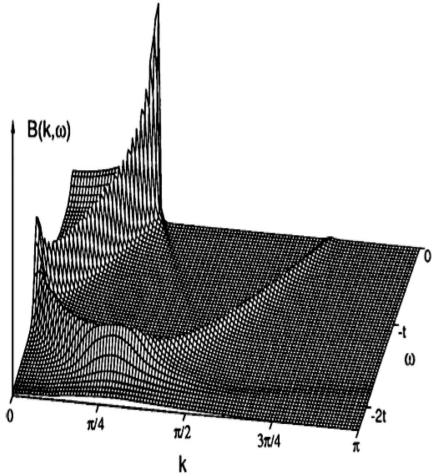}
\caption{\label{figKarloRHB} The one-electron removal (RHB) full spectral function from Ref. \cite{Karlo2}, for $(U/t)=\infty$ at quarter filling $n=0.5$ and $m \rightarrow 0$. Note the overall agreement with Fig. (\ref{figspecU100}), for example the $c$- and the $s$-branch line features, and the Fermi peak singularity. Moreover, the $c$-branch line feature fades away as it approaches $3k_F$.  \vspace{0.5cm}}
\end{center}
\end{figure}

The spectral weight distribution for the entire ($k,\omega$) plane for arbitrary values of ($U/t$), $n$ and magnetization $m \rightarrow 0$, was obtained by the use of the expressions derived in the previous section. Our results for $(U/t)=100$ should be very similar to other results valid in the large ($U/t$) limit. In Fig. (\ref{figspecU100}), we plot the spectral function for ($U/t$) equal to $100$ and in Fig. (\ref{figspecU4p9}) for ($U/t$) equal to $4.9$, respectively. The former case should then be compared with the corresponding Fig. (\ref{figKarloRHB}), originally presented in Ref. \cite{Karlo2} and valid for $(U/t) \rightarrow \infty$ only. That reference uses properties of the Hubbard model unique for the infinite repulsion case, and does not use the pseudofermion representation {\it per se}. However, the representation used in that reference is related to ours since in the $(U/t) \rightarrow \infty$ limit, the $c0$ pseudofermion becomes the "spinless fermion" of that reference. The spin part of that reference is described by the 1D Heisenberg spin hamiltonian. Here we use it as a reference for the validity of our results.

The features of the "large $(U/t)$" spectral function obtained by using the pseudofermion representation are described in the caption of Fig. (\ref{figspecU100}). It is a verification that the "Basic" transition is a good approximation to the total spectral weight of the large ($U/t$) spectral function, by comparison with Fig. (\ref{figKarloRHB}). Considering other transitions (for example, the previously described "Exotic" transition) will only modify the total spectral weight very slightly. For example, these other transitions will not bring about new features to the overall spectral function, but add small corrections to the already existing features and ultimately make so that the exact sum rule will be satisfied. This is the reason of not having any tick marks on the $z$-axis of the figures presented here: considering more transitions could, however slightly, shift the total weight. The general shape is however directly proportional to the probability of finding the created electronic hole at momentum $k$ and energy $\omega$. As a final remark, we note that our result that the 2P "background" contribution is indeed very small, is confirmed by the studies of Refs. \cite{Karlo1}-\cite{Karlo3}. 

The main difference between the full spectral function of the large ($U/t$) case and of the intermediate ($U/t$) case, is the increase of the velocity $v_{s1}$. This increase introduces spin related excitation energies significantly departed from the zero energy level. Since we always have that $v_{c0} \geq v_{s1}$, the border line covers the entire $s1$ band, but only a segment (symmetrical around zero) of the $c0$ band. Due to the larger value of $v_{s1}$ for the intermediate ($U/t$) case, as compared to its value for large ($U/t$), the border line extends to higher energy excitation values, than for large ($U/t$) case. Our results for $(U/t)=4.9$ are presented in Fig. (\ref{figspecU4p9}).

\section{The LHB spectral weight}

In this section we present the full one electron spectral functions for the LHB, as obtained by use of Eqs. 
(\ref{2P}), (\ref{borde1}), (\ref{borde2}), (\ref{alphabr}), (\ref{speclutt}), and (\ref{0P}), respectively. Furthermore, we present the momentum and ($U/t$) dependence of the $\alpha$-branch line exponents, given by $\zeta_0 -1= -1+\sum_{\alpha\nu} 2\Delta_{\alpha\nu}$, where $2\Delta_{\alpha\nu}=\sum_{\iota=\pm} 2 \Delta_{\alpha\nu}^{\iota}$ is defined by Eq. (\ref{sumdelta}), and $2 \Delta_{\alpha\nu}^{\iota}$ is defined in section (\ref{LHBtrans}). We also plot the regions in the ($k,\omega$) plane where the contributions to the one electron spectral function generates a finite spectral weight. These regions were obtained by the defining equations for $k$ and $\omega$, respectively, presented in section (\ref{LHBtrans}). With an eye to the applications of the theory, presented in chapter (\ref{applica}), we have chosen the intermediate value of ($U/t$) to be equal to $5.61$ and not $4.9$ as in the RHB case.

The one electron addition spectral function is described by creation of one $c0$ pseudofermion and the appearance of one extra $s1$ pseudofermion hole. The LHB "Basic" transition is described in section (\ref{LHBtrans}), from which we find that there is a smaller number of branch lines than in the RHB case.

\begin{figure}
\subfigure{\includegraphics[width=7cm,height=5cm]{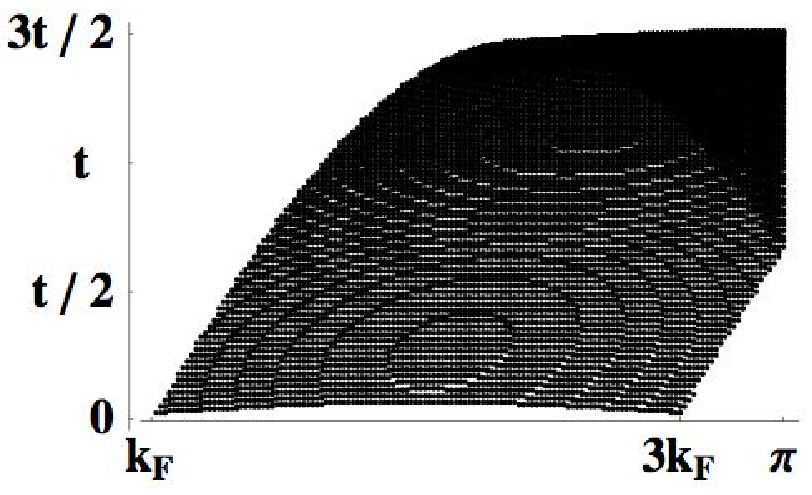}}
\subfigure{\includegraphics[width=7cm,height=5cm]{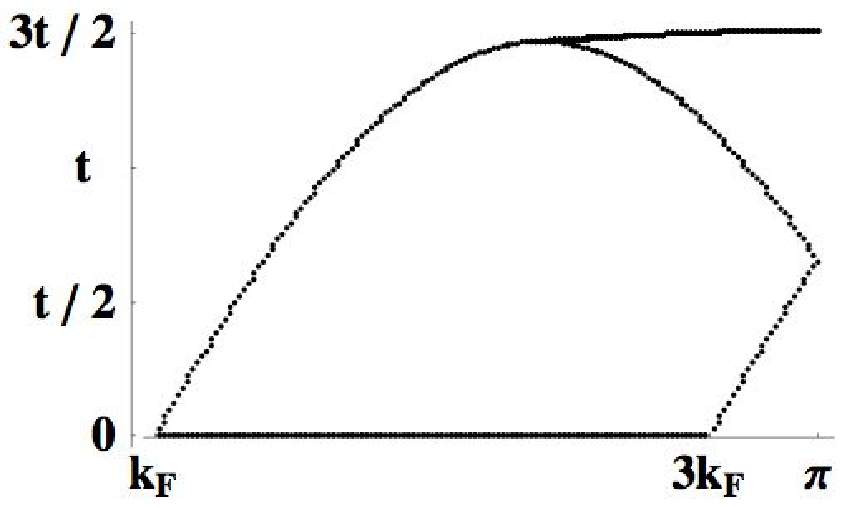}}
\caption{\label{fig2DLHBU100} The region of the ($k,\omega$) plane with a finite spectral weight from the 2P contribution (left) and the branch lines and the border line (right), respectively, for the one electron addition band (LHB) with $(U/t)=100$, $n=0.59$ and $m \rightarrow 0$. The $s1$ band is nearly dispersionless, in analogy with the large ($U/t$) case for the RHB. There is no finite spectral weight for $0 < k < k_F$, as both branch lines originate at $k=k_F$ and extend into regions with larger $k$. The border line is nearly flat, due to having $v_{s1} \approx 0$. \vspace{0.5cm}} 
\end{figure}

This results in part from the subtle effect that the $s1$ pseudofermion hole is not created at the expense of a $s1$ pseudofermion.  The number $N_{s1}$ remains constant under this transition. The contrast to the RHB case described in the previous section can be described by the fact that the $s1$ pseudofermion current is zero for all values of the $s1$ canonical hole momenta $q_{s1}$ different from $q_{Fs1} = \pm k_{F\downarrow} \approx k_F$. For positive values of the momentum $k$, we hence have one c-branch and one s-branch in total. 

\begin{figure}
\subfigure{\includegraphics[width=7cm,height=5cm]{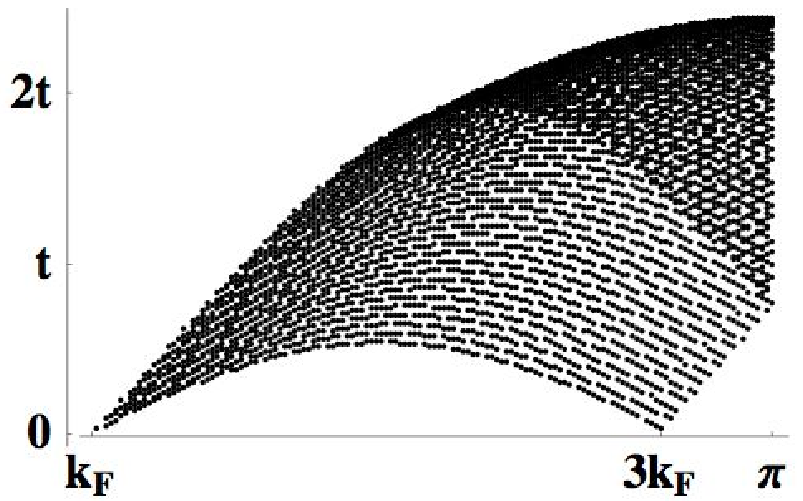}}
\subfigure{\includegraphics[width=7cm,height=5cm]{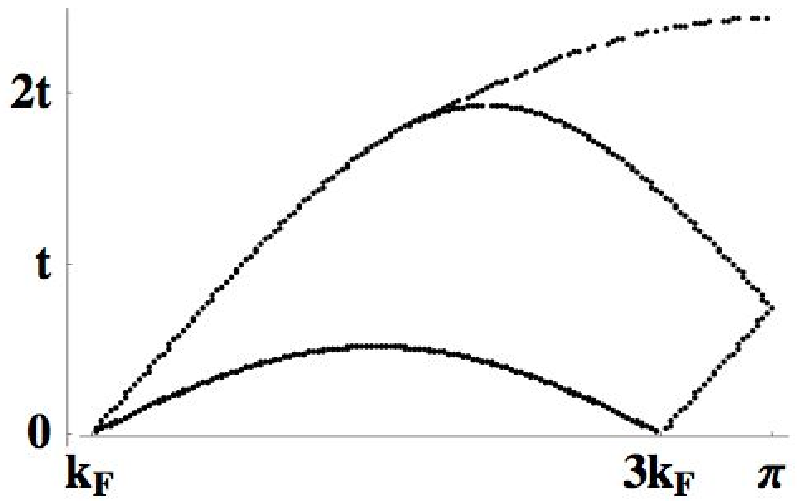}}
\caption{\label{fig2DLHBU5p61} The region of the ($k,\omega$) plane with a finite spectral weight from the 2P contribution (left) and the branch lines and the border line (right), respectively, for the one electron addition band (LHB) with $(U/t)=5.61$, $n=0.59$ and $m \rightarrow 0$. The basic topology is the same as for the $(U/t)=100$ case, but with a larger value for $v_{s1}$, which influences the extension of the s-branch line and the border line onto higher values of $\omega$, as compared to the $(U/t)=100$ case. \vspace{0.5cm}}
\end{figure}

The general domains of finite LHB spectral weight is given in Fig. (\ref{fig2DLHBU100}) for $(U/t)=100$ and in Fig. (\ref{fig2DLHBU5p61}) for $(U/t)=5.61$. Note that the $c0$ total bandwidth $\epsilon_{c0} (\pm \pi) - \epsilon_{c0} (0) = 4t$ is independent of ($U/t$). Since a $c0$ pseudofermion is created in the "Basic" LHB transition, having canonical momentum values between $-\pi$ and $-2k_F$ and between $2k_F$ and $\pi$, respectively, we see that the maximum excitation energy of the $c0$ creation is governed by $\epsilon_{c0} (\pm \pi)$. This value can be read off of Fig (\ref{figEc0}), and increases for decreasing ($U/t$). This explains the difference in the scale of the energy axis in Figs. (\ref{fig2DLHBU100}) and (\ref{fig2DLHBU5p61}). The s-branch is described by momentum values $k=2k_F - q_{s1}$ and runs through values between $k_F$ and $3k_F$, respectively, and not between $0$ and $k_F$ as in the RHB case described in the previous section. 

\begin{figure}
\subfigure{\includegraphics[width=7cm,height=5cm]{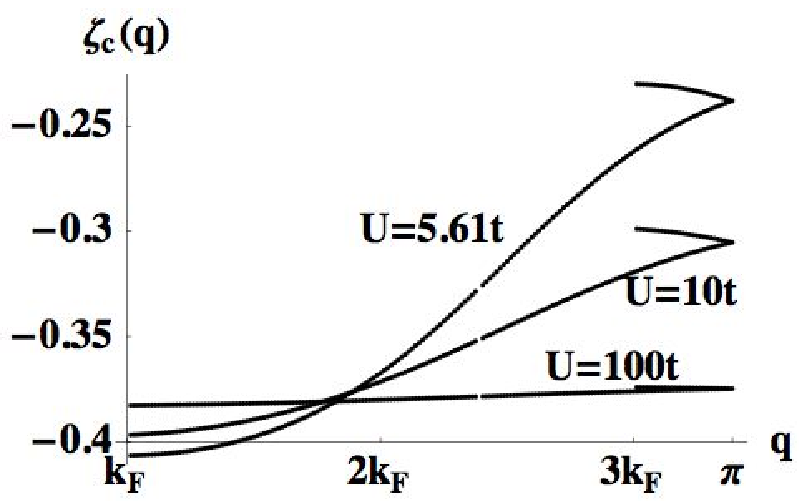}} 
\subfigure{\includegraphics[width=7cm,height=5cm]{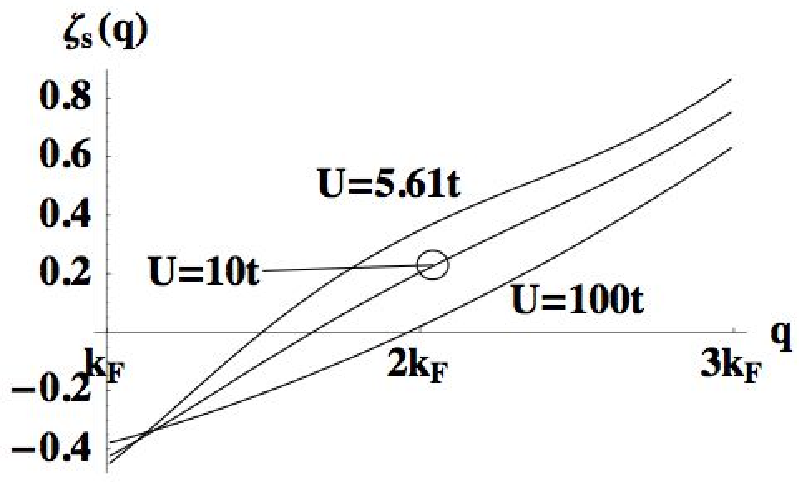}}
\caption{\label{figexpoLHB} The value of the exponents for the c-branch line (left) and the s-branch line (right) for the LHB, for various values of ($U/t$), $n=0.59$ and $m \rightarrow 0$. The value of the c-branch exponent for $(U/t)=100$ is almost constant. Note that the exponent for the c-branch line segment for momentum values $q>\pi$ is folded into the first Brillouin zone (the exponent with canonical momentum $q=\pi + k_F$ is folded back into the Brillouin zone at the small break of the continuous line given by $\zeta_c (q)$, visible between $2k_F$ and $3k_F$). The c-branch line exponents start from $q=k_F$ with roughly the same slope, but with an increasing value of the curvature for decreasing values of ($U/t$), and thus the values of $\zeta_c (q)$ increases for decreasing ($U/t$). The s-branch line exponent is negative for momentum values close to $k_F$, but grows monotonously with $q$ and becomes positive after some specific value of $q$ and remains positive for the remainder of the branch line. We would thus expect the s-branch line feature of the full spectral function to vanish as this value of $q$ is reached and passed. Note that $\zeta_s (q)$ depends almost linearly on $q$ for $(U/t)=10$, and for smaller values of this ratio the dependence is mainly concave, whilst for larger values of this ratio, it is mainly convex. \vspace{0.5cm}}
\end{figure}

The values of the branch line exponents are plotted in Fig. (\ref{figexpoLHB}), for various values of ($U/t$). Similarily to that case, we are always restricting ourselves to momentum values $k$ such that $0 < k < \pi$, which explains the apparent double valuedness of the c-branch exponent. The two c-branch lines, corresponding to the two possible values of $\iota_{s1}=\pm$, maps onto one single continuous line by only allowing positive values of $k$ inside the Brillouin zone, in the following way: The c-branch line feature with $\iota_{s1}=+$ ranges from $k=-\pi-k_F$ to $k=-3k_F$ (for negative $c0$ pseudofermion canonical momentum values) and from $k=k_F$ to $k=\pi-k_F$ (for positive $c0$ pseudofermion canonical momentum values), respectively. Similarily, the c-branch line feature with $\iota_{s1}=-$ ranges from $k=-\pi+k_F$ to $k=-k_F$ (for negative $c0$ pseudofermion canonical momentum values) and from $k=3k_F$ to $k=\pi+k_F$ (for positive $c0$ pseudofermion canonical momentum values), respectively.

For example, in the $(U/t)=5.61$ curve of Fig. (\ref{figexpoLHB}), we see the value of the c-branch line exponent for the $\iota_{s1}=+$ subbranch, for $k$ ranging between $k_F$ and $\pi-k_F$. The other subbranch ($\iota_{s1}=-$), $k$ ranging between $3k_F$ and $\pi+k_F$, starts at $3k_F$ with a value of the exponent roughly around $-0.23$ and then decreases as $k$ increases beyond $k=\pi$. The value of the exponent at $k=\pi+k_F$ (backfolded to momentum $\pi-k_F$) is roughly equal to the value of the exponent belonging to the subbranch with $\iota_{s1}=+$ at the same momentum value.

\begin{figure}
\subfigure{\includegraphics[width=7cm,height=7cm]{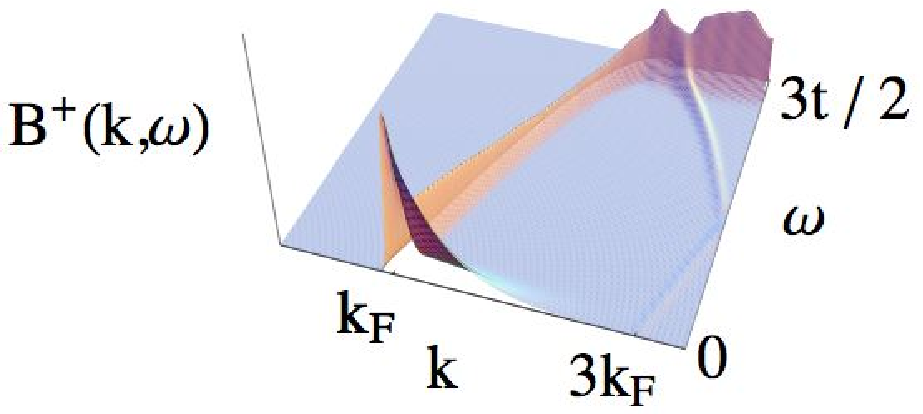}}
\hspace{1.0cm}
\subfigure{\includegraphics[width=7cm,height=7cm]{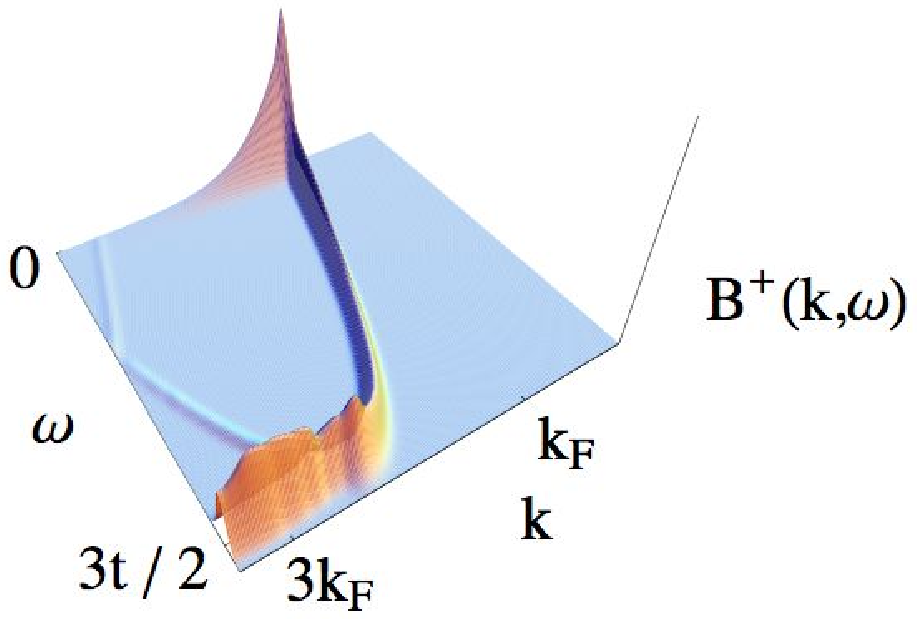}}
\caption{\label{figLHBU100} The one-electron addition (LHB) full spectral function, for $(U/t)=100$, $n=0.59$ and $m \rightarrow 0$, as viewed from two different angles. The strongest divergency of the spectral function occurs at the zero energy Fermi point $k=k_F$. The c-branch originating from this point has a smoothly decreasing spectral weight and is almost vanishing at the other end (towards the zero energy point at $k=3k_F$). The origin of the large rim of the border line is discussed in the text of this section. However, this line seems here a little bit rugged which is due to the numerically sensitive calculations as well as to limitations in our "constant cutoff" and LHB "Basic" transition approximations. \vspace{0.5cm}}
\end{figure}

\begin{figure}
\subfigure{\includegraphics[width=7cm,height=7cm]{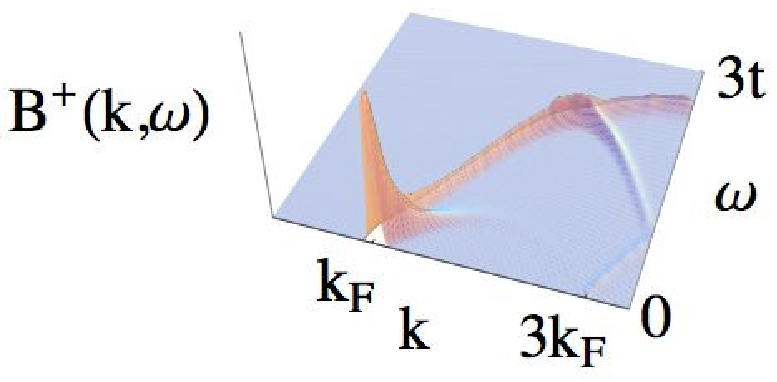}}
\hspace{1.0cm}
\subfigure{\includegraphics[width=7cm,height=7cm]{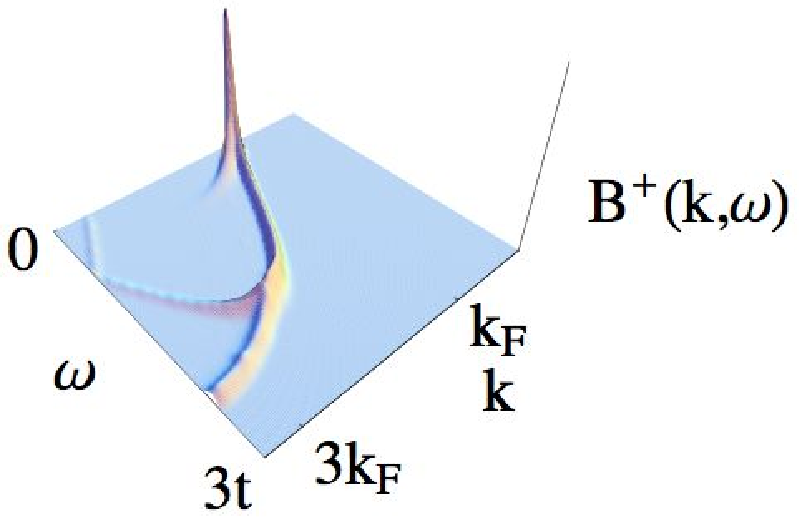}}
\caption{\label{figLHBU5p61} The one-electron addition (LHB) full spectral function, for $(U/t)=5.61$, $n=0.59$ and $m \rightarrow 0$, as viewed from two different angles. The non zero dispersion of the $s1$ band introduces a curved border line, as well as a curved s-branch line. The latter is vanishing due to the positiveness of the exponent $\zeta_s$, as confirmed by Fig. (\ref{figexpoLHB}). The exponent for the Fermi point contribution is even smaller here than in the $(U/t)=100$ case, explaining the strong peak at the point ($k,\omega$)=($k_F,0$). \vspace{0.5cm}}
\end{figure}

We note that the s-branch exponents are monotonously increasing with $k$, becoming larger than zero for a large segment of the total branch line. We are thus expecting that the spectral weight of the s-branch line will vanish completely as we travel along the line from $k=k_F$ towards $k=3k_F$. This characteristic behavior of the s-branch line is indeed verified in Figs. (\ref{figLHBU100}) and (\ref{figLHBU5p61}), where the fading away of this line feature is evident. 

Another characteristic feature of the LHB is the importance of the border lines, which in general carry much more spectral weight than their RHB counterparts. This effect has been attributed to a van-Hove singularity \cite{Karlo1}-\cite{Karlo3} in the strong coupling limit.

The exponents of the Fermi point singularities are $\zeta_F \approx -0.889$ for $(U/t)=100$ and $\zeta_F \approx -0.965$ for $(U/t)=5.61$. The strong negative exponent in both cases motivates the high peak at the point ($k,\omega$)=($k_F,0$) for both values of ($U/t$). For the c-branch line, however, we have that $\zeta_c$ grows with decreasing ($U/t$), producing a weaker spectral weight as compared to larger values of this ratio.


\begin{figure}
\begin{center}
\includegraphics[width=7cm,height=7cm]{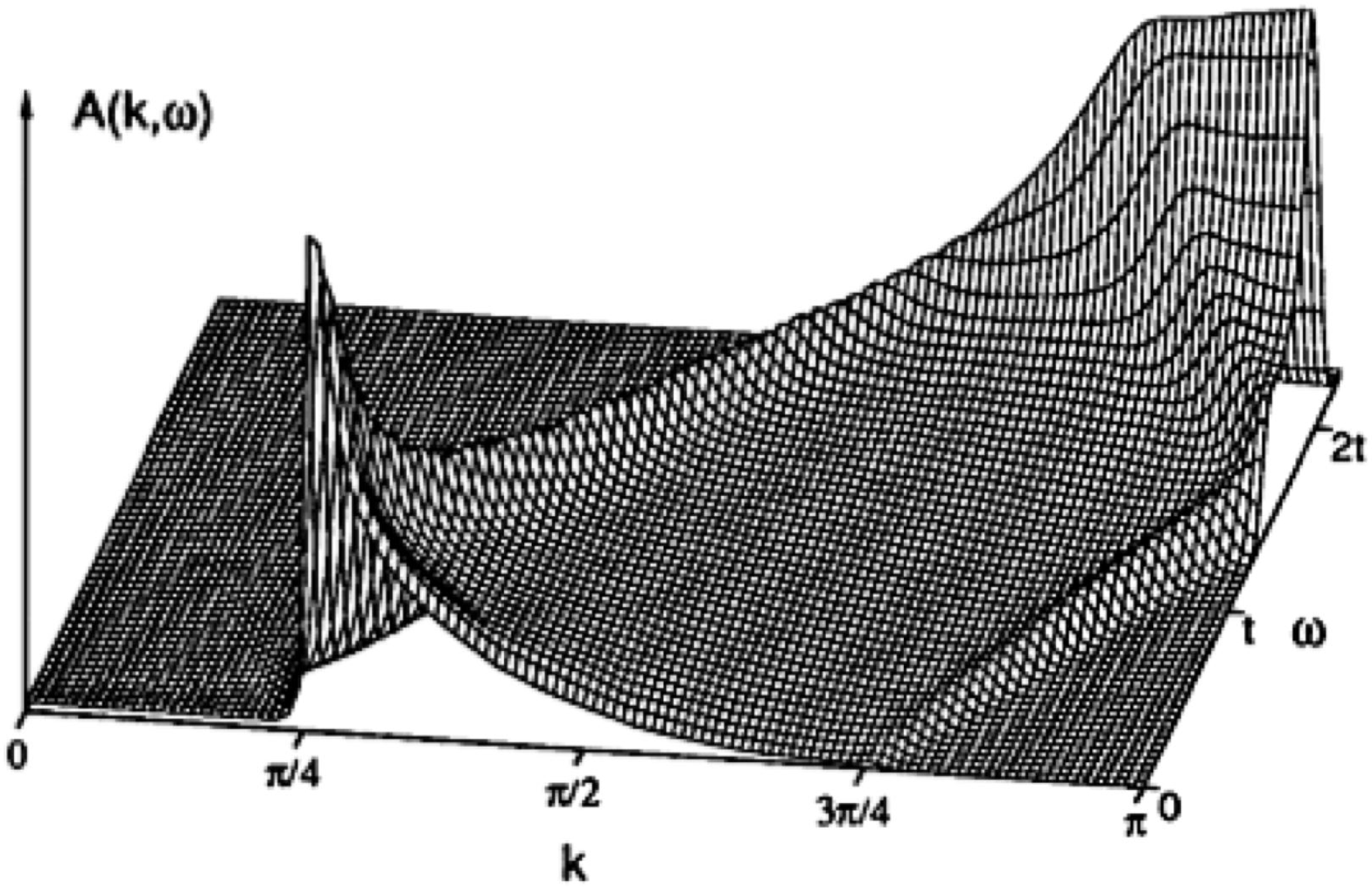}
\caption{\label{figKarloLHB} The one-electron addition (LHB) full spectral function from Ref. \cite{Karlo2}, for $(U/t)=\infty$ at quarter filling $n=0.5$ and $m \rightarrow 0$. This figure should be compared to Fig. (\ref{figLHBU100}). We see that the characteristic spectral features shown here is also present in Fig. (\ref{figLHBU100}). For example, the pronounced border line singularity, the $c0$ pseudofermion branch line, $s1$ pseudpfermion branch line, and the strong divergency at the point $(k,\omega)=(k_F,0)$, are all features accounted for in both figures.
 \vspace{0.5cm}}
\end{center}
\end{figure}

The spectral function of Fig. (\ref{figLHBU100}) should be compared with that of Ref. \cite{Karlo2}, given in Fig. (\ref{figKarloLHB}). In this reference, there is no division of the different types of contributions leading to the total spectral function, but all types of final states fall into the same mathematical treatment, in contrast to the pseudofermion method. This means that there is no clear division between the contributions of the branch lines and the border lines, for example. However, there is also a positive effect of this: there are no troublesome crossover regions in which it is not clear exactly which type of contribution should be valid. For the c-branch line, in the vicinity of the zero velocity point, we have that this contribution superposes on the border line contribution. Moreover, due to the flatness of the $s1$ band, it is not clear how important the Luttinger contribution will be for the border line, since it can be argued that the Luttinger liquid region increases as the dispersion relations becomes flatter, and thus "more linear". These small uncertainties cause the rugged appearance of the border line of the pseudofermion method, which is a numerical feature totally absent in the method of Ref. \cite{Karlo2}. The reasoning behind this ruggedness in the $(U/t)=100$ case implies that for lower values of ($U/t$), this numerical effect would be less pronounced, as the $s1$ dispersion becomes less flat and hence the Luttinger contribution more confined. This is verified in Fig. (\ref{figLHBU5p61}), where virtually all of this ruggedness is non existent.

\section{Density dependent exponents and dispersions}
\label{densexp}

In this section we will briefly discuss the filling dependence of the $\alpha$-branch line exponents, without plotting the full spectral function. As before, these exponents were calculated by use of the defining equations in sections (\ref{RHBtrans}) and (\ref{LHBtrans}), respectively. When discussing these exponents, we will exclusively focus on the same branch lines as in the previous sections of this chapter. Also, the energy dispersions presented here were obtained from Eq. (\ref{energybands}).

\begin{figure}
\begin{center}
\includegraphics{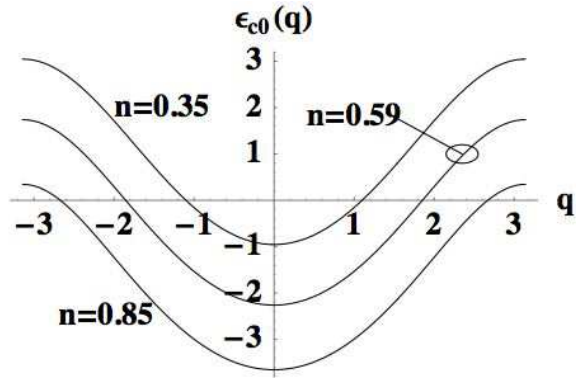}
\caption{\label{figEc0fill} The $c0$ pseudofermion dispersion relations for various values of the electronic density $n$, $(U/t)=10$ and $m \rightarrow 0$. According to the definition of the Fermi momentum, $q_{Fc0}=2k_F = \pi n$, we see that the Fermi momenta approaches the limiting values for the effective Brillouin zone as $n \rightarrow 1$, and goes to zero as $n$ goes to zero. The bandwidth is constant and equals $4t$ independently of the electron density. \vspace{0.5cm}}
\end{center}
\end{figure}

\begin{figure}
\begin{center}
\includegraphics{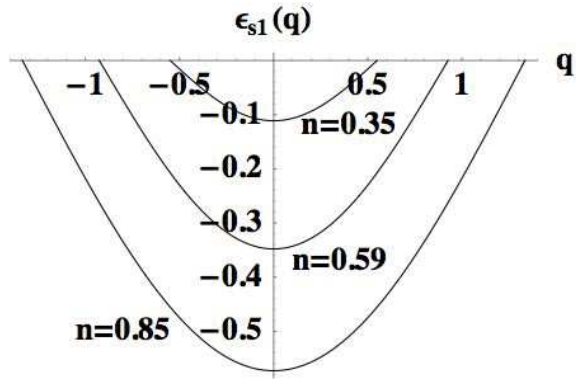}
\caption{\label{figEs1fill} The $s1$ pseudofermion dispersion relations for various values of the electronic density $n$, $(U/t)=10$ and $m \rightarrow 0$. The band shrinks with decreasing $n$, with a decreasing bandwidth, and with Fermi points $\pm q_{Fs1}=\pm k_F= \pm \pi n / 2 \rightarrow 0$ as $n \rightarrow 0$ \vspace{0.5cm}}
\end{center}
\end{figure}

\begin{figure}
\begin{center}
\includegraphics{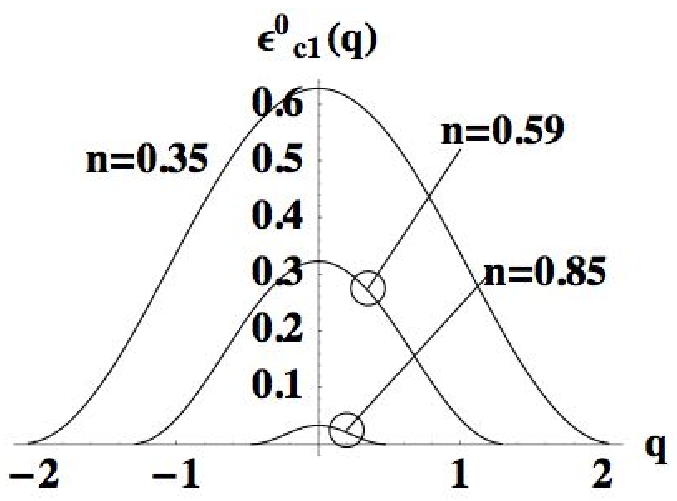}
\caption{\label{figEc1fill} The $c1$ pseudofermion dispersion relations for various values of the electronic density $n$, $(U/t)=10$ and $m \rightarrow 0$. Even though we do not consider final states with finite occupancies of $c1$ pseudofermions, the dispersion is included here for completeness. Both the energy bandwidth and the values of the canonical momenta at the effective Brillouin zone boundaries decrease as $n \rightarrow 1$. This is basically an effect of the diminishing number of doubly occupied, and empty, rotated electron sites, respectively, in this limit. \vspace{0.5cm}}
\end{center}
\end{figure}

All of the figures presented here are valid for $(U/t)=10$ and $m \rightarrow 0$. In Figs. (\ref{figEc0fill})-(\ref{figEc1fill}), we present the $c0$, $s1$, and $c1$ pseudofermion dispersion relations for $n=0.35, 0.59,$ and $0.85$ respectively. 

For the RHB, the dispersing quantum objects are $c0$ and $s1$ pseudofermion holes. Thus, as $n$ decreases, so does the value of the Fermi momentum, and hence the $c0$ and $s1$ pseudofermion holes will have a smaller canonical momentum range in which to disperse. This means that the characteristic peak at the Fermi point ($k,\omega$)=($k_F,0$) approaches the zero momentum and zero energy corner in the ($k,\omega$) plane, as $n \rightarrow 0$.  The minimum value of the $\alpha\nu=c0,s1$ dispersion relations, namely $\epsilon_{\alpha\nu}(0)$, approaches zero in this limit. Thus, the $\alpha$-branch lines will also shrink towards ($k,\omega$)=($0,0$).

For $n \rightarrow 1$, the situation is reversed: the $c0$ and $s1$ pseudofermion holes will have an increasingly larger canonical momentum range in which to disperse. As before, the position of the peak at ($k,\omega$)=($k_F,0$) will move as $n$ varies. The $\alpha$-branch lines will extend over increasingly larger portions of the ($k,\omega$) plane, as the domains of the pseudofermion energies and canonical momenta increases with increasing $n$.

For the LHB, the situation is somewhat different since in this case the dispersive quantum objects correspond to the $c0$ pseudofermion (and not the pseudofermion hole), and the $s1$ pseudofermion hole, respectively. For the $c0$ pseudofermion, we have that for decreasing values of $n$, the domain in which the $c0$ pseudofermion can disperse {\it increases}. For very low densities, the added $c0$ pseudofermion will have almost the entire $c0$ band "for itself", and will thus yield a more extended c-branch line as for higher densities. We remember that the $c0$ pseudofermion domain of dispersion in this case corresponds to canonical momentum values in the ($-\pi , -\pi n$) and the ($\pi n , \pi$) domains, respectively. The s-branch line, however, follows the same general behavior as in the RHB case, with its branch line feature shrinking towards the point ($0,0$).

\begin{figure}
\subfigure{\includegraphics[width=7cm,height=5cm]{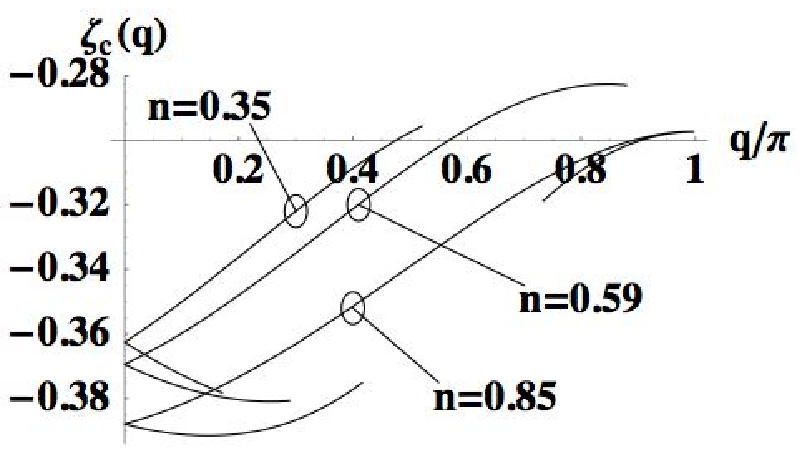}} 
\subfigure{\includegraphics[width=7cm,height=5cm]{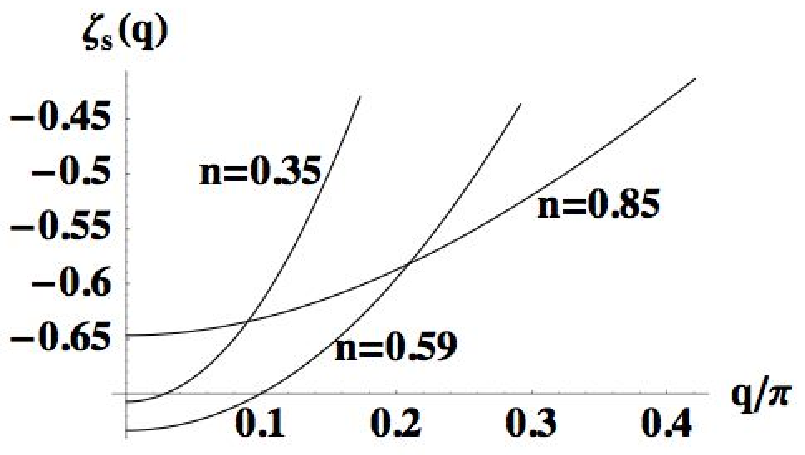}}
\caption{\label{figexpofillRHB} The exponents for the c-branch line (left) and the s-branch line (right) for the RHB, for various values of $n$, $(U/t)=10$ and $m \rightarrow 0$. For a fixed momentum value $q=q_0$, we see that $\zeta_c (q_0)$ is increasing as $n$ decreases (i.e. by following a vertical line upwards). Hence in general, the c-branch line feature diverges more slowly as the density decreases. Note however, that the value of this exponent at the Fermi point $k_F / \pi= n / 2$, does not alter significantly between different densities. An interesting effect occurs at $n=0.85$ for the c-branch line: the value of the exponent at the Fermi point is larger than the corresponding value at momenta between 0 and $k_F$, in contrast to the values of the other exponents of the c-branch line. For the s-branch line, no linear-type trend can be deduced, as the intermediate density value produces the smallest exponent for small momentum values. For any $n$, $\zeta_s$ is always negative however increasing as $q$ increases from $0$ towards $k_F$. As with the c-branch line, the exponent for the s-branch line does not vary significantly at the corresponding Fermi points $k_F / \pi= n / 2$. \vspace{0.5cm}}
\end{figure}

Once again, the situation is reversed in the opposite limit, $n \rightarrow 1$. Here, the added $c0$ pseudofermion will only be able to disperse along the "wings" near the canonical momentum values $q=\pm \pi$ (corresponding to the shrinking regions for which $\epsilon_{c0}(q) > 0$). According to the sum rule of Eq. (\ref{sumrule}), the total spectral weight of the LHB vanishes in this limit, transferring its weight to the UHB. 

\begin{figure}
\subfigure{\includegraphics[width=7cm,height=5cm]{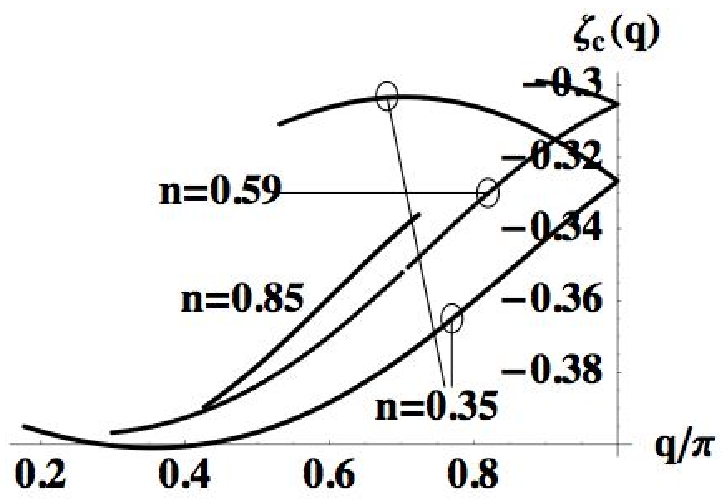}} 
\subfigure{\includegraphics[width=7cm,height=5cm]{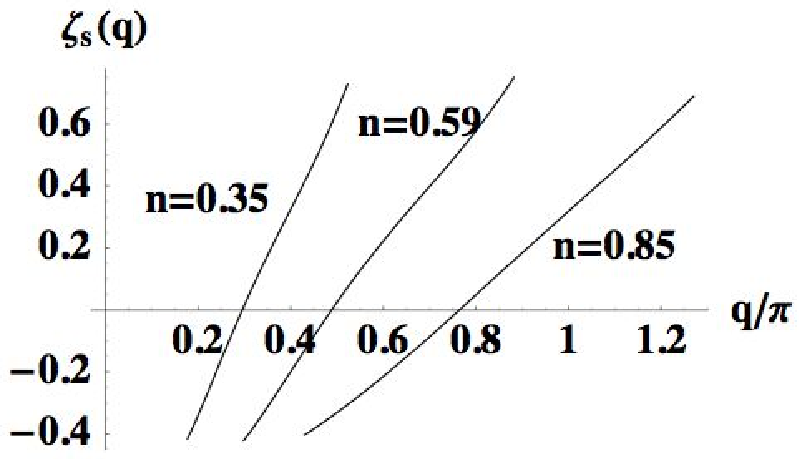}}
\caption{\label{figexpofillLHB} The exponents for the c-branch line (left) and the s-branch line (right) for the LHB, for various values of $n$, $(U/t)=10$ and $m \rightarrow 0$. For the c-branch line, the exponent is always negative and not varying significantly with $n$, roughly larger than $-0.40$ and smaller than $-0.30$. For the s-branch line however, as $n$ decreases, the region for which  $\zeta_s (q) < 0$ shrinks, and thus we expect that the s-branch line becomes less and less significant. Note that the length of the s-branch line shrinks as $n$ decreases, whilst we have the opposite dependence on the filling for the c-branch line. In conclusion, as $n$ decreases, the weight in the vicinity of the s-branch line decreases whilst the weight in the vicinity of the c-branch line does not.
\vspace{0.5cm}}
\end{figure}

Note that in all of these cases, the region in the ($k,\omega$) plane with contributions from the border line, is governed by the value of $v_{s1}$. Irrespectively if we study the RHB or the LHB, all canonical momentum values in the $s1$ band such that $\epsilon_{s1}(q) \leq 0$ have a corresponding canonical momentum value in the $c0$ band, such that the pseudofermion group velocities are equal to each other.

The $\alpha$-branch line exponents are plotted in Figs. (\ref{figexpofillRHB}) and (\ref{figexpofillLHB}). Their dependence on the filling $n$ is discussed in the corresponding captions. Generally, for the RHB, we have that as $n$ decreases, so does the total spectral weight, in accordance with the sum rule. Indeed, the value of the negative c-branch line exponent increases with decreasing $n$. However, the s-branch line exponent does not. The conclusion of this is that a study of the branch line exponents alone, is not sufficient to characterize the behavior of the branch line spectral feature. As $n$ increases we have that the RHB c-branch line feature exhibits a stronger divergent behavior as $n \rightarrow 1$, even though this effect is not "dramatic".  

For the LHB, we have an interesting effect for the s-branch line. Namely, as the momentum increases from $k=k_F$, the exponent for the s-branch line becomes positive at a certain momentum value larger than $k_F$. However, this momentum value approaches $k_F$ as $n$ decreases. Hence, the weight of this branch line must also decrease as $n$ decreases. We do not detect any similar effect for the c-branch line exponent. It seems reasonable to assume that the disappearence of the spectral weight in the LHB case as $n \rightarrow 1$ is in the pseudofermion picture linked with the disappearence of a dynamical $c0$ branch, rather than to the values of the branch line exponents themselves.

In this chapter, we have studied the behavior of the $\alpha$-branch line exponent only. For a complete understanding of the filling dependence of the total $\alpha$-branch line spectral behavior, we need in addition to study the corresponding behavior of the pre-factors.

\setcounter{chapter}{5}
\setcounter{section}{5}

\chapter{Applications - Experimental Spectral Weight}
\label{applica}

\section{The organic compound TTF-TCNQ}

We will in this section devote some attention to the organic material "Tetrathiafulvalene Tetracyanoquinodimethane", abbreviated TTF-TCNQ. For temperatures above the broken symmetry state (linked to a Peierls transition as further described in this section), it is characterized as a metallic "charge transfer salt" consisting of linear stacks of planar molecules. In the subsequent sections, some general properties of this material will be discussed conveying the reasons for why it constitutes a reasonable quasi 1D material allowing it to be compared with the theoretical results obtained so far. It is not our intention to explain in greater detail the rich physical literature that exists regarding TTF-TCNQ, nor to make a "from first principles" derivation of its physical properties. This is outside the scope of this thesis report. However, the interested reader could use the references given in this and in the subsequent section for a deeper study of the properties of TTF-TCNQ.

\begin{figure}
\hspace{1.0cm}
\subfigure{\includegraphics[width=6cm,height=8cm]{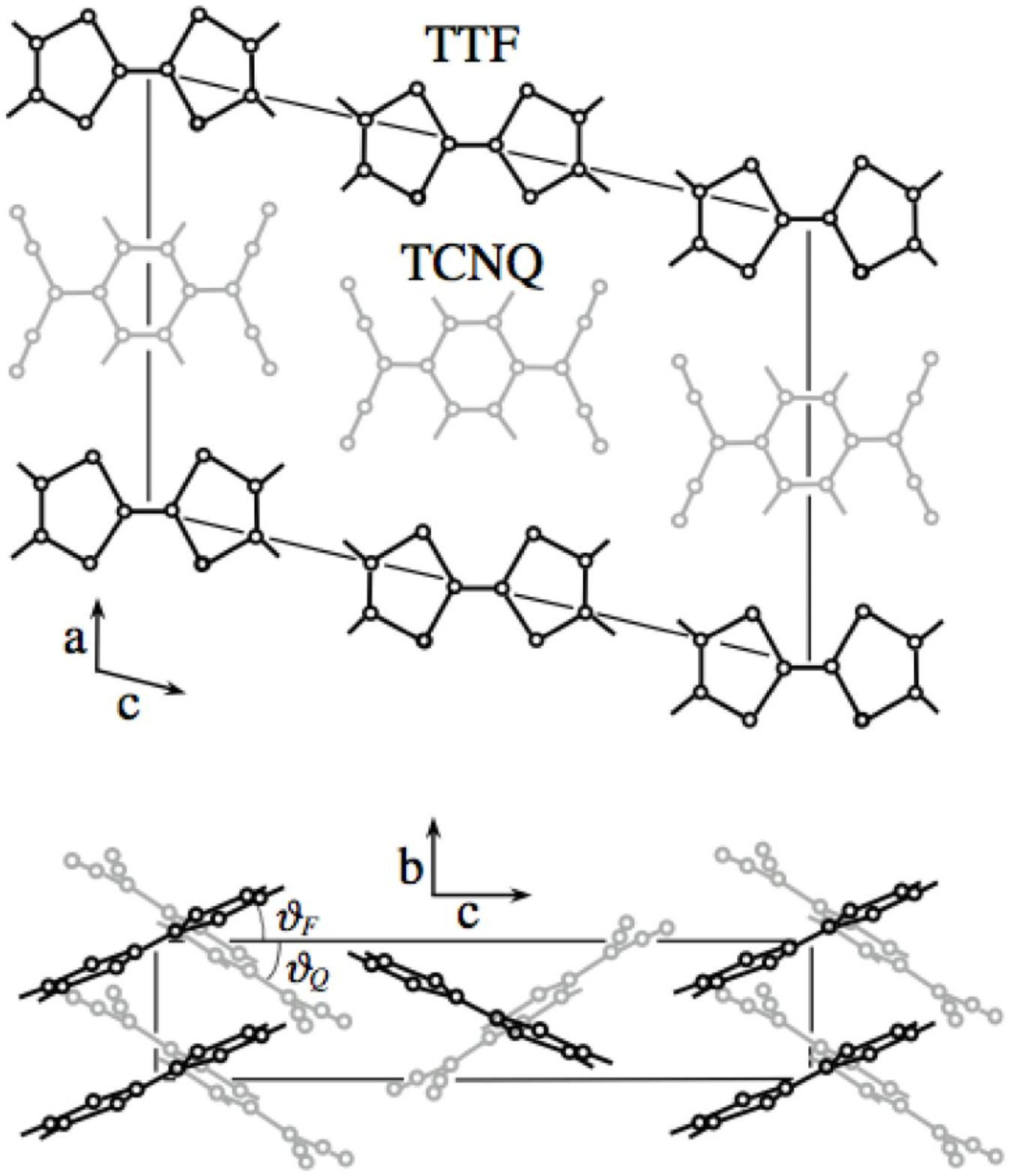}}
\hspace{2.0cm}
\subfigure{\includegraphics[width=5cm,height=8cm]{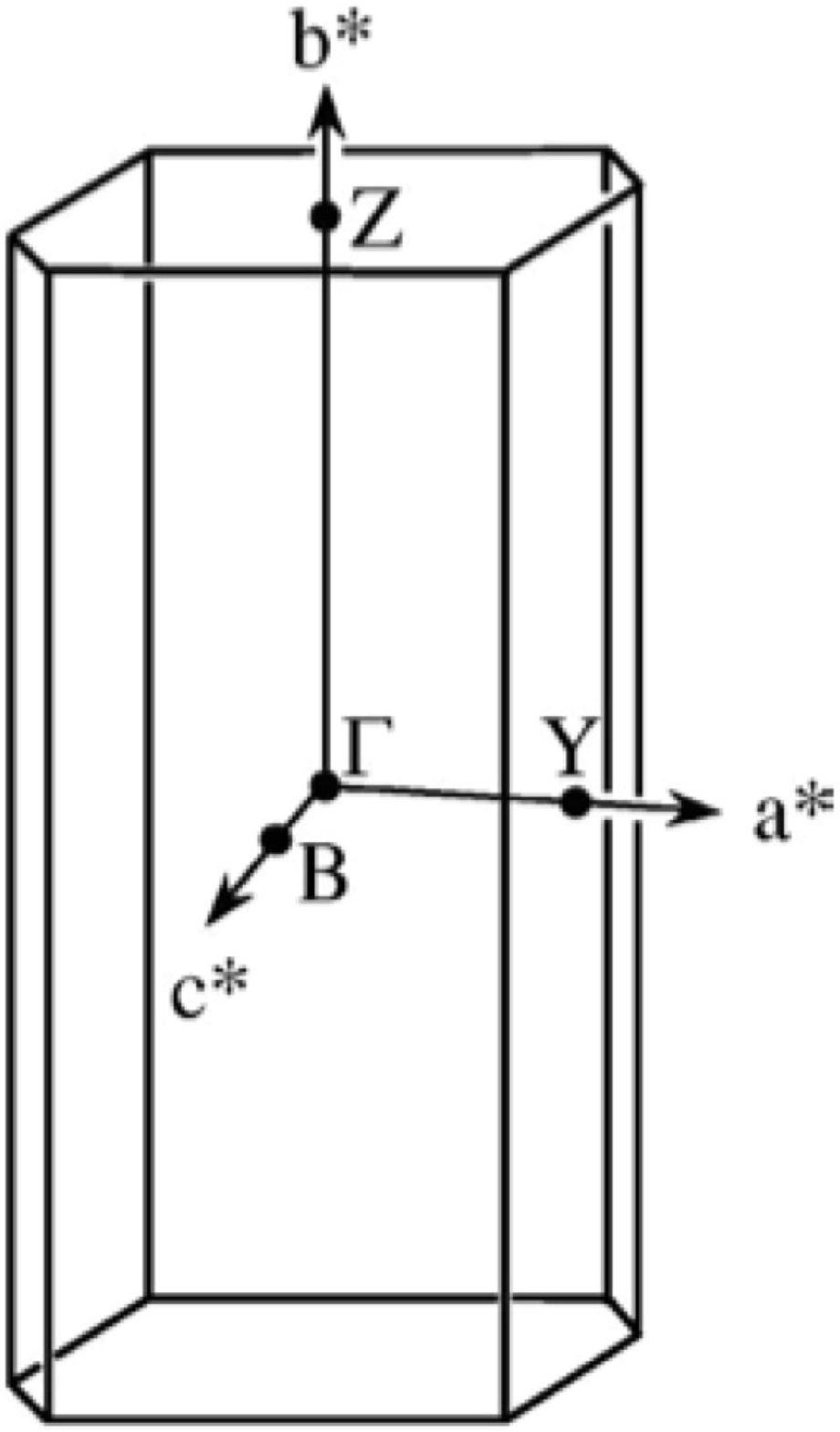}}
\caption{\label{figcrystal} The crystal structure of TTF-TCNQ. The angles about the $\bm{a}$ axis, as measured with respect to the $\bm{c}$ axis, ensures maximum covalent bonding along the stack direction, hence making the electronic conductivity strongly anisotropic. As a result, the electrical conductivity in the $\bm{b}$ direction is three orders of magnitude larger than in the other directions. The Brillouin zone (right) shows the high symmetry points in the corresponding reciprocal space directions. Thus, we measure electronic momentum along the $\bm{b}^*$ direction, identifying the centre of the Brillouin zone $\Gamma$, as the zero momentum point. \vspace{0.5cm}}
\end{figure}

The charge transfer occurs between the two types of molecules, i.e. between the stacks: an approximate $0.59$ electrons per molecule is transferred from the TTF to the TCNQ molecule, which drives the stacks metallic. In the metallic phase, the electrical conductivity is about three orders of magnitude larger in the intra-strack direction (for both molecules) than in other directions \cite{ttftcnq1}, a property attributed to the crystal structure \cite{ttftcnq01} \cite{ttftcnq02}. This is manifested through $\pi$-type orbitals overlapping in the conduction direction, i.e. overlapping with neighboring molecules belonging to the same stack. The high conductivity, as compared to inorganic metals, has been a key motivator for scientific investigations since the 1970's. Actually, TTF-TCNQ is one of the most celebrated organic conductors which have been widely examined since its discovery \cite{ttftcnq11} mainly regarding its electronic conductivity and optical properties, which have been thoroughly investigated in for example Refs. \cite{ttftcnq2}-\cite{ttftcnq6}. For our purposes, the material makes a good candidate for a "one dimensional electronic system" regarding the one electron removal spectral properties of the metallic phase. The crystal structure of TTF-TCNQ, along with its Brillouin zone, is depicted in Fig. (\ref{figcrystal}).

The non Drude behavior of the metallic regime is described in Ref. \cite{ttftcnq7}, a regime reached above the critical temperature $T_P=54$ K. This critical temperature is about half the value of the predicted mean-field weak coupling value, a deviation attributed to strong 1D fluctuations \cite{claessen1}. Below this temperature a charge density wave builds up in the TCNQ chains, eventually turning the system into an insulator at 38 K (the corresponding temperature for the TTF stacks is 49 K \cite{ttftcnq7}). The critical temperature is linked with a Peierls transition, where an electronic gap opens up due to the molecule lattice displacements \cite{ttftcnq8}-\cite{ttftcnq10}. This means that even though the theoretical model developed here refers to zero temperature, we have to study the material at temperatures higher than the critical temperature in order to reach the metallic phase, for which our theory is valid.

The charge transfer between the molecules shifts the intra-molecular density of electrons by $n_{TCNQ}=0.59$ for TCNQ and thus by $n_{TTF}=2-0.59=1.41$ for TTF, i.e. with a hole density of $0.59$. Thus, due to the particle-hole symmetry of our model, it suffices to study systems with a density of $0.59$. In this way, "removing an electron" (RHB) for the TCNQ translates into "adding a hole" for the TTF (LHB), and hence the spectral weight of both transitions can be mapped onto the {\it same} ($k,\omega$) region, with $\omega \leq 0$. We must note however, that the transfer integrals for the individual stacks are different from each other. 

The fact that Coloumb interaction plays a key role for the electronic structure of TTF-TCNQ is not a new claim \cite{ttftcnq02} \cite{torrance}, and from physical properties other than the photo emission spectrum, one would expect that $U \approx 4t$, with a slightly higher value for this ratio for TTF, than for TCNQ. By comparing with the photo emission spectrum, it is a simple fitting procedure (with the pseudofermion energy dispersions as the fitting functions) to deduce the values of ($U/t$) yielding the best match between experimental and theoretical results \cite{claessen1} (note that the $c0$ pseudofermion bandwidth is constant and equal to $4t$). In the following, we shall use the results of Refs. \cite{ourprl} and \cite{carmclaes}, where an almost perfect agreement between the theoretical and the experimental spectral weight is found for $t=0.40$ eV and $U=1.96$ eV for TCNQ, and $t=0.35$ eV and $U=1.96$ eV for TTF, respectively. We note that these values of the Hubbard parameters produce an exact fit with the pseudofermion energy bandwidths. We thus obtain effective values $(U/t)=4.9$ for TCNQ and $(U/t)=5.61$ for TTF, respectively, which furthermore coincide with the findings of Ref. \cite{ttftcnq7} (and will be further demonstrated below). 

One last note should be made on the subject of "dimensional crossover", i.e. the phenomenon that the quasi 1D material, for some reason and in some regime, changes its physical behavior to start acting like, for example, a 2D layer. In Ref. \cite{zwick6} it is confirmed that the intermolecular transfer integral is on the order of $5$ meV $\approx k_B T_P$ in the $\Gamma B$ direction.  This means that at sufficiently low temperatures, we will start to see hopping between identical molecules belonging to {\it different} stacks. Moreover, below the Peierls transition we will also have important contributions to the dynamics of the system from electron-phonon interactions. These considerations force us to only study the system at temperatures $T>T_P$.

\section{ARPES experiments on TTF-TCNQ}

ARPES is an abbreviation for "angle resolved photoelectron spectroscopy" and basically stands for the experimental procedure equivalent to what a theoretician would call "one electron removal" \cite{arpes}. The basic physical considerations regarding the ARPES technique will not be accounted for here, other than just stating that the energy and direction of the photoemitted electron defines the quantum state of the material sample. Thus, varying these parameters, it is possible to obtain a full energy and momentum map. The ARPES technique does not demand a certain environment or sample temperature {\it per se}, and the experiments can be conducted at zero magnetic fields. Usually, this technique is employed for probing the shape of the Fermi surface of the sample, however in our case this technique is used to study properties of the bulk material itself. 

To motivate this, we use the reasoning of F. Zwick {\it et al} \cite{zwick}. In this reference, special concern is taken regarding surface effects: surface sensitiveness to radiation damage, aging of cleaved surfaces, eventual perturbations on the bulk charge balance and lattice periodicity due to surface effects, possibility of an insulating surface, and so forth. The "passing" of these tests for TTF-TCNQ allows us to conclude that the ARPES measurements actually do give us information on the bulk properties of the material. Some of the references that F. Zwick {\it et al} use in this analysis include Refs. \cite{zwick1}-\cite{zwick7}, with the main conclusion that cleaved surfaces of TTF-TCNQ are "highly ordered and retain the periodicity of the bulk" up to the penetration depth used by the ARPES technique. By keeping the material sample at an ambient temperature of 150 K, F. Zwick {\it et al} \cite{zwick} measure the spectral weight along the $\bm{b}$ axis, as well as along the perpendicular $\bm{a}$ axis. The last measurement serves the purpose of excluding this direction as a conductive direction, showing no dispersive behavior. The $\Gamma$ point of the Brillouin zone is the natural zero momentum point, with the spectral weight in the $\bm{b}$ direction symmetrical around this point. Further on, Zwick goes on discussing the temperature dependence of the spectra, and how the size of the Peierls gap fits with various theoretical models. The interested reader should note that the spectral weight distribution of TTF-TCNQ indeed does have a very interesting temperature dependence, which is summarized for example in Refs. \cite{claessen1} \cite{zwick}. However, the conclusions are expected: the experimental results do not fit with a Fermi liquid description (an example of this is the absence of the usual metallic Fermi edges of the spectral weight; instead a complete supression of the quasi particle weight is found), neither do they fit with a strong electron-phonon interaction description. Due to the incoherency of the spectral weight and the interaction dependent singular lines in the mapped ($k,\omega$) plane, one should use 1D correlated fermion models in order to explain the experimentally obtained spectral features. 

But there is one feature not explicitly touched in that reference, a feature upon which our theory triumphs or fails completely: the separation of the charge-type $c0$ excitations and the spin-type $s1$ excitations, usually referred to as the "spin-charge separation" \cite{boson3}. In contrast to standard Luttinger liquid theory, where one studies the spin-charge separation for low lying excitations only, here this separation occurs for {\it finite} energy excitations. In recent studies of the spectral behavior of the 1D Hubbard model, such separate charge and spin excitations were identified \cite{spectralattempt1} \cite{spectralattempt2}. In our language, however, this separation is manifested through the different types of contributions which lead to the full spectral function, with the branch lines and the border lines as the most obvious features. 

\begin{figure}
\begin{center}
\includegraphics[width=11cm,height=8.5cm]{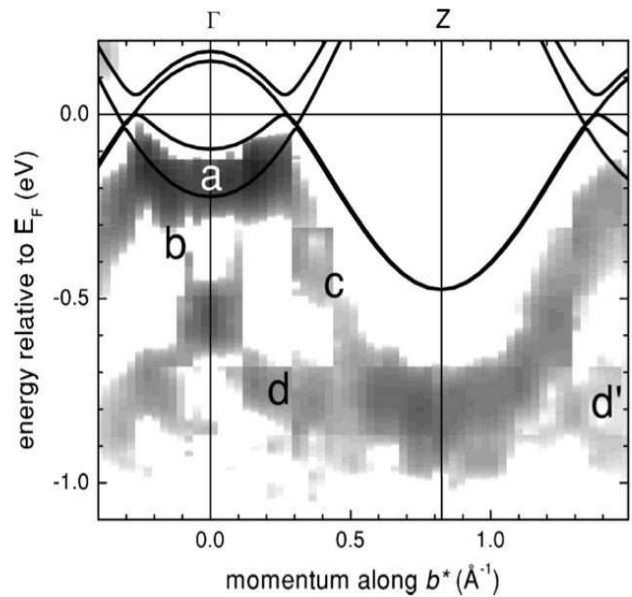}
\caption{\label{figklasband} A density plot of the obtained ARPES dispersions of TTF-TCNQ according to Ref. \cite{ttftcnq1}, together with a density functional band prediction regarding the singular line features (of the same reference). We see major quantitative and qualitative discrepancies, for example the mismatch between the lines $a$, $b$ and $c$ and the experimentally obtained data. Also, we see that the band theory is incapable to reproduce the experimental features labeled $d$ and $d'$, which is perhaps the most serious failure of this method.  \vspace{0.5cm}}
\end{center}
\end{figure}

From an experimental point of view, we find that Ref. \cite{ttftcnq1} presents ARPES measurements on TTF-TCNQ with details of the spectral features not previously reported. In that reference, the ARPES procedure involved a momentum and energy resolution of $0.07$ \AA$^{-1}$ and $60$ meV, respectively, and with an ambient temperature of $T=61$ K. Moreover, a comparison with the predictions of density functional band theory is made. Unfortunately, the line predictions of this method fails completely, producing non physical band dispersions (i.e. predicting spectral weight in the ($k,\omega$) plane along lines with the wrong bandwidth) and missing some experimentally proven spectral features all together. This is illustrated in Fig. (\ref{figklasband}). 

\begin{figure}
\begin{center}
\includegraphics[width=11cm,height=8.5cm]{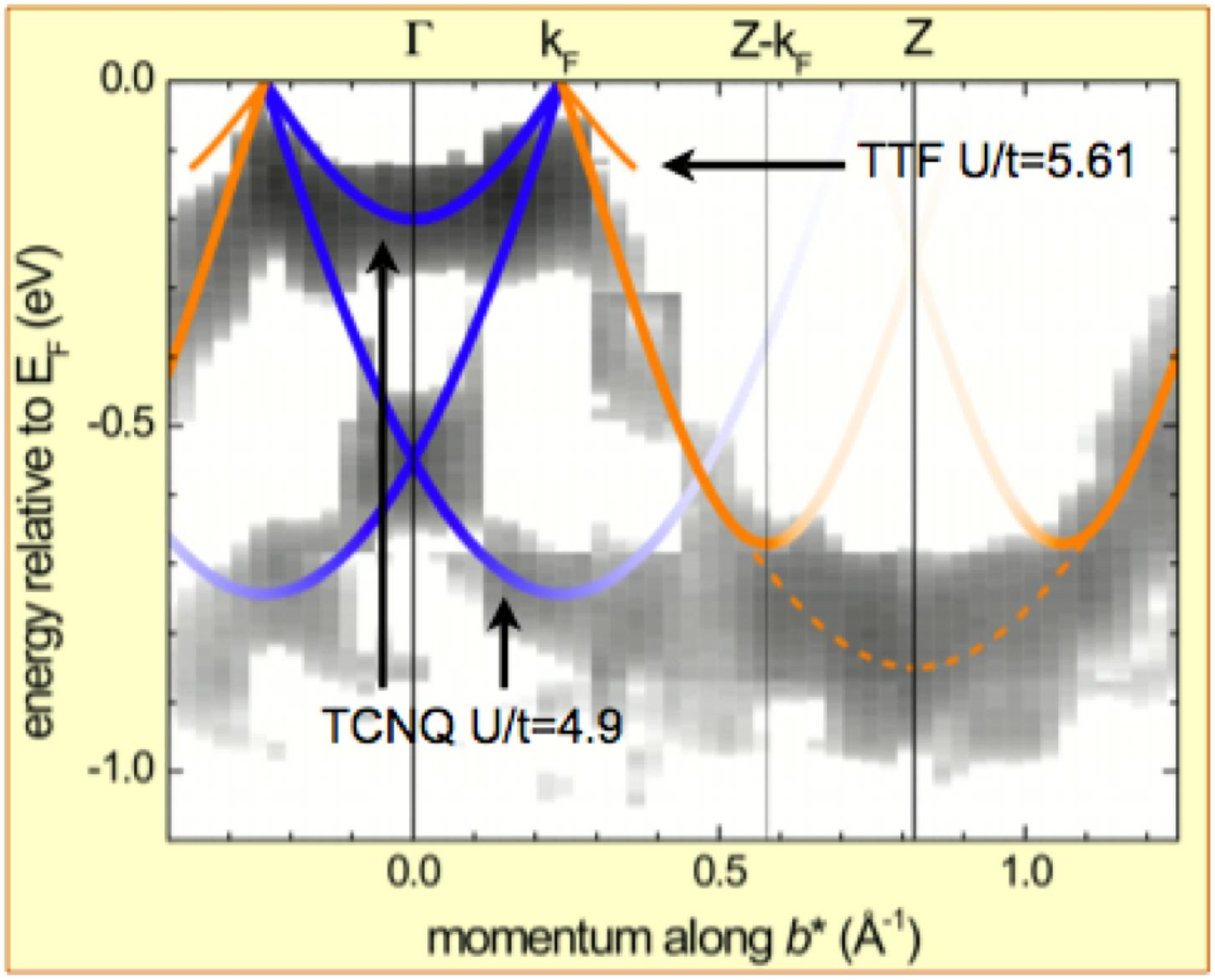}
\caption{\label{figtheend} A density plot of the obtained ARPES dispersions of TTF-TCNQ according to Ref. \cite{ttftcnq1}, together with the line features of the pseudofermion representation of the 1D Hubbard model. The line between $-k_F$ and $+k_F$ encircling the symmetry point $\Gamma$, as well as the lines emerging from $\pm k_F$ and extending to higher values of $\vert \omega \vert$, find their exact equivalence in the RHB s-branch line and c-branch line theoretical features, respectively, with $(U/t)=4.9$, see Fig. (\ref{fig2DRHBU4p9}). The other line shapes, originating at $+k_F$ and fading away as we follow the dispersion towards the symmetry point $Z$, are exactly matched by the LHB s-branch line and c-branch line theoretical features, respectively, with $(U/t)=5.61$. Note that the latter s-branch vanishes quite rapidly as we depart from the Fermi momentum. The dashed line between the symmetry points $Z-k_F$ and $Z+k_F$ is nothing but the $(U/t)=5.61$ LHB border line. The LHB features are adopted from Fig. (\ref{fig2DLHBU5p61}). TTF has a hole concentration of $0.59$, and thus TCNQ has an electronic density of $0.59$. Due to the particle-hole symmetry of our model, the spectral weight of both transitions can be mapped onto the {\it same} ($k,\omega$) region, with $\omega \leq 0$. \vspace{0.5cm}}
\end{center}
\end{figure}

\begin{figure}
\begin{center}
\includegraphics[width=10cm,height=10cm]{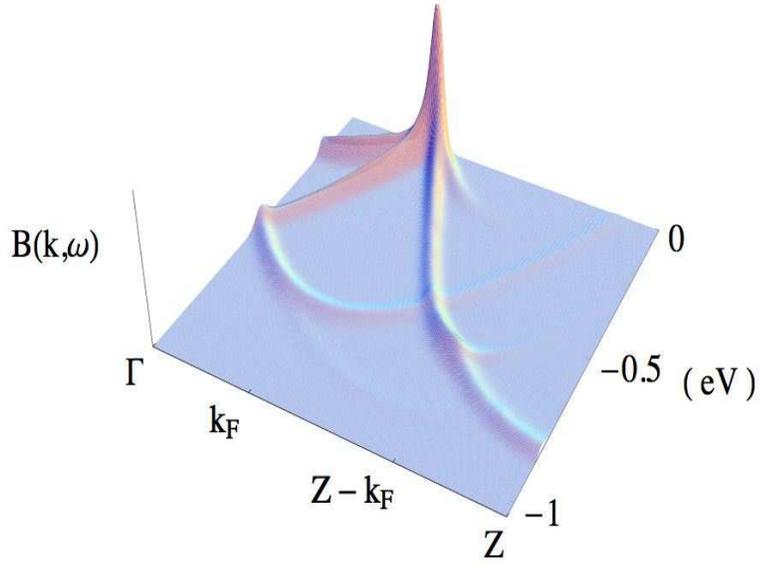}
\caption{\label{figfull} Theoretical spectral weight for TTF-TCNQ, obtained by use of Eqs. (\ref{2P}), (\ref{borde1}), (\ref{borde2}), (\ref{alphabr}), (\ref{speclutt}), and (\ref{0P}), respectively. This figure is nothing but the superposition of the two spectral weight distributions already given in Figs. (\ref{figspecU4p9}) and (\ref{figLHBU5p61}), respectively. We expect the singular features of this spectral function to be visible in suitable photo emission studies of TTF-TCNQ. This figure is used to obtain Fig. (\ref{densityspec}), which indeed confirms the agreement between theoretical and experimental spectral weights. \vspace{0.5cm}}
\end{center}
\end{figure}

\begin{figure}
\begin{center}
\includegraphics[width=10cm,height=10cm]{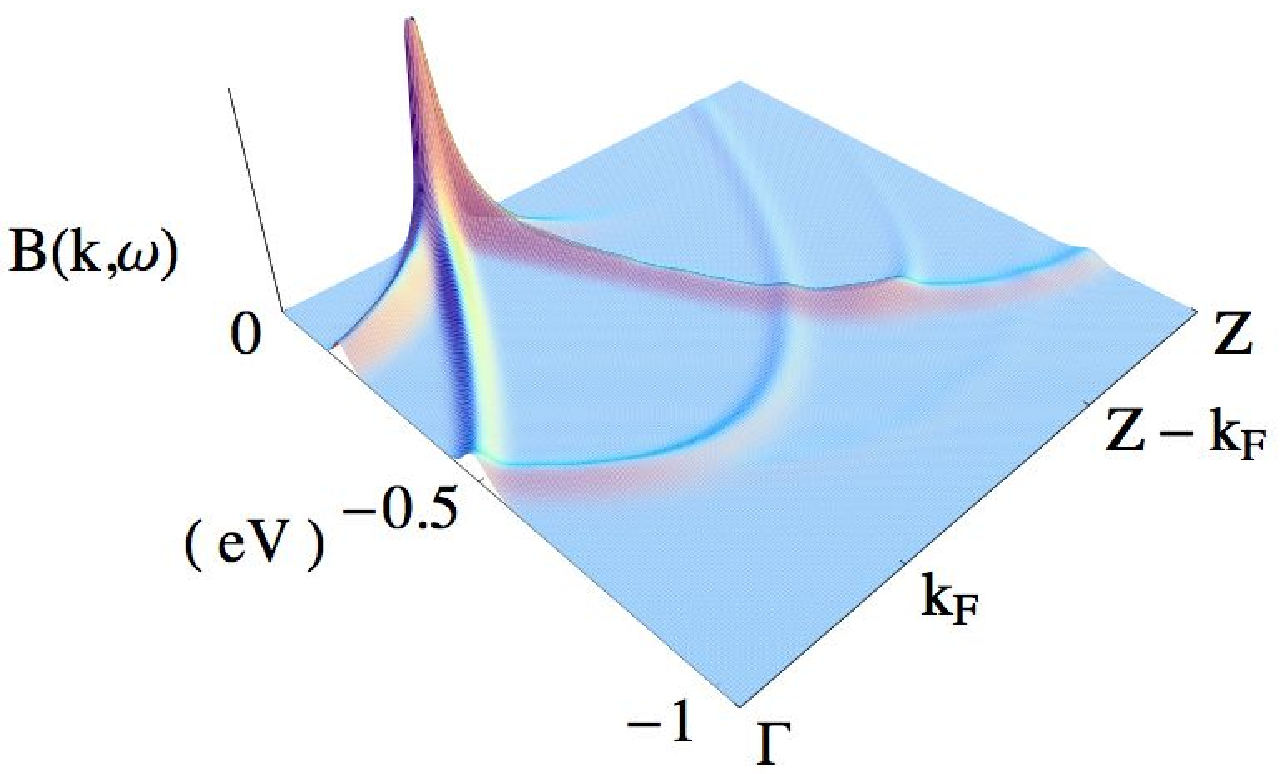}
\caption{\label{figfull2} Theoretical spectral weight for TTF-TCNQ, obtained by use of Eqs. (\ref{2P}), (\ref{borde1}), (\ref{borde2}), (\ref{alphabr}), (\ref{speclutt}), and (\ref{0P}), respectively. This figure is identical to Fig. (\ref{figfull}) but shows the spectral weight distribution from a different angle. \vspace{0.5cm}}
\end{center}
\end{figure}

We have already discussed and shown in section (\ref{theorweight}), the various line features that the pseudofermion description of the Hubbard model produces. In Fig. (\ref{figtheend}) we show the same grey scale density plot of the experimental spectral weight of Ref. \cite{ttftcnq1}, but now fitted with the characteristic branch lines and border line features of our pseudofermion theory. These line features are exactly the same as those given in section (\ref{theorweight}) for the relevant values of ($U/t$), but with the TTF spectra folded onto negative energies, illustrating the donor-acceptor relationship between TTF (donor, holes) and TCNQ (acceptor, electrons). Note that a different choice of ($U/t$) for any of the two materials would produce line features not corresponding to the experimental line features. Now, it is straightforward to invert the spectral function of Fig. (\ref{figLHBU5p61}), from positive to negative energies, and add it to the spectral function of Fig. (\ref{figspecU4p9}), to produce the full theoretical spectral weight corresponding to the ARPES experimental result, Figs. (\ref{figfull}) and (\ref{figfull2}).

\begin{figure}
\begin{center}
\includegraphics[width=10cm,height=10cm]{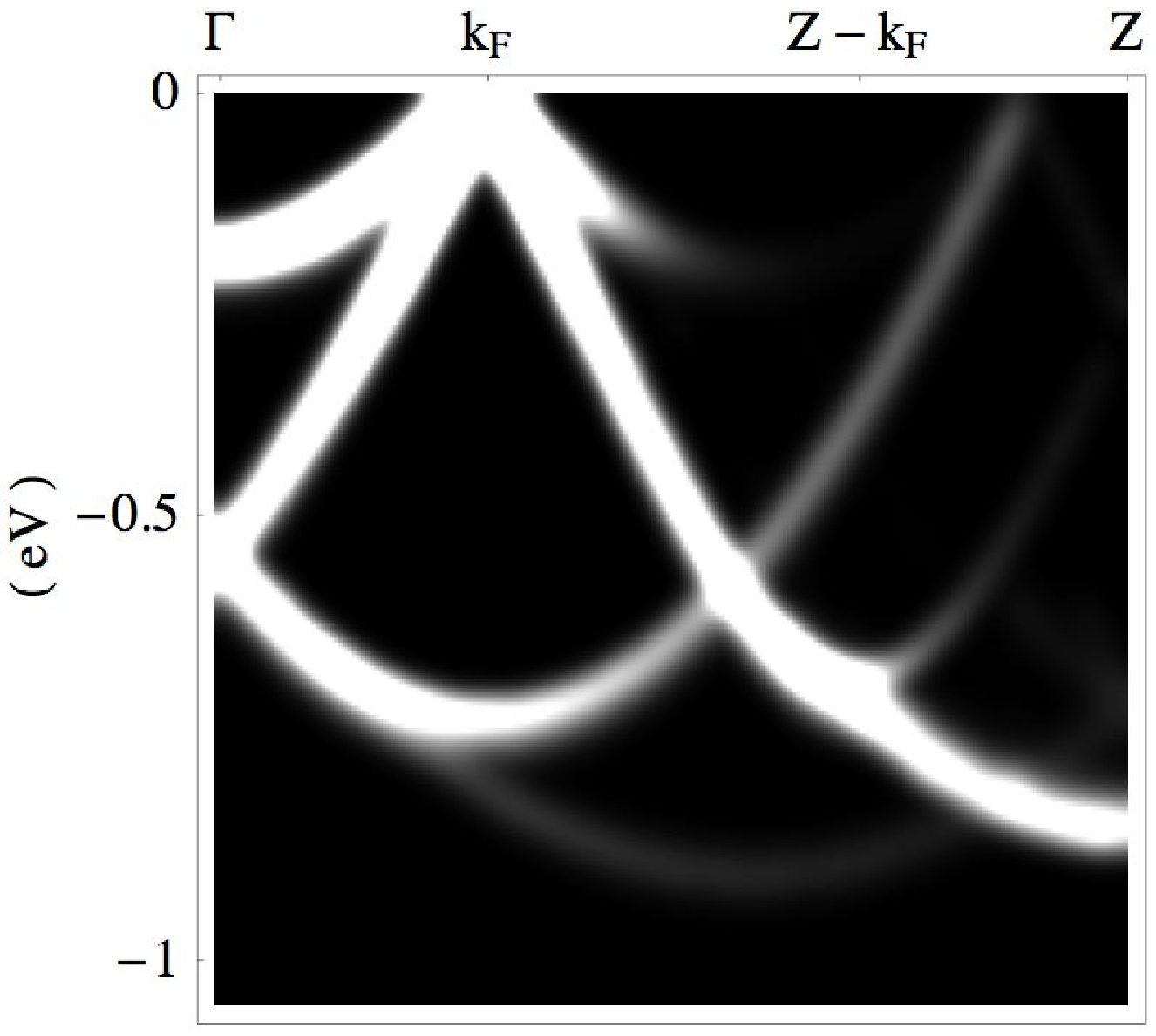}
\caption{\label{densityspec} Theoretical density plot of the spectral weight for TTF-TCNQ. Technically, it is this figure which should be compared with the experimentally obtained ARPES dispersions. We see that we have an almost complete agreement between this figure and Fig. (\ref{figtheend}). Important features include the two charge type excitations both originating at $k=k_F$, one related to the TCNQ stack and one to the TTF stack. They both fade away at higher momentum values. In contrast, the TCNQ spin type excitation carries much more spectral weight than its TTF counterpart, with the opposite relationship for the border lines. \vspace{0.5cm}}
\end{center}
\end{figure}

Lastly, we present the density plot corresponding to the spectral function provided in Figs. (\ref{figfull}) and (\ref{figfull2}), which is provided in Fig. (\ref{densityspec}). These plots lets us identify the types of excitations responsible for the total spectral weight of the system, hence chartering previously unknown territory. The dominant line shapes are thus due to separate charge type and spin type excitations for all relevant excitations, not just the low lying ones. Moreover, the relatively large spectral weight quite deep inside the band can here be identified with a "velocity resonance effect": the border line is the line where the two types of excitations propagate with the same group velocity.

\setcounter{chapter}{6}
\setcounter{section}{6}

\chapter{Conclusions and Discussion}
\label{conclusion}

The pseudofermion dynamical theory reported in this thesis describes the quantum objects that diagonalize the normal ordered 1D Hubbard hamiltonian. In the pseudofermion representation, the scatterers and scattering centers are $\eta$-spin and spin zero objects, and hence the S-matrix describing the scattering events between them is merely a complex number: it is given by the phase shift of the ground state $\rightarrow$ final state quantum mechanical excitation. Thus, the scattering events of these quantum objects are of zero energy forward scattering type \cite{josescatt}. 

The form of the S-matrix, being a one dimensional matrix, is crucial to the development of the dynamical theory. This ultimately results from the "diagonal" form of the pseudofermion anticommutation relations. The $\alpha\nu$ pseudofermion or pseudofermion hole S-matrix fully controls the one electron matrix elements between the ground state and excited energy eigenstates through these anticommutation relations. Indeed, the anticommutator can solely be expressed in terms of the S-matrix, as shown in section (\ref{humbadu}). The form of the pseudofermion S-matrix constitutes an important new result of this thesis report and of Ref. \cite{CarmBoziPedro}. 

The studies of Ref. \cite{NPB04} showed that the various quantum numbers introduced by the Takahashi string hypothesis describe occupancies of pseudoparticles. In this reference, the original electrons were related to the "rotated electrons" via a unitary transformation $\hat{V}(U/t)$, as described in chapter (\ref{themodel}). The double occupancy, the single $\sigma$-spin occupancy ($\sigma=\uparrow,\downarrow$) and the no occupancy number of the rotated electrons are good quantum numbers for all values of ($U/t$). The separated charge and spin degrees of freedom of the rotated electrons give rise to the pseudoparticles. The related pseudofermion description differs from the pseudoparticle description by a shift in the discrete momentum values of order ($1/L$), which are associated with the scattering phase shifts due to the ground state $\rightarrow$ final state transitions \cite{CarmBoziPedro}. 

The pseudoparticles have residual energy interactions which makes them unsuitable for the development of a dynamical theory. Indeed, the residual energy interaction prevents a pseudoparticle wave-function factorization. However, for the pseudofermions, we have no such residual energy interactions, which indeed allows for such a factorization \cite{wavefcnfact}. This wave-function factorization is valid for the normal ordered 1D Hubbard model at all values of the energy ratio ($U/t$), filling $n$ and magnetization $m$.

In contrast to the usual low-energy Luttinger liquid theory, the theory reported here allows us to categorize a separation of the charge type degrees of freedom and the spin type degrees of freedom at a finite energy excitation scale. However, in the low energy elementary excitation regime, the results of the conformal field theory \cite{crit6} \cite{crit61} coincide with the dynamical pseudofermion theory, as demonstrated in Ref. \cite{equival}. 

The pseudofermion dynamical theory presented in this thesis report, was originally inspired by the $(U/t) \gg 1$ methods of Refs. \cite{OgShib}-\cite{Karlo3}, where the spectral properties of the 1D Hubbard model for $(U/t) \rightarrow \infty$ were studied. For arbitrary ($U/t$), the dynamical theory allows us to calculate general closed-form analytical expressions of the finite energy one electron spectral weight distributions of a 1D correlated system (with on-site electronic repulsion). This derivation is done in detail in chapters (\ref{dynamics}) and (\ref{onelecspec}) and constitutes important new contributions to the understanding of the spectral properties of the 1D Hubbard model. This work was also presented in Ref. \cite{Carmspec2}.

The canonical pseudoparticle-pseudofermion transformation involves a momentum shift $Q_{\alpha\nu}^{\Phi} (q) / L$. This shift is zero for the original ground state, for which the $\alpha\nu=c0,s1$ pseudofermions have well defined Fermi points $\iota q_{F\alpha\nu}$ (where $\iota=\pm$ denotes the left and the right Fermi point, respectively). The ($k,\omega$) dependent exponents of the theory are then described in terms of the canonical momentum shifts of these Fermi points associated with a finite-number electron excitation.

The dynamical theory applied to the case of one-electron excitations is presented in this thesis report for the cases of one-electron removal (RHB) and one-electron lower Hubbard band addition (LHB). The closed form expressions are explicitly derived in chapter (\ref{onelecspec}) and plotted in chapter (\ref{theorweight}). 

The spectral properties of the RHB and the LHB cases can be categorized according to different types of {\it contributions} (this categorization is also possible for the one electron upper Hubbard band addition but is not presented here). These contributions, corresponding to different regions of the ($k,\omega$) plane, each have different sets of ($k,\omega$) dependent exponents and pre-factors. By this classification, we are able to identify practically all features of the spectral weight of the 1D Hubbard model, in terms of $c0$ and $s1$ pseudofermion or pseudofermion hole excitations. An example is the border line, which is found here and in Ref. \cite{Carmspec2}, to generate divergent spectral features when the $c0$ pseudofermion or pseudofermion hole and the $s1$ pseudofermion hole propagate with equal group velocity.

On a microscopical level, for each of the "contributions", the $\alpha\nu=c0,s1$ particle-hole towers of states give rise to an orthogonal catastrophe. For any such tower of states, we can associate an exponent, which in the case of being negative produces a power-law type decay of the spectral weight as we move away from the base of the tower. The spectral weight distribution associated with these particle-hole processes is controlled by the value of the exponent but also by the value of the pre-factor, both of which are constant for one specific tower of states. The pseudofermion dynamical theory is capable of explicitly calculating both of these quantities, for the entire ($k,\omega$) plane.

Thus on the theoretical side, we are able to predict the position and origin of the one electron spectral singular features of a 1D correlated metal. However, we are also able to connect our theoretical predictions to experimental results. The singular behavior of the spectral function, as predicted by the explicitly calculated values of the relevant exponents, leads to a spectral weight distribution which should be detectable by photo emission and / or photo absorption experiments. It turns out that within the approximation of only considering the leading order elementary processes to the RHB and the LHB one-electron spectral weight, we are able to reproduce, for the whole energy bandwidth, the experimental spectral distributions found for the organic compound TTF-TCNQ by high-resolution ARPES. These new results are presented in chapter (\ref{applica}) and in Ref. \cite{ourprl}. The TTF-TCNQ high-resolution ARPES experiments were reported in Refs. \cite{claessen1} \cite{carmclaes} \cite{ttftcnq1}. In conclusion, the dynamical theory presented here allows for an understanding of the elementary quantum processes that give rise to the spectral features of TTF-TCNQ. 

With the advent of new experimental techniques that allow for a high-resolution study of the spectral features of quasi-1D materials, the pseudofermion dynamical theory has yet many challenges ahead of itself. One of the most exciting recent experimental setups is the "optical lattice" in which ultra-cold fermions are trapped in a potential well, forming a "real" 1D quantum chain. These systems can be described by the 1D Hubbard model, with the electrons replaced by ultra-cold fermionic atoms. Even though some preliminary results already have been reported \cite{sexy5}-\cite{sexy52}, this technique is still at its infancy. However, it does promise an unprecedented control over the necessary parameters (such as the on-site Coloumb repulsion $U$ and transfer amplitude $t$), enabling the high-resolution measurements necessary for a complete understanding of these materials. The theoretical spectral weight expressions obtained in this thesis report should be taken into account when characterizing the experimental spectral features obtained through this method. Indeed, this experimental setup comprises one of the most exciting future applications of the pseudofermion dynamical theory.

In conclusion, we see that the pseudofermion dynamical theory is a suitable theory for the study of the spectral properties of the 1D Hubbard model, yielding results in good agreement with the behavior of the one electron spectral function in the $(U/t) \rightarrow \infty$ limit \cite{Karlo1}-\cite{Karlo3}, and the low-lying elementary excitation limit \cite{equival} \cite{crit6} \cite{crit61} and with experimental results \cite{claessen1} \cite{carmclaes} \cite{ttftcnq1}.

\end{document}